\documentclass{aa}  
\usepackage{natbib}
\usepackage{graphicx}
\usepackage{txfonts}
\usepackage{hyperref}
\usepackage{xcolor}
\usepackage{upquote} % To have the correct quotes in verbatim mode, so that copy/paste works for ADQL commands
\usepackage{xspace}
\usepackage{ulem}  % for \sout{}
\usepackage{multirow}

\def\GaiaSrcId#1{\texttt{\small Gaia DR3 #1}}
\def\GaiaSrcIdWithoutDR3#1{\texttt{\small #1}}
\def\GaiaSrcIdInCaption#1{{\tiny Gaia DR3 #1}}
\def\GaiaSrcIdInCaptionWithoutDR3#1{{\tiny #1}}

\def\Gaia{\textit{Gaia}\xspace}
\def\gmag{\ensuremath{G}\xspace}
\def\gbp{\ensuremath{G_\mathrm{BP}}\xspace}
\def\grp{\ensuremath{G_\mathrm{RP}}\xspace}
\def\absMeanG{\ensuremath{\mathrm{M}_G}\xspace}

\def\BPminusRP{\ensuremath{G_\mathrm{BP} - G_\mathrm{RP}}\xspace}

\def\phaseEclPrimary{\ensuremath{\varphi_\mathrm{ecl,1}}\xspace}
\def\durationEclPrimary{\ensuremath{w_\mathrm{ecl,1}}\xspace}
\def\depthEclPrimary{\ensuremath{d_\mathrm{ecl,1}}\xspace}
\def\phaseEclSecondary{\ensuremath{\varphi_\mathrm{ecl,2}}\xspace}
\def\durationEclSecondary{\ensuremath{w_\mathrm{ecl,2}}\xspace}
\def\depthEclSecondary{\ensuremath{d_\mathrm{ecl,2}}\xspace}
\def\Aell{\ensuremath{A_\mathrm{ell}}\xspace}
\def\phaseell{\ensuremath{\mu_\mathrm{ell}}\xspace}
\def\eccProxy{\ensuremath{e_\mathrm{proxy}}\xspace}
\def\eccProxyError{\ensuremath{\varepsilon(e_\mathrm{proxy})}\xspace}
\def\deltaPhi{\ensuremath{\Delta_{0.5}(\varphi_\mathrm{ecl})}\xspace}

\def\BPRPexcess{{BP$+$RP flux excess factor}\xspace}
\def\OgleI{\ensuremath{I}\xspace}
\def\OgleV{\ensuremath{V}\xspace}

\begin{document}

   \title{\Gaia Data Release 3}

   \subtitle{The first \Gaia catalogue of eclipsing binary candidates}

  \author{% Core team
          N.~Mowlavi\inst{\ref{Sauverny},\ref{Ecogia}}\fnmsep\thanks{Corresponding author: N. Mowlavi
(\href{mailto:Nami.Mowlavi@unige.ch}{\tt Nami.Mowlavi@unige.ch})}
          \and B.~Holl\inst{\ref{Sauverny},\ref{Ecogia}}
          \and I.~Lec\oe ur-Ta\"ibi\inst{\ref{Ecogia}}
          % Larger EB team
          \and F.~Barblan\inst{\ref{Sauverny}}
          \and A.~Kochoska\inst{\ref{Villanova}}
          \and A.~Pr\v{s}a\inst{\ref{Villanova}}
          \and T. Mazeh\inst{\ref{Israel}}
          % Classif team
          \and L.~Rimoldini\inst{\ref{Ecogia}}
          \and P.~Gavras\inst{\ref{RHEA}}
          % Geneva CU7 / DPCG teams
          \and M.~Audard\inst{\ref{Sauverny},\ref{Ecogia}}
          \and G.~Jevardat de Fombelle\inst{\ref{Ecogia}}
          \and K.~Nienartowicz\inst{\ref{Ecogia},\ref{Sednai}}
          % Others
          \and P.~Garc\'{i}a-Lario\inst{\ref{ESAC}}
          % On request by himself to be the last one 
          \and L.~Eyer\inst{\ref{Sauverny},\ref{Ecogia}}
%N. Mowlavi, B. Holl, I. Lecœur-Taïbi, F. Barblan, A. Kochoska, A. Prsa, T. Mazeh, L. Rimoldini, P. Gavras, M. Audard, G. Jevardat de Fombelle, K. Nienartowicz, P. Garcia-Lario, L. Eyer
         }         
  \authorrunning{Mowlavi et al.}
  
  \institute{Department of Astronomy, University of Geneva, Chemin Pegasi 51, 1290 Versoix, Switzerland \label{Sauverny}
             \and
             Department of Astronomy, University of Geneva, Ch. d'Ecogia 16, 1290 Versoix, Switzerland\label{Ecogia}
             \and
             Villanova University, Dept. of Astrophysics and Planetary Science, 800 Lancaster Ave, Villanova PA 19085, USA\label{Villanova}
             \and
             School of Physics and Astronomy, Tel Aviv University, Tel Aviv 6997801, Israel\label{Israel}
             \and
             RHEA for European Space Agency (ESA), Camino bajo del Castillo, s/n, Urbanizacion Villafranca del Castillo, Villanueva de la Ca\~{n}ada, 28692 Madrid, Spain\label{RHEA}
             \and
             Sednai Sarl, 1204 Geneva, Switzerland\label{Sednai}
             \and
             European Space Agency (ESA), European Space Astronomy Centre (ESAC), Camino bajo del Castillo, s/n, Urbanizacion Villafranca del Castillo, Villanueva de la Ca\~{n}ada, 28692 Madrid, Spain\label{ESAC}
            }

   \date{Received September XX, 2022; accepted September XX, 2022}

% \abstract{}{}{}{}{} 
% 5 {} token are mandatory
 
  \abstract
   {\Gaia Data Release 3 (DR3) provides a number of new data products that complement the early DR3 made available two years earlier, among which the first \Gaia catalogue of eclipsing binary candidates containing 2\,184\,477 sources with brightnesses from a few magnitudes to 20~mag in the \Gaia \gmag-band and covering the full sky.
   }
   {We present the catalogue, describe its content, provide tips for its usage, estimate its quality, and show illustrative samples.
   }
   {Candidate selection is based on the results of variable object classification performed within the \Gaia Data Processing and Analysis Consortium, further filtered using eclipsing binary-tailored criteria based on the \gmag-band light curves.
    To find the orbital period, a large ensemble of trial periods is first acquired using three distinct period search methods applied to the cleaned \gmag light curve of each source. 
    The \gmag light curve is then modelled with up-to two Gaussians and a cosine for each trial period.
    The best combination of orbital period and geometric model is finally selected using Bayesian model comparison based on the BIC.
    A global ranking metric is provided to rank the quality of the chosen model between sources.
    The catalogue is restricted to orbital periods larger than 0.2 days.
   }
   {About 530\,000 of the candidates are classified as eclipsing binaries in the literature as well, out of $\sim$600\,000 available crossmatches, and 93\% of them have published periods compatible with the \Gaia periods.
    Catalogue completeness is estimated to be between 25\% and 50\%, depending on the sky region, relative to the OGLE4 catalogues of eclipsing binaries towards the Galactic Bulge and the Magellanic Clouds.
    The analysis of an illustrative sample of $\sim$400\,000 candidates with significant parallaxes shows properties in the observational Hertzsprung-Russell diagram as expected for eclipsing binaries.
    The subsequent analysis of a sub-sample of detached bright candidates provides further hints for the exploitation of the catalogue.
    The orbital periods, light curve model parameters, and global rankings are all published in the catalogue with their related uncertainties where applicable.
   }
   {This \Gaia DR3 catalogue of eclipsing binary candidates constitutes the largest catalogue to date in number of sources, sky coverage, and magnitude range.
   }


   \maketitle
%
%-------------------------------------------------------------------

%==================================================================
\section{Introduction}
\label{Sect:introduction}

Most stars are in binary systems and a fraction of them appear to an observer as eclipsing.
These eclipsing systems allow us, under certain conditions, to determine fundamental parameters of stars, such as mass and radius, together with the orbital parameters.
They are stringent tests for stellar evolution when the two stars are in wide systems, while they are laboratories for many physical processes when the two stars interact with one another.
Some eccentric systems can also serve as a test of the theory of general relativity thanks to the determination of their apsidal motion.
In addition, when one of the components is oscillating and provides suitable conditions to perform asteroseismology, the system provides an independent determination of stellar parameters and tests for the asteroseismic scaling relations.
Clearly, eclipsing binaries are exceptionally interesting objects for astronomy. 
Still, the number of well studied cases is relatively small.
For example, the catalogue of well studied systems presented by \citet{Southworth_2015} contains 170%
\footnote{
305 binaries on Aug. 30, 2022, see \url{https://www.astro.keele.ac.uk/jkt/debcat}.
}
binaries, based on an initial compilation of 45 eclipsing binaries by \citet{Andersen_1991}. 
%Before being able to perform these detailed studies, these systems need to be identified, with a minimal characterisation so that a selection of promising cases can be achieved.

With the advent of large-scale multi-epoch ground-based photometric surveys, pioneered by the microlensing-search tailored `Exp\'erience pour la recherche d'objets sombres' \cite[EROS1][]{RenaultAubourgBareyre98}, the `Massive compact halo object' experiment \citep[MACHO][]{AlcockAllsmanAlves97}, and the `Optical gravitational lensing experiment' \citep[OGLE2][]{UdalskiSzymanskiKaluzny92}, the opportunities to find eclipsing binaries increased dramatically.
The precursor of \Gaia, HIPPARCOS, already provided an all-sky survey of eclipsing binaries \citep{Hipparcos_1997}.
The number of eclipsing binaries was rather limited, about 900 among 11\,597 detected variables (from 118\,218 monitored stars), yet it contained $\sim$30\% new candidates.
Before this \Gaia Data Release 3 (DR3), the largest catalogue specifically dedicated to eclipsing binaries comes from the OGLE4 survey team with the publication of 40\,204 sources in the Large Magellanic Cloud (LMC) and 8\,401 sources in the Small Magellanic Cloud \citep[SMC,][]{PawlakSoszynskiUdalski_etal16}, and 450\,598 sources towards the Galactic Bulge \citep{SoszynskiPawlakPietrukowicz_etal16}.
In parallel, multiple other large-scale multi-epoch surveys provide additional opportunities with automated classification of their variable stars.
Such is the case, for example, for (number of eclipsing binaries given in parenthesis)
the Trans-Atlantic Exoplanet Survey \citep[TRES;][773]{TRES_ECL_DEVOR_2008},
the All Sky Automated Survey \citep[ASAS;][1055 and 180, respectively]{Pojmanski_2002,ASAS_KEPLER_VAR_PIGULSKI_2009},
the Lincoln Near-Earth Asteroid Research survey \citep[LINEAR;][2700]{LINEAR_VAR_PALAVERSA_2013},
the EROS2 survey \citep[][$\sim$45\,600]{KimProtopapasBailerJones_etal14},
the CATALINA survey \citep[][23\,312]{CATALINA_VAR_DRAKE_2017},
the Asteroid Terrestrial-impact Last Alert System survey \citep[ATLAS;][$\sim$110\,000]{HeinzeTonryDenneau_etal18},
or
the Zwicky Transient Facility survey \citep[][$\sim$350\,000]{ChenWangDeng_etal20}.
The catalogue of variable stars made available by the American Association of Variable Star Observers (AAVSO) through their Variable Star Index (VSX) database also provides a wealth of data for the study of eclipsing binaries \citep{COMP_VAR_VSX_2019}.

% ASAS\_KEPLER\_VAR\_PIGULSKI\_2009 \citep{ASAS_KEPLER_VAR_PIGULSKI_2009}\\
% ASAS\_VAR\_POJMANSKI\_2002 \citep{ASAS_VAR_POJMANSKI_2002}\\
% ATLAS\_VAR\_HEINZE\_2018 \citep{HeinzeTonryDenneau_etal18} \\
% CATALINA\_VAR\_DRAKE\_2017 \citep{CATALINA_VAR_DRAKE_2017} \\
% EROS2\_VAR\_KIM\_2014 \citep{KimProtopapasBailerJones_etal14} \\
% GAIA\_ECL\_RYBIZKI\_2018 (Internal!) \\
% HIPGAIA\_VAR\_BRANDT\_2018 \\
% INTEGRAL\_VAR\_ALFONSOGARZON\_2012 \\
% LINEAR\_VAR\_PALAVERSA\_2013 \citep{LINEAR_VAR_PALAVERSA_2013}\\
% OGLE in BLG \citep{SoszynskiPawlakPietrukowicz_etal16} \\
% OGLE in MCs \citep{PawlakSoszynskiUdalski_etal16} \\
% TRES\_ECL\_DEVOR\_2008 \citep{TRES_ECL_DEVOR_2008} \\
% VVV\_VAR\_BRAGA\_2019 (this is about new type II cepheids!)\\
% EBs in TESS \citep{PrsaKochoskaConroa_etal22}\\
% COMP\_VAR\_VSX\_2019 \citep[][AAVSO's International Variable Star Index (VSX)]{COMP_VAR_VSX_2019}

Space missions dedicated to exoplanet search provide another source of data for the study of eclipsing binaries.
Their strengths come from the continuous, high-cadence observation on long time scale, combined to the high photometric precision that can be obtained from space.
Catalogues dedicated to eclipsing binaries from these missions include, for example, \citet{KEPLER_ECL_KIRK_2016} for Kepler (2878 candidates including ellipsoidal variables) and \citet{PrsaKochoskaConroa_etal22} from the Transiting Exoplanet Survey Satellite (TESS; 4584 eclipsing binaries).
They, however, are limited in terms of sky coverage and/or brightness range.

% KEPLER\_ECL\_KIRK\_2016 \citep{KEPLER_ECL_KIRK_2016}\\
% KEPLER\_VAR\_DEBOSSCHER\_2011 \\
% TESS

The \Gaia space mission from the European Space Agency (ESA) offers a new opportunity to study eclipsing binaries.
Launched at the end of 2013, this all-sky survey started its nominal mission in July 2014 \citep{Gaia_PrustiDeBruijneBrown_etal16}.
Among the strong points of the mission for variability analysis, we can mention, in addition to its well-known astrometric capabilities, the large dynamical range reached in stellar brightness, from a few magnitudes to fainter than 20~mag, the specific scanning law leading to irregularly sampled time series, and the quasi-simultaneity (within tens of seconds) of the observations in \gmag photometry, \gbp and \grp spectrophotometry, and RVS (Radial Velocity Spectrometer) spectroscopy.
Data products based on 34 months of astrometry and photometry data have been released in the early Data Release 3 \citep[EDR3 in Dec. 3, 2020;][]{Gaia_BrownVallenariPrusti_etal21,Riello_etal21}.
These have been complemented with numerous additional data products in DR3 \citep[June 13, 2022;][]{DR3-DPACP-185}, including variability catalogues for more than ten million variable objects \citep{2022arXiv220606416E}.

This paper presents the first \Gaia catalogue of eclipsing binaries, published as part of \Gaia DR3.
It is the largest such catalogue to date, with more than two million candidates.
A balance was reached between completeness and purity.
The selection of the eclipsing binaries starts with the classification of variable objects performed within the Gaia Processing and Analysis Consortium (DPAC) as described in \citet{DR3-DPACP-165}, followed by a specific eclipsing binary module that automatically selects a geometric two-Gaussian model \citep[see][]{MowlaviLecoeurHoll_etal17} and orbital period based on the \gmag light curves, after which a final filtering step on various statistics parameters is made.
The \gbp and \grp time series were not used.
The eclipsing binary processing pipeline is described in Sect.~\ref{Sect:catalogue}.
In particular, the section describes candidate selection, orbital period search, the two-Gaussian model used to fit the morphology of the \gmag light curves, and the procedure implemented to automate the selection of the best model and orbital period, as well as to derive uncertainties for the determined parameters.
Section~\ref{Sect:catalogue} also details the content of the catalogue.
Recommendations for catalogue exploitation using published parameters are given in Sect.~\ref{Sect:catalogue_usage}.
%In particular, it is shown how the morphological model parameters can be used to identify various types of eclipsing binaries.
The quality of the catalogue is then addressed in Sect.~\ref{Sect:quality}, with an estimate of catalogue completeness and an investigation of the new \Gaia candidates.
Illustrative samples of candidates with good parallaxes are presented in Sect.~\ref{Sect:overview}, with a specific application to the period--eccentricity analysis of bright candidates.
Section~\ref{Sect:conclusions} ends the main body of the text with a summary and conclusions.

Additional content is presented in four appendices.
Appendix~\ref{Appendix:twoGaussianModel} analyses the various types of two-Gaussian models used to fit the eclipsing binary light curves.
Appendix~\ref{Appendix:eccentricty} elaborates on the eccentricity proxy that can be derived from the light curve.
Appendix~\ref{Appendix:additionalFigures} presents additional figures referenced in the main body of the text.
Appendix~\ref{Appendix:acknowledgements} completes the acknowledgments.

%==================================================================
\section{The catalogue}
\label{Sect:catalogue}

\begin{figure*}
  \centering
  \includegraphics[width=\linewidth]{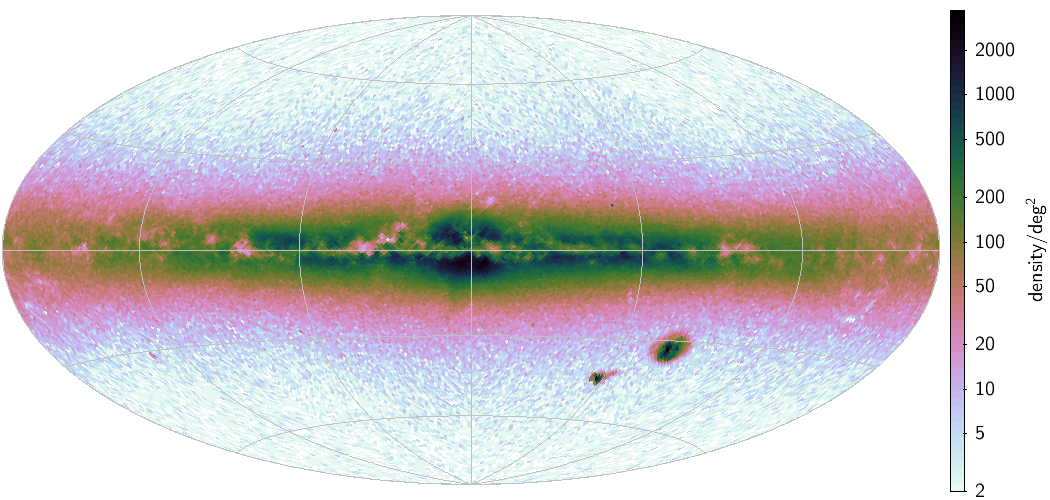}
  \caption{Sky density map of the \Gaia catalogue of eclipsing binaries, in Galactic coordinates, colour-coded according to the colour scale shown on the right of the figure.
          }
\label{Fig:sky}
\end{figure*}

%\begin{figure}
%  \centering
%  \includegraphics[width=\linewidth]{Figures/figSky_Nobs.png}
%  \caption{Sky distribution (Galactic coordinates) of the number $N_\gmag$ of field-of-view observations selected for variability analysis in the \gmag light curves of the \Gaia eclipsing binaries, colour-coded according to the colour scale shown on the right of the figure.
%          }
%\label{Fig:sky_Nobs}
%\end{figure}

\begin{figure}
  \centering
  \includegraphics[trim={0 0 0 42},clip,width=\linewidth]{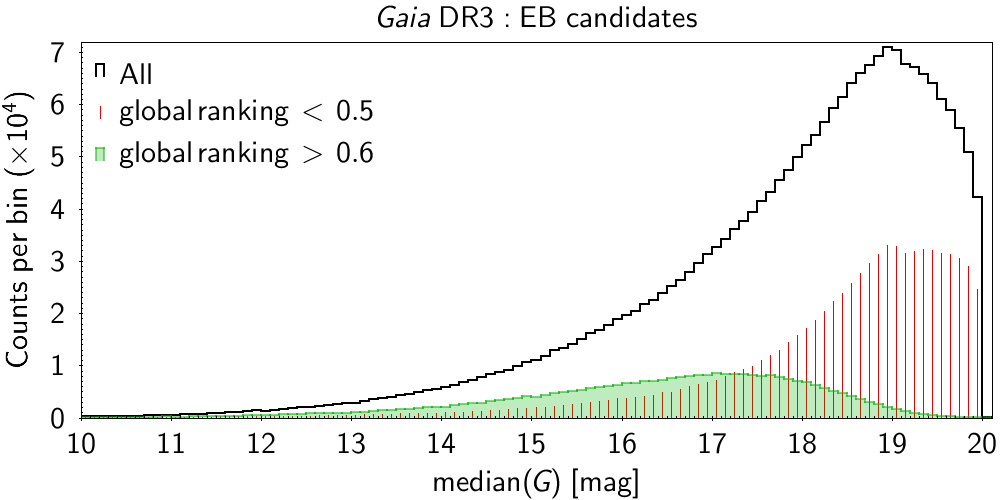}
  \caption{Distribution of \gmag magnitude of the full sample (black histogram) and of the samples with global ranking larger than 0.6 (filled green histogram) and smaller than 0.5 (red spiked histogram).
           The abscissa scale is truncated at the lower side for better visibility.
          }
\label{Fig:histo_gmag}
\end{figure}

The 2\,184\,477 sources published in table \texttt{gaiadr3.vari\_eclipsing\_binary} (under \texttt{Variability} in the \Gaia archive) constitute the \Gaia DR3 catalogue of eclipsing binaries.
The candidates were selected considering a mixture of various criteria with the goal of reaching a relatively good degree of completeness while limiting the level of contamination.
The list of sources in this catalogue is essentially the same as the list of variables identified as eclipsing binaries in the general \Gaia DR3 classification table \texttt{vari\_classifier\_result} \citep[variability type \texttt{ECL}, for details see][]{DR3-DPACP-165}.
Small differences nevertheless exist between the two tables.
Nineteen sources are present in the classification table but are not in the catalogue of eclipsing binaries.
Periods and light curve characterisation are thus not available for these sources.
Conversely, the catalogue of eclipsing binaries contains 140 candidates not listed in the classification table due to a post-processing step of the classification table that modified the label of a small fraction of sources.
In this paper, we restrict the analysis to the catalogue of eclipsing binaries.

From the two million eclipsing binary candidates, 86\,918 have further been processed within the DPAC to derive orbital solutions.
The results are published in table \texttt{gaiadr3.nss\_two\_body\_orbit} (under \texttt{Non-single stars} in the \Gaia archive), with \texttt{nss\_solution\_type='EclipsingBinary'}.
We refer to \citet{DR3-DPACP-179} for a presentation of that table.
In addition, 155 of them have combined photometric + spectroscopic solutions (identified in the table with \texttt{nss\_solution\_type='EclipsingSpectro'}).
We refer to \citet{GaiaArenouBabusiaux_etal22} for further information.

The distribution on the sky of the eclipsing binary candidates from the catalogue is shown in Fig.~\ref{Fig:sky}.
The \gmag light curves contain between 16 and 259 cleaned field-of-view measurements, depending on the sky position according to the \Gaia scanning law.% (Fig.~\ref{Fig:sky_Nobs}).
For each candidate, an orbital period is provided in the catalogue, together with a geometrical characterisation of its \gmag light curve and a global ranking that ranges from 0.4 to 0.84 (Eqs.~\ref{Eq:FVU} and \ref{Eq:global_ranking} in Sect.~\ref{Sect:catalogue_method}), where a higher value indicates a better light curve characterisation.
Figure~\ref{Fig:histo_gmag} gives the \gmag magnitude distribution for the full catalogue (in black) and for the sub-samples with the highest (>0.6, in green) and lowest (<0.5, in red) global rankings.

The eclipsing binary pipeline is presented in Sects.~\ref{Sect:pipeline_input} to \ref{Sect:catalogue_selection}.
The input to the pipeline is shortly described in Sect.~\ref{Sect:pipeline_input}.
The geometrical characterisation of the light curves is detailed in Sect.~\ref{Sect:catalogue_method}, and our post-pipeline selection criteria is presented in Sect.~\ref{Sect:catalogue_selection}.
The content of the catalogue is summarised in Sect.~\ref{Sect:catalogue_content}.

%------------------------------------------------------------------
\subsection{Eclipsing binary pipeline input}
\label{Sect:pipeline_input}

The eclipsing binary module that generated the candidates published here are part of the variability pipeline consisting of several stages described in \cite{2017arXiv170203295E,2022arXiv220606416E}.
After a general variability detection performed on all \Gaia sources, variable source candidates go through a classification stage \citep{DR3-DPACP-165}.
Sources classified as eclipsing binaries are then fed to our eclipsing binary module.

Not all sources initially classified as eclipsing binaries are published in DR3.
An initial selection keeps only sources that are brighter than 20~mag in \gmag, that have at least sixteen cleaned field-of-view measurements in their \gmag light curves, and for which the skewness in the \gmag time series is larger than $-0.2$.
This constitutes $\sim$20 million sources.
The eclipsing binary pipeline then processes the \gmag light curves (as described in Sect.~\ref{Sect:catalogue_method}), and a final selection further filters out sources according to period and folded light curve properties (see Sect.~\ref{Sect:catalogue_selection}).

%------------------------------------------------------------------
\subsection{Light curve characterisation}
\label{Sect:catalogue_method}

For each eclipsing binary candidate, a geometric model of its \gmag-band light curve is constructed by fitting to the cleaned \gmag-band time series up to two Gaussians and one cosine.
The Gaussian components aim at modelling the geometrical light curve shape of the eclipses and the cosine component that of an ellipsoidal-like variability.
The model serves the purposes of characterising the geometry of the light curve, of selecting the most probable orbital period based purely on the photometry, and of providing a ranking among all sources.

The `two-Gaussian' model is introduced in Sect.~\ref{Sect:twoGaussModel}, and its derived parameters is described in Sect.~\ref{Sect:derivedParams}.
The period search method is then presented in Sect.~\ref{Sect:periodSearch}, followed  in Sect.~\ref{Sect:paramsUncertaintyEst} by the procedure to estimate the uncertainty on these parameters.
Our final per-source light curve model selection strategy is given in Sect.~\ref{Sect:modelSelection}.

%- - - - - - - - - - - - - - - - - - - - - - -
\subsubsection{Two-Gaussian model parameters}
\label{Sect:twoGaussModel}

\begin{figure}
  \centering
  \includegraphics[width=0.8\linewidth]{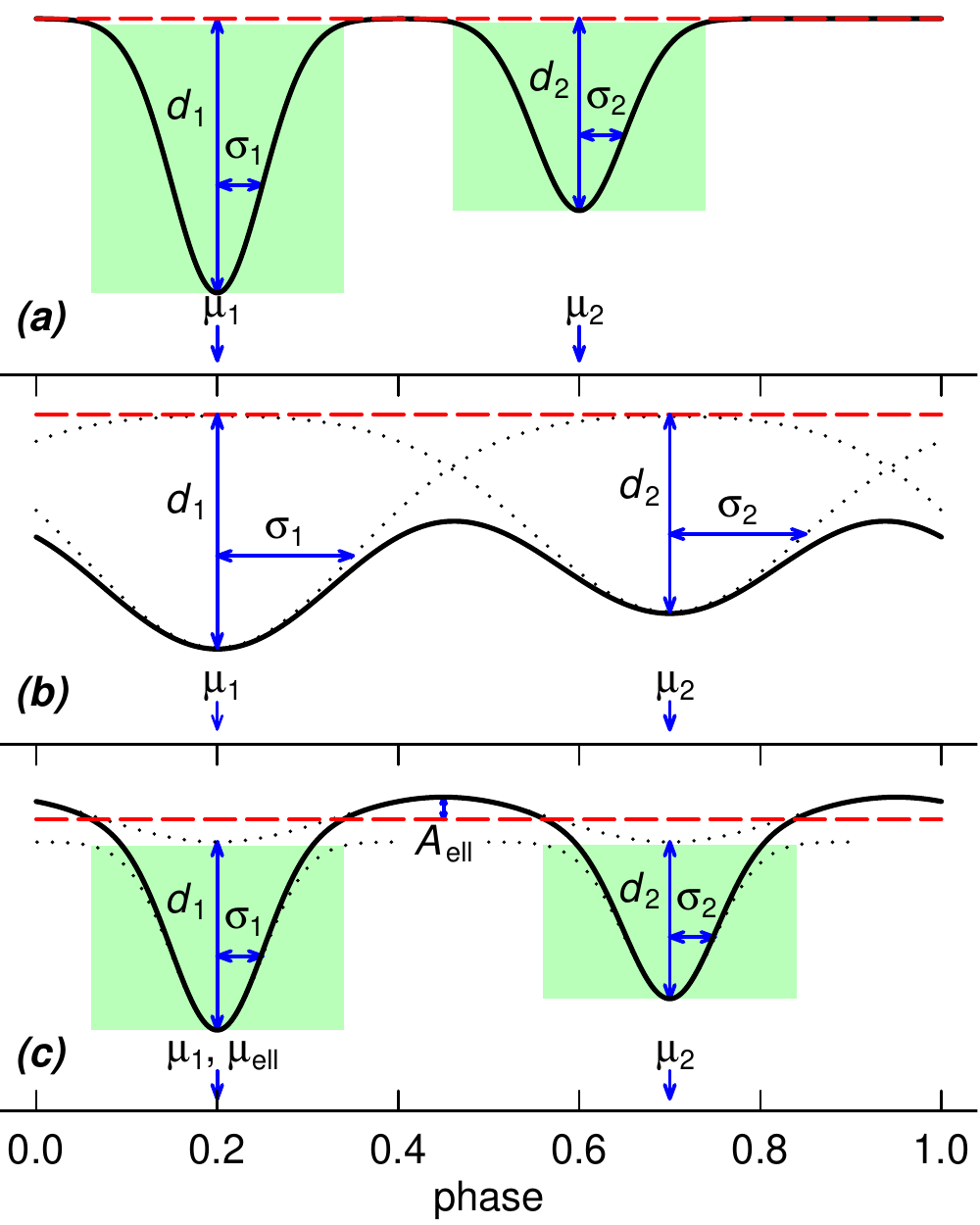}
  \caption{Schematic representation of the two-Gaussian model parameters used in Eq.~(\ref{Eq:twoGaussianModel}) to fit folded light curves of eclipsing binaries.
           The ordinate represents magnitude in reverse order.
           Three cases are shown with their primary eclipses (arbitrarily) located at phase 0.2.
           Case \textit{(a)} illustrates the modelling of a well-detached eccentric system using two non-overlapping Gaussians.
           Case \textit{(b)} shows a very tight circular system modelled with two overlapping Gaussians.
           Case \textit{(c)} represents a tight circular system with an out-of-eclipse ellipsoidal variation modelled with a cosine component.
           The red dashed horizontal line in each panel indicates the value of the constant $C$ in Eq.~\ref{Eq:twoGaussianModel}.
           The green areas delimit the eclipse durations.
           The thin black dotted lines in the middle and bottom panels show the individual Gaussian and/or cosine components of the two-Gaussian models.
           The thick black solid lines show the resulting two-Gaussian models.
          }
\label{Fig:twoGaussianModel}
\end{figure}

%---------
\begin{table*}
\caption{Types of geometric models fitting the \gmag light curves.
         The columns give the model type, the number of model parameters, the ranking used in model prioritization, a description of the model type, and the number of sources of the given type in the \Gaia DR3 table \texttt{vari\_eclipsing\_binary}.
        }
\centering
\begin{tabular}{l c c l r}
\hline\hline
Model type & Num. of & Model & Description & Num. of \\
           & params  & rank  &             & sources \\
\hline
\hline
\tiny{\texttt{TWOGAUSSIANS}}                                 & 8 & 6 &  Two Gaussians & 1\,587\,926 \\
\tiny{\texttt{TWOGAUSSIANS\_WITH\_ELLIPSOIDAL\_ON\_ECLIPSE1}}& 9 & 5 &  Two Gaussians $+$ cosine\tablefootmark{a} aligned on Gaussian 1 & 389\,725 \\
\tiny{\texttt{TWOGAUSSIANS\_WITH\_ELLIPSOIDAL\_ON\_ECLIPSE2}}& 9 & 4 &  Two Gaussians $+$ cosine\tablefootmark{a} aligned on Gaussian 2 & 85\,400 \\
\textit{Total with two Gaussians} & & & & \textit{2\,063\,051} \\
\hline
\tiny{\texttt{ONEGAUSSIAN}}                                  & 5 & 3 &  One Gausian. & 36\,984 \\
\tiny{\texttt{ONEGAUSSIAN\_WITH\_ELLIPSOIDAL}}               & 6 & 2 &  One Gaussian $+$ cosine\tablefootmark{a} & 48\,215 \\
\textit{Total with one Gaussian}  & & & & \textit{85\,199} \\
\hline
\tiny{\texttt{ELLIPSOIDAL}}                                  & 4 & 1 &  A cosine\tablefootmark{a} & 36\,227\\
\hline
\multicolumn{4}{l}{\textit{All}} & \textit{2\,184\,477} \\
\hline
\hline
\end{tabular}
\vskip -2mm
\tablefoot{\tablefoottext{a}{Cosine function with half the orbital period}
          }
\label{Tab:modelTypes}
\end{table*}

%---------
\begin{table*}
\caption{Data fields in the \Gaia DR3 table of eclipsing binaries (\Gaia DR3 table \texttt{vari\_eclipsing\_binary}), with their units (col.~2), the mathematical symbol used in this paper (if used, col.~3), and a short description (col.~4).
}
\centering
\begin{tabular}{l c c l}
\hline\hline
Data field name & Unit & Symb & Description \\
%
%\hline\hline
%\multicolumn{2}{l}{Table \texttt{vari\_classifier\_result}} \\
\hline
\tiny{\texttt{source\_id}}                           & -- & & Unique source identifier of the EB candidate \\
\tiny{\texttt{model\_type}}                          & -- & & Geometric model type fitting the \gmag light curve (Table~\ref{Tab:modelTypes}) \\
\tiny{\texttt{num\_model\_parameters}}               & -- & & Number of free parameters of the geometric model \\
\tiny{\texttt{global\_ranking}}                      & -- & & Number between 0 (worst) and 1 (best) \\
\tiny{\texttt{reduced\_chi2}}                        & -- & $\chi^2_{\rm red}$  & Reduced $\chi^2$ of the geometric model fit \\
\tiny{\texttt{frequency}}                            & day$^{-1}$ & $f_\mathrm{orb} = 1/P_\mathrm{orb}$ & Orbital frequency of the EB \\
\tiny{\texttt{frequency\_error}}                     & day$^{-1}$ & $f_\mathrm{orb,err}$ & Uncertainty on the orbital frequency \\
\tiny{\texttt{reference\_time}}                      & BJD\tablefootmark{a} & $T_0$ & Reference time for the geometric model fit \\
\hline
\tiny{\texttt{geom\_model\_reference\_level}}        & mag & $C$ & Magnitude reference level of geometric model \\
\tiny{\texttt{geom\_model\_reference\_level\_error}} & mag & & Uncertainty on \tiny{\texttt{geom\_model\_reference\_level\_error}} \\
\tiny{\texttt{geom\_model\_gaussian1\_phase}}        & -- & $\mu_1$ & Phase of Gaussian 1\tablefootmark{b} \\
\tiny{\texttt{geom\_model\_gaussian1\_phase\_error}} & -- & & Uncertainty on \tiny{\texttt{geom\_model\_gaussian1\_phase\_error}}\tablefootmark{b} \\
\tiny{\texttt{geom\_model\_gaussian1\_sigma}}        & -- & $\sigma_{1}$ & Width (standard deviation, in phase) of  Gaussian 1\tablefootmark{b} \\
\tiny{\texttt{geom\_model\_gaussian1\_sigma\_error}} & -- & & Uncertainty on \tiny{\texttt{geom\_model\_gaussian1\_sigma\_error}}\tablefootmark{b} \\
\tiny{\texttt{geom\_model\_gaussian1\_depth}}        & mag & $d_{1}$ & Depth of Gaussian 1\tablefootmark{b} \\
\tiny{\texttt{geom\_model\_gaussian1\_depth\_error}} & mag & & Uncertainty on \tiny{\texttt{geom\_model\_gaussian1\_depth\_error}}\tablefootmark{b} \\
\tiny{\texttt{geom\_model\_gaussian2\_phase}}        & -- & $\mu_2$ & Phase of Gaussian 2\tablefootmark{c} \\
\tiny{\texttt{geom\_model\_gaussian2\_phase\_error}} & -- & & Uncertainty on \tiny{\texttt{geom\_model\_gaussian2\_phase\_error}}\tablefootmark{c} \\
\tiny{\texttt{geom\_model\_gaussian2\_sigma}}        & -- & $\sigma_{2}$ & Width (standard deviation, in phase) of  Gaussian 2\tablefootmark{c} \\
\tiny{\texttt{geom\_model\_gaussian2\_sigma\_error}} & -- & & Uncertainty on \tiny{\texttt{geom\_model\_gaussian2\_sigma\_error}}\tablefootmark{c} \\
\tiny{\texttt{geom\_model\_gaussian2\_depth}}        & mag & $d_{2}$ & Depth of Gaussian 2\tablefootmark{c} \\
\tiny{\texttt{geom\_model\_gaussian2\_depth\_error}} & mag & & Uncertainty on \tiny{\texttt{geom\_model\_gaussian2\_depth\_error}}\tablefootmark{c} \\
\tiny{\texttt{geom\_model\_cosine\_half\_period\_amplitude}} & mag & \Aell & Amplitude (half peak-to-peak) of the cosine component \\
                                                      & & & \hspace{2mm} with half the period of the geometric model\tablefootmark{d} \\
\tiny{\texttt{geom\_model\_cosine\_half\_period\_amplitude\_error\!\!}} & mag & & Uncertainty on \tiny{\texttt{geom\_model\_cosine\_half\_period\_amplitude}}\tablefootmark{d} \\

\tiny{\texttt{geom\_model\_cosine\_half\_period\_phase}} & mag & \phaseell & Phase of the cosine component \\
                                                      & & & \hspace{2mm} with half the period of the geometric model\tablefootmark{e} \\
\tiny{\texttt{geom\_model\_cosine\_half\_period\_phase\_error\!\!}} & mag & & Uncertainty on \tiny{\texttt{geom\_model\_cosine\_half\_period\_phase}}\tablefootmark{e} \\

\hline
\tiny{\texttt{derived\_primary\_ecl\_phase}}         & -- & \phaseEclPrimary & Phase location at geometrically deepest point\tablefootmark{b} \\
\tiny{\texttt{derived\_primary\_ecl\_phase\_error}}  & -- & $\varepsilon(\phaseEclPrimary)$ & Uncertainty on \tiny{\texttt{derived\_primary\_ecl\_phase}}\tablefootmark{b} \\
\tiny{\texttt{derived\_primary\_ecl\_duration}}        & -- & \durationEclPrimary & Phase duration of deepest eclipse in phase fraction\tablefootmark{b} \\
\tiny{\texttt{derived\_primary\_ecl\_duration\_error}} & -- & $\varepsilon(\durationEclPrimary)$ & Uncertainty on \tiny{\texttt{derived\_primary\_ecl\_duration}}\tablefootmark{b} \\
\tiny{\texttt{derived\_primary\_ecl\_depth}}        & mag & \depthEclPrimary & Depth of deepest eclipse\tablefootmark{b}. \\
\tiny{\texttt{derived\_primary\_ecl\_depth\_error}} & mag & & Uncertainty on \tiny{\texttt{derived\_primary\_ecl\_depth}}\tablefootmark{b} \\
\tiny{\texttt{derived\_secondary\_ecl\_phase}}         & -- & \phaseEclSecondary & Phase location at geometrically second deepest point\tablefootmark{c} \\
\tiny{\texttt{derived\_secondary\_ecl\_phase\_error}}  & -- & $\varepsilon(\phaseEclSecondary)$ & Uncertainty on \tiny{\texttt{derived\_secondary\_ecl\_phase}}\tablefootmark{c} \\
\tiny{\texttt{derived\_secondary\_ecl\_duration}}        & -- & \durationEclSecondary & Phase duration of second deepest eclipse in phase fraction\tablefootmark{c} \\
\tiny{\texttt{derived\_secondary\_ecl\_duration\_error}} & -- & $\varepsilon(\durationEclSecondary)$ & Uncertainty on \tiny{\texttt{derived\_secondary\_ecl\_duration}}\tablefootmark{c} \\
\tiny{\texttt{derived\_secondary\_ecl\_depth}}        & mag & \depthEclSecondary & Depth of second deepest eclipse\tablefootmark{c} \\
\tiny{\texttt{derived\_secondary\_ecl\_depth\_error}} & mag & & Uncertainty on \tiny{\texttt{derived\_secondary\_ecl\_depth}}\tablefootmark{c} \\
\hline
\end{tabular}
\tablefoot{\tablefoottext{a}{Referenced time given in barycentric JD in TCB - 2455197.5 day.}\\
           \tablefoottext{b}{Null if no Gaussian component in the model.}\\
           \tablefoottext{c}{Null if only one Gaussian component in the model.}\\
           \tablefoottext{d}{Null if no cosine component with half the period of the geometric model.}\\
           \tablefoottext{e}{Null if no cosine component with half the period of the geometric model. Equal to one of the \texttt{geom\_model\_gaussian*\_phase} and associated error if model type contains ``\texttt{\_WITH\_ELLIPSOIDAL*}''.}
          }
\label{Tab:dataFields}
\end{table*}

The geometrical model fitted to the \gmag light curve consists of up to two Gaussians and a cosine.
The model can contain any combination of these three components, not all necessarily present.
It is called a `two-Gaussian' model irrespective of the number of components it eventually contains.
A full description of the model is given in \citet{MowlaviLecoeurHoll_etal17}, to which we refer for more details.
We here summarise the model components and associated parameters.

A Gaussian component $k$ is defined as
\begin{equation}
  f_{\mathrm{Gauss},k}(\varphi) = d_k \; exp{\left(-\frac{(\varphi- {\mu_k})^2}{2\,\sigma_k^2} \right)}\;,
\label{Eq:Gaussian}
\end{equation}
where $\varphi$ is the orbital phase, i.e. (observation time $-$ reference time $T_0$) modulo (orbital period), and $\mu_k$, $d_k$, $\sigma_k$ are the Gaussian parameters (phase location of the centre, depth in magnitude, and width in phase, respectively) of the first ($k=1$) and second ($k=2$) Gaussian, when present.
A schematic representation of a model with two Gaussians mimicking a well detached binary system is shown in the top panel of Fig.~\ref{Fig:twoGaussianModel}, while the middle panel illustrates the case of a tighter system modelled with two overlapping Gaussians.
We note that there are not always two Gaussians in the models and, when there are two, the first Gaussian is not necessarily the deepest of the two.

When a Gaussian component is included, its mirror functions at phases below zero and above one are automatically added to take into account the contribution of the tails of the Gaussian function from adjacent phases due to the periodicity of the eclipses \citep[see Eq.~(2) of][]{MowlaviLecoeurHoll_etal17}.
This is necessary for a correct inclusion of wide Gaussians.

The cosine component, when present, has a period equal to half of the orbital period.
It is given by
\begin{equation}
    f_\mathrm{cos}(\varphi) = A_\mathrm{ell} \; \cos [4\pi (\varphi-\mu_\mathrm{ell})] \;,
\label{Eq:sine}
\end{equation}
where $A_\mathrm{ell}$ is the amplitude of the cosine function.
If there are any Gaussian components, $\mu_\mathrm{ell}$ is either equal to $\mu_1$ or $\mu_2$, depending on whether the cosine is centred on the first or second Gaussian component, respectively.
If the model contains only a cosine, $\mu_\mathrm{ell}$ is fitted to the data as an independent parameter.

When all the components are present, the model writes
\begin{equation}
    f(\varphi) = C + f_{\mathrm{Gauss},1}(\varphi) + f_{\mathrm{Gauss},2}(\varphi) + f_\mathrm{cos}(\varphi) \;,
\label{Eq:twoGaussianModel}
\end{equation}
where $C$ is the reference level.
The list of model types according to the number of components, and the number of parameters for each model type are summarised in Table~\ref{Tab:modelTypes}.
We note that this model is adequate to represent eccentric systems only in the absence of a cosine component, and that reflection, which would be described with a cosine component with a period equal to the orbital period, is not included in this first \Gaia catalogue of eclipsing binaries.

All parameters necessary to reconstruct the geometric model are published in the catalogue.
They are summarised in Table~\ref{Tab:dataFields}.
The orbital period is given as a frequency (to which a frequency uncertainty can be associated, see Sect.~\ref{Sect:paramsUncertaintyEst}).
The model component parameters are given in field names prepended with ``\texttt{geom\_model\_}''.
The reference time used for phase folding is also published.

%- - - - - - - - - - - - - - - - - - - - - - -
\subsubsection{Derived geometric model parameters}
\label{Sect:derivedParams}

In addition to the `two-Gaussian' model parameters, several parameters are derived from the geometric model and published in the catalogue in field names prepended with ``\texttt{derived\_}'' (see Table ~\ref{Tab:dataFields}).
These derived parameters give eclipse characteristics (phase location, phase duration, depth) based on the geometric model as given by Eq.~(\ref{Eq:twoGaussianModel}).
The deepest and second deepest eclipse information are stored in the `primary' and `secondary' eclipse fields, respectively.
We remind that the underlying Gaussian model components 1 and 2 have no specific order.

Derived eclipse parameters are only provided in association with a Gaussian component.
A dip in the folded light curve that results from a cosine component and that has no associated Gaussian does not have derived eclipse parameters.
Therefore, models containing a cosine and a Gaussian, for example, only have one set of derived eclipse parameters.
Only the ``\texttt{derived\_primary\_}*'' fields are then filled in the catalogue.
Likewise, purely cosine models have no derived parameters.

The derived eclipse phase locations are obtained by starting at the centre of the Gaussian (1 or 2) and identifying the closest zero-derivative (flat) point in the light-curve, which is not necessarily located at the same positions as the centres of the Gaussians if they are not offset by 0.5 in phase or when there is an ellipsoidal component.
The derived eclipse depth is defined as the distance between the model value at the derived primary or secondary eclipse phase and the brightest model value, and the derived eclipse duration in phase is defined %at a magnitude depth of 2\% within eclipse $k$ (Eq.~\ref{Eq:Gaussian}) 
as $5.6\,\sigma_k$, $\sigma_k$ being defined in Eq.~(\ref{Eq:Gaussian}), with a maximum of 0.4 \citep[see ][]{MowlaviLecoeurHoll_etal17}.
These last two definitions equally apply for models with and without an ellipsoidal component.

%- - - - - - - - - - - - - - - - - - - - - - -
\subsubsection{Period search}
\label{Sect:periodSearch}

The orbital period is obtained in two steps.
First, a list of up to twenty candidate periods is established from the \gmag light curve as described in this section.
Two-Gaussian models are then fitted to the light curve for each of these periods, and the best model is selected as described in Sect.~\ref{Sect:modelSelection}.
The period of this best model is the orbital period published in the catalogue together with the best model parameters.

Due to the variety of eclipsing binary light curve geometries, we combined the results of three different period search methods to identify the list of candidate periods.
The three methods are the Generalised Least-Squares \citep{HeckManfroidMersch_85,CummingMarcyBulter99,ZechmeisterKurster09}, the Phase Dispersion Minimisation \cite[PDM;][]{Jurkevich71,Stellingwerf78,SchwarzenbergCzerny97} and the String Length \citep{LaflerKinman65,BurkeRollandBoy70} methods.
The choice for these three different methods is based on earlier internal tests on HIPPARCOS \citep{Hipparcos_1997} eclipsing binaries showing the largest correct period recovery to be found in the union of this ensemble.
The unweighted procedure has been used in all cases because the observations in the eclipses are fainter, and their uncertainties consequently larger, than their corresponding out-of-eclipses values, and they would therefore be down-weighted in a weighted procedure.
Periodograms are computed using these three methods in the frequency range between 0.005 and 15 d$^{-1}$ (spanning 1.6~h to 200~d) using a fixed frequency step of $10^{-5}$~d$^{-1}$.

The two most significant peaks in each of the three periodograms are then gathered in a list of candidate frequencies, to which half and twice their values are added for all three methods, as well as one third and four times their values for Generalised Least-Squares.
In this way, a set of twenty candidate periods are constructed, some of which might of course overlap between the different methods.

%- - - - - - - - - - - - - - - - - - - - - - -
\subsubsection{Model parameter uncertainties}
\label{Sect:paramsUncertaintyEst}

Due to the often low duty-cycle of eclipsing signals (e.g. down to an adopted minimum of three observations in eclipse), estimation of the uncertainties in our models can be inherently imprecise.
As formal errors from the least-squares fit do not capture any modelling errors, we opted the jackknife method to get a sense of the uncertainties around our best-fit solution parameters.

For this data release, we implemented a Jackknife method with non-robust mean and variance estimates \citep{WallJenkins03}.
Essentially this means that, in order to estimate the uncertainties of the best fit model parameters $\vec{p}$ (including frequency, reference level, and derived parameters) of a source with $N$ observations $X_{i=1\rightarrow N}$, we re-fit the model $N$-times, where each time one of the observations $X_i$ is left out.
Generally, for each re-fit, this recovers a similar, but not identical, parameter solution $\vec{p_i}$ of which the variance
\begin{equation}
\label{eq:jkSigmaParamEst}
(\vec{\sigma}^*)^2 = \frac{N-1}{N} \sum^N_{i=1} \left( \vec{p_i} - \frac{1}{N}\sum^N_{i=1}\vec{p_i} \right)^2
\nonumber
\end{equation}
\citep[Eq.~6.20 in][]{WallJenkins03} is used to populate the uncertainty estimate.
Because instances of the $N$ jackknife re-fits can cause non-convergence, a minimum of 30\% converged solutions was required to estimate the uncertainties.
If more than 70\% of the re-fits failed, the model is rejected from the list of model candidates for the given source (see Sect.~\ref{Sect:modelSelection}).
Even though most Jackknife solutions converged, some included some wildly large values, which is reflected in some of the published uncertainties.
Alternatively, the Jackknife samples showed in some cases too little variation for a good uncertainty estimate, resulting in some near-zero uncertainty estimates.
We intend to improve upon that in DR4 by implementing a more robust estimate of the variance. 

The Jackknife method described above allows to estimate the uncertainties of not only the geometric model parameters, but also of the frequency, reference level, and derived parameters.
These uncertainties are generally more informative (and larger) than the formal errors obtained from a simple linear covariance estimation at the best-fit parameter set, because the latter does not include any modelling errors and assumes that observation uncertainties are correctly estimated.

As the frequency is among the most important parameters, we applied more stringent checks and bounds on its estimated Jackknife uncertainty.
We set \texttt{frequency\_error} = MAX( \texttt{frequency\_error}, 0.001/\texttt{time\_duration\_g\_fov}) where \texttt{time\_duration\_g\_fov} is the duration between the first and last observations, as published in the \texttt{gaiadr3.vari\_summary} table.
Additionally we identified that for \texttt{frequency\_error} $\times$ \texttt{time\_duration\_g\_fov} > 0.6, no correct period is recovered in our literature cross-match.
Therefore, all models with a value above this limit have been rejected.
These lower and upper bounds on the frequency uncertainty $f_\mathrm{orb,err}$ correspond to, respectively, 0.1\% and 60\% phase deviations%
\footnote{
The phase deviation at the last cycle of an observation due a shift $\delta P$ in period is given by $\delta P \, \Delta T / P^2$ where $\Delta T$ is the observation duration% (see also Eq.~\ref{Eq:fracPlit})
.
Expressed in terms of a frequency shift $\delta f$, the phase deviation writes $\delta f \, \Delta T$.
}
at the last cycle of the observations with the given period $P_\mathrm{orb}=1/f_\mathrm{orb}$.

The uncertainties on all model parameters are published in the catalogue in field names appended with ``\texttt{\_error}''.
They include the geometrical model parameters as well as the orbital frequency and derived parameters.

\subsubsection{Model selection strategy}
\label{Sect:modelSelection}

For each of the up-to-twenty candidate periods identified in Sect.~\ref{Sect:periodSearch}, seven two-Gaussian models are fitted to the \gmag light curve by considering all possible combinations of the two-Gaussian components, including a simple constant model in order to do a proper model comparison against a non-variable model.
This results in a list of up to 140 model candidates per source, considering the six model types listed in Table~\ref{Tab:modelTypes} and the additional constant model.
The models are then cleaned and sorted according to their Bayesian Information Criterion (BIC) score \citep[][Eq.~3.54]{FeigelsonBabu12}, which allows to compare model fits for all combinations of the candidate periods and geometric models, and the best model is selected.
These steps are each briefly described in the next paragraphs.

In the first, cleaning step, models having component parameters that we deem non-physical are removed from the list of model candidates.
Visual inspection of earlier iterations of our pipeline on \Gaia data revealed that the geometric model parameters may model features that we deem non-physical.
This is the case when two Gaussian components are too close to each other.
We therefore remove a model from the list of model candidates if the derived primary and secondary eclipse locations are distant by less than 0.08 in phase to avoid  stacking Gaussians on the same eclipse.
We also remove models with one Gaussian if its width is larger than 0.4 in phase, as well as models with one Gaussian and a cosine component if the Gaussian width is larger than 0.4 in phase (as such wide Gaussian is partially degenerate with the ellipsoidal component).
The pipeline also checks the uncertainties of the geometric model parameters, and rejects models that have uncertainties larger than 10~mag for the reference level ($C$ in Eq.~\ref{Eq:twoGaussianModel}) or for the cosine amplitude ($A_\mathrm{ell}$), or larger than one for one for the phase locations ($\mu_1, \mu_2$) or widths ($\sigma_1, \sigma_2$) of the Gaussians.
No condition is given on the uncertainties of the Gaussian depths as this quantity can be unconstrained for well-detached systems with narrow eclipses.

After this first pruning of models, we order the list of remaining model candidates by their BIC score.
In the adopted BIC convention, a higher BIC score identifies a better model fit to the data taking into account the number of free parameters in each model and giving a higher weight to models that have a smaller number of parameters.
We then retain all models that have a BIC score within 30 of the highest BIC score.
All these model candidates are considered to be equivalently good at this point.
This list is then filtered according to several exclusion criteria. 
We remove the constant model that was added to the list of models, if it remains in the list of model candidates, models that have a phase coverage less than 0.6 (the phase coverage is computed by binning the phase-folded data in ten bins and counting the fraction of filled bins), and models that have less than three observations in an eclipse.
%\NM{The list described in the SRS is much longer, should we mention all of them here? \textbf{Andrej's suggestion:} yes, perhaps tabulate exclusions.  $\rightarrow$ Berry, could you check this?} \NM{Should also check that all these were indeed applied to the Cycle 3 data}.
%The model with the longest period in the list of model candidates remaining after the application of these filters is finally selected as the best model.
%\NM{These are from the SRS, to confirm that this is what has finally been used}
If multiple models survive at this point, a pre-defined model ranking is used to select the model with the highest rank according to the model ranking indicated in Table~\ref{Tab:modelTypes}.
It must be noted that this model ranking inevitably introduces priors in the model selection.
For example, circular systems with two equal-depth eclipses will be favored over eccentric systems displaying only one eclipse (these two cases differ by a factor of two in their orbital periods).
%[if wanted we can remove the new model ranking column and just refer to the current order in the table used as ranking?]}
If no candidate model remains in the final list, the source is removed from the catalogue.

%------------------------------------------------------------------
\subsection{Post-pipeline source filtering and model ranking} %{Candidate selection}
\label{Sect:catalogue_selection}

%\NM{This section is difficult and tricky to write given the complex multi-dimensional filters that we applied, sometimes even based on different runs using different outlier rejection operators. To see how this section could be completed/improved...}

%The identification of eclipsing binaries in a sample of 2.7 billion stars \NM{Replace with the number of variable candidates if given in the main CU7 paper} is not a trivial task, especially for detached systems with a sparse time series as results from the \Gaia scanning law. 
%In this first \Gaia catalogue of eclipsing binaries, the candidates were selected considering a mixture of various criteria with the goal of reaching a good degree of completeness while limiting the level of contamination.
%The final list derives from the results of the classification pipeline \citep[see][]{DR3-DPACP-165} combined with criteria based on photometric properties, on the list of frequencies obtained from the period search methods described in Sect.~\ref{Sect:catalogue_method}, and on properties of the folded light curve.
 
% Commented out as it was moved to section about variability pipeline input 
%We used the classification probabilities provided by two sets of meta-classifiers, one using multi-class classifiers and another one using one-versus-rest classifiers \citep[see][]{DR3-DPACP-165}.
%Thresholds were used in these meta-classifier results, combined with additional criteria given below.

In this first \Gaia catalogue of eclipsing binaries, the output of the pipeline underwent a large variety of verification and validation checks that led to the application of additional filters outlined here.
% \NM{These first filters are moved earlier in the pipeline input description}
%We considered only sources that are brighter than 20~mag in \gmag, that have at least sixteen measurements in their \gmag light curves, and for which the skewness in the \gmag time series is larger than -0.2.
The first concerns the periods found in the time series, requiring that the internal second best model (see Sect.~\ref{Sect:modelSelection}) must have a period compatible with the one found in the best model (i.e. with period ratios equal to 0.5, 1, 1.5 or 2).
Additional criteria further consider the Abbe value on the folded light curves in combination with various frequency limits and %ratios and
global ranking criteria.
%(\NM{\sout{quantitative criterion?}}), \sout{the phase coverage} (\NM{\sout{quantitative criterion?}}) \sout{, and the signal-to-noise of the \gmag time series} (\NM{\sout{quantitative criterion?}})) \NM{\sout{This last condition should logically go in Sect. 2.1; however, if it is done at this stage in the pipeline, it may affect the highest BIC value and should this be kept here}}.
Finally, sources with periods smaller than 0.2~d were removed because of the larger occurrence in DR3 of aliases at these small periods.

%\NM{Moved here from up as it compares sources, while above we compared models for a given source}
In order to compare the models of all sources in the catalogue, a global ranking is computed based on the Fraction of Variance Unexplained (FVU).
This quantity is defined as the ratio of the variance of the residuals to the variance of the signal, and is given by
\begin{equation}
%  FVU =  \frac{\displaystyle \frac{1}{N_\gmag} \sum_{i=1}^{N_\gmag} \; (\gmag_{\mathrm{obs},i} - \gmag_{\mathrm{model},i})^2}
%              {\displaystyle \frac{N_\gmag-1}{N_\gmag} \; var(\gmag_\mathrm{obs})}\;.
  FVU =  \frac{\displaystyle \sum_{i=1}^{N_\gmag} \; (\gmag_{\mathrm{obs},i} - \gmag_{\mathrm{model},i})^2}
              {\displaystyle \sum_{i=1}^{N_\gmag} \; (\gmag_{\mathrm{obs},i} - \overline{\gmag})^2}\;.
\label{Eq:FVU}  
\end{equation}
In this equation, $\gmag_{\mathrm{obs},i}$ is the $i$th measurement of the ${N_\gmag}$ observations in the \gmag time series, $\gmag_{\mathrm{model},i}$ is the value of the model at that time, and $\overline{\gmag}$ the mean \gmag magnitude.
A global ranking that ranges between zero and one is then derived using a linear transformation of the base ten logarithm of the FUV, given by
\begin{equation}
  \texttt{global\_ranking} = 0.11 \; [3.45 - \log_{10}(FVU)] \;\; 
\label{Eq:global_ranking}
\end{equation}
The constants in this equation are empirically derived to map the $\log(FVU)$ values in the range from zero to one.

Our last source filter uses this global ranking.
Only sources with a global ranking larger than 0.4 are published in the catalogue.

%------------------------------------------------------------------
\subsection{Catalogue content}
\label{Sect:catalogue_content}

The data fields published in the catalogue are listed in Table~\ref{Tab:dataFields}.
They include the orbital frequency, the geometrical model parameters of the \gmag-band light curve, the parameters derived from the model, the uncertainties on these parameters, and the global ranking.
Orbital frequencies are published rather than orbital periods for consistency with the internal model parameterisation and subsequent uncertainty estimates by the Jackknife method.

The model type is one of the six possible combinations of two Gaussian and a cosine functions.
They are listed in Table~\ref{Tab:modelTypes}, together with the number of sources present in the catalogue for each type.
All model parameters are named with a prefix \texttt{`geom\_'}.
The numbering of the first and second Gaussians follows the order of dip detection in the pipeline, and does not necessarily correspond to an order where the deepest Gaussian would be Gaussian one and the shallowest Gaussian would be Gaussian two. 

The two-Gaussian model represents a purely geometrical description of the light curve morphology and is not intended to model the physical properties of the binary system.
From the two-Gaussian model, however, an estimate of the phase locations, durations and depths of the primary and secondary eclipses are derived by identifying the deepest and second deepest dips, respectively, in the model light curve (see Sect.~\ref{Sect:derivedParams}).
These quantities are published in the catalogue with data field names prefixed with \texttt{`derived\_'}.

As mentioned in Sect.~\ref{Sect:paramsUncertaintyEst}, the current uncertainty estimation is not robust against outlying samples in the Jackknife method, and thus can lead to arbitrarily high uncertainties in some cases.
This explains the presence in the table of unrealistically large estimates of the errors on some parameters.
Besides, values above 3.4E38 have been converted to NULL values, as they cannot fit in a numeric float type in the database.
As a result, there are 1131~sources which have NULL values for {\small{\texttt{geom\_model\_gaussian1\_depth\_error}}},
824~sources for {\small{\texttt{geom\_model\_gaussian2\_depth\_error}}},
776~sources for {\small{\texttt{derived\_primary\_ecl\_depth\_error}}},
and 1145~sources for {\small{\texttt{derived\_secondary\_ecl\_depth\_error}}}, despite the presence for these same sources of non-NULL values for the quantities to which the errors are associated.

%------------------------------------------------------------------
\section{Catalogue usage}
\label{Sect:catalogue_usage}

%This is the first \Gaia release of eclipsing binaries that covers a comprehensive set of candidates within the filtering criteria summarised in Sect.~\ref{Sect:catalogue_selection}.
%In this section, we provide insight into the data, starting with the two-Gaussian model parameters in Sect.~\ref{Sect:catalogue_usage_model}.
%The global ranking is then addressed in Sect.~\ref{Sect:catalogue_usage_ranking}, and the orbital period in Sect.~\ref{Sect:catalogue_usage_period}.

%------------------------------------------------------------------
\subsection{Light curve models}
\label{Sect:catalogue_usage_model}

%-----
\begin{figure}
  \centering
  \includegraphics[trim={40 145 0 70},clip,width=\linewidth]{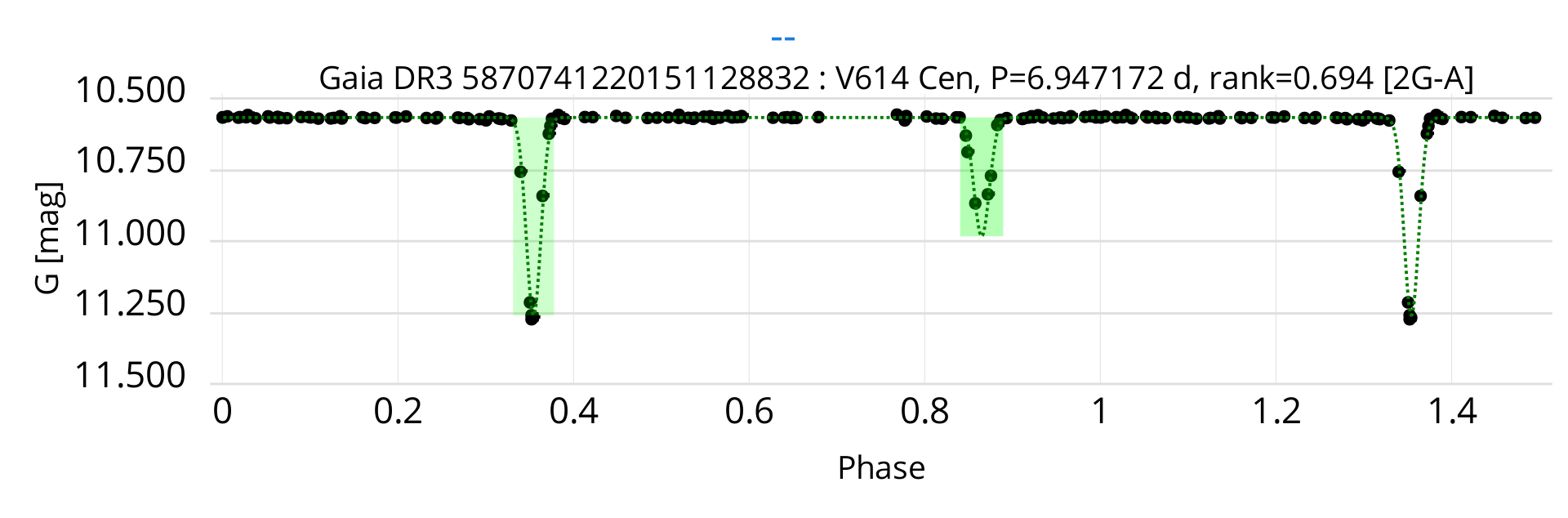}
  \vskip -0.5mm
  \includegraphics[trim={40 45 0 60},clip,width=\linewidth]{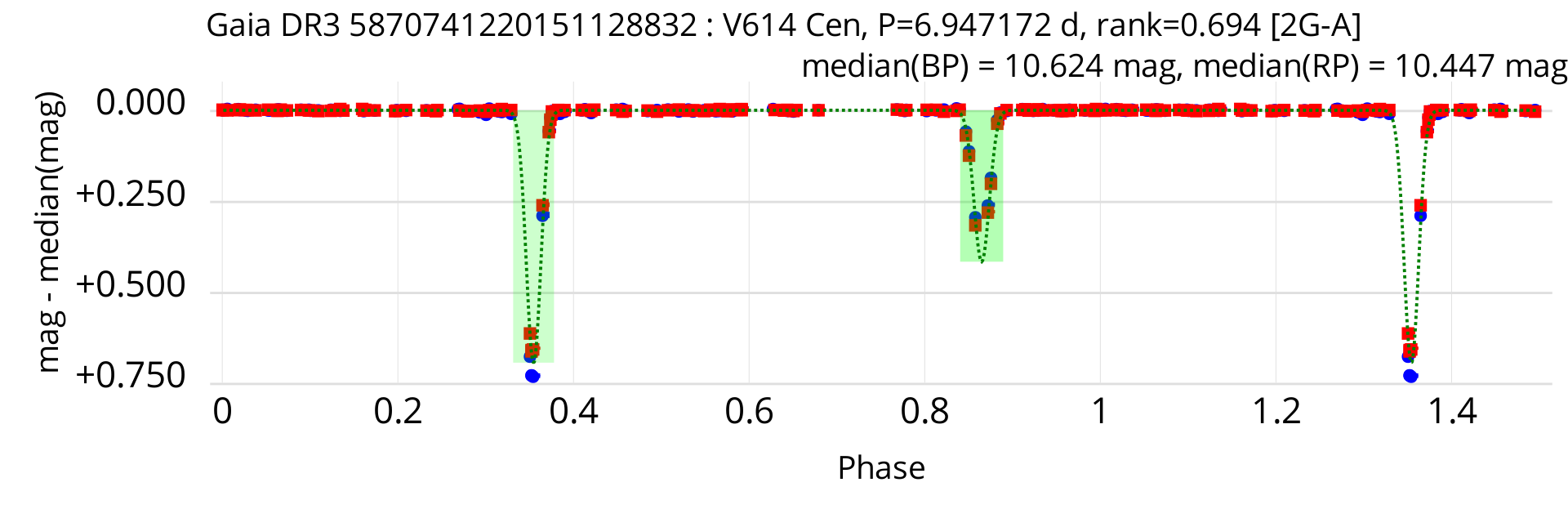}
  \caption{Folded light curves of V614~Ven.
           \textbf{Top panel:} \gmag light curve.
           The two-Gaussian model is superposed in dotted line.
           The green areas indicate the derived eclipse durations.
           \textbf{Bottom panel:} \gbp and \grp light curves, shifted by a value equal to their respective median magnitudes as written in the top of the panel.
           The dotted model and green areas shown in the bottom panel are the ones from the \gmag light curve, shifted to match zero median magnitude.
           The \Gaia period, global ranking and the light curve classification (in brackets; see text) are given in title of the figure after the \Gaia DR3 ID and GCVS name.
           }
\label{Fig:lcs_example_2GA}
\end{figure}
%-----

%-----
\begin{figure}
  \centering
  \includegraphics[trim={40 145 0 74},clip,width=\linewidth]{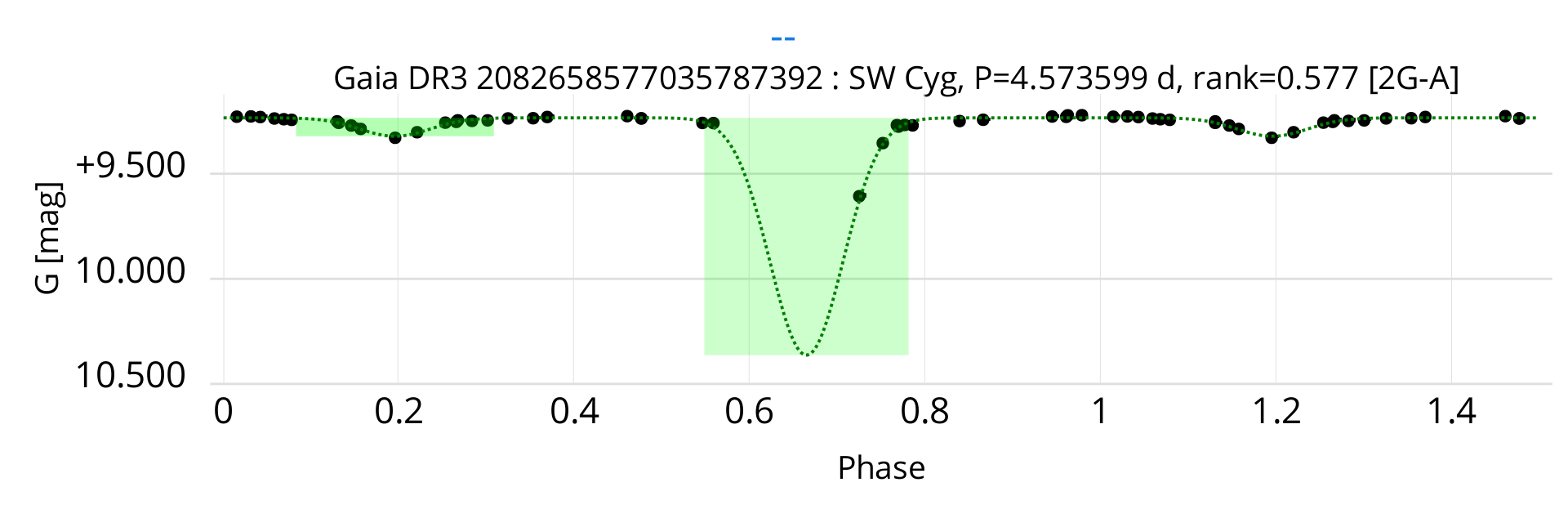}
  \vskip -0.5mm
  \includegraphics[trim={40 145 0 74},clip,width=\linewidth]{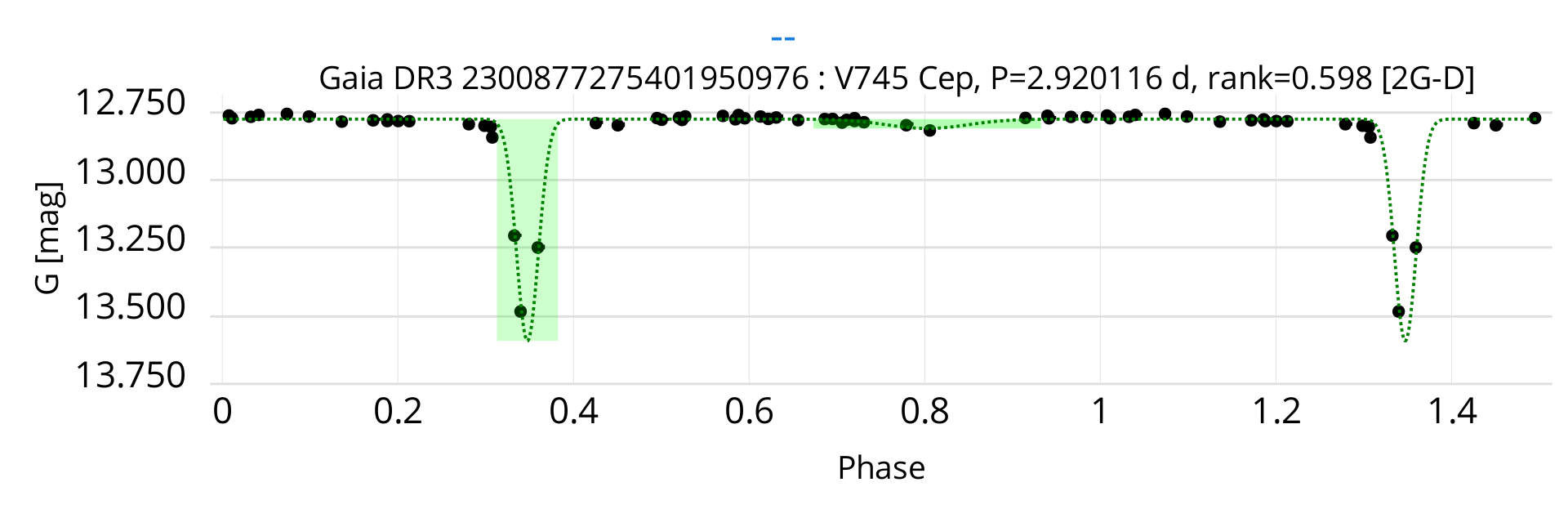}
  \vskip -0.5mm
  \includegraphics[trim={40 145 0 74},clip,width=\linewidth]{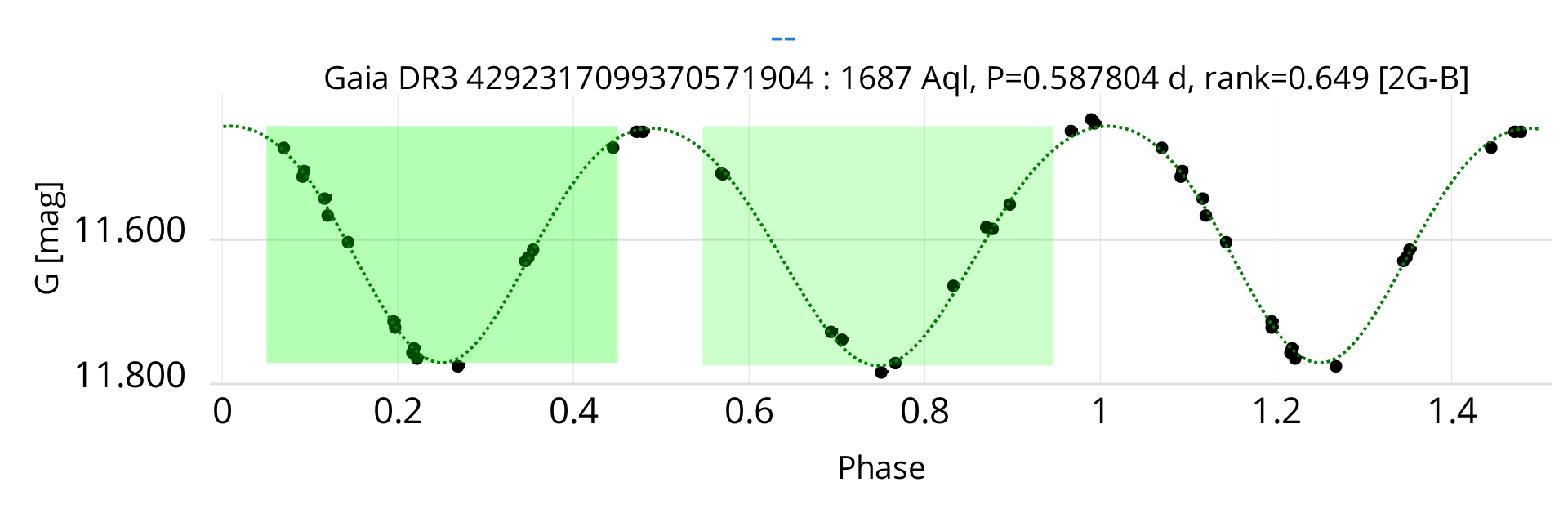}
  \vskip -0.5mm
  \includegraphics[trim={40 145 0 74},clip,width=\linewidth]{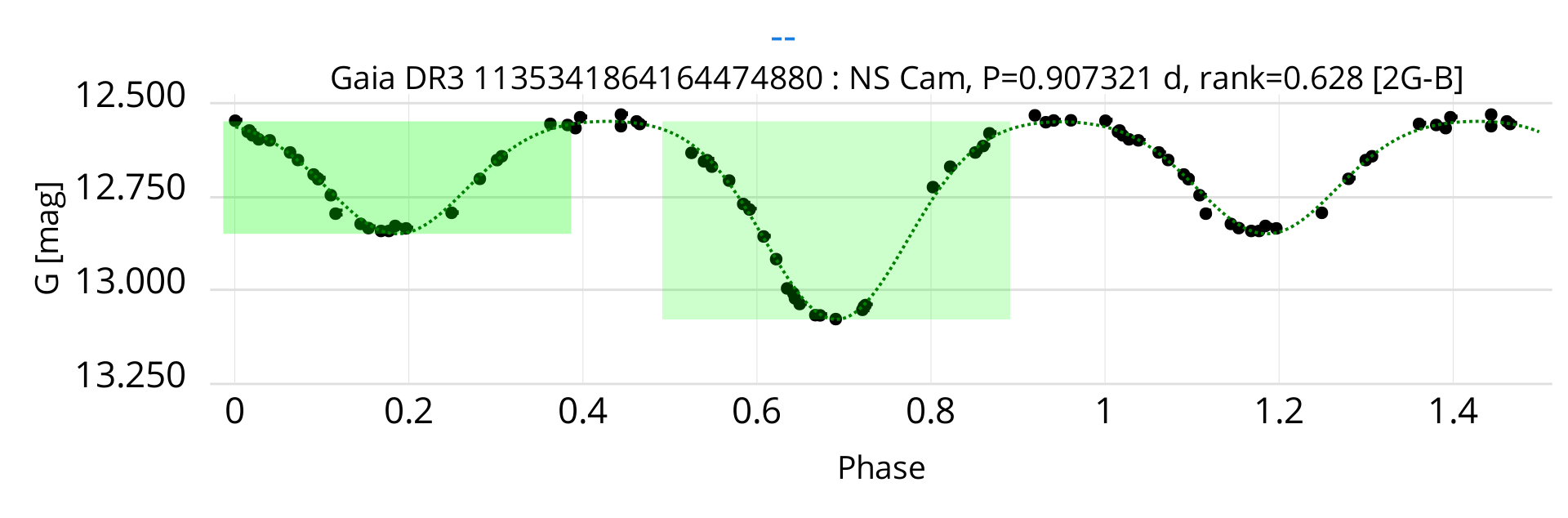}
  \vskip -0.5mm
  \includegraphics[trim={40 45 0 74},clip,width=\linewidth]{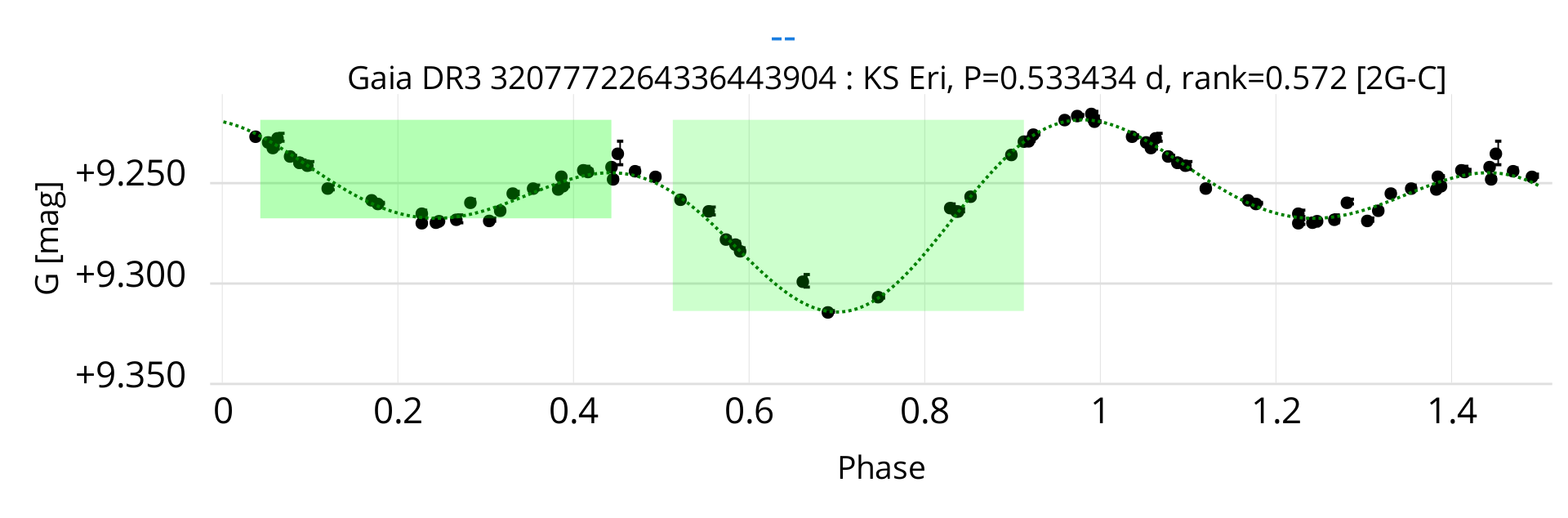}
  \caption{Same as top panel of Fig.~\ref{Fig:lcs_example_2GA}, but for additional eclipsing binaries for which the \gmag light curves are modelled with only two Gaussians.
           From top to bottom: SW~Cyg, V745~Cep, 1687~Aql, NS~Cam, and KS~Eri.
          }
\label{Fig:lcs_examples_2G}
\end{figure}
%-----
% 1927062769572959744, 
% 5870741220151128832, 2082658577035787392, 4292317099370571904, 1135341864164474880, 3207772264336443904, 2300877275401950976

%-----
\begin{figure}
  \centering
  \includegraphics[trim={0 80 0 42},clip,width=\linewidth]{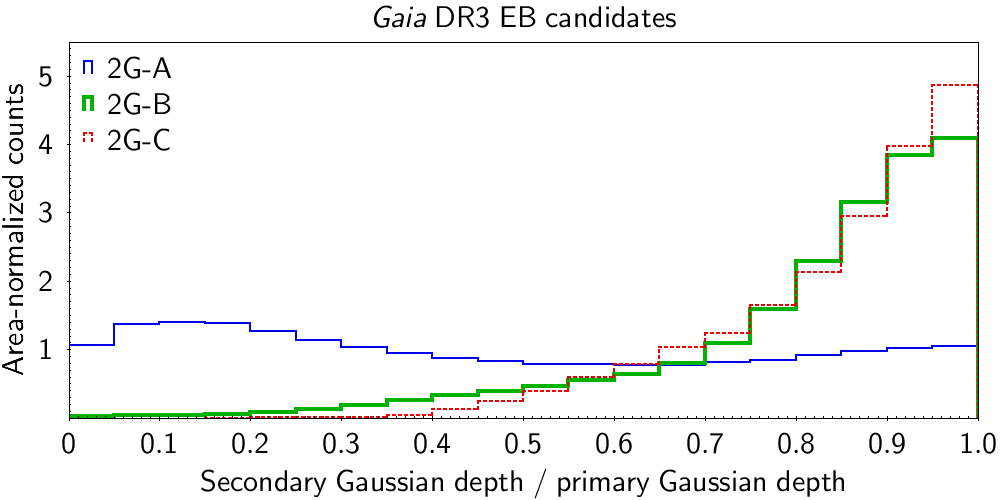}
  \vskip -0.5mm
  \includegraphics[trim={0 80 0 42},clip,width=\linewidth]{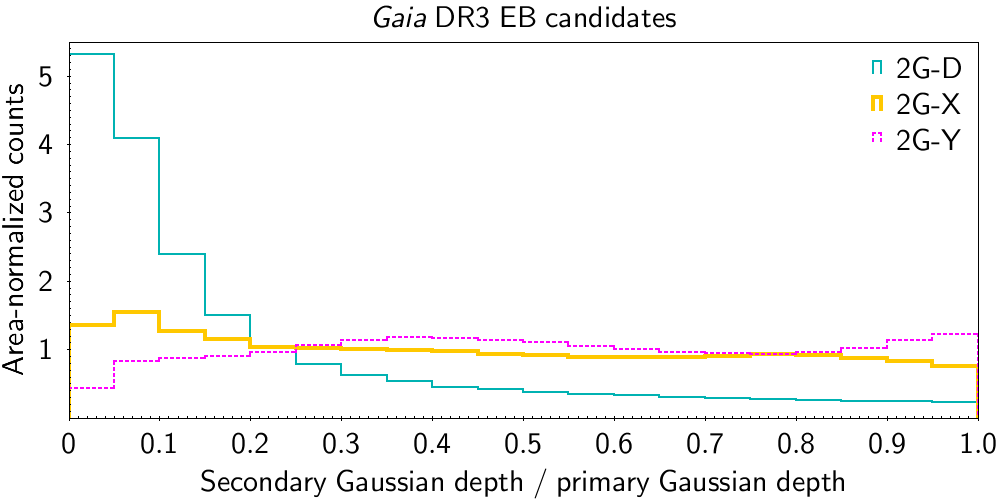}
  \vskip -0.5mm
  \includegraphics[trim={0 0 0 42},clip,width=\linewidth]{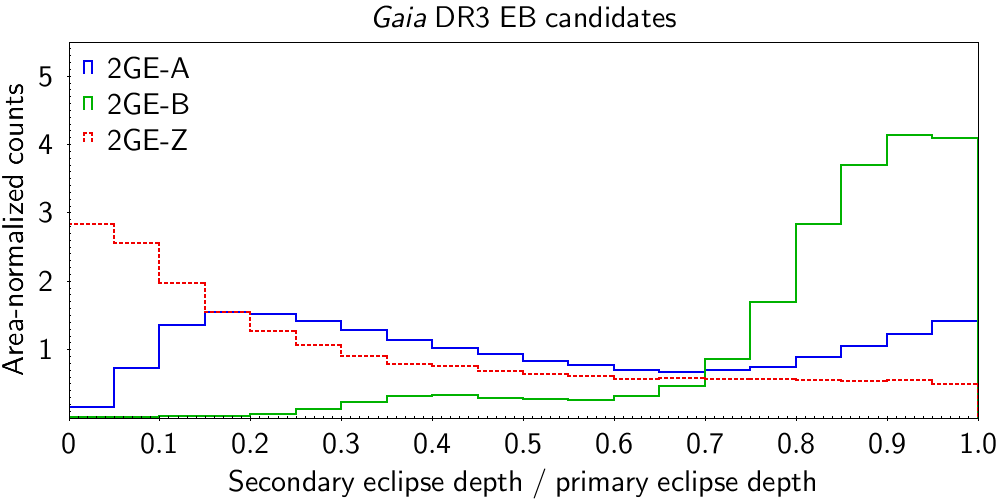}
  \caption{Distribution of the derived eclipse depths ratio (secondary over primary) for the various samples having two Gaussians in their light curve models without (top and middle panels) or with (bottom panel) an ellipsoidal component, as labeled in the panels (see text).
           The histograms are area-normalized.
          }
    \label{Fig:histo_depthRatios_2G}
\end{figure}
%-----

The automated procedure that processes the data of the two million \Gaia eclipsing binary candidates finds the best two-Gaussian model fit to the \gmag light curves.
As stressed in Sect.~\ref{Sect:catalogue_content}, the model represents a purely geometrical description of the light curve morphology.
The model parameters are not necessarily linked to physical properties of the binary system despite a good description of light curve geometry, due, for example, to a lack of phase coverage, spurious feature identifications in the light curves, or potential wrong period determination
The model parameters can, however, in a large number of cases, inform on the physical properties of the eclipses (depth, duration, eccentricity) and the ellipsoidal variability (amplitude).

A detailed analysis of the light curve models is presented in Appendix~\ref{Appendix:twoGaussianModel}.
In that appendix, the light curves are classified in samples that have two, one, or no Gaussian components, with a naming convention starting with 2G, 1G, or 0G, respectively, and with a letter E added when an ellipsoidal component is present.
In addition, groups 2G and 2GE are further sub-classified depending on model parameters by post-fixing the group name with -A, -B, -C, -D, -X, -Y and -Z.
The definition of the groups are given in Table~\ref{Tab:sample_definition} of the appendix.
This basic classification is only meant to guide the user on the catalogue content and various types of light curve morphologies.
In this section, we present examples of known eclipsing binaries in each of these groups, all available in the catalogue of \citet{AvvakumovaMalkovKniazev13}.

%-----
\begin{figure}
  \centering
  \includegraphics[trim={40 145 0 70},clip,width=\linewidth]{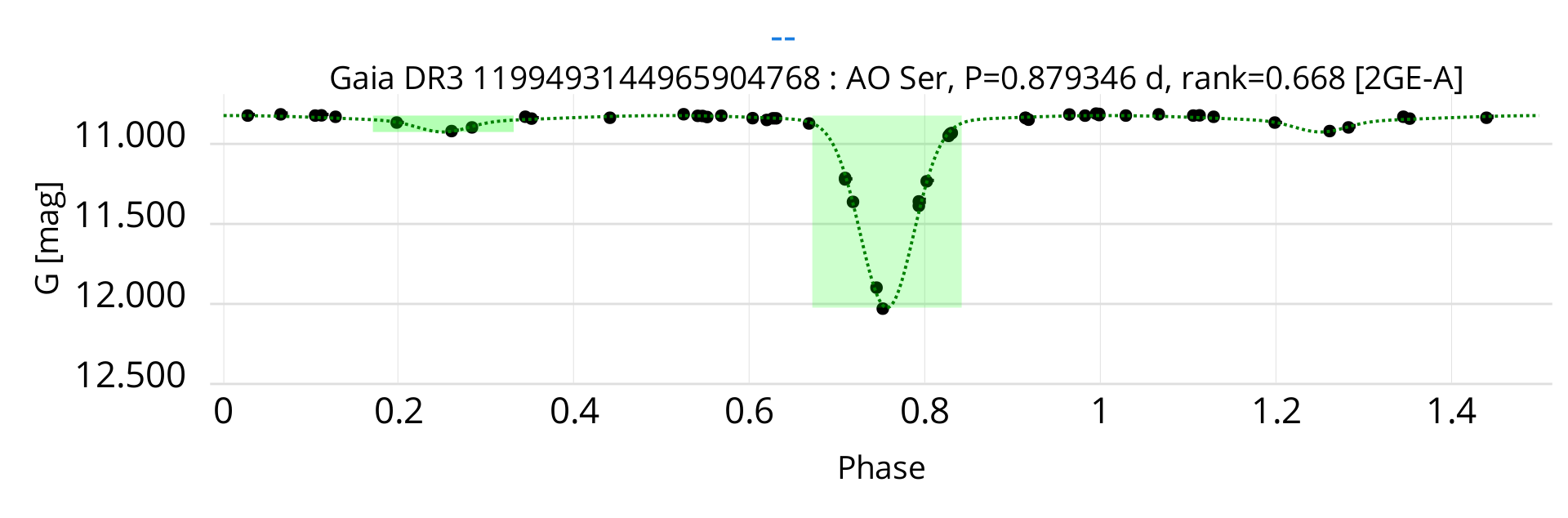}
  \vskip -0.5mm
  \includegraphics[trim={40 145 0 70},clip,width=\linewidth]{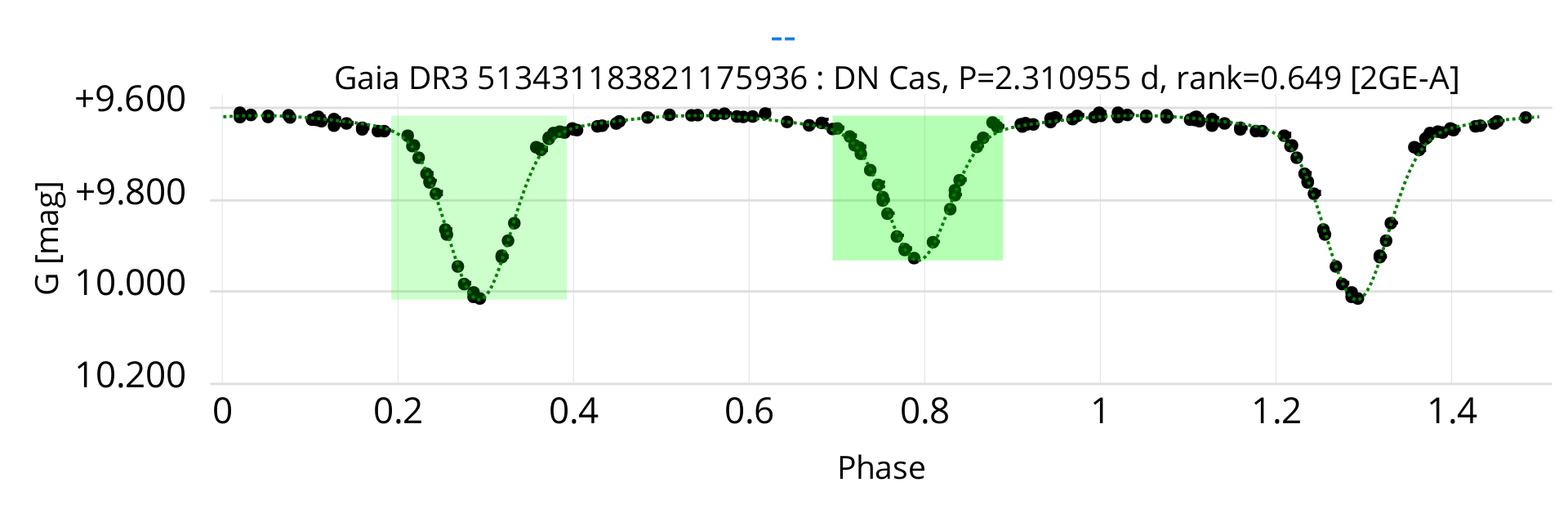}
  \vskip -0.5mm
  \includegraphics[trim={40 145 0 70},clip,width=\linewidth]{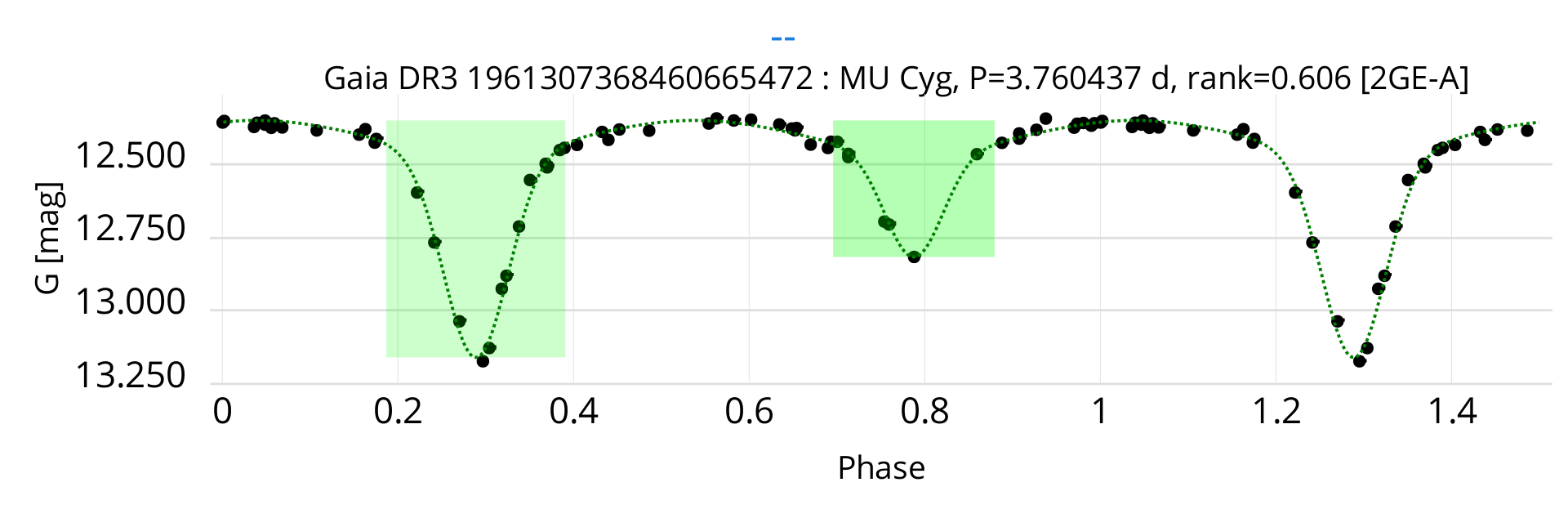}
  \vskip -0.5mm
  \includegraphics[trim={40 45 0 70},clip,width=\linewidth]{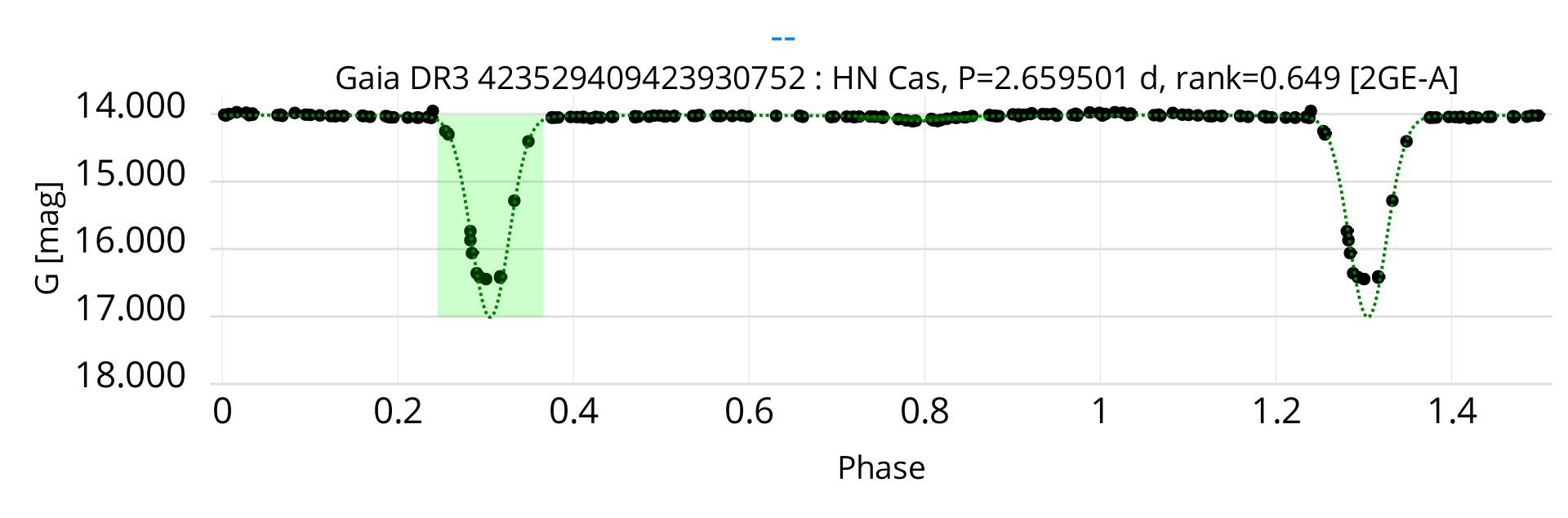}
  \caption{Same as top panel of Fig.~\ref{Fig:lcs_example_2GA}, but for sources with light curves modelled with two Gaussians and a cosine of small to medium amplitude.
           From top to bottom: AO~Ser, DN~Cas, MU~Cyg and HN~Cas.
          }
\label{Fig:lcs_examples_2GE_A}
\end{figure}
%-----
% 1199493144965904768, 513431183821175936, 1961307368460665472, 423529409423930752

%\begin{figure}
%  \centering
%  \includegraphics[trim={0 0 0 0},clip,width=\linewidth]{Figures/figHisto_depthRatios_2GE_ABZ.png}
%  \caption{Same as Fig.~\ref{Fig:histo_depthRatios_2G}, but for the derived eclipse depths ratio (secondary over primary) for the various samples of sources with two Gaussians and an ellipsoidal variability  in their light curves.
%          }
%    \label{Fig:histo_depthRatios_2GE}
%\end{figure}

%-----
\begin{figure}
  \centering
  \includegraphics[trim={40 145 0 70},clip,width=\linewidth]{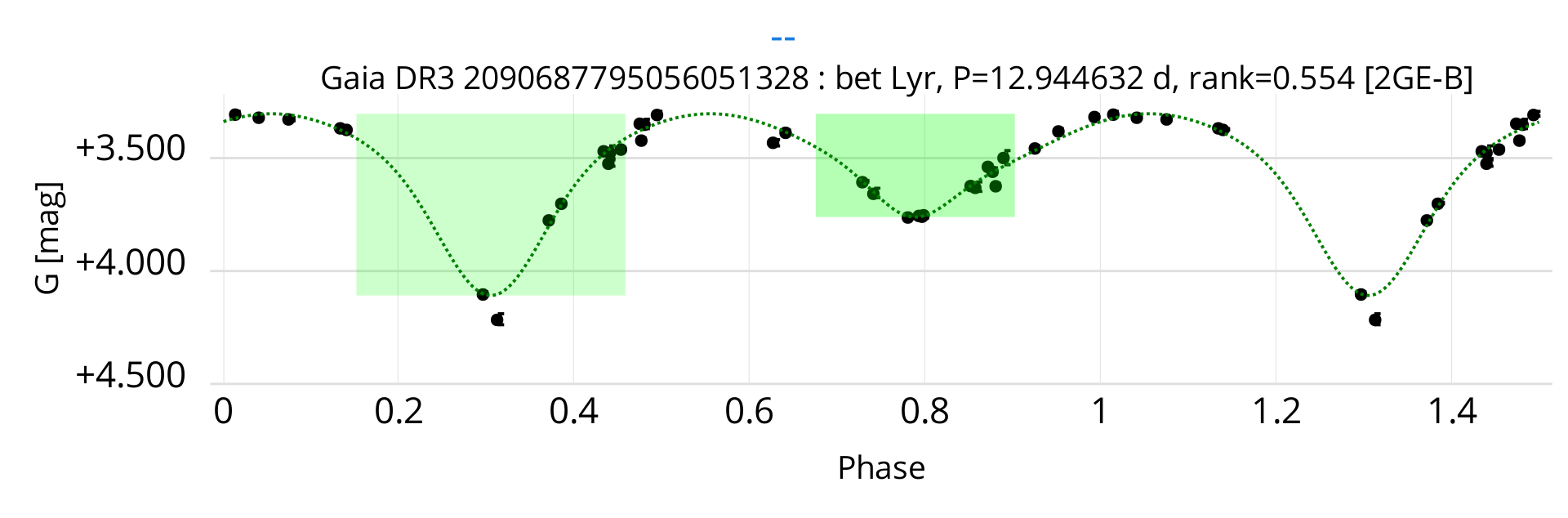}
  \vskip -0.5mm
  \includegraphics[trim={40 45 0 70},clip,width=\linewidth]{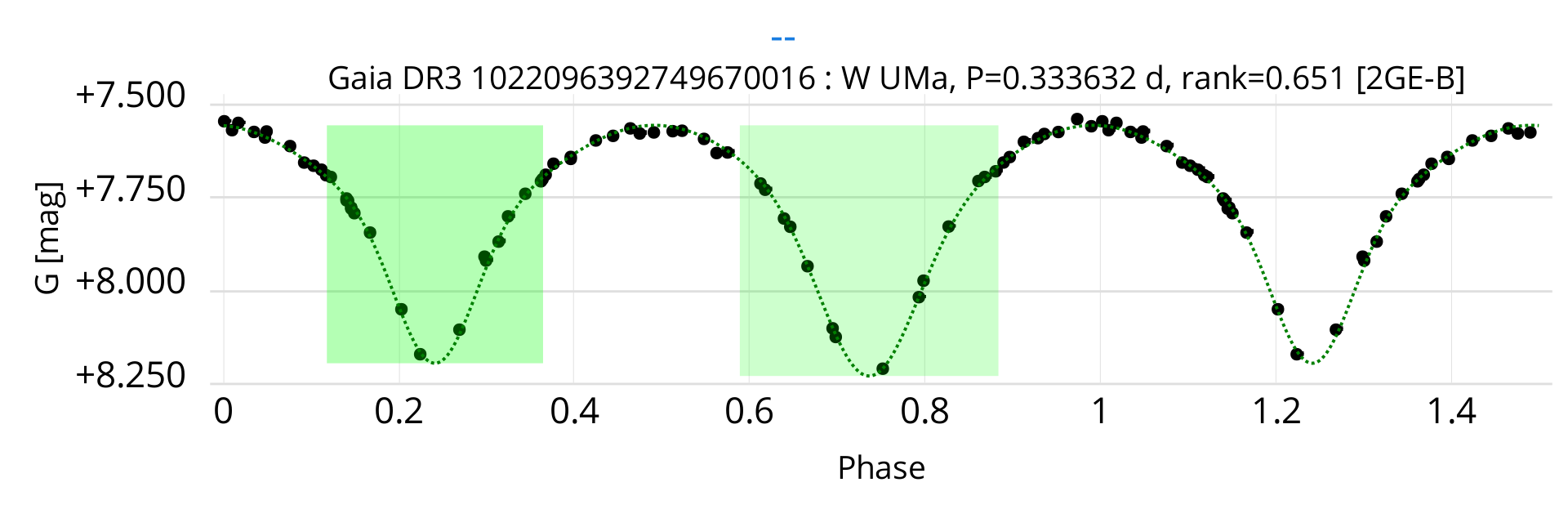}
  \caption{Same as top panel of Fig.~\ref{Fig:lcs_example_2GA}, but for sources with light curves modelled with two Gaussians and a cosine of large amplitude.
           From top to bottom: $\beta$~Lyr and W~UMa.
           }
\label{Fig:lcs_examples_2GE_B}
\end{figure}
%-----
% 2090687795056051328, 1022096392749670016

%-----
\begin{figure}
  \centering
  \includegraphics[trim={40 145 0 70},clip,width=\linewidth]{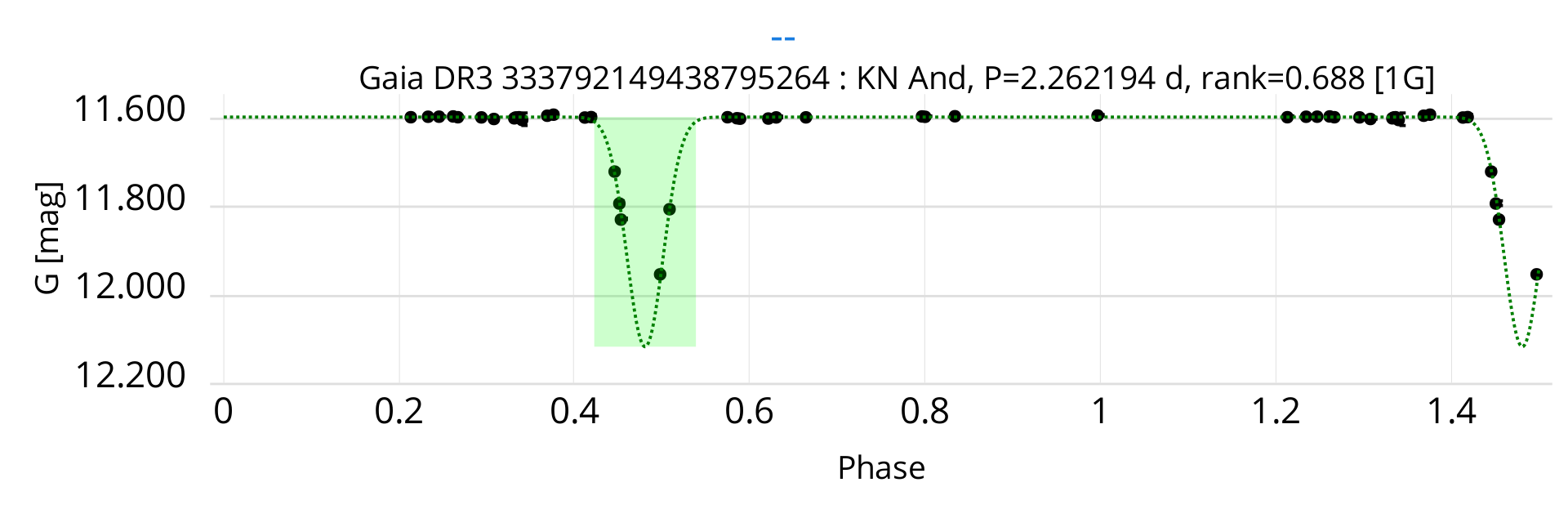}
  \vskip -0.5mm
  \includegraphics[trim={40 145 0 70},clip,width=\linewidth]{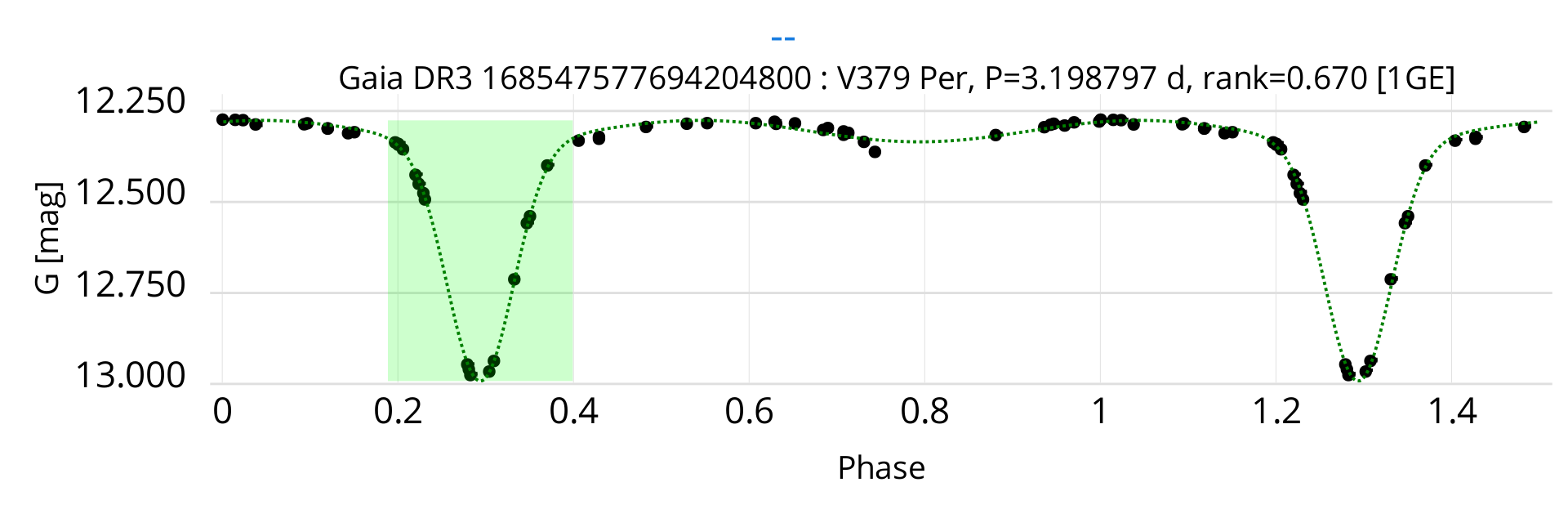}
  \vskip -0.5mm
  \includegraphics[trim={40 145 0 70},clip,width=\linewidth]{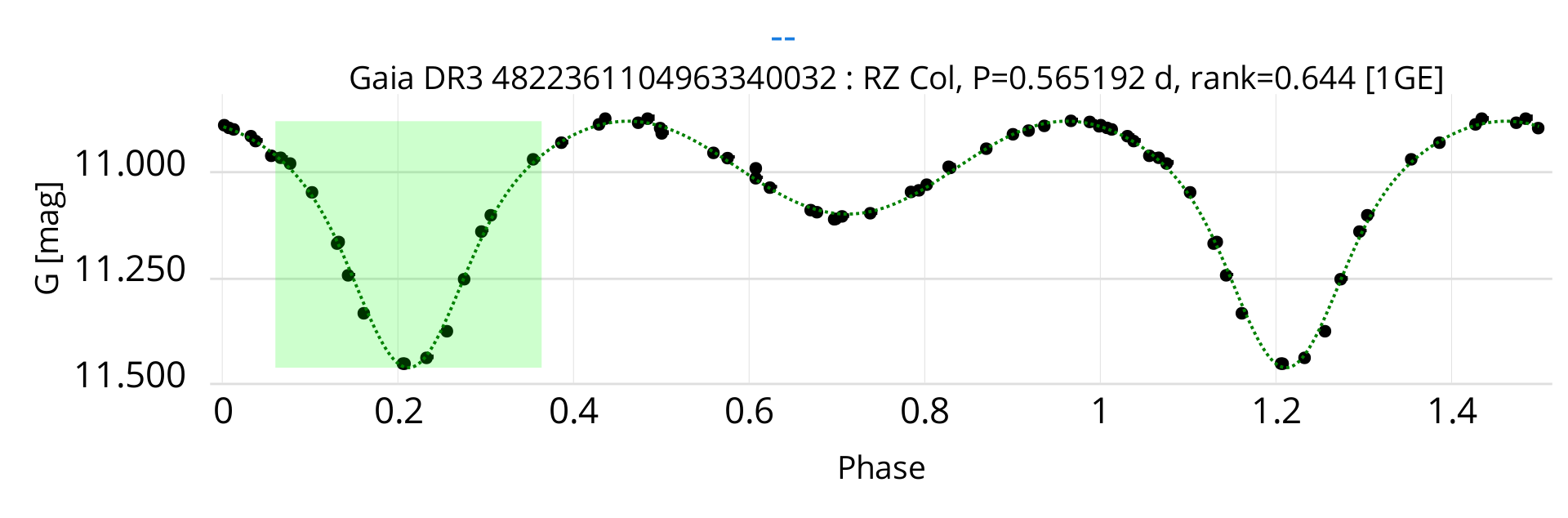}
  \vskip -0.5mm
  \includegraphics[trim={40 45 0 70},clip,width=\linewidth]{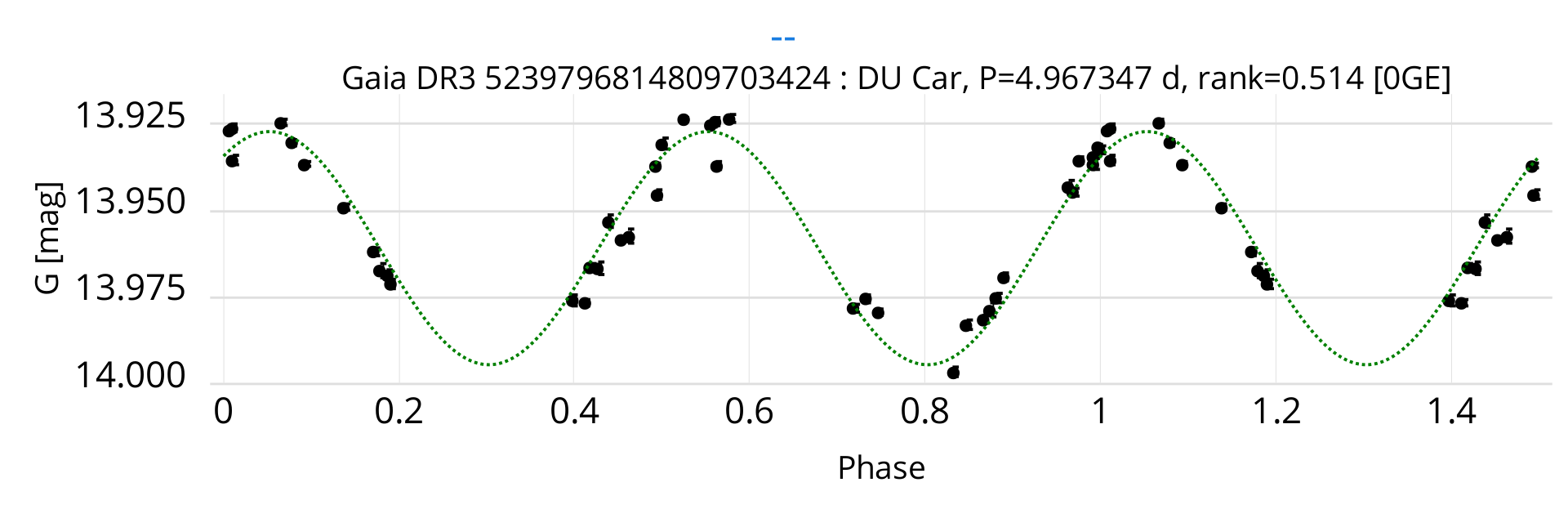}
  \caption{Same as top panel of Fig.~\ref{Fig:lcs_example_2GA}, but for sources with light curves modelled with only one Gaussian (KN~And, top panel), one Gaussian and a cosine (V379~Per in second panel and RZ~Col in third panel) or only a cosine (DU~Car, bottom panel).
           No green area is displayed for eclipses that are not modelled with a Gaussian component in the light curve model.
            }
\label{Fig:lcs_examples_1G_1GE_0G}
\end{figure}
%-----
% 333792149438795264, 168547577694204800, 4822361104963340032, 5239796814809703424

%-----
\begin{figure}
  \centering
  \includegraphics[trim={40 145 0 70},clip,width=\linewidth]{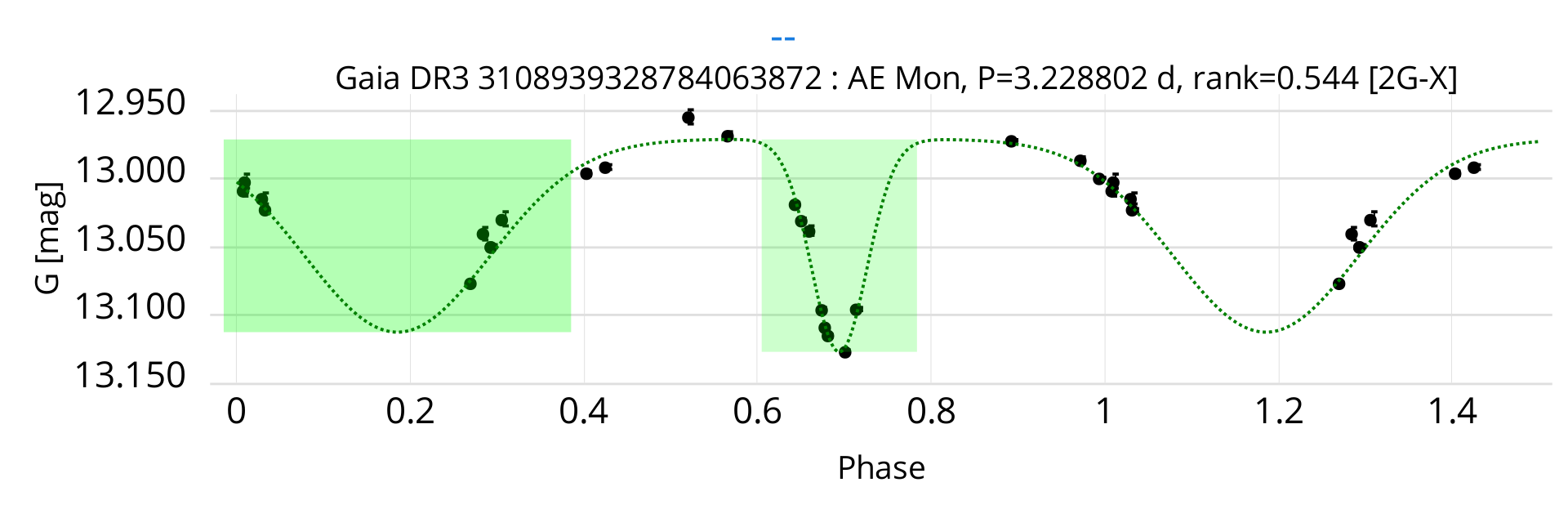}
  \vskip -0.5mm
  \includegraphics[trim={40 145 0 70},clip,width=\linewidth]{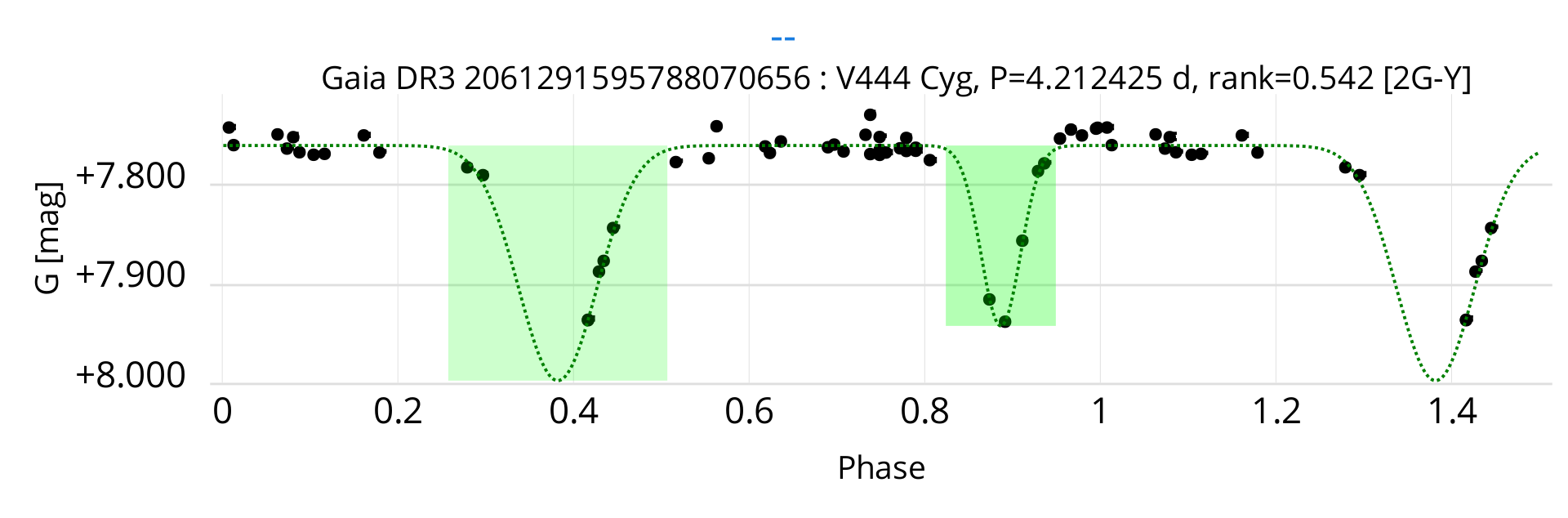}
  \vskip -0.5mm
  \includegraphics[trim={40 45 0 70},clip,width=\linewidth]{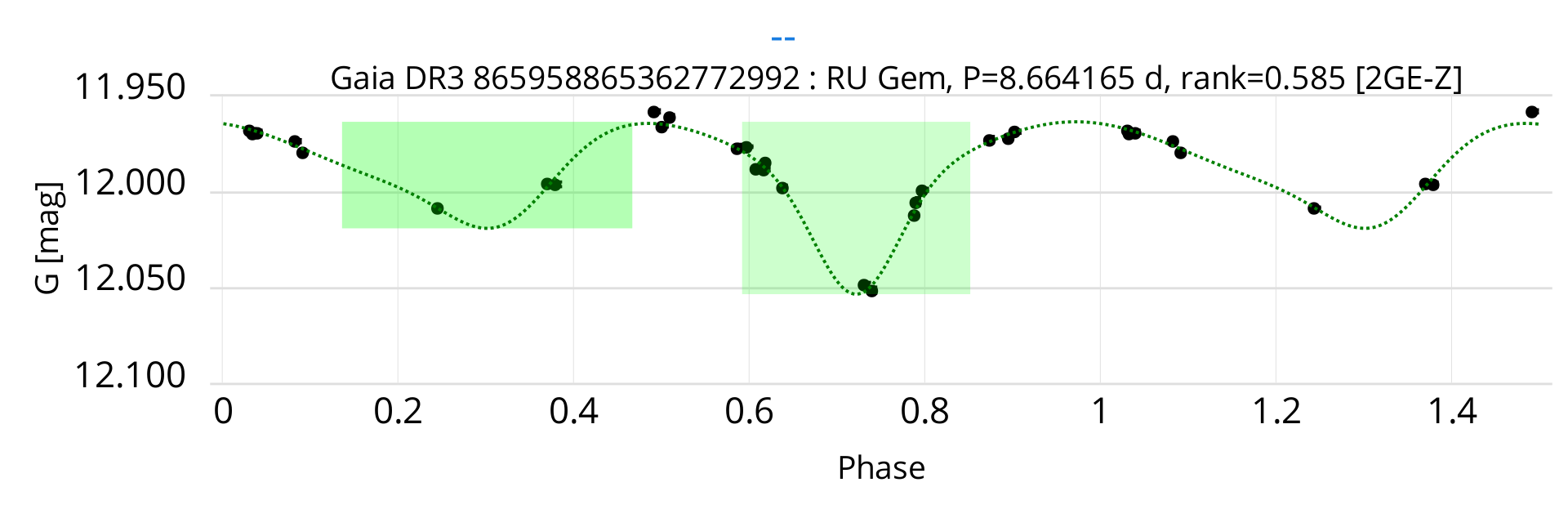}
  \caption{Same as top panel of Fig.~\ref{Fig:lcs_example_2GA}, but for additional sources with light curves modelled with at least two Gaussians.
           From top to bottom: AE~Mon, V444~Cyg and RU~Gem.
           }
\label{Fig:lcs_examples_2G_special}
\end{figure}

The overwhelming majority of \gmag light curves are modelled with two Gaussians (94\% of sources in the catalogue, see Table~\ref{Tab:modelTypes}).
Among these, two-third have strictly two Gaussian components.
The most obvious eclipsing binary configuration whose light curve can be modelled in this way is that of well-detached systems, with constant out-of-eclipse light.
The two Gaussians have similar widths, but not necessarily similar depths.
In Appendix~\ref{Appendix:twoGaussianModel}, they define Sample 2G-A (285\,320 candidates).
The \gmag folded light curve of V614~Ven in this sample is displayed in the top panel of Fig.~\ref{Fig:lcs_example_2GA}.
We remind that only the \gmag data have been used to produce the results published in the DR3 catalogue of eclipsing binaries.
The \gbp and \grp time series are nevertheless available in the DR3 \Gaia archive.
The \gbp and \grp folded light curves of V614~Ven are shown in the bottom panel of Fig.~\ref{Fig:lcs_example_2GA}.

Tighter systems in which one or both stars fill their Roche lobes can also display light curves reminiscent of detached systems \citep[e.g.,][]{Pojmanski_2002,PaczynskiSzczygielPilecki_etal06}, and hence be found in Sample 2G-A.
This can, for example, happen when the star that fills its Roche lobe is much fainter than its companion such that the induced ellipsoidal variability is below detection limit (depending on instrument photometric precision).
The secondary eclipse would then also be much shallower than the primary eclipse.
They typically characterise Algo-type binaries, which are understood to result from a past mass-transfer episode.
Algol itself is not available in \Gaia DR3 due to its brightness (2.1~mag in $V$), but the example of SW~Cyg, a A2Ve+KI system \citep{Malkov20}, is given in the top panel of Fig.~\ref{Fig:lcs_examples_2G}.\footnote{
The \gbp and \grp light curves of SW~Cyg as well as of the other binaries mentioned in this section are displayed in Appendix~\ref{Appendix:additionalFigures}.
}
The absence of detected out-of-eclipse variability does thus not necessarily imply a well-detached system.
The variety of binary configuration in Sample~2G-A is also attested by the depth ratio distribution  shown in blue in the top panel of Fig.~\ref{Fig:histo_depthRatios_2G}.
The histogram covers all values from close-to-zero to one, with two main peaks, one at small ratios below 0.2, and another at depth ratios close to one.

Some light curves are modelled with a very narrow primary Gaussian and a wide secondary.
In the sub-classification presented in Appendix~\ref{Appendix:twoGaussianModel}, they are gathered in Sample~2G-D.
The secondary Gaussians of these cases are, on the mean, much shallower than their primary Gaussian, as shown in Fig.~\ref{Fig:histo_depthRatios_2G} (second panel, cyan histogram).
When the primary eclipse is very narrow, the detection of the secondary eclipse may be challenging, due for example to insufficient measurements in the eclipse and/or too shallow secondary eclipse.
The probability that the pipeline fails to correctly detect the secondary, or that the orbital period is incorrect, is thus much greater than for Sample~2G-A candidates.
The second example in Fig.~\ref{Fig:lcs_examples_2G} displays a case in Sample~2G-D, V745~Cep classified as a semi-detached system in \citet{AvvakumovaMalkovKniazev13}, where both Gaussians correctly identify the eclipses.

%If the Gaussian widths (and depths) are very different, the candidate light curve is classified in a separate sample called 2G-D (see Appendix~\ref{Appendix:twoGaussianModel} for details).
%The probability that their secondary Gaussian does not represent a real eclipse or that their orbital period is incorrect is much greater than for candidates in Sample~2G-A, and they %must be investigated on a case-by-case basis.
%The sixth example in Fig.~\ref{Fig:lcs_examples_2G} displays a case in Sample 2G-D where both Gaussians correctly identify the real eclipses.
%It is V745~Cep, classified as a semi-detached system in \citet{AvvakumovaMalkovKniazev13}.
%The depth ratio distribution of the 285\,320 candidates in Sample 2G-A is shown in blue in the top panel of Fig.~\ref{Fig:histo_depthRatios_2G}.
%It shows two peaks, one at depth ratios close to one, and one at small ratios below 0.2.
%The 111\,820 candidates in Sample 2G-D (cyan histogram in the middle panel) complete the 2G-A distribution at the lowest depth ratios below 0.2.
%The depth ratio distribution of the 162\,630 candidates in Sample 2GE-A, shown in the bottom panel in Fig.~\ref{Fig:histo_depthRatios_2G} (blue histogram), follows the same pattern as the distribution of Sample 2G-A.

Tight systems are generally modelled with two Gaussians and a cosine to account for the ellipsoidal out-of-eclipse variability.
These light curves belong to either Sample~2GE-A or 2GE-B in Appendix~\ref{Appendix:twoGaussianModel}, depending on the amplitude of the ellipsoidal variability.
Sample~2GE-A (162\,630 sources) contains candidates with small to medium amplitudes of $2\,\Aell<0.11$~mag, while Sample~2GE-B (265\,276 sources) has $2\,\Aell>0.11$~mag.
Sample 2GE-A is similar to Sample 2G-A except for the additional cosine component.
Four such examples are shown in Fig.~\ref{Fig:lcs_examples_2GE_A}, with increasing ellipsoidal amplitude (relative to primary eclipse depth) from top to third case, and with a total eclipse in the fourth case.
The two famous eclipsing binaries $\beta$-Lyr and W~UMa, the prototypes of the classical EB- and EW-type eclipsing binaries, respectively, belong to Sample~2GE-B.
Their light curves are shown in Fig.~\ref{Fig:lcs_examples_2GE_B}.

Very tight systems, including semi-detached systems with large ellipsoidal variability, in-contact systems, or systems with a common envelope, have their light curves modelled in several ways using a two-Gaussian model.
The most common way consists of two wide overlapping Gaussians of similar width.
They form Sample~2G-B containing 834\,093 sources.
The Gaussians are located at a phase separation of about 0.5 from each other.
The majority of them have similar eclipse depths, as seen by the green histogram in Fig.~\ref{Fig:histo_depthRatios_2G} (top panel).
1687~Aql is such an example, shown in the third panel of Fig.~\ref{Fig:lcs_examples_2G}.
An example in Sample~2G-B with significantly unequal eclipse depths, NS~Cam, is shown in the fourth panel.

A small fraction of candidates modelled with two wide overlapping Gaussians have non-equal Gaussian widths.
This feature can model asymmetries in the light curves of tight systems.
They form Sample~2G-C (24\,081 sources).
Their eclipse depth ratio distribution is very similar to that of Sample 2G-B (red dotted histogram in Fig.~\ref{Fig:histo_depthRatios_2G}, top panel).
An example is given in the bottom panel of Fig.~\ref{Fig:lcs_examples_2G} with KS~Eri, a binary system displaying the O'Connell effect. 

In less than 4\% of the DR3 eclipsing binary candidates, the light curve is modelled without a second Gaussian.
They belong to samples 1G and 1GE depending on whether the model contains or not a cosine component.
The lack of a secondary Gaussian can be due to several reasons.
One of them is the lack of eclipse phase coverage.
Such is the case for KN~And (Fig.~\ref{Fig:lcs_examples_1G_1GE_0G}, top panel) and V379~Per (second panel).
The absence of a second Gaussian can also be due to the presence of a cosine component that models by itself the secondary eclipse.
This is the case for RZ~Col shown in the third panel of Fig.~\ref{Fig:lcs_examples_1G_1GE_0G}.

Finally, a single cosine my be sufficient to model a light curve.
They form Sample~0GE (36\,227 sources).
DU~Car illustrates an example in Fig.~\ref{Fig:lcs_examples_1G_1GE_0G} (bottom panel).

About one fifth of the $\sim$2 million sources that contain two Gaussians in their light curve model do not fall in one of the above categories 2G-A, 2G-B, 2G-C, 2G-D, 2GE-A, 2GE-B, 1G, 1GE or 0GE.
They form three additional categories, 2G-X, 2G-Y and 2GE-Z, depending on their model parameters.
% They form three additional categories, depending on whether their secondary Gaussian is significantly wider than their primary Gaussian (Sample~2G-X) or narrower (Sample~2G-Y), or if they have a cosine and a separation of their derived eclipse locations significantly differing from $\sim$0.5 in phase (Sample~2G-Z).
We refer to Appendix~\ref{Appendix:twoGaussianModel} for more details.
The probability that their model components reflect physical configurations of the eclipsing binaries is much lower than for the other groups, and they are to be investigated on a case-by-case basis.
Example of nevertheless correct cases in each of these three samples, and where the \Gaia period agrees with literature period, are shown in Fig.~\ref{Fig:lcs_examples_2G_special}. 

In conclusion, the two-Gaussian model provides a powerful tool to study the two million eclipsing binary candidates published in \Gaia DR3.
The classification provided in Appendix~\ref{Appendix:twoGaussianModel} gives some insight into the type of binary system, keeping in mind that each group defined in that appendix contains a variety of different light curve morphologies.
In addition, there is an inherent degeneracy in light curve morphology between different types of binary systems that makes it impossible to discriminate between them solely based on \gmag photometry.
The case of detached and semi-detached systems was mentioned above.
From the 119 semi-detached systems listed in \citet{Malkov20}, 95 are present in the DR3 catalogue and 74 have \Gaia periods compatible within 5\% with the values gathered by that author.
Among these 74 sources, 60 have an ellipsoidal component in their two-Gaussian model (36 in Sample~2GE-A, 19 in 2GE-B and five in 1GE), and 14 do not (five in 2G-A, six in 2G-B, one in 1G, and two in 2G-X).  
%
% In the list of semi-detached systems of Malkov 2020 (119 systems): 74/95 have similar Gaia-Malkov periods, 5 of which are classified as 2G-A. In increasing order of ecl_sec_depth/ecl_prim_depth, they are:
% 2082658577035787392  A2Ve+[KI]
% 1413786483748744960  A0V+G4V
% 211807141136171136   B7V+A5V
% 2213360032485424768  B2V+A0IV
% 5499415974230271488  B0.5+B0.5-3

%------------------------------------------------------------------
\subsection{Global ranking}
\label{Sect:catalogue_usage_ranking}

\begin{figure}
  \centering
  \includegraphics[trim={0 0 0 42},clip,width=\linewidth]{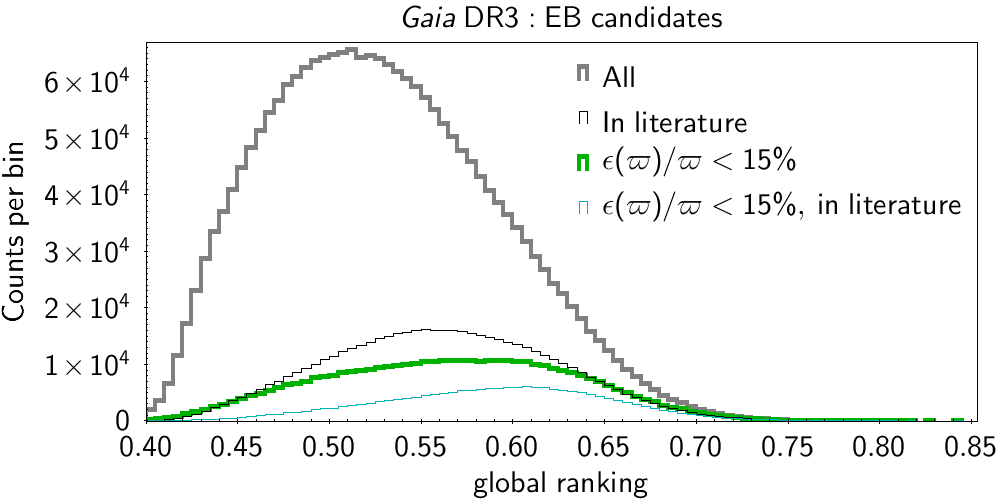}
  \caption{Distribution of the global ranking of DR3 EB candidates.
           The full sample is displayed in thick grey, with the candidates therein that have a cross-match with known EBs in the literature shown in thin black.
           The sample with positive parallax uncertainties better than 15\% is displayed in thick green, with the candidates therein with a literature cross-match shown in thin cyan.
          }
\label{Fig:histo_globalRanking}
\end{figure}

\begin{figure}
  \centering
  \includegraphics[trim={0 0 0 42},clip,width=1\linewidth]{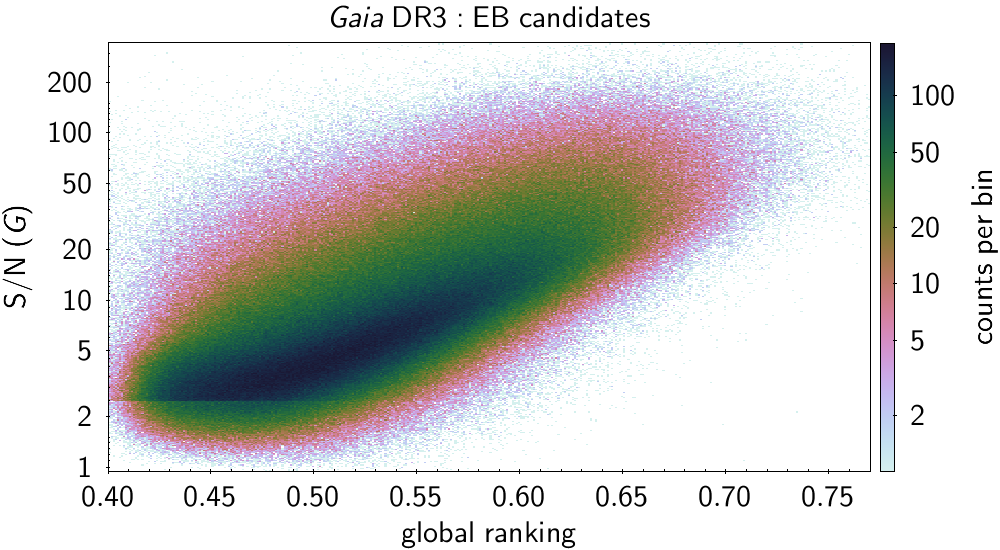}
  \caption{Density map of the signal-to-noise of the \gmag time series (standard deviation of the measurements over root mean square of their uncertainties) of all eclipsing binary candidates versus their global ranking. 
           The density in the map is colour coded according to the colour scale shown on the right of the figure.
          }
\label{Fig:snG_GlobalRanking}
\end{figure}

\begin{figure}
  \centering
  \includegraphics[trim={0 80 0 42},clip,width=1.0\linewidth]{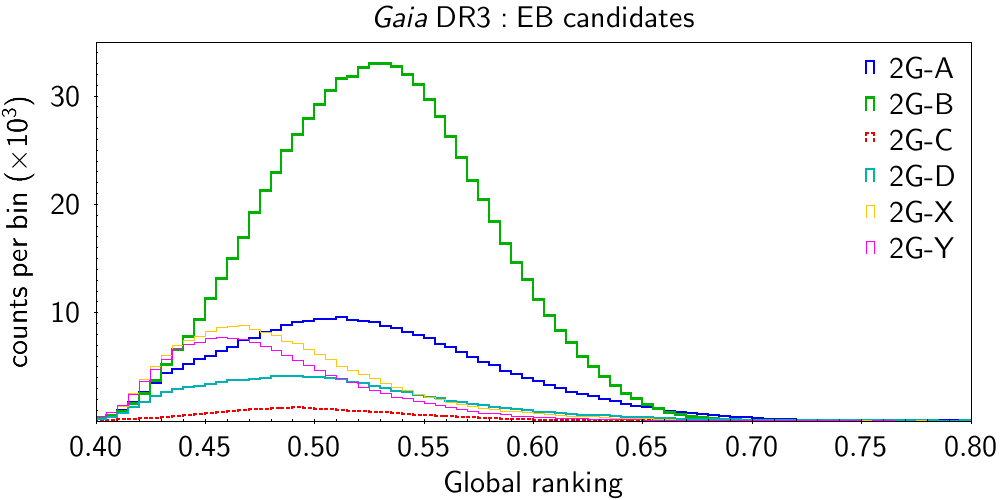}
  \vskip -0.5mm
  \includegraphics[trim={0 0 0 42},clip,width=1.0\linewidth]{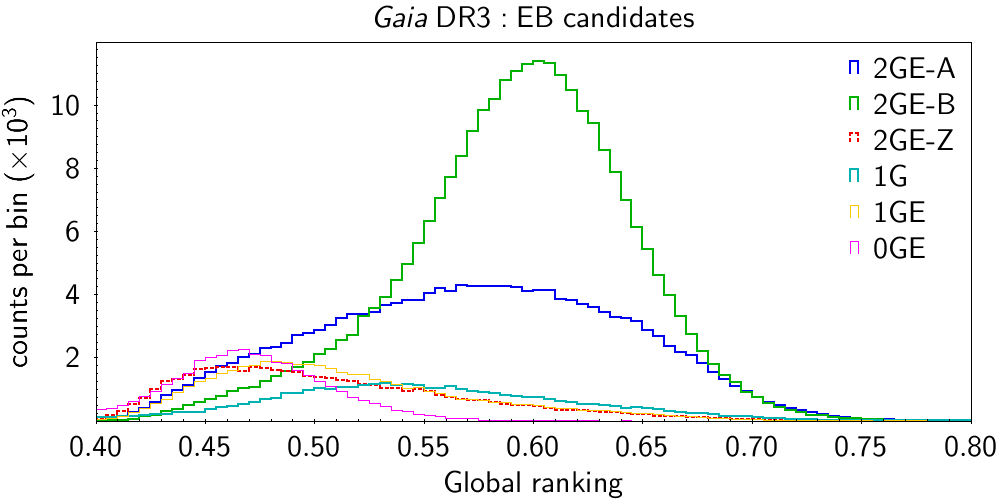}
  \caption{Same as Fig.~\ref{Fig:histo_depthRatios_2G}, but for the global ranking.
           The histograms of the samples are displayed in two panels as labeled in the panels.
          }
\label{Fig:histo_globalRanking_samples}
\end{figure}

The global ranking is directly linked to the fraction of the variance unexplained by the two-Gaussian model through Eq.~(\ref{Eq:global_ranking}).
As such, it informs on the reliability of a candidate to be an eclipsing binary,
a larger global ranking corresponding to a better fit to the light curve, and hence to a more reliable eclipsing binary candidate.
A poor global ranking, however, does not necessarily imply a false detection, as it relies on the assumption that the functions included in the model can adequately describe the light curve of an eclipsing binary.
The two-Gaussian model will fail to recognise an eclipsing binary if some physics dominating the shape of the light curve is not modelled by these functions.
Such would be the case, for example,
for ellipsoidal variables on an eccentric orbit including heartbeat stars,
%for binaries having a total eclipse (which translates in a flat light curve at the bottom of an eclipse),
or for close binaries featuring a reflection effect (which translates in a cosine function with a period equal to the orbital period).
Sources in the catalogue with a low global ranking will therefore need additional investigation to confirm and characterise their binary nature.
Sources with a high global ranking, on the other hand, have a high probability to be eclipsing binaries.

The distribution of the global ranking is shown in Fig.~\ref{Fig:histo_globalRanking} for the full catalogue (black histogram).
It ranges from 0.40 to 0.84, with a maximum of the distribution around 0.51.
Candidates with low global rankings are, on average, fainter than the ones with high global rankings.
This is illustrated by the green and red histograms in Fig.~\ref{Fig:histo_gmag}, where the sample with rankings larger than 0.6 (filled green histogram) peaks around 17.3~mag, while sources with rankings less than 0.5 (red hatched histogram) are located at much fainter magnitudes around 19~mag.
It mainly results from the fact that faint sources have larger epoch \gmag uncertainties than bright sources, which in turn generally leads to poorer eclipsing binary light curve characterisation, and hence lower rankings.
Figure~\ref{Fig:snG_GlobalRanking}, which plots the signal-to-noise ratio in \gmag (\texttt{std\_dev\_over\_rms\_err\_mag\_g\_fov} in the \Gaia archive) versus global ranking, overall supports this explanation.

The histograms of the global ranking for the various samples discussed in the previous section are shown in Fig.~\ref{Fig:histo_globalRanking_samples}.
The largest global rankings, on the mean, are found in samples whose model components have a higher probability to represent physical features (eclipses and ellipsoidal variability).
These are samples 2G-A and 2G-B without an ellipsoidal component (respectively blue and green distributions in the top panel of Fig.~\ref{Fig:histo_globalRanking_samples}), and samples 2GE-A and 2GE-B with an ellipsoidal component (respectively blue and green distributions in the bottom panel).
We note that the presence of an ellipsoidal component leads to a higher global ranking, on the mean, as the ellipsoidal variability is well defined by a cosine (in circular orbits).
Noticeable is the histogram of Sample 2G-C (dashed red histogram in the top panel) that peaks to lower global rankings than the 2G-A, 2G-B, 2GE-A and 2GE-B samples, despite their generally good light curves.
Global ranking-limited selections should take this into account to avoid their exclusion.
%These are characterised by ellipsoidal variability with large amplitudes, especially in Sample 2GE-B (see Fig.~\ref{Fig:histo_Aell} in Appendix~\ref{Appendix:twoGaussianModel}), which is generally well described by the two-Gaussian model.
Samples 2G-X, 2G-Y and 2GE-Z, on the other hand, containing light curve models with components that are predominantly unrelated with physical features, have small rankings.
%Samples 1GE and 0GE also have predominantly small rankings.

Additional insight in the global ranking distributions is provided in Sect.~\ref{Sect:catalogue_usage_summary} of Appendix~\ref{Appendix:twoGaussianModel}.
Figure~\ref{Fig:rank_magG_samples}, in particular, shows the distribution of the global ranking versus \gmag magnitude for each sample.

%------------------------------------------------------------------
\subsection{Orbital periods}
\label{Sect:catalogue_usage_period}

%-------
\begin{figure}
  \centering
  \includegraphics[trim={0 74 0 30},clip,width=1.0\linewidth]{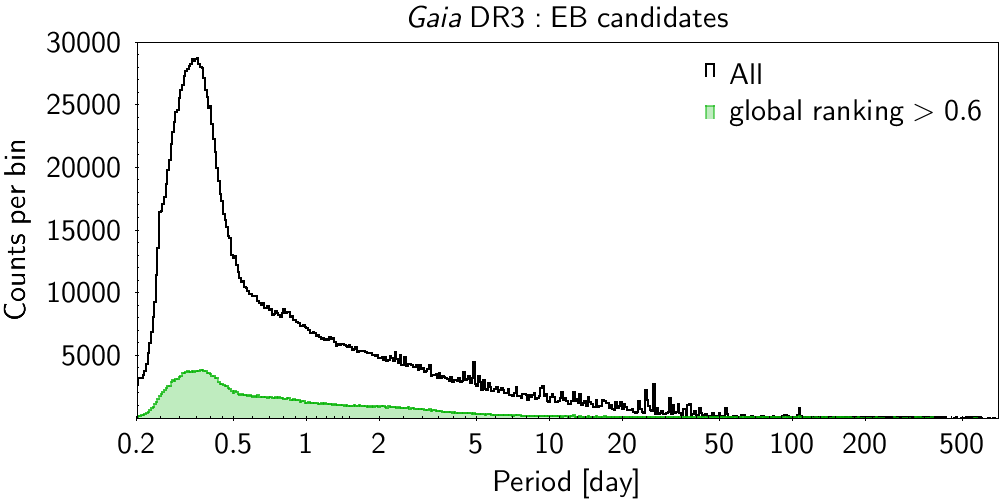}
  \vskip -0.5mm
  \includegraphics[trim={0 0 0 44},clip,width=1.0\linewidth]{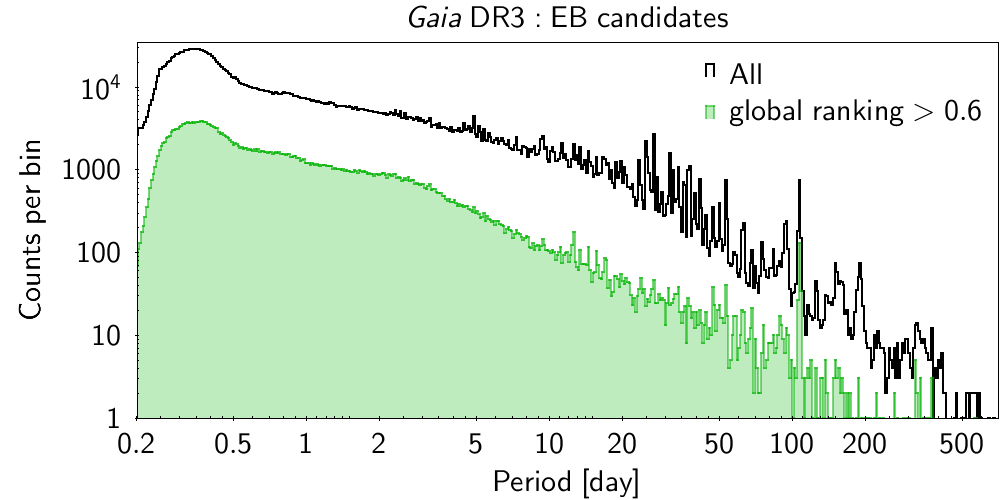}
  \caption{Distribution of the orbital periods of the DR3 eclipsing binary candidates.
           The full sample is displayed in black line, and the sample with global ranking larger than 0.6 in filled green.
           The top panel shows the number of counts per bin on a linear scale, while the bottom panel shows them on a logarithmic scale.
          }
\label{Fig:histoPeriod}
\end{figure}
%-------

%-------
\begin{figure}
  \centering
  \includegraphics[trim={0 0 0 42},clip,width=\linewidth]{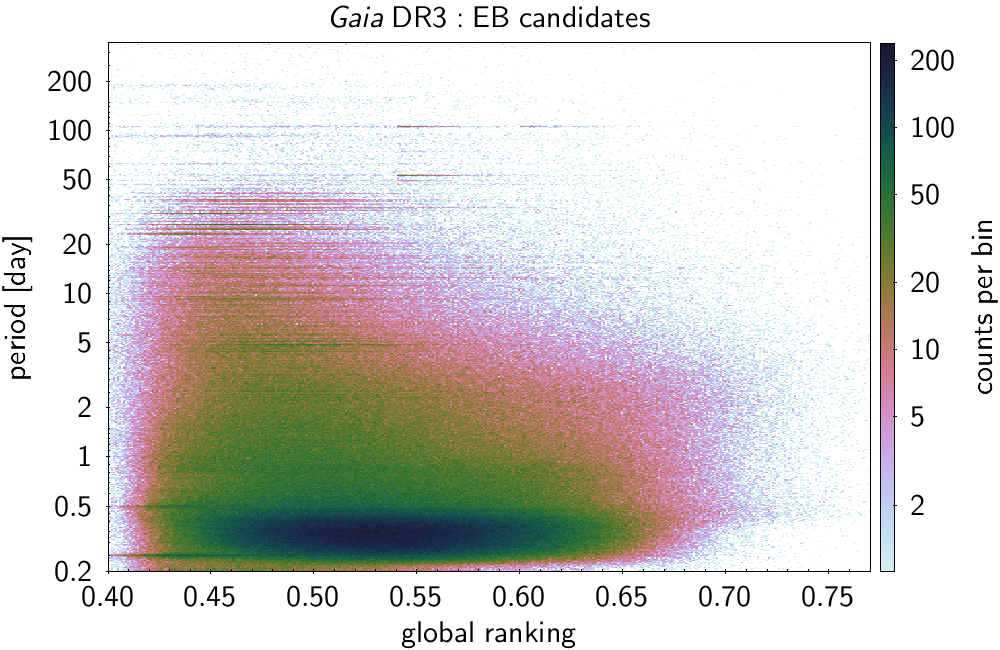}
  \caption{Density map of the orbital period versus global ranking of the DR3 eclipsing binary candidates.
          }
\label{Fig:periodVsGR}
\end{figure}
%-------

%-------
\begin{figure}
  \centering
  \includegraphics[trim={0 74 0 42},clip,width=\linewidth]{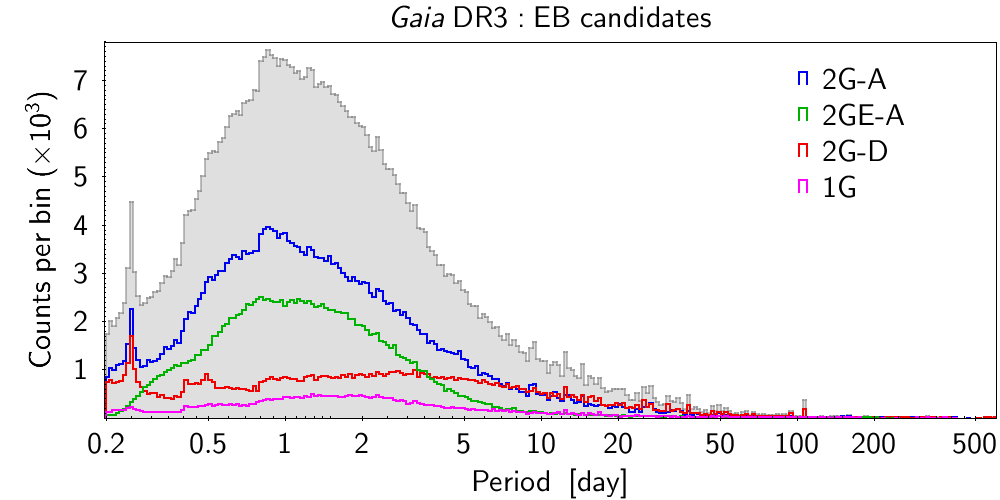}
  \vskip -0.5mm
  \includegraphics[trim={0 74 0 44},clip,width=\linewidth]{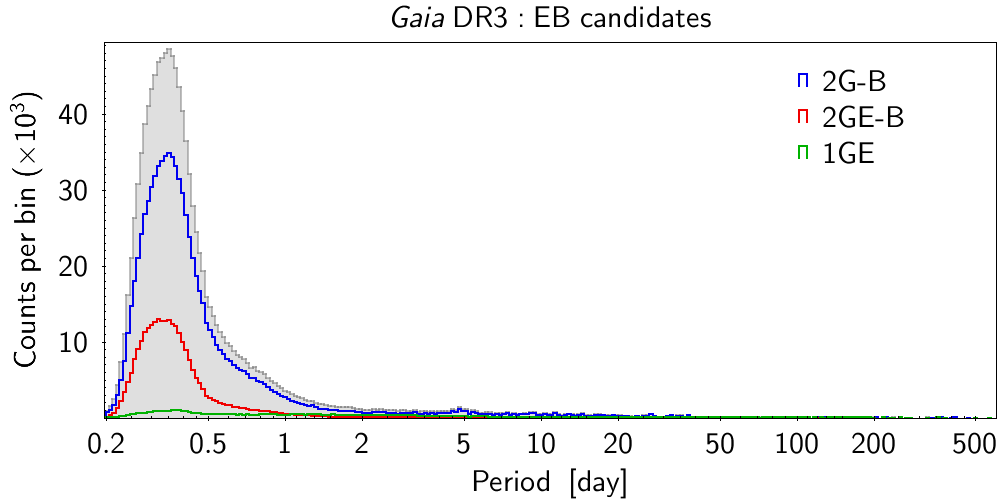}
  \vskip -0.5mm
  \includegraphics[trim={0 74 0 44},clip,width=\linewidth]{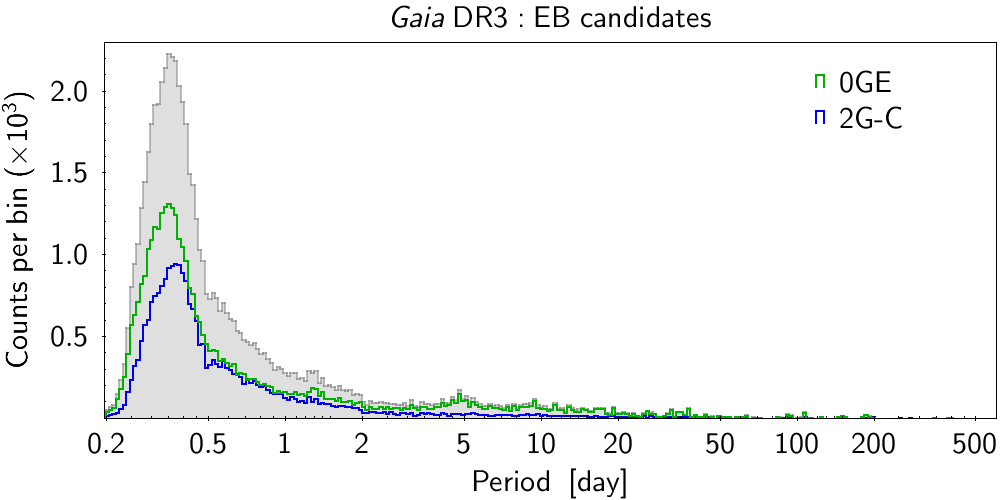}
  \vskip -0.5mm
  \includegraphics[trim={0 0 0 44},clip,width=\linewidth]{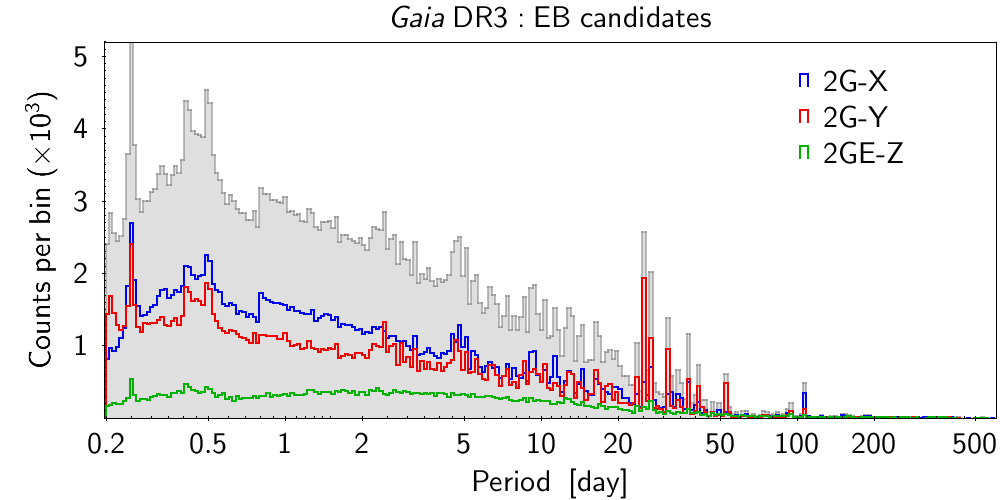}
  \caption{Period distributions of various samples according to their two-model parameters as labeled in the panels (see text).
           The filled grey histograms represent the combined samples in each panel.
          }
    \label{Fig:histo_period_samples}
\end{figure}
%-------

The distribution of the orbital periods of the full sample is shown in Fig.~\ref{Fig:histoPeriod} (black histogram), on a linear scale for the number of sources in the top panel and on a logarithmic scale in the bottom panel.
Peaks are observed at specific periods above about five days, and especially above 20 days.
These structures result from aliases and other period search artefacts inherent to the \Gaia mission properties such as the scanning law and orientation \citep{DR3-DPACP-164}.
They are much reduced in the sample with global rankings larger than 0.60, as shown by the green filled histogram in Fig.~\ref{Fig:histoPeriod}.
Figure~\ref{Fig:periodVsGR} gives a more detailed view of the period distribution versus global rankings.
At global rankings larger than 0.54, unexpected peaks are visible mainly at periods longer than 30 days, and more specifically around, in decreasing order of importance, 53.7~d, 107.2~d and 34.1~d.
These structures are much less present at global rankings below 0.54.
Instead, structures are visible between 4 and 50 days.
At global rankings below $\sim$0.5, the 6-hour alias due to the spacecraft rotation and its related 12-hour alias become significant.
The reader is referred to \citet{DR3-DPACP-164} for a general discussion on the structures and aliases present in DR3 period distributions.

The period distributions for the various samples discussed in Sect.~\ref{Sect:catalogue_usage_model} are shown in Fig.~\ref{Fig:histo_period_samples} (solid lines).
The samples are grouped in three categories.
Wider systems (on the mean) are shown in the top panel (samples 2G-A, 2GE-A, 2G-D, 1G).
The periods span all values, with a peak at around one day and an extended tail above twenty days.
The observed period distributions reflect the real distribution of these mainly detached eclipsing binaries convolved with the (complex) selection function resulting from the \Gaia eclipsing binary identification, period determination, and light curve modelling procedures.
We also note the presence of the alias peak at the six-hour rotation period.
The excess is predominant in Sample~2G-D containing very narrow primary Gaussians (red histogram), while it is absent in Sample~2GE-A (green histogram) where the presence of a small- to medium-amplitude ellipsoidal component ($2\,\Aell<0.11$~mag) in the light curve model better constrains the period.

The periods of tighter systems are shown in the second (samples 2G-B, 2GE-B, 1GE) and third (samples 0GE, 2G-C) panels.
They have narrower distributions than the ones of well-detached systems, peaking at $\sim$0.35~days (second and third panels in Fig.~\ref{Fig:histo_period_samples}).
The tighter systems include tighter detached, contact, and ellipsoidal systems.
No excess is observed at $\sim$0.25~d in the period distributions of these systems.

The last category shown in the bottom panel gathers samples 2G-X, 2G-Y and 2GE-Z whose light curve model components are not necessary linked to physical features of the binary system.
It contains about one fifth of the full catalogue
The period distributions reveal much more complex structures. 
Many peaks are observed over the full range of periods, with a predominance at 0.25~days and above twenty days.
These distributions support the conclusions drawn in Sect.~\ref{Sect:catalogue_usage_model} that ask for a confirmation of their periods and the nature of the eclipsing binaries in these samples.

Additional insight in the period distributions is provided in Sect.~\ref{Sect:catalogue_usage_summary} of Appendix~\ref{Appendix:twoGaussianModel}.
Figure~\ref{Fig:period_magG_samples}, in particular, shows the distribution of the periods versus \gmag magnitude for each sample.

%==================================================================
\section{Catalogue quality}
\label{Sect:quality}

%-------
\begin{table}
\caption{Surveys cross-matched with the \Gaia DR3 catalogue of eclipsing binaries.
         The first column gives the survey.
         The second column gives the percentage of cross-matched sources belonging to that survey.
         The third column gives the catalogue label(s) used by \citet{DR3-DPACP-177}, on which we based our cross-matches.
         The literature references corresponding to the catalogue labels are given in their Table~1.
        }
\centering
\begin{tabular}{l r l}
\hline
Survey & \% & Catalogues\\
\hline
ZTF       &  42\% & \small{\texttt{ZTF\_PERIODIC\_CHEN\_2020}}  \\
OGLE4     &  17\% & \small{\texttt{OGLE4\_VAR\_OGLE\_2019}} (mainly), \\
          &       & \small{\texttt{OGLE4\_LMC\_ECL\_OGLE4\_2017}}, \\
          &       & \small{\texttt{OGLE4\_SMC\_ECL\_OGLE4\_2017}}, \\
          &       & \small{\texttt{OGLE4\_BLG\_RRL\_SOSZYNSKI\_2019}}, \\
          &       & \small{\texttt{OGLE4\_GSEP\_VAR\_SOSZYNSKI\_2012}} \\
ASAS-SN   &  14\% & \small{\texttt{ASASSN\_VAR\_JAYASINGHE\_2019}} \\
ATLAS     &  10\% & \small{\texttt{ATLAS\_VAR\_HEINZE\_2018}} \\
CATALINA  &   8\% & \small{\texttt{CATALINA\_VAR\_DRAKE\_2014}}, \\
          &       & \small{\texttt{CATALINA\_VAR\_DRAKE\_2017}} \\
PS1       &   5\% & \small{\texttt{PS1\_RRL\_SESAR\_2017}} \\
Other     &   4\% & Various other sources \\
\hline
\end{tabular}
\label{Tab:XM_catalogues}
\end{table}
%-------

%-------
\begin{table}
\caption{Some statistics on period comparison between the \Gaia DR3 catalogue of eclipsing binaries and literature data.
         The number of sources with parallaxes better than 10\% are indicated in italics below the parent sample.
         See text for a description of the table.
        }
\centering
\begin{tabular}{l r r | r r}
\hline
\hline
\multirow{2}{*}{XMs} & \multirow{2}{*}{All}   & \multirow{2}{*}{with $P_\mathrm{lit}$} & \multicolumn{2}{c}{$P_\mathrm{lit} / P_\mathrm{\Gaia}$} \\
                     &                        &                                        &                 $\simeq$1 &       $\simeq$(0.5, 1, 2) \\
\hline
All                  &               606\,393 &               600\,902 &                455\,821 &                513\,523 \\
                     &\textit{\small 166\,951}&\textit{\small 144\,890}& \textit{\small 123\,232}& \textit{\small 143\,245}\\
\hline
EB                   &               527\,779 &               527\,526 &                451\,575 &                489\,926 \\
                     &\textit{\small 145\,053}&\textit{\small 144\,890}& \textit{\small 121\,408}& \textit{\small 134\,949}\\
\hline
non-EB               &                78\,614 &                73\,376 &                    4246 &                 23\,597 \\
                     & \textit{\small 21\,898}& \textit{\small 19\,286}&     \textit{\small 1824}&     \textit{\small 8296}\\
\hline
\end{tabular}
\label{Tab:XM_statistics}
\end{table}
%-------

%-------
\begin{figure}
  \centering
  \includegraphics[trim={0 0 0 42},clip,width=\linewidth]{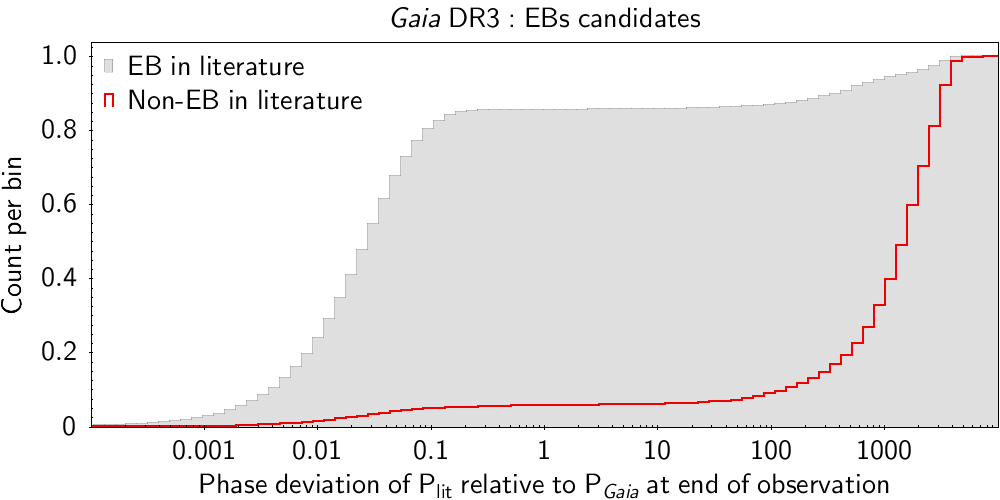}
    \caption{Cumulative distribution of the phase deviation $r_\mathrm{P,lit}$ at the end of the observation obtained when adopting the literature period $P_\mathrm{lit}$ instead of the \Gaia period $P_\mathrm{Gaia}$ (see Eq.~\ref{Eq:fracPlit}).
             The sample of \Gaia eclipsing binary candidates that are also classified as eclipsing binaries in the literatures is shown by the filled grey histogram, and the sample which has literature cross-matches but with a classification other than eclipsing binary in the literature is shown by the red histogram.
            }
    \label{Fig:histoFracRecovery_cumul}
\end{figure}
%-------

%-------
\begin{figure}
  \centering
  \includegraphics[trim={0 74 0 42},clip,width=1.0\linewidth]{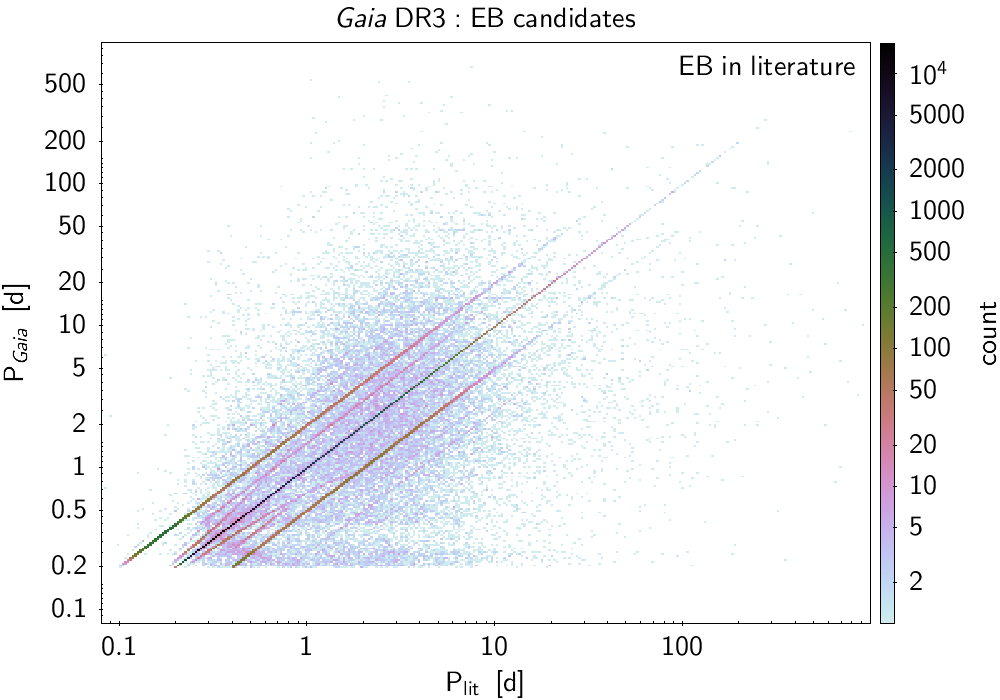}
  \vskip -0.5mm
  \includegraphics[trim={0 0 0 44},clip,width=1.0\linewidth]{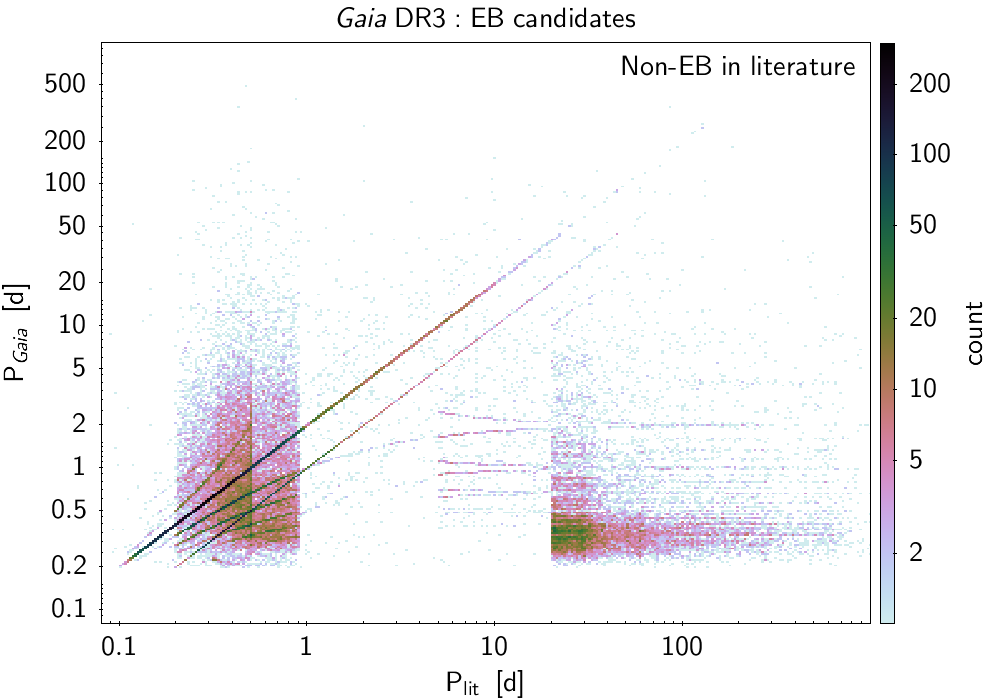}
  \caption{\Gaia period versus literature period for all \Gaia DR3 eclipsing binary candidates that have a cross-match in the literature.
           \textbf{Top panel:} Candidates also identified as eclipsing binaries in the literature.
           \textbf{Bottom panel:} Candidates not classified as eclipsing binaries in the literature.
           The imprints of literature catalogues are visible in the distributions of $P_\mathrm{lit}$ (see text).
          }
    \label{Fig:PlitPgaia}
\end{figure}
%-------

%-------
%\begin{figure}
%  \centering
%  \includegraphics[trim={0 74 0 0},clip,width=1.0\linewidth]{Figures/figPlitPgaia_ogle4_LMC.png}
%  \vskip -0.5mm
%  \includegraphics[trim={0 74 0 44},clip,width=1.0\linewidth]{Figures/figPlitPgaia_ogle4_SMC.png}
%  \vskip -0.5mm
%  \includegraphics[trim={0 0 0 44},clip,width=1.0\linewidth]{Figures/figPlitPgaia_ogle4_BLG.png}
%  \caption{Same as Fig.~\ref{Fig:PlitPgaia}, but with OGLE4 periods of eclipsing binaries towards the LMC (to ppanel), SMC (middle panel), and the Galactic Bulge (bottom panel)
%          }
%    \label{Fig:PoglePgaia}
%\end{figure}
%-------

%-----------
\begin{figure}
  \centering
  \includegraphics[trim={0 77 0 42},clip,width=1.0\linewidth]{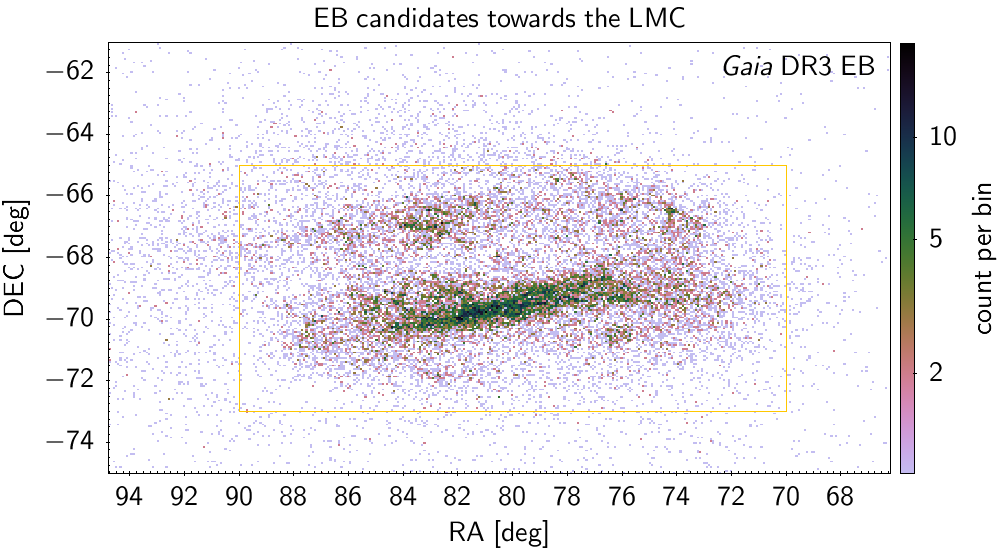}
  \vskip -0.25mm
  \includegraphics[trim={0 77 0 42},clip,width=1.0\linewidth]{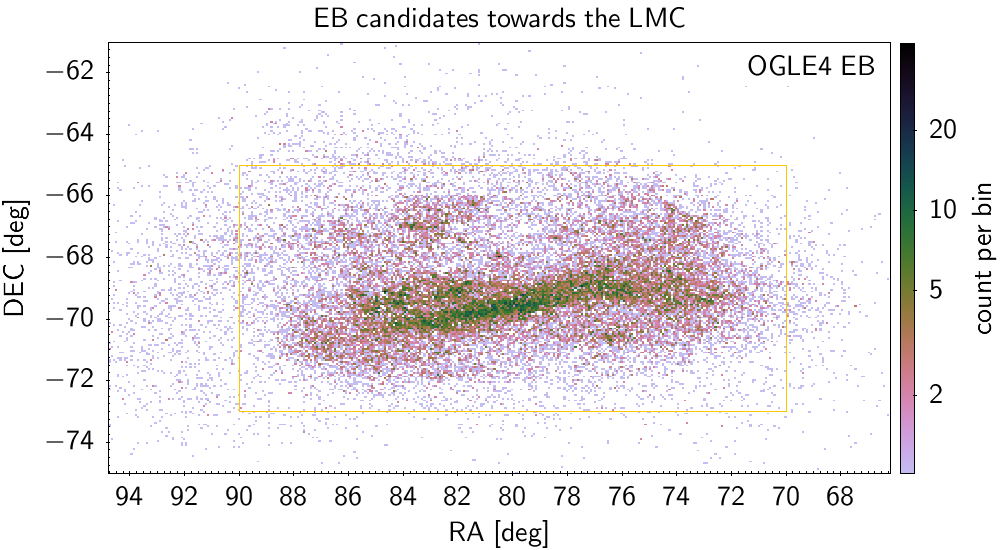}
  \vskip -0.25mm
  \includegraphics[trim={0 77 0 42},clip,width=1.0\linewidth]{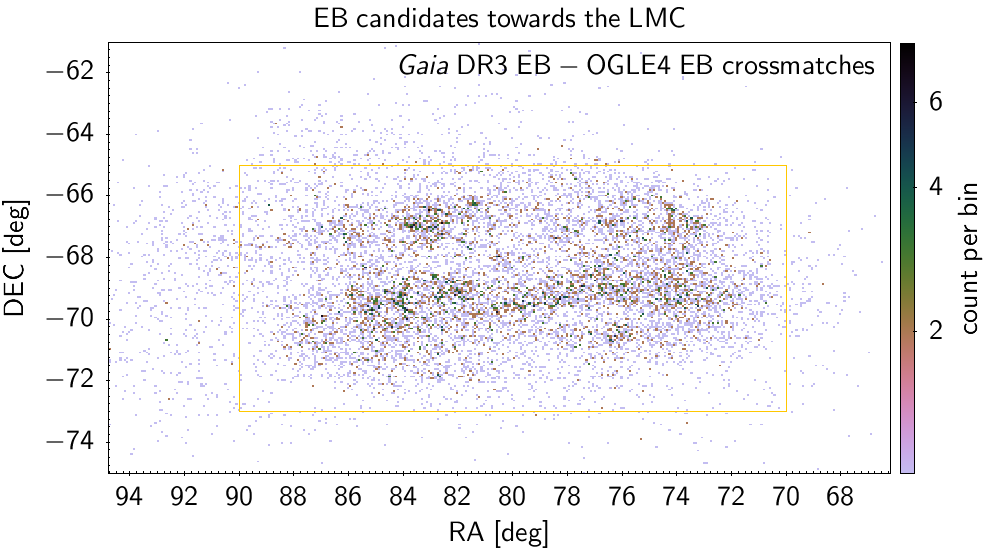}
  \vskip -0.25mm
  \includegraphics[trim={0 0 0 42},clip,width=1.0\linewidth]{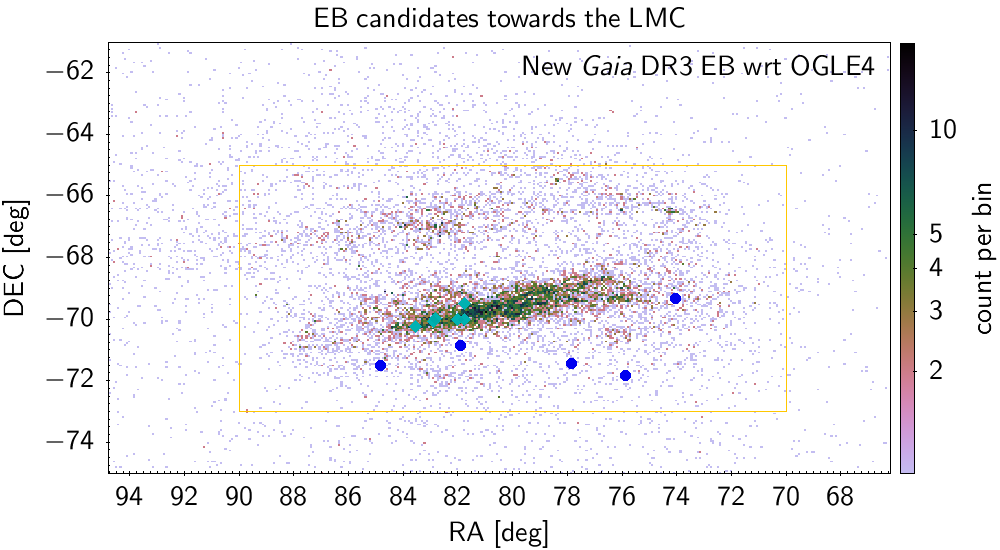}
  \caption{Sky distributions (density maps) in equatorial coordinates of eclipsing binary candidates around the LMC.
           The panels show, from top to bottom, \Gaia DR3 candidates, OGLE4 candidates, the \Gaia-OGLE4 crossmatches (two arc seconds radius), and the new \Gaia candidates with respect to OGLE4.
           Sources highlighted by filled circles in the bottom panel have their \gmag light curves displayed in Figs.~\ref{Fig:lcs_hasNotOgle4_outOfBar} and \ref{Fig:lcs_hasNotOgle4_inBar}.
           They are located in the bar of the LMC for the cyan markers and outside the bar for the blue markers. 
           The orange area delineates the sub-region of the sky used in the text to compute the fraction of \Gaia new candidates towards the LMC.
          }
    \label{Fig:skyLMC_DR3versusOgle4}
\end{figure}

\begin{figure}
  \centering
  \includegraphics[trim={0 77 0 42},clip,width=1.0\linewidth]{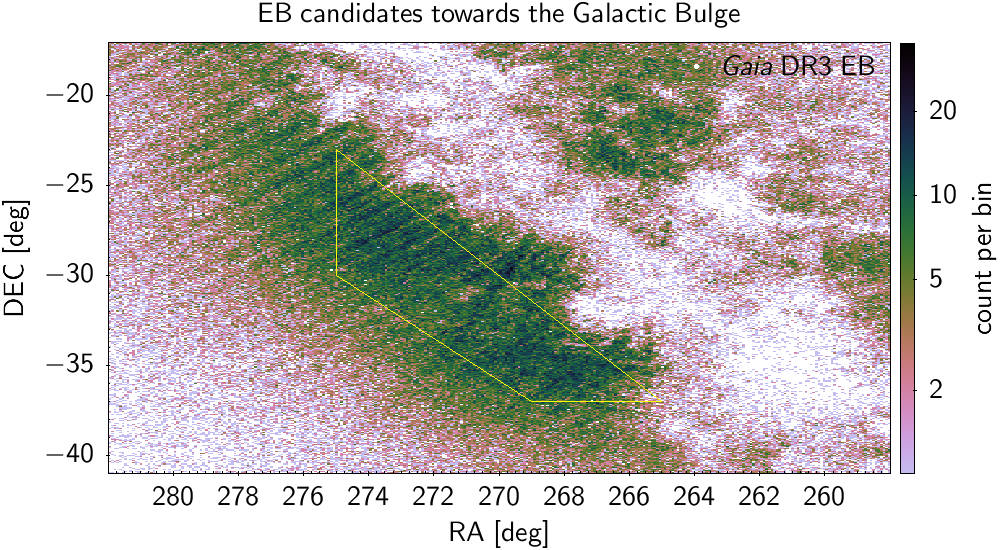}
  \vskip -0.25mm
  \includegraphics[trim={0 77 0 42},clip,width=1.0\linewidth]{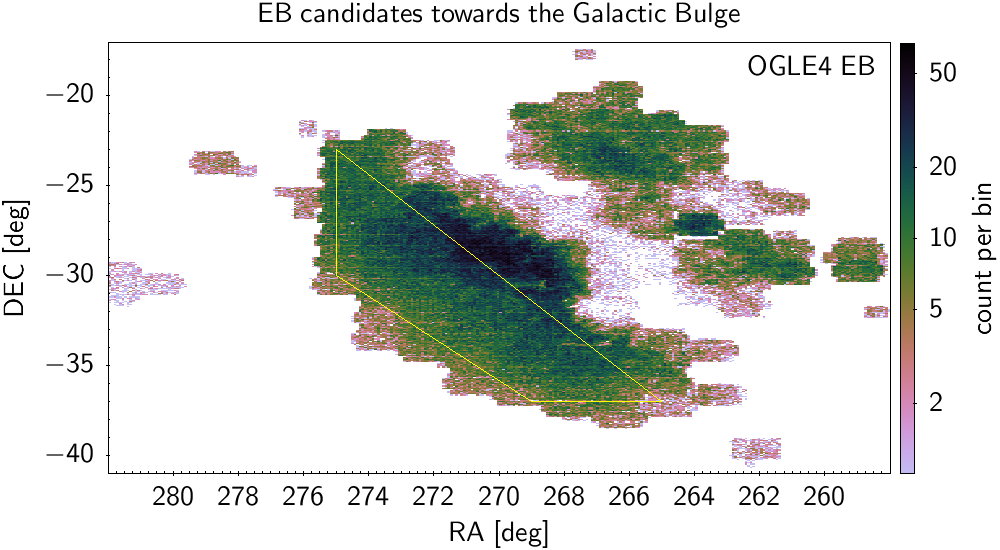}
  \vskip -0.25mm
  \includegraphics[trim={0 77 0 42},clip,width=1.0\linewidth]{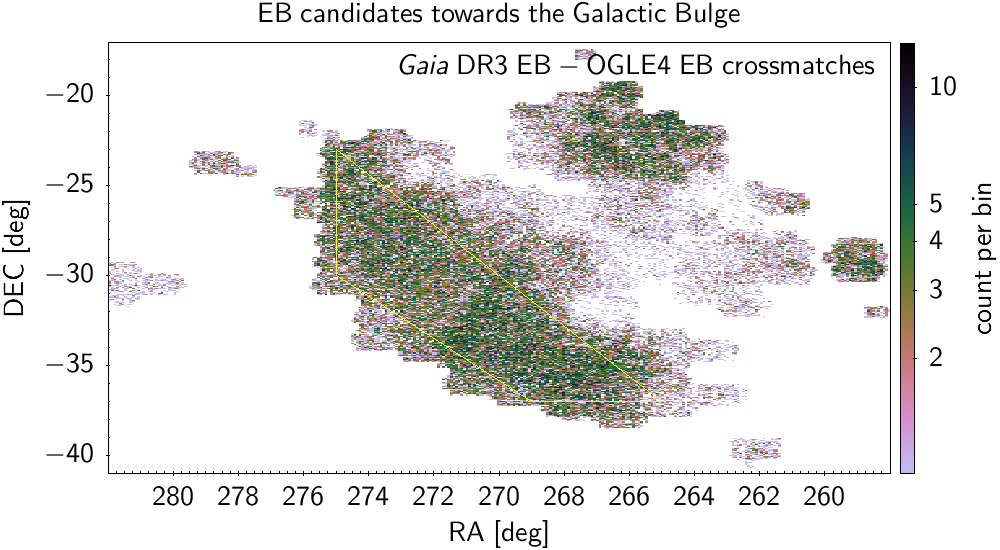}
  \vskip -0.25mm
  \includegraphics[trim={0 0 0 42},clip,width=1.0\linewidth]{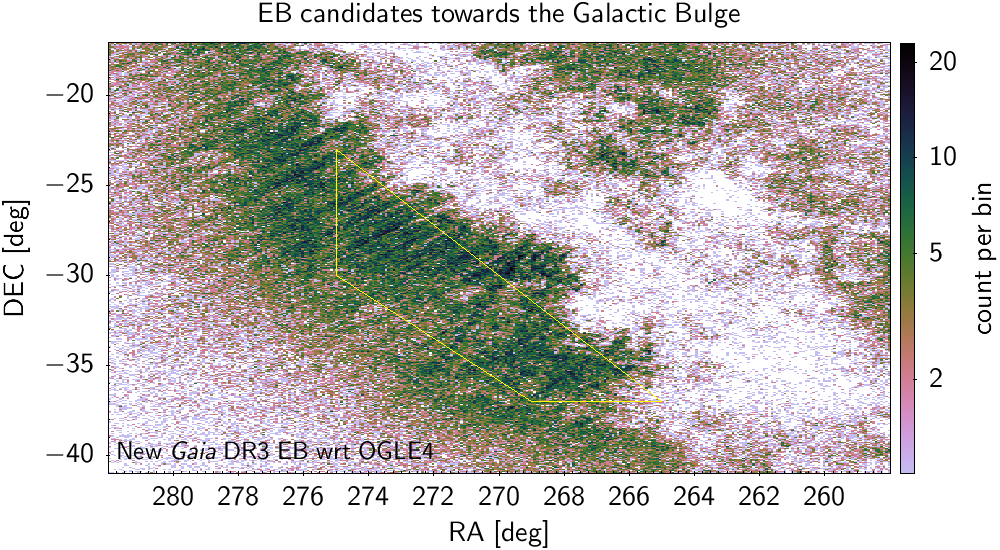}
  \caption{Same as Fig.~\ref{Fig:skyLMC_DR3versusOgle4}, but towards the Galactic Bulge.
          }
    \label{Fig:skyBLG_DR3versusOgle4}
\end{figure}

\begin{figure}
  \centering
  \includegraphics[trim={0 0 0 42},clip,width=1.0\linewidth]{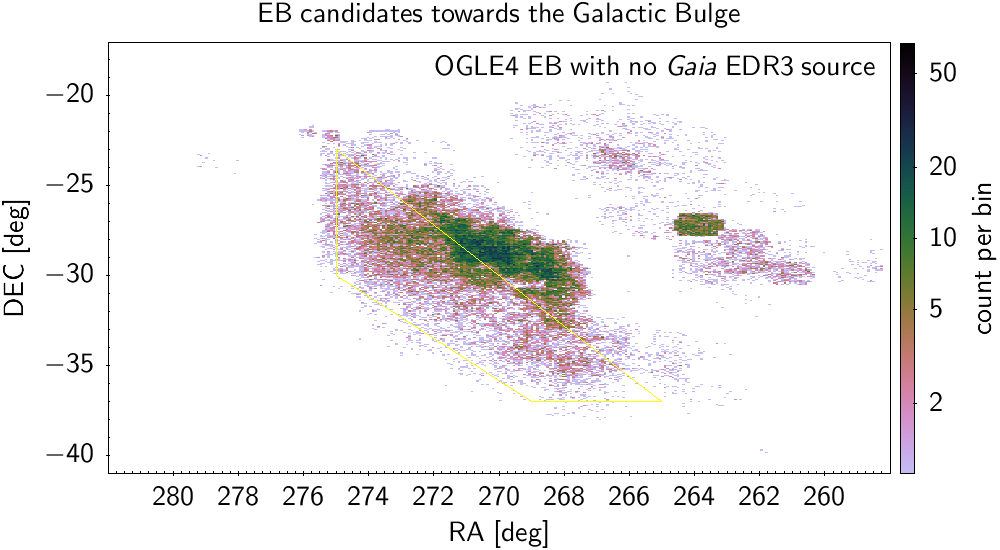}
  \caption{Same as Fig.~\ref{Fig:skyLMC_DR3versusOgle4}, but for OGLE4 sources in the Galactic Bulge that have no source counterpart in the \Gaia DR3 archive.
          }
    \label{Fig:skyBLG_Ogle4_notInDR3}
\end{figure}
%-----------

%-----------
\begin{table}
\caption{Number of OGLE4 eclipsing binaries towards the LMC, SMC and Galactic Bulge.
         The number of cross-matches with the full \Gaia DR3 sources are given in the second to fourth rows of numbers, and the ones with the \Gaia DR3 catalogue of eclipsing binaries in magnitude-limited samples are given in the last four rows, with their percentages relative to the number of OGLE4-DR3 crossmatches given below the relevant rows. 
        }
\centering
\begin{tabular}{l r r r r}
\hline\hline 
                        & \hspace{5mm}  LMC &  SMC &  Bulge &    All \\
\hline
OGLE4                   & 40204 & 8401 & 425193 & 473798 \\[3pt]
%\hline
\multicolumn{5}{l}{\textit{Cross-matches with DR3}}          \\
\hspace{2mm} All        & 40023 & 8392 & 371902 & 420317 \\
\hspace{2mm} \gmag < 20 & 35392 & 7843 & 315523 & 358758 \\
\hspace{2mm} \gmag < 19 & 21900 & 5796 & 174984 & 202680 \\[3pt]
%\hline
%\hline
\multicolumn{5}{l}{\textit{Cross-matches with DR3 catalogue of eclipsing binaries}}          \\
\hspace{2mm} \gmag < 20 & 13390 & 3780 &  83885 & 101055 \\
                        &  37\% & 48\% &   26\% &   28\% \\
\hspace{2mm} \gmag < 19 &  9317 & 2940 &  61557 &  73814 \\
                        &  42\% & 50\% &   35\% &   36\% \\
\hline
\end{tabular}
\label{Tab:OGLE4}
\end{table}
%-----------

%-----------
\begin{table}
\caption{Number of \Gaia DR3 eclipsing binaries candidates in selected sky areas of the LMC and Galactic Bulge. % shown in Figs.~\ref{Fig:skyLMC_DR3versusOgle4} and \ref{Fig:skyBLG_DR3versusOgle4}, respectively.
         The first dataset (columns 2-3) includes all global rankings, while the second set (columns 4-5) consists of sources with \small{\texttt{global\_ranking}}\,>\,0.50.
        }
\centering
\begin{tabular}{l | r r | r r}
\hline\hline
                 & \multicolumn{2}{c|}{All rankings} & \multicolumn{2}{c}{Ranking $>0.5$}\\
                 & LMC$^a$ & Bulge$^b$ & LMC$^a$ & Bulge$^b$ \\
\hline
\Gaia DR3 EBs    & 26020 & 96199 & 8358 & 31446 \\
in OGLE4         & 12123 & 33469 & 6912 & 20197 \\
not in OGLE4     & 13897 & 62730 & 1446 & 11249 \\
\% new           &  53\% &  65\% & 17\% &  35\% \\
\hline
\end{tabular}
\vskip -2mm
\tablefoot{\tablefoottext{a}{Within the rectangular sky region shown in Fig.~\ref{Fig:skyLMC_DR3versusOgle4}.}\\
           \tablefoottext{b}{Within the polygon sky region shown in Fig.~\ref{Fig:skyBLG_DR3versusOgle4}.}
           }
\label{Tab:newWrtOgle4}
\end{table}
%----------- 

%-----------
\begin{figure}
  \centering
  \includegraphics[trim={0 77 0 42},clip,width=1.0\linewidth]{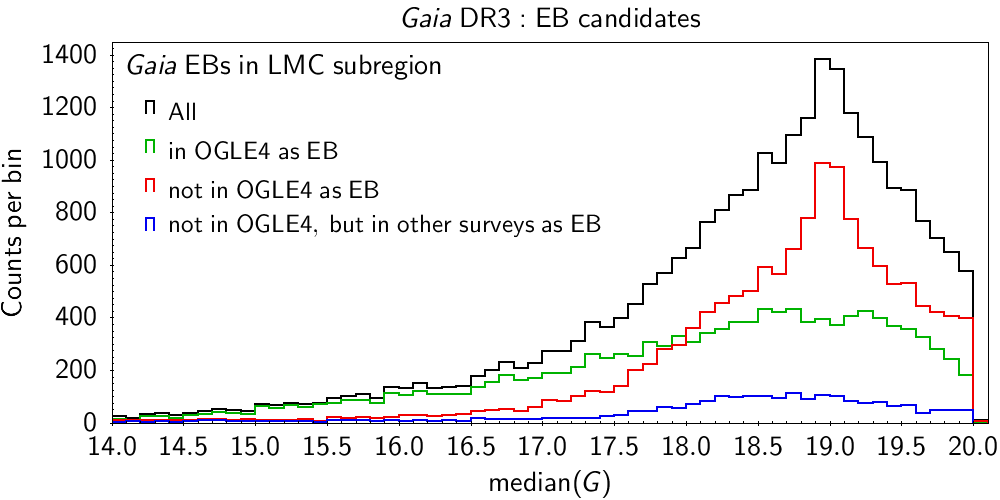}
  \vskip -0.25mm
  \includegraphics[trim={0 0 0 42},clip,width=1.0\linewidth]{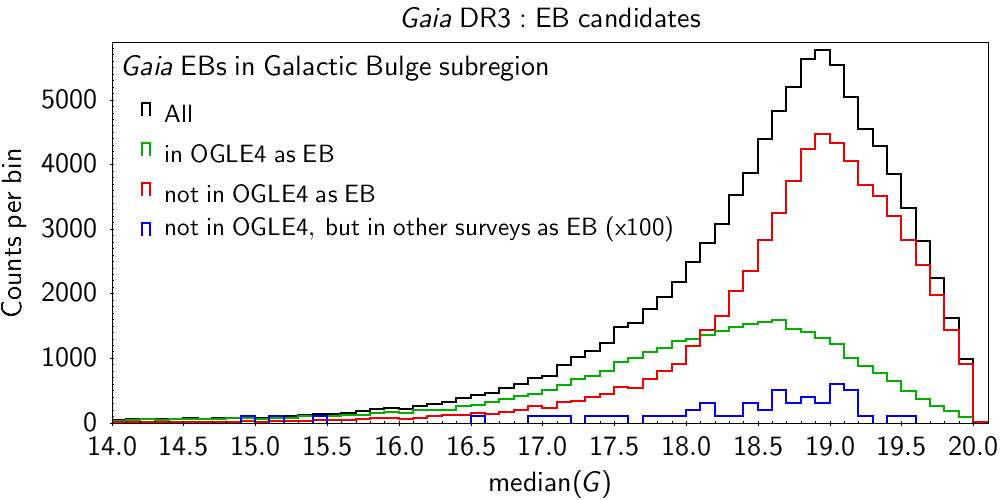}
  \caption{Median \gmag magnitude distributions of \Gaia eclipsing binaries in specific regions of the sky.
           \textbf{Top panel:} Sky region towards the LMC shown in Fig.~\ref{Fig:skyLMC_DR3versusOgle4}.
           \textbf{Bottom panel:} Sky region towards the Galactic Bulge shown in Fig.~\ref{Fig:skyBLG_DR3versusOgle4}.
           The black histograms show all \Gaia DR3 eclipsing binary candidates in the given sky area.
           The sources among them that have or do not have a cross-match (two arc second radius) with the OGLE4 catalogue of eclipsing binaries are shown by the green and red histograms, respectively.
           The blue histograms are the distributions of sources with no OGLE4 cross-match, but which have a cross match with eclipsing binary candidates identified in other surveys.
           The abscissa range is truncated on the bright side for better visibility.
          }
    \label{Fig:histo_medianMag}
\end{figure}
%-----------

%-----------
\begin{figure*}
  \centering
\includegraphics[trim={0  81 115 50},clip,height=0.2776\linewidth]{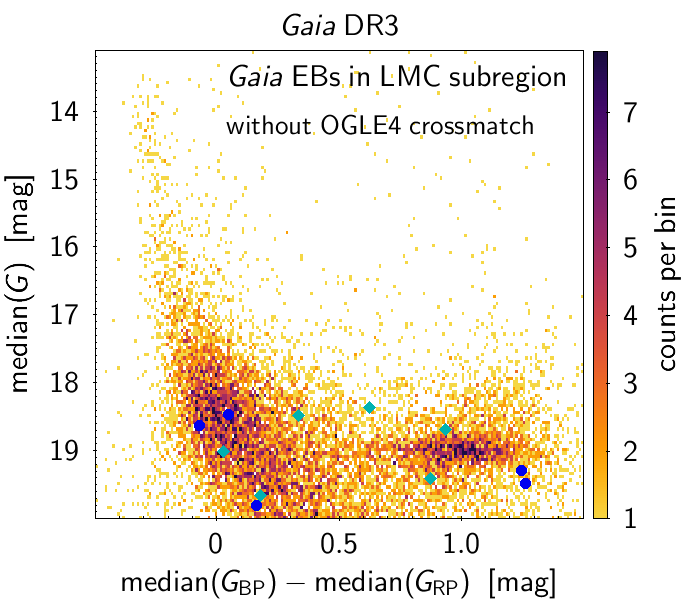}%
\includegraphics[trim={93 81 115 50},clip,height=0.2776\linewidth]{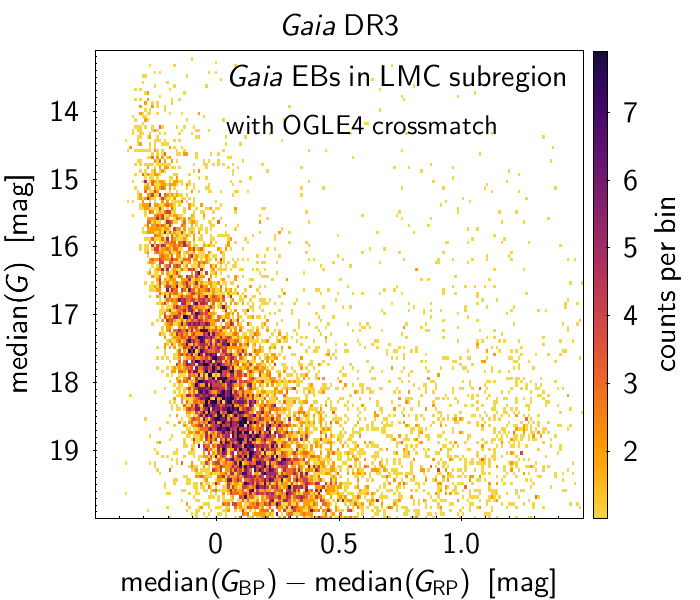}%
\includegraphics[trim={93 81   0 50},clip,height=0.2776\linewidth]{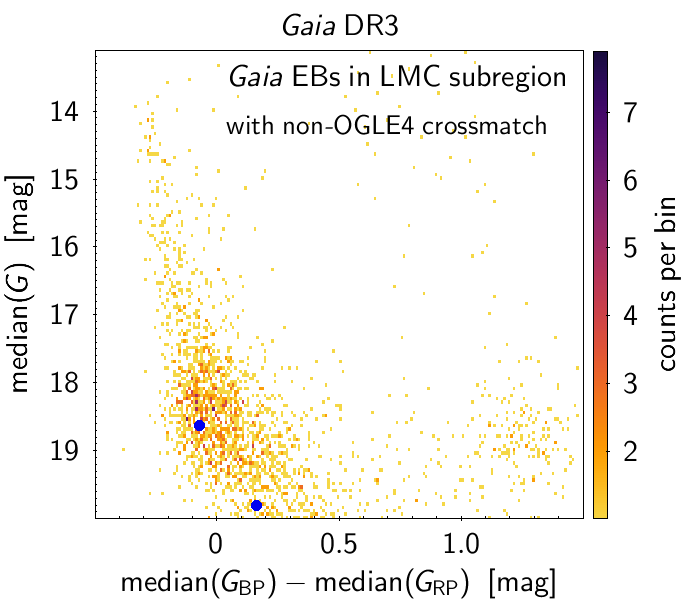}
\includegraphics[trim={0   0 115 50},clip,height=0.1775\linewidth]{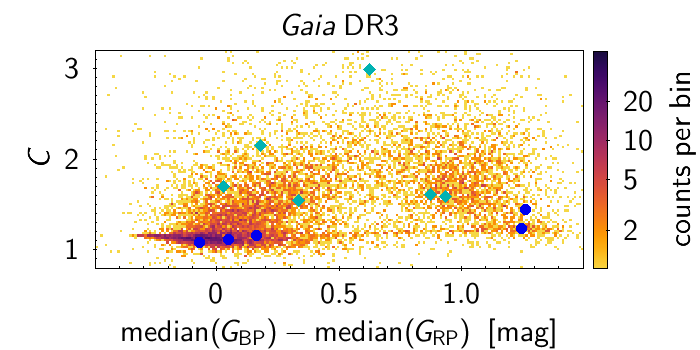}%
\includegraphics[trim={93  0 115 50},clip,height=0.1775\linewidth]{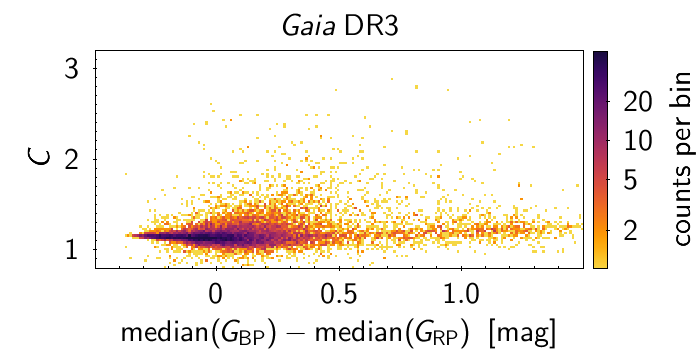}%
\includegraphics[trim={93  0   0 50},clip,height=0.1775\linewidth]{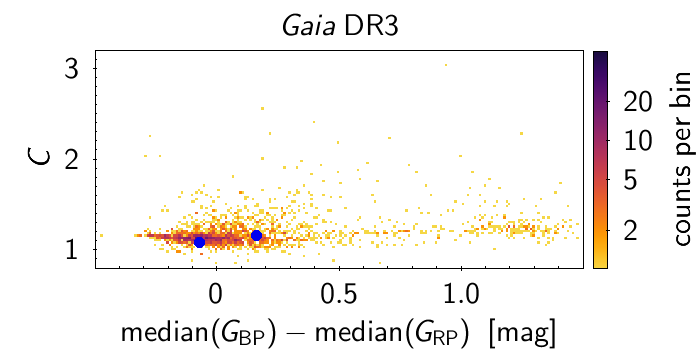}
  \caption{Colour-magnitude (top row) and \BPRPexcess versus colour (bottom row) diagrams of the \Gaia eclipsing binaries in the sky area towards the LMC shown in Fig.~\ref{Fig:skyLMC_DR3versusOgle4}.
           \textbf{Left panels:} New \Gaia eclipsing binary candidates with respect to the OGLE4 catalogue of LMC eclipsing binaries.
           \textbf{Middle panels:} \Gaia sources having a crossmatch (two arc-second radius) with the OGLE4 catalogue.
           \textbf{Right panels:} \Gaia candidates having no crossmatch with the OGLE4 candidates but having a crossmatch with EB candidates in other surveys.
           The colour of each bin is proportional to the number of counts per bin according to the colour scales shown on the right for each row.
           Sources highlighted by filled circles have their \gmag light curves displayed in Figs.~\ref{Fig:lcs_hasNotOgle4_outOfBar} and \ref{Fig:lcs_hasNotOgle4_inBar}.
           They are located in the bar of the LMC for blue circle markers and outside the bar for cyan diamond markers. 
           The axes ranges have been truncated for better visibility.
          }
    \label{Fig:CM_LMC}
\end{figure*}
%-----------

%-----------
\begin{figure}
  \centering
  \includegraphics[trim={40 145 0 70},clip,width=\linewidth]{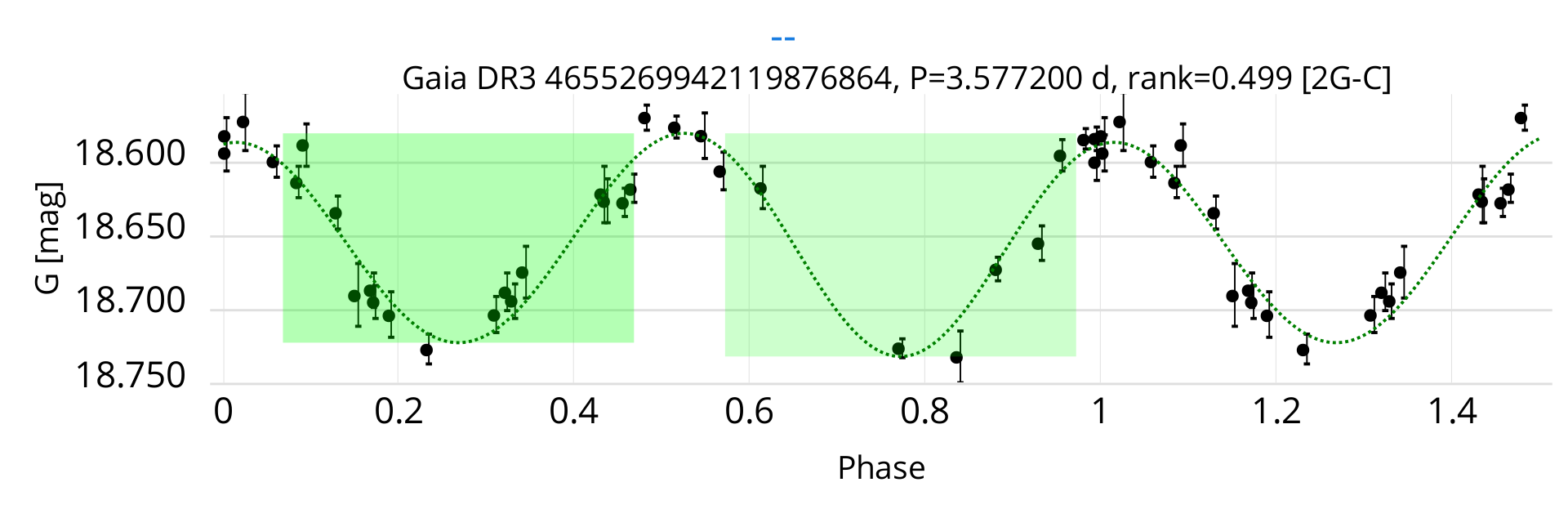}
  \vskip -0.5mm
  \includegraphics[trim={40 145 0 70},clip,width=\linewidth]{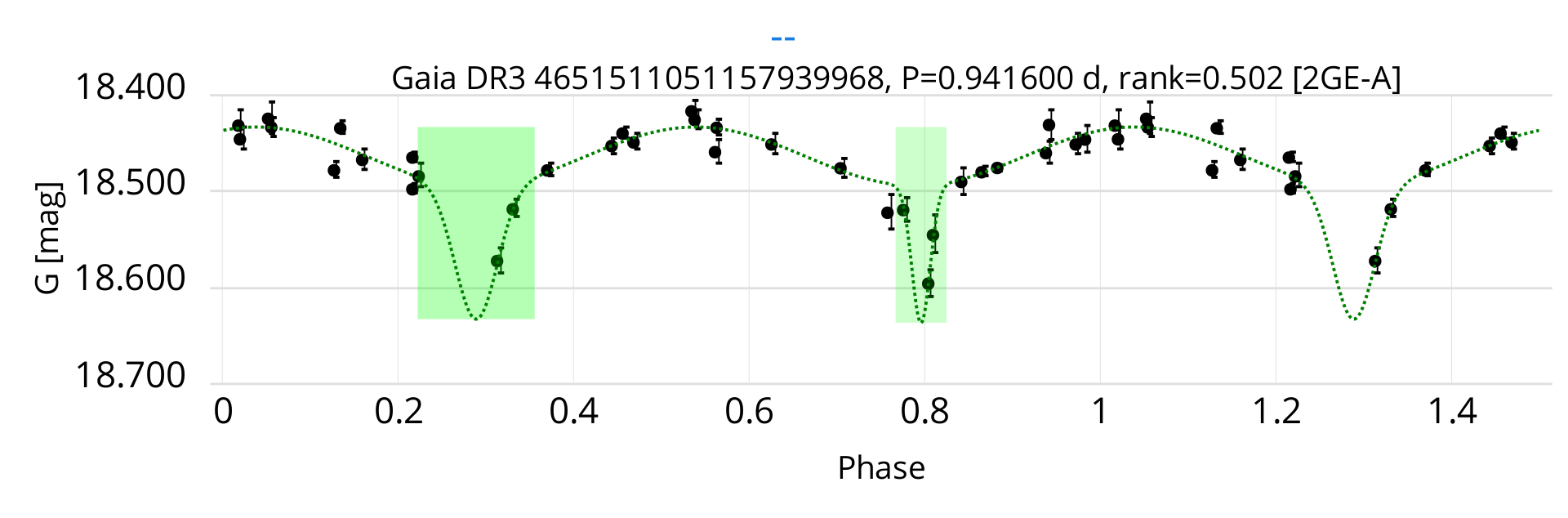}
  \vskip -0.5mm
  \includegraphics[trim={40 145 0 70},clip,width=\linewidth]{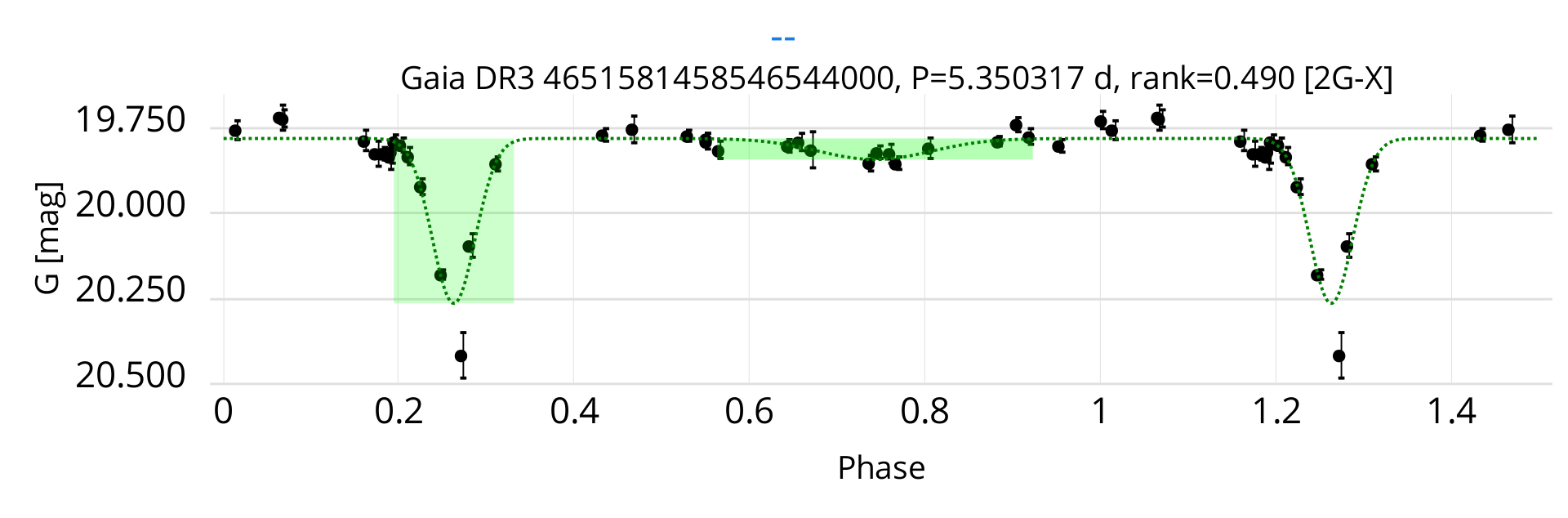}
  \vskip -0.5mm
  \includegraphics[trim={40 145 0 70},clip,width=\linewidth]{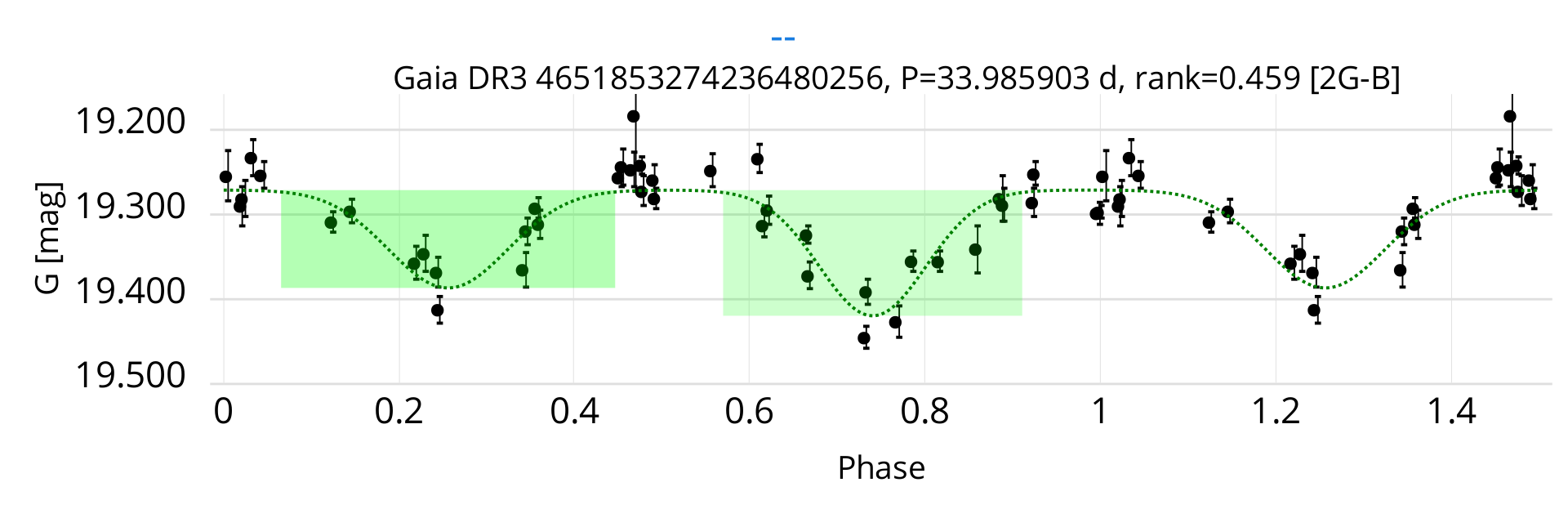}
  \vskip -0.5mm
  \includegraphics[trim={40 45 0 70},clip,width=\linewidth]{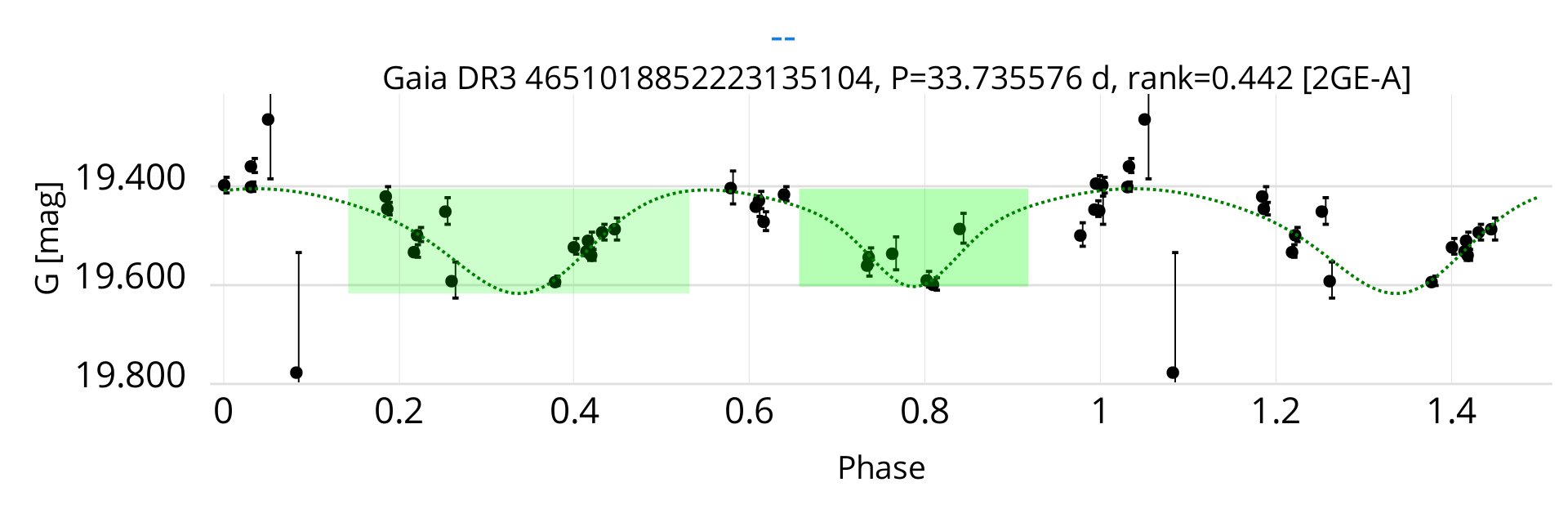}
    \caption{Same as top panel of Fig.~\ref{Fig:lcs_example_2GA}, but for five \Gaia candidates outside the LMC bar that are not present in the OGLE4 catalogues of eclipsing binaries, sorted with increasing \BPminusRP colour from top to bottom.
             The sources are \GaiaSrcIdInCaption{4655269942119876864, 4651511051157939968, 4651581458546544000, 4651853274236480256, 4651018852223135104}.
             Their sky positions are identified in Fig.~\ref{Fig:skyLMC_DR3versusOgle4}.
             %Their locations in the colour-magnitude diagram and in the BP\,RP excess factor versus colour diagrams are indicated in Fig.~\ref{Fig:CM_LMC}.
            }
\label{Fig:lcs_hasNotOgle4_outOfBar}
\end{figure}
%-----------
% -- out of bar:
% source_id==4655269942119876864L || source_id==4651511051157939968L || source_id==4651581458546544000L || source_id==4651853274236480256L || source_id==4651018852223135104L
%-----------
% 4655269942119876864, 4651511051157939968, 4651581458546544000, 4651853274236480256, 4651018852223135104

%-----------
\begin{figure}
  \centering
  \includegraphics[trim={40 145 0 70},clip,width=\linewidth]{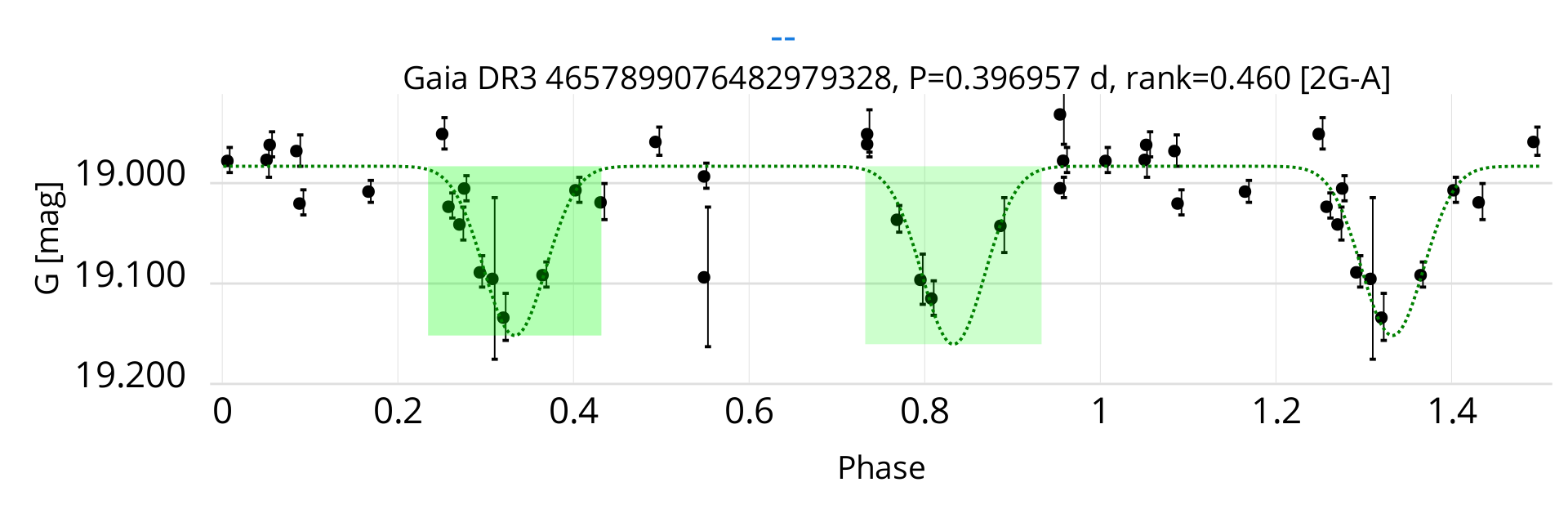}
  \vskip -0.5mm
  \includegraphics[trim={40 145 0 70},clip,width=\linewidth]{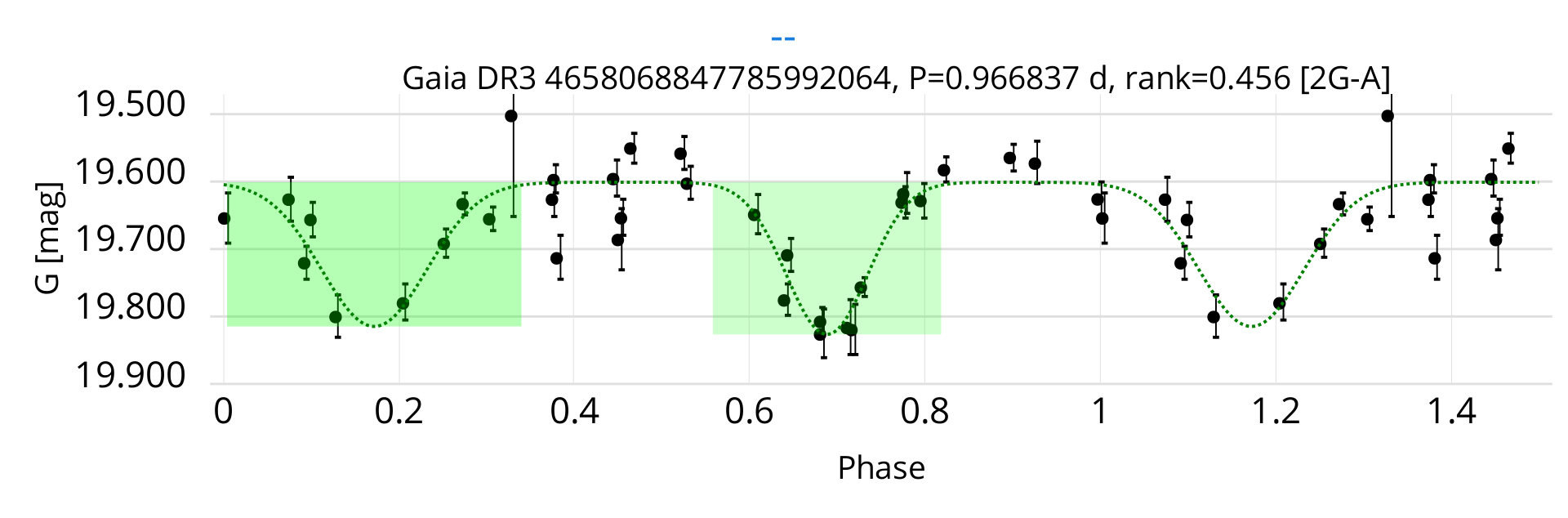}
  \vskip -0.5mm
  \includegraphics[trim={40 145 0 70},clip,width=\linewidth]{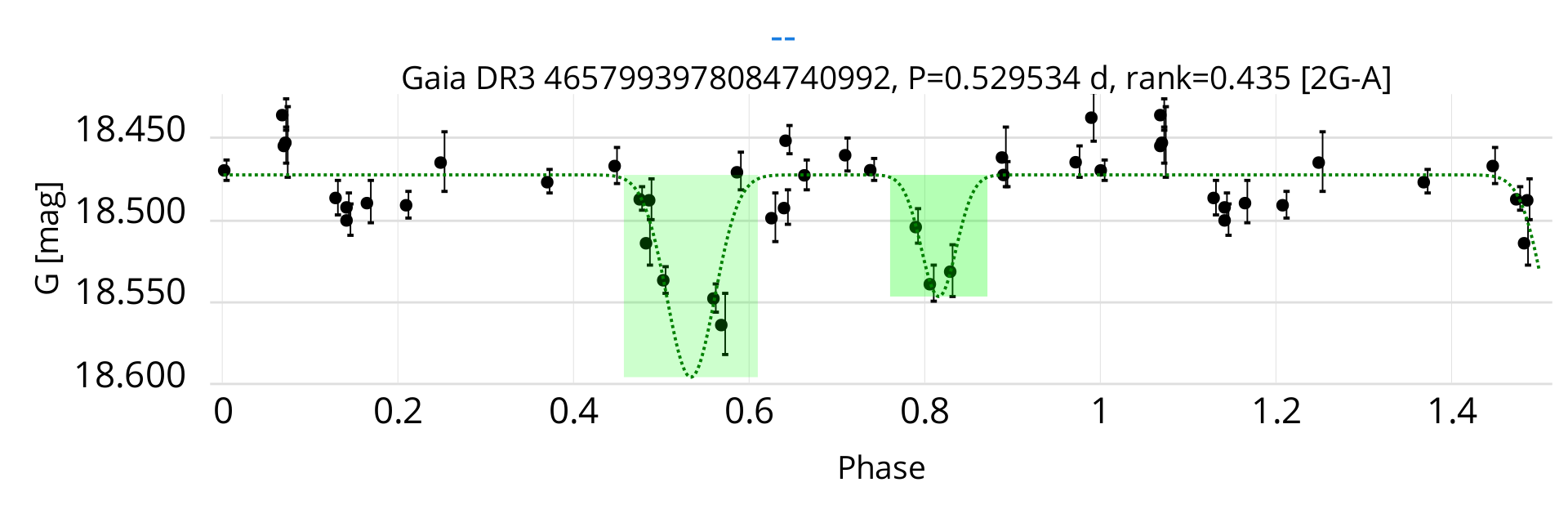}
  \vskip -0.5mm
  \includegraphics[trim={40 145 0 70},clip,width=\linewidth]{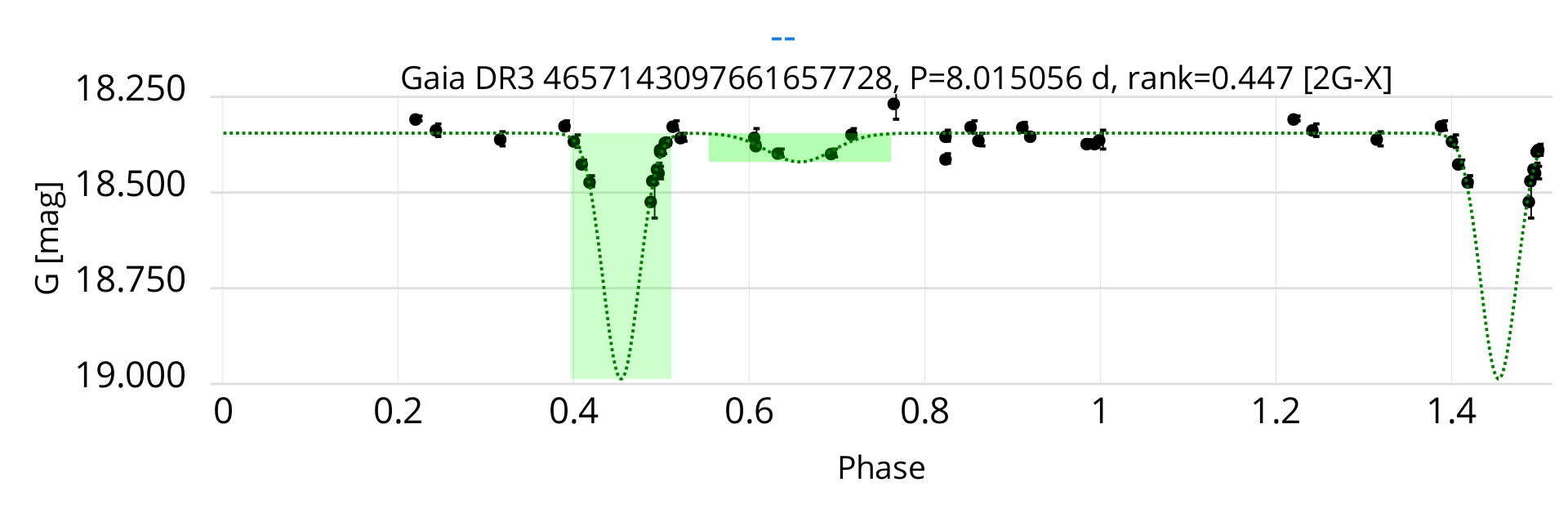}
  \vskip -0.5mm
  \includegraphics[trim={40 145 0 70},clip,width=\linewidth]{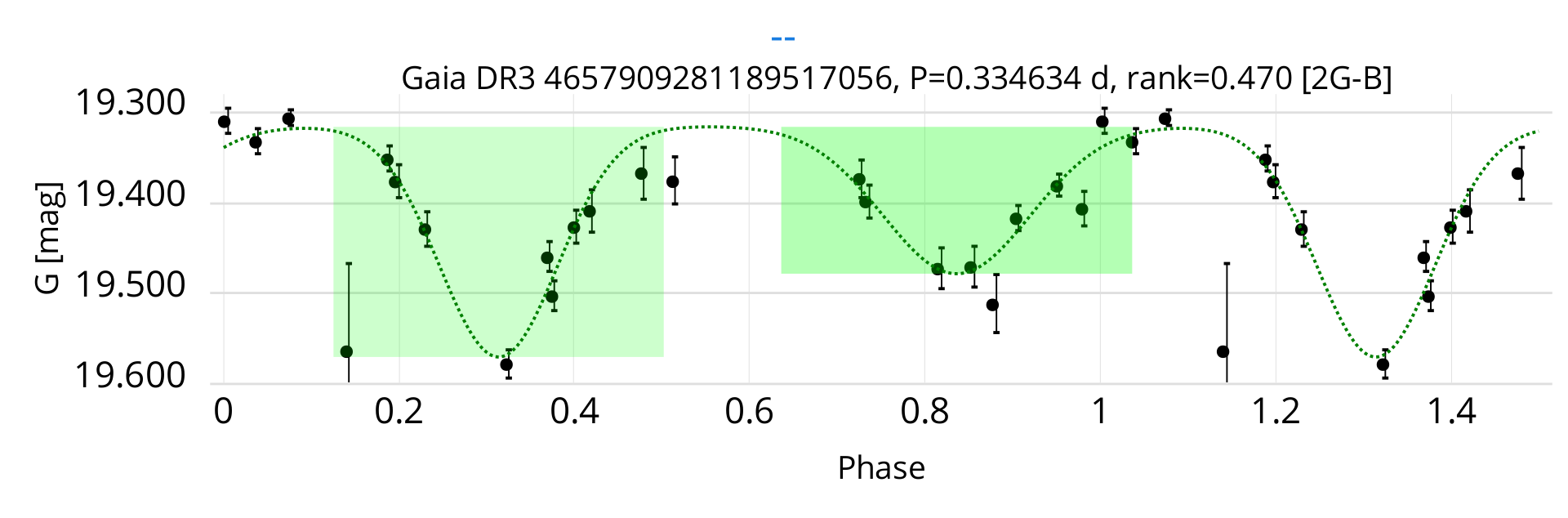}
  \vskip -0.5mm
  \includegraphics[trim={40 45 0 70},clip,width=\linewidth]{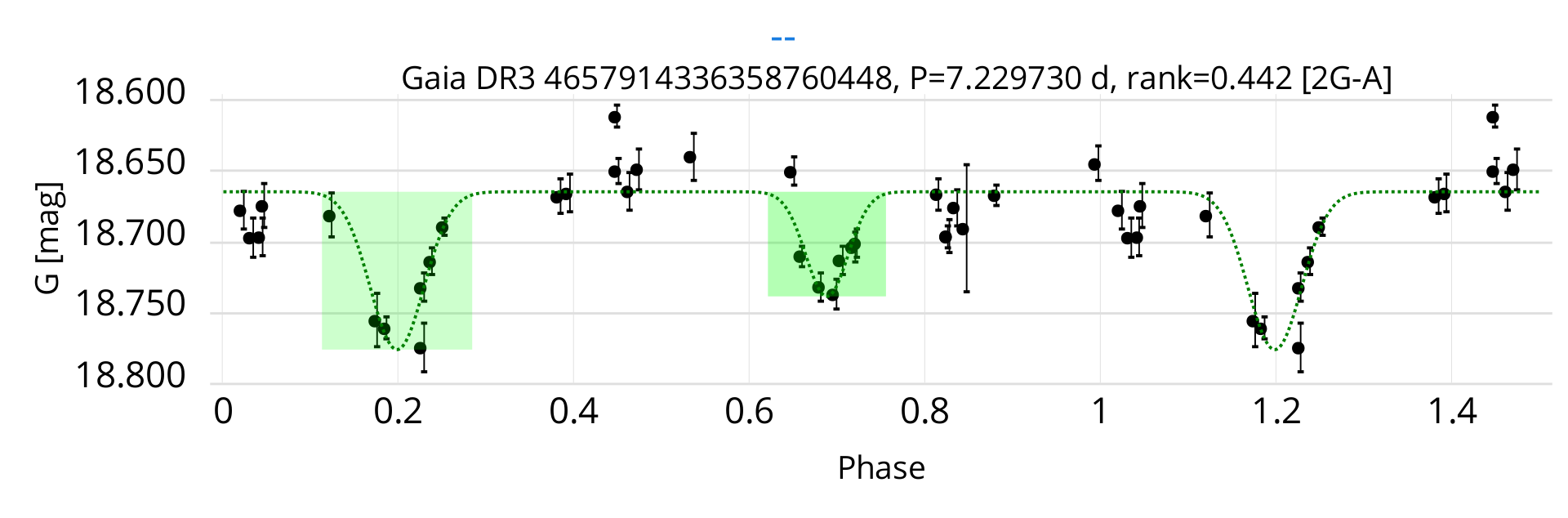}
    \caption{Same as Fig.~\ref{Fig:lcs_hasNotOgle4_outOfBar}, but for six  \Gaia candidates in the LMC bar.
             The sources are \GaiaSrcIdInCaption{4657899076482979328, 4658068847785992064, 4657993978084740992, 4657143097661657728, 4657909281189517056, 4657914336358760448}.
             %, as shown in Fig.~\ref{Fig:skyLMC_DR3versusOgle4}.
             %Their locations in the colour-magnitude diagram and in the BP\,RP excess factor versus colour diagrams are indicated in Fig.~\ref{Fig:CM_LMC}.
            }
\label{Fig:lcs_hasNotOgle4_inBar}
\end{figure}
% -- in bar:
% source_id==4657899076482979328L || source_id==4658068847785992064L || source_id==4657993978084740992L || source_id==4657143097661657728L || source_id==4657909281189517056L || source_id==4657914336358760448L
%-----------
% 4657899076482979328, 4658068847785992064, 4657993978084740992, 4657143097661657728, 4657909281189517056, 4657914336358760448

%-----------
\begin{figure}
  \centering
  \includegraphics[trim={0 77 0 42},clip,width=1.0\linewidth]{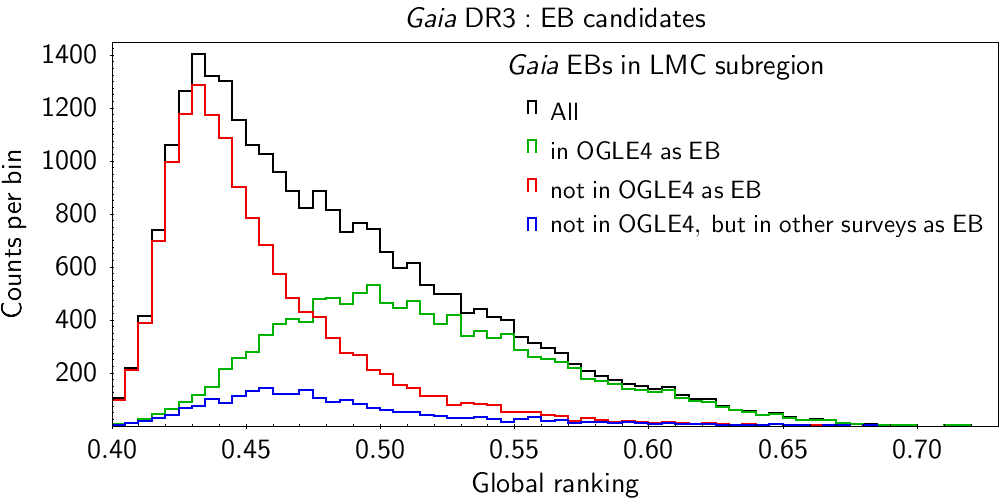}
  \vskip -0.25mm
  \includegraphics[trim={0 0 0 41},clip,width=1.0\linewidth]{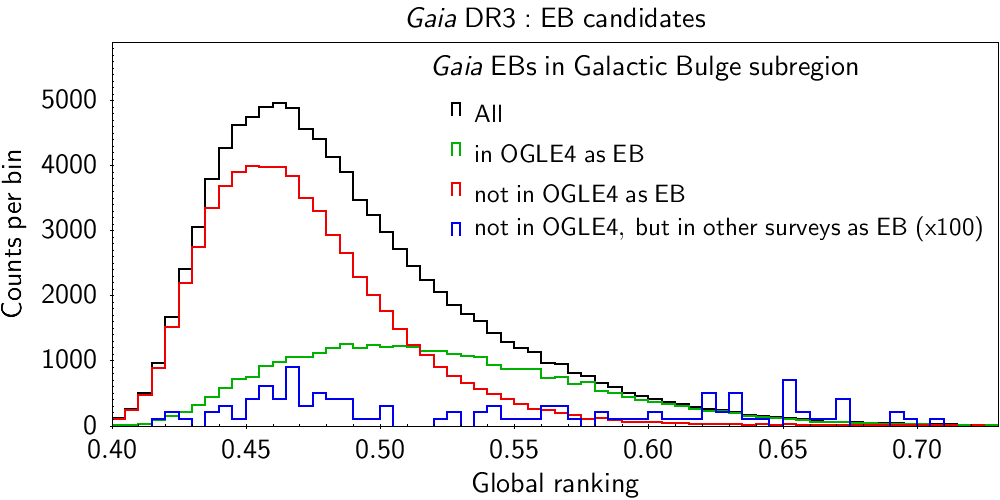}
  \caption{Same as Fig.~\ref{Fig:histo_medianMag}, but for the global ranking. 
          }
    \label{Fig:histo_rankingOgle4}
\end{figure}
%-----------

%-----------
\begin{figure}
  \centering
  \includegraphics[trim={0 75 0 42},clip,width=1.0\linewidth]{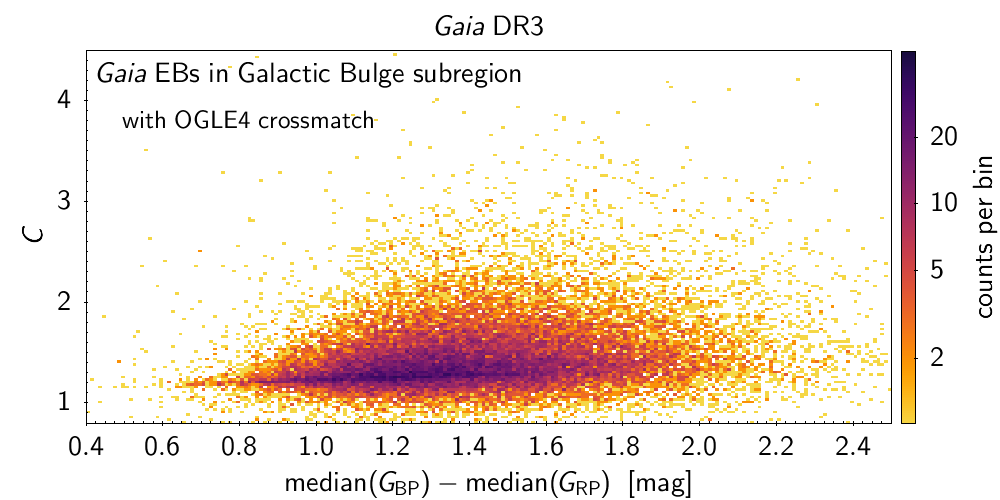}
  \vskip -0.22mm
  \includegraphics[trim={0 0 0 50},clip,width=1.0\linewidth]{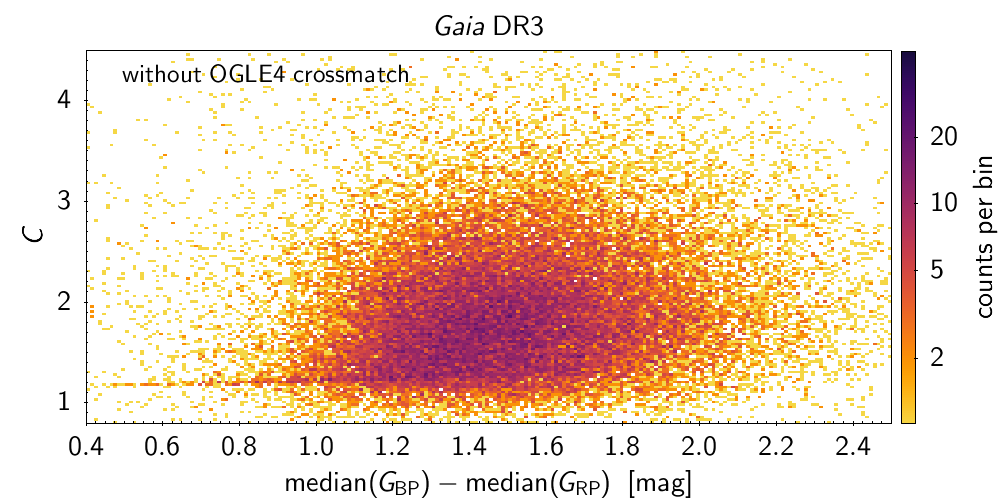}
  \caption{\BPRPexcess versus colour of the \Gaia eclipsing binaries in the sky area towards the Galactic Bulge shown in Fig.~\ref{Fig:skyBLG_DR3versusOgle4}.
           \textbf{Top panel:} \Gaia candidates having OGLE4 crossmatches.
           \textbf{Bottom panel:} \Gaia candidates having no OGLE4 crossmatch.
           The axes ranges have been truncated for better visibility.
          }
    \label{Fig:bprpExcess_BLG}
\end{figure}
%-----------

We assess the quality of our catalogue by comparison of our results with literature data, based on the \Gaia DR3 cross-matches presented in \citet{DR3-DPACP-177}.
For the \Gaia DR3 catalogue of eclipsing binaries, there are 606\,393 cross-matches.
The main surveys and number of cross-matched sources are listed in Table~\ref{Tab:XM_catalogues}.
The largest number of cross-matches relates to the ZTF survey (42\%), then OGLE4 (17\%), ASAS-SN (14\%), ATLAS (10\%), CATALINA (8\%), and PS1 (5\%).
The remaining 4\% cross-matches come from a variety sources not detailed here.

The statistics of the \Gaia DR3 cross matches with the literature are reported in Table~\ref{Tab:XM_statistics}.
The first two-row set of rows (labeled 'All' in the XMs column) gives the statistics for the sample of all cross-matches, irrespective of whether the source is classified as an eclipsing binary in the literature or not.
The table lists the number of sources, the number of sources that have a period reported in the literature, and the number of sources for which the literature period is compatible with the \Gaia period, either directly (1:1 ratio, see Sect.~\ref{Sect:quality_period}) or within a factor of one or two (1:1, 1:2 or 2:1 ratios).
The second two-row set (labeled 'EB') then provides the same statistics, but only for the subsample of cross-matches that are also classified as eclipsing binaries in the literature.
The last two-row set (labeled 'non-EB') finally gives the statistics for the complementary subsample of cross-matches that are classified in the literature in a variability type other than eclipsing binary.

Table~\ref{Tab:XM_statistics} shows that the great majority (87\%) of the \Gaia DR3 eclipsing binaries cross-matched with literature data are also identified in the literature as eclipsing binaries.
This is a good score given the fact that classification of large catalogues is performed through automated procedures, a process that necessarily introduces a fraction of wrong classifications that will impact the comparison between two independent catalogues.
Among the non-EB crossmatches, we note that the \Gaia eclipsing binary candidates cross-matched with non-eclipsing binaries in the literature include 1205 candidates classified as ellipsoidal variables in OGLE4.

We first compare in Sect.~\ref{Sect:quality_period} our periods with the ones found in the literature.
The questions of completeness and purity are then addressed in Sects.~\ref{Sect:quality_completeness} and \ref{Sect:quality_newGaia}, respectively.

%------------------------------------------------------------------
\subsection{Orbital periods}
\label{Sect:quality_period}

Almost all \Gaia DR3 eclipsing binary candidates that have a cross-match in the literature also have a period published in the literature (99\% of them, see Table~\ref{Tab:XM_statistics}), allowing a direct comparison with our periods.
To do so, for any given source, we evaluate the phase deviation $r_\mathrm{P,lit}$ at the end of the observation obtained when adopting the literature period $P_\mathrm{lit}$ instead of the \Gaia period $P_\mathrm{Gaia}$.
This is computed by multiplying the relative difference between the literature and \Gaia periods with the number of cycles during the observation, and is given by
\begin{equation}
  r_\mathrm{P,lit} = \frac{|P_\mathrm{\Gaia} - P_\mathrm{lit}|}{P_\mathrm{lit}} \, \frac{\Delta T}{P_\mathrm{\Gaia}} \; ,
\label{Eq:fracPlit}
\end{equation}
where $\Delta T$ is the duration of the \gmag light curve.
Its cumulative distribution is shown in grey filled histogram in Fig.~\ref{Fig:histoFracRecovery_cumul} for cross-matches that are classified as eclipsing binaries in the literature.
More than 85\% of the sources have a phase deviation of less than 0.5 at the last cycle of their observation.
The histogram also shows that when this is not the case, $r_\mathrm{P,lit}$ is much larger than one, indicating a significant difference between the \Gaia and literature periods.
We therefore consider the \Gaia and literature periods to be equal when $r_\mathrm{P,lit}<1$.
The number of such sources is reported in the fourth column in Table~\ref{Tab:XM_statistics}.
If we also include sources with \Gaia periods that are half or twice the literature periods (replacing $P_\mathrm{lit}$ by $0.5\,P_\mathrm{lit}$ or $2\,P_\mathrm{lit}$ in Eq.~(\ref{Eq:fracPlit})), the percentage of sources having compatible \Gaia and literature data increases to 93\% (fifth column in the table).

In contrast, less than 6\% of sources not classified as eclipsing binaries in the literature have equal \Gaia and literature periods (red histogram in Fig.~\ref{Fig:histoFracRecovery_cumul} and fourth column in Table~\ref{Tab:XM_statistics}).
Interestingly, this number increases to 32\% when considering compatible periods  ($P_\mathrm{lit} / P_\mathrm{\Gaia} \simeq (0.5, 1, 2)$ in the table).
This can easily be understood if the sources have sinusoidal-like light curves.
The detected period can then easily be a factor of two the orbital period if it is an eclipsing binary or ellipsoidal variable, and a survey may pick either one of these periods.

The comparison between \Gaia and literature periods is shown in Fig.~\ref{Fig:PlitPgaia}.
The upper panel displays the periods of the crossmatches classified as eclipsing binaries in both \Gaia DR3 and the literature.
They are distributed as expected, with the presence of (mainly) $P_\mathrm{Gaia}$:$P_\mathrm{lit}$ = 1:2, 2:1 and 2:3 ratios in addition to the overwhelming 1:1 cases.
Alias features are also seen.
The distribution of the \Gaia eclipsing binary candidates crossmatched with sources classified as non eclipsing binaries in the literature, on the other hand, reveals the imprints of the underlying literature catalogues in the distributions of $P_\mathrm{lit}$ (bottom panel in the figure).
At literature periods below one day, we see the imprint of $\sim$31\,600 sources from \texttt{\small PS1\_RRL\_SESAR\_2017} \citep{SesarBranimirHernitschek_etal17}, while the main contribution at literature periods above twenty days comes from $\sim$17\,200 sources from \texttt{\small ATLAS\_VAR\_HEINZE\_2018} \citep{HeinzeTonryDenneau_etal18}.
The former catalogue targets RR~Lyrae variables, while the cross-matches in the latter catalogue were assigned the tailored `OMIT' classification type in \citet{DR3-DPACP-177} to gather sources whose classification in the literature were considered to be `too generic, uncertain, or with insufficient variability characterisation' \citep[see][]{DR3-DPACP-177}.
In \citet{HeinzeTonryDenneau_etal18}, they are mainly assigned, in decreasing order of number of crossmatches with our eclipsing binaries, the types NSINE (pure sine wave fit, but noisy data), SINE, or MSINE (modulated sine wave).
These crossmatches not classified as eclipsing binaries in the literature or considered uncertain by \citet{DR3-DPACP-177} are, however, a minority in the full sample of crossmatches.

In summary, the \Gaia periods are compatible with literature periods in about 85\% of cases.
This includes cases where the literature period is twice or half the \Gaia period.

%------------------------------------------------------------------
\subsection{Completeness of the \Gaia catalogue}
\label{Sect:quality_completeness}

To estimate the completeness of our catalogue, we compare it with the OGLE4 catalogues of eclipsing binaries available, which are available for the Large (LMC) and small (SMC) Magellanic Clouds \citep{PawlakSoszynskiUdalski_etal16} and for the Galactic Bulge \citep{SoszynskiPawlakPietrukowicz_etal16}.
The sky distribution of the \Gaia DR3 eclipsing binaries towards the LMC and Galactic Bulge are displayed in the top panels of Figs.~\ref{Fig:skyLMC_DR3versusOgle4} and \ref{Fig:skyBLG_DR3versusOgle4}, respectively, and the distributions of the OGLE4 eclipsing binaries in the second panels.
Sources in common in \Gaia DR3 and OGLE4 catalogues are shown in the third panels.

Two steps are achieved in order to estimate the completeness of the \Gaia catalogue relative to the OGLE4 catalogues.
We first restrict the OGLE4 catalogues to sources present in the full \Gaia DR3 archive, with a cross-match search radius of one arcsecond.
The statistics are given in Table~\ref{Tab:OGLE4}.
\Gaia cross-matches are found for practically all OGLE4 eclipsing binaries in the LMC and SMC, but for only 87\% of the OGLE4 sources in the Galactic Bulge (420\,321/473\,798 in Table~\ref{Tab:OGLE4}).
The 13\% OGLE4 sources from the Bulge that are not in the \Gaia archive are all very red faint sources with OGLE \OgleI magnitudes mainly between 18 and 20.5~mag.
Their sky distribution is shown in Fig.~\ref{Fig:skyBLG_Ogle4_notInDR3}.

We then limit the OGLE4 samples to sources brighter than 20~mag in \gmag to comply with the input magnitude selection of the \Gaia eclipsing binaries (see Sect.~\ref{Sect:pipeline_input}).
The final OGLE4 samples contain 35\,392, 7843 and 315\,523 sources in the LMC, SMC and Galactic Bulge%
\footnote{
Most OGLE4 eclipsing binaries in the Magellanic Clouds are on the main sequence.
In contrast, in the Galactic Bulge, they are red sources.
This is reflected in their \OgleI, \OgleV and \gmag magnitudes shown in Fig.~\ref{Fig:histo_mags_Ogle_DR3} in Appendix~\ref{Appendix:additionalFigures}.
}%
, respectively (see Table~\ref{Tab:OGLE4}).

From these OGLE4 samples, 28\% are present in the \Gaia catalogue of eclipsing binaries  (36\% had we considered 19~mag as the faintest limit for both OGLE4 and \Gaia catalogues) .
The recovery rates are larger in the Magellanic Clouds than in the Bulge, as detailed in Table~\ref{Tab:OGLE4}, reaching 48\% in the SMC while being 26\% in the Bulge.
An investigation of the 72\% missing OGLE4 sources reveals that $\sim$45\% were excluded from the initial selection (see Sect.~\ref{Sect:pipeline_input}, with $\sim$40\% not being classified as eclipsing binaries and another $\sim$5\% having less than sixteen measurements in their \gmag light curves, mainly in the Bulge).
The remaining $\sim$27\% of missing sources were further filtered out from the final selection procedure (Sect.~\ref{Sect:catalogue_selection}).

A small fraction of the missing OGLE4 eclipsing binaries that were not classified as eclipsing binaries in \Gaia DR3 are present in other variability tables published in DR3 (tables \texttt{gaiadr3.vari\_*} in the \Gaia archive).
They consist of
2195 short time-scale variables,
426 binary candidates with a compact companion,
384 long-period variables,
89 main-sequence oscillators,
32 rotation modulation variables,
31 Cepheids,
and one Active Galactic Nucleus.

In summary, the completeness of the \Gaia catalogue of eclipsing binaries amounts to between 25\% and 50\% depending on the sky region, when compared to the OGLE4 catalogues of eclipsing binaries.
The missing OGLE4 sources were excluded from the \Gaia catalogue at candidate selection steps in our processing pipeline.
A significant increase in the number of eclipsing binary candidates is thus expected for the next \Gaia release, DR4.

%------------------------------------------------------------------
\subsection{New \Gaia candidates}
\label{Sect:quality_newGaia}

We investigate in the section the \Gaia eclipsing binary candidates that are not present in the OGLE4 catalogues of eclipsing binaries, using the LMC and Galactic Bulge regions as test cases.
For this purpose, two sky areas well covered by the OGLE4 surveys are defined towards these regions:
the rectangle area
$70^\mathrm{o}$\,<\,$\mathrm{RA}$\,<\,$90^\mathrm{o}$ and
$-73^\mathrm{o}$\,<\,$\mathrm{DEC}$\,<\,$-65^\mathrm{o}$ towards the LMC,
and the parallelogram area with corners
(RA, DEC) =
$(269^\mathrm{o}, -37^\mathrm{o})$,
$(275^\mathrm{o}, -30^\mathrm{o})$,
$(275^\mathrm{o}, -23^\mathrm{o})$,
$(265^\mathrm{o}, -37^\mathrm{o})$ towards the Galactic Bulge.
They are shown by the orange borders in Figs.~\ref{Fig:skyLMC_DR3versusOgle4} and \ref{Fig:skyBLG_DR3versusOgle4}, respectively.
The statistics on \Gaia and OGLE4 sources in these regions are summarised in Table~\ref{Tab:newWrtOgle4}.
There are 26\,020 \Gaia eclipsing binary candidates towards the LMC and 96\,199 sources towards the Galactic Bulge.
More than half of them are new relative to the OGLE4 catalogues (53\% in the LMC and 65\% in the Galactic Bulge, see Table~\ref{Tab:newWrtOgle4}).

The sky distribution of the new \Gaia candidates towards the LMC is shown in the bottom panel of Fig.~\ref{Fig:skyLMC_DR3versusOgle4}.
They are mostly concentrated in the bar, where the sky density of stars is highest.
Their magnitude distribution peaks around 19~mag in \gmag (red histogram in the top panel of Fig.~\ref{Fig:histo_medianMag}), similarly to the magnitude distribution of the full \Gaia sample in the defined sky area (black histogram).
In contrast, the magnitude distribution of the \Gaia--OGLE4 crossmatch sample reveals a plateau between $\sim$18.5 and $\sim$19.5~mag (green histogram).
The origin of this plateau is unclear, as the full OGLE4 sample in the defined sky area shows a continuously increasing distribution of \gmag up to 20~mag (not shown here).
We checked that the new faint sources are not contaminated by the potential presence of a nearby brighter eclipsing binaries.

The distribution of the new \Gaia candidates towards the LMC in the colour-magnitude diagram is shown in the upper-left panel of Fig.~\ref{Fig:CM_LMC}, together, in the upper-middle panel, with the distribution of the \Gaia candidates that have an OGLE4 crossmatch.
The new \Gaia candidates are seen to lie not only on the main sequence, as is the case for the OGLE4 crossmatches, but also, for one third of them, on the red side of the diagram at $\BPminusRP > 0.6$~mag.
Such red candidates are much less abundant in the cross-matched sample.
We also note in this regard the larger colour dispersion observed at the faint end of the main sequence for the new \Gaia candidates (upper-left panel in Fig.~\ref{Fig:CM_LMC}) compared to the narrower colour dispersion for the crossmatched sample (upper-middle panel).
The correctness of the red colours must, therefore, be checked as the \gbp and \grp values of these faint sources may be affected by residual background estimates and/or multiple source blending in the BP and RP spectra.
The \BPRPexcess $C$ available in the \Gaia archive (\texttt{phot\_bp\_rp\_excess\_factor} in the \texttt{gaia\_source} table) provides a handy tool for this purpose (assuming \gmag is correct).
This quantity, notated $C$ by these authors, evaluates the excess of the integrated BP and RP fluxes in comparison with the \gmag flux \citep{Riello_etal21}.
It is shown versus \BPminusRP in the bottom-left panel of Fig.~\ref{Fig:CM_LMC}.
While many sources are seen to have an excess factor between 1.1 and 1.2, as expected, a non-negligible fraction of them have excess factors significantly above 1.2.
This is particularly true for the red sources.
These large excess factors lead to unreliable \gbp and/or \grp magnitudes, and can thus be at the origin of both the large colour dispersion observed at the faint end of the main sequence and the presence of an excess of red sources at $\gmag \simeq 19$~mag.
We note that the value of $C$ of the crossmatched sources have much cleaner distribution around the expected values (bottom-middle panel of Fig.~\ref{Fig:CM_LMC}).

The \gmag values, on the other hand, should be reliable, within the uncertainties expected at the faint magnitudes.
A visual inspection of the \gmag light curves of the new \Gaia candidates provides confidence that at least one third of them are genuine eclipsing binaries.
% In topcat: (inLMC_regionForCompleteness && isEclInLit && !cat_OGLE4) -> 2370 -> 2370/13897 = 0.170
In fact, about 17\% of these new \Gaia candidates towards the LMC with no OGLE4 crossmatch are classified as eclipsing binaries in other surveys, using crossmatches from \citet{DR3-DPACP-177}.
They were mainly identified from the EROS2 survey by \citet{KimProtopapasBailerJones_etal14}.%, and some from, among other catalogues, the ASAS-SN survey \citep{ASASSN} [Only 64, compared to >2000 in EROS2 -> no mention of it].

Eleven examples of good light curves from the \Gaia new candidates not in the OGLE4 catalogue are shown in Figs.~\ref{Fig:lcs_hasNotOgle4_outOfBar} and \ref{Fig:lcs_hasNotOgle4_inBar}.
Figure~\ref{Fig:lcs_hasNotOgle4_outOfBar} shows sources located outside the bar of the LMC.
Two of them, sources \GaiaSrcId{4651581458546544000} and \GaiaSrcId{4655269942119876864}, are also referenced as eclipsing binaries in EROS2.
The light curves in Fig.~\ref{Fig:lcs_hasNotOgle4_inBar} are from sources located in the bar of the LMC.
Their sky positions are indicated in Fig.~\ref{Fig:skyLMC_DR3versusOgle4}, in blue for the sources outside the LMC bar and in cyan for those inside the bar.
All eleven light curves show a variety of geometric morphologies typical of eclipsing binaries.

The global rankings of the new \Gaia candidates (with respect to OGLE4) towards the LMC are, on the mean, much smaller than the ones of the \Gaia--OGLE4 crossmatches.
This is seen from the global ranking distributions shown in the top panel of Fig.~\ref{Fig:histo_rankingOgle4}, in red for the new candidates and in green for the crossmatch sample.
Since the majority of these new \Gaia candidates are faint and lie in crowded regions of the LMC, the eclipsing binary signature in the \gmag light curves can be mingled with variability of non-astrophysical origin.
This would explain their low global rankings.
The fraction of new \Gaia candidates is much smaller if we limit the samples to sources with larger global rankings.
In a sample limited to global rankings larger than 0.5, for example, the fraction of new \Gaia candidates towards the LMC is three times less than in the full sample (17\% new candidates with respect to OGLE4, compared to 53\% considering all global rankings, see Table~\ref{Tab:newWrtOgle4}).
It is interesting to note that the global ranking distribution of the \Gaia candidates not in OGLE4 but identified as eclipsing binaries in other surveys (blue histogram in the Fig.~\ref{Fig:histo_rankingOgle4}) peaks at values between those of the new candidates and those of the \Gaia--OGLE4 crossmatch sample.
The distribution in the colour-magnitude diagram of the new \Gaia candidates is shown in the upper-right panel of Fig.~\ref{Fig:CM_LMC}, and their {\BPRPexcess}s in the lower-right panel.

The situation of the new \Gaia candidates in the Galactic Bulge with respect to OGLE4 is very similar to the situation described above for the LMC, with the additional observation that the sky density is much higher in the Galactic Bulge than in the LMC, and the sources are much redder due to heavy extinction.
As a result, $C$ reaches very large values, above three, as shown in Fig.~\ref{Fig:bprpExcess_BLG}.
This is particularly true for the new \Gaia candidates having no OGLE4 crossmatch (bottom panel in the figure) as compared to the sample with OGLE4 crossmatch (top panel).
Among these new \Gaia candidates compared to OGLE4, only very few have a detection as an eclipsing binary in other surveys.
Their magnitude and global ranking distributions are shown in blue in the bottom panels of Figs.~\ref{Fig:histo_medianMag} and \ref{Fig:histo_rankingOgle4}, respectively (note the one hundred times amplification factor compared to the distributions of the other samples).
The larger sky crowdedness towards the Galactic Bulge may explain the larger fraction of new \Gaia candidates in that region of the sky (65\% compared to 53\% towards the LMC, see Table~\ref{Tab:newWrtOgle4}).
This fraction is still 35\% in a sample limited to \texttt{global ranking}\;>\;0.5 towards the Galactic Bulge.

In summary, more than half of the \Gaia eclipsing binaries are new discoveries, the percentage being larger in crowded regions than in less dense regions of the sky.
They generally have low global rankings and large {\BPRPexcess}s, requiring to be cautious when using their \gbp and \grp magnitudes.
They, however, show genuine \gmag light curves of eclipsing binaries in many cases.

%==================================================================
\section{Illustrative samples with good parallaxes}
\label{Sect:overview}

%-----------
\begin{figure}
  \centering
  \includegraphics[trim={0 0 0 42},clip,width=\linewidth]{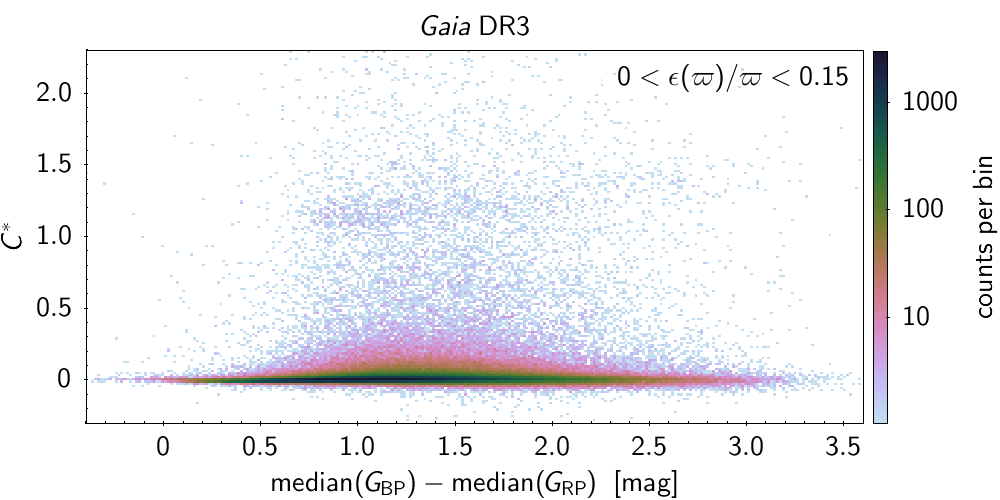}
  \caption{Corrected \BPRPexcess versus colour of \Gaia DR3 eclipsing binaries having positive parallax uncertainties better than 15\%.
           Median values of \gbp, \grp and \gmag are used in all quantities.
           The axes ranges have been truncated for better visibility.
          }
    \label{Fig:bprpExcessCorrected_par15perc}
\end{figure}
%-----------

%-----------
\begin{figure}
  \centering
  \includegraphics[trim={0 0 0 42},clip,width=\linewidth]{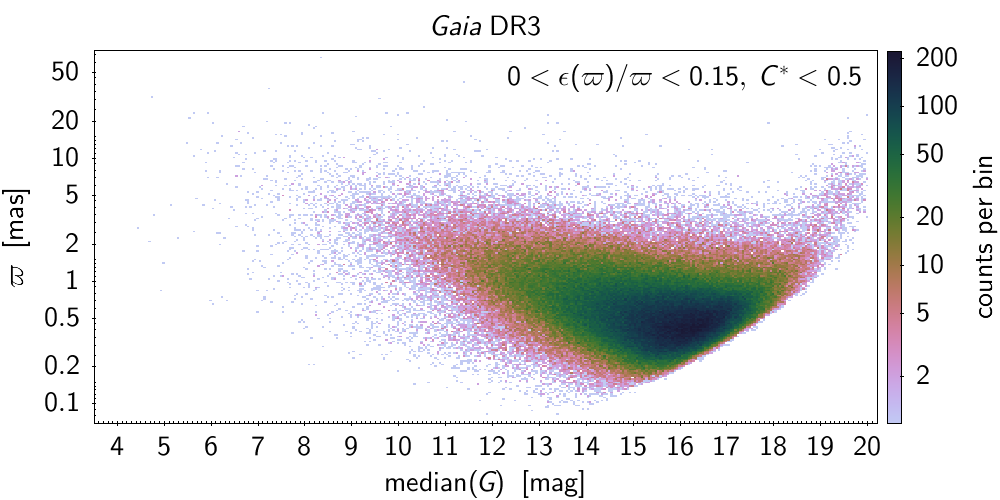}
  \caption{Parallax in milliarcsec versus median(\gmag) of \Gaia DR3 eclipsing binaries having positive parallax uncertainties better than 15\% and corrected \BPRPexcess smaller than 0.5.
          }
    \label{Fig:parallaxVsG}
\end{figure}
%-----------

%-----------
\begin{figure}
  \centering
  \includegraphics[trim={0 80 0 50},clip,width=\linewidth]{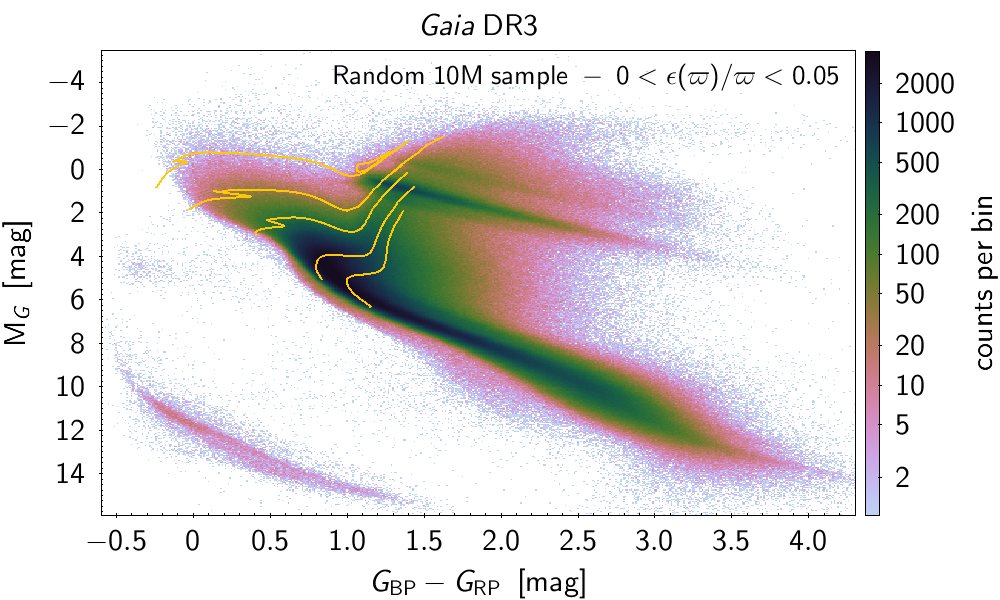}
  \vskip -0.5mm
  \includegraphics[trim={0 80 0 50},clip,width=\linewidth]{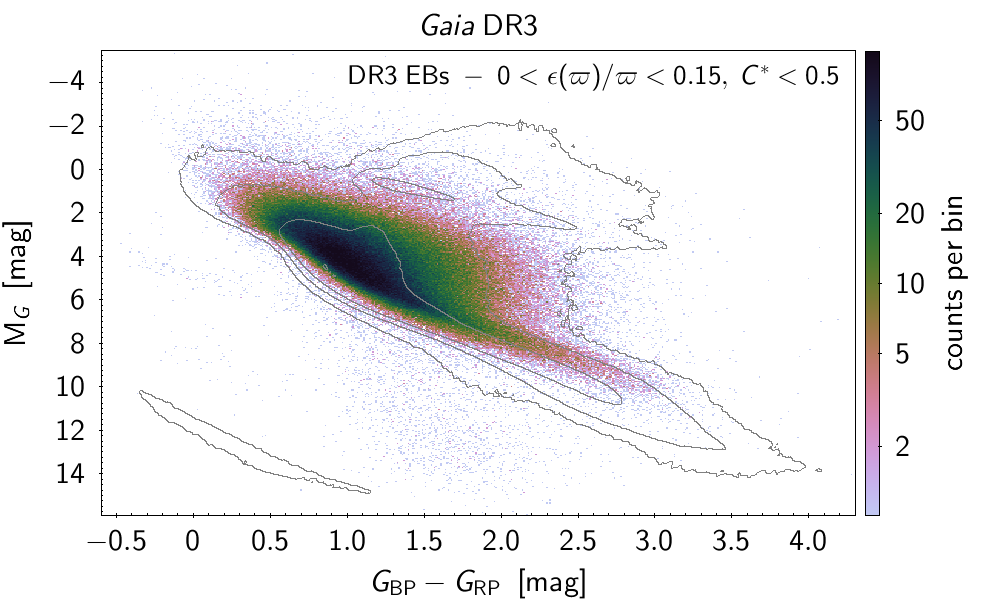}
  \vskip -0.5mm
  \includegraphics[trim={0 80 0 50},clip,width=\linewidth]{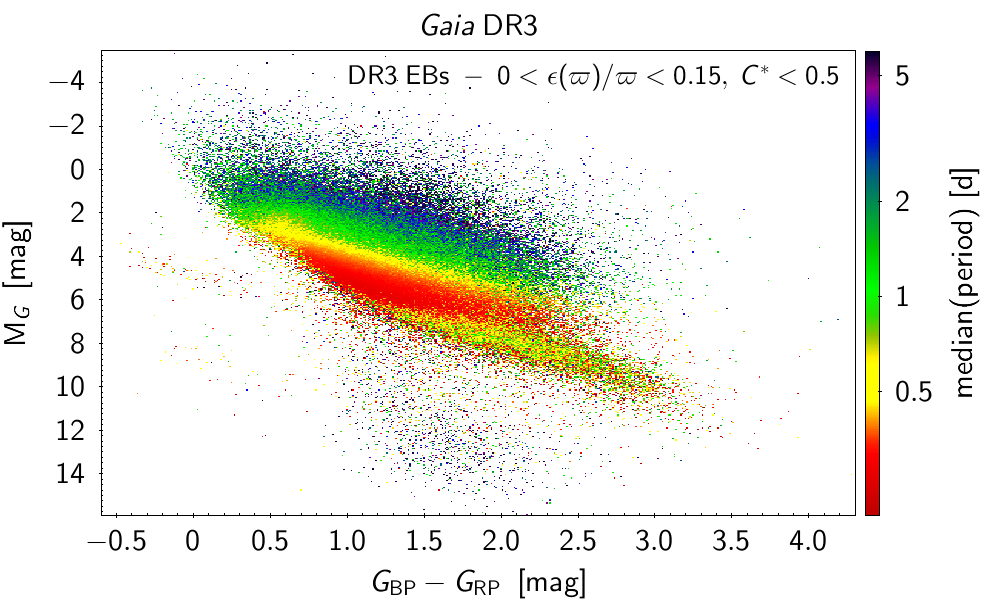}
  \vskip -0.5mm
  \includegraphics[trim={0 0 0 50},clip,width=\linewidth]{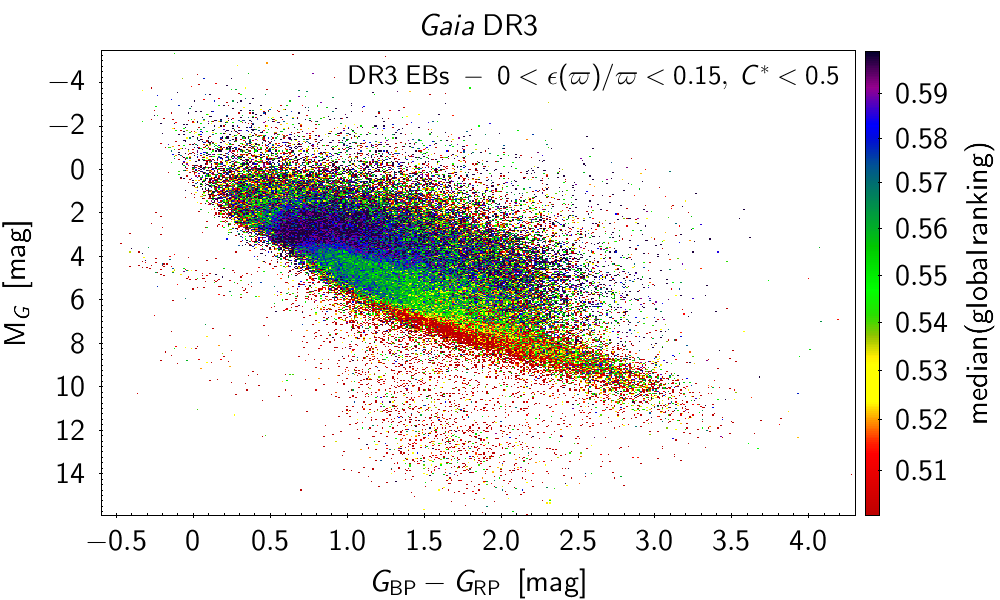}
  \caption{Observational HR diagrams of \Gaia DR3 stars with good parallaxes.
           \textbf{Top panel:} Density map of a random sample of ten million stars with parallax uncertainties better than 5\% and additional conditions on the number of measurements and source image quality (see text).
           The yellow lines are evolution tracks of, from bottom to top, 0.8, 1, 1.5 and 2~M$_\odot$ solar-metacclicity stellar models from \citet{EkstromGeorgyEggenberger_etal12}. 
           \textbf{Second panel:} Density map of \Gaia eclipsing binary candidates with parallax uncertainties better than 15\% and corrected {\BPRPexcess}s less than 0.5.
           Contour lines (logarithmic scale) of the sample shown in the top panel are drawn in grey.
           \textbf{Third panel:} Same as second panel, but colour-coded with the median value of the orbital period in each bin.
           \textbf{Bottom panel:} Same as second panel, but colour-coded with the median value of the global ranking in each bin.
           Median values of the \gbp, \grp and \gmag cleaned time series are used in all panels except in the top one, where mean values are used due to the unavailability of median values for all sources in \Gaia DR3.
           The colours in the figures are coded according to the colour scales on the right of each panel.
           The ranges of the axes and colour scales are truncated for better visibility.
          }
\label{Fig:obsHR_15perc}
\end{figure}
%-----------

%-----------
\begin{figure}
  \centering
  %\includegraphics[trim={0 80 0 0},clip,width=\linewidth]{Figures/figObsHR_EBs_15perc_rank0p6.png}
  %\vskip -0.5mm
  \includegraphics[trim={0 0 0 50},clip,width=\linewidth]{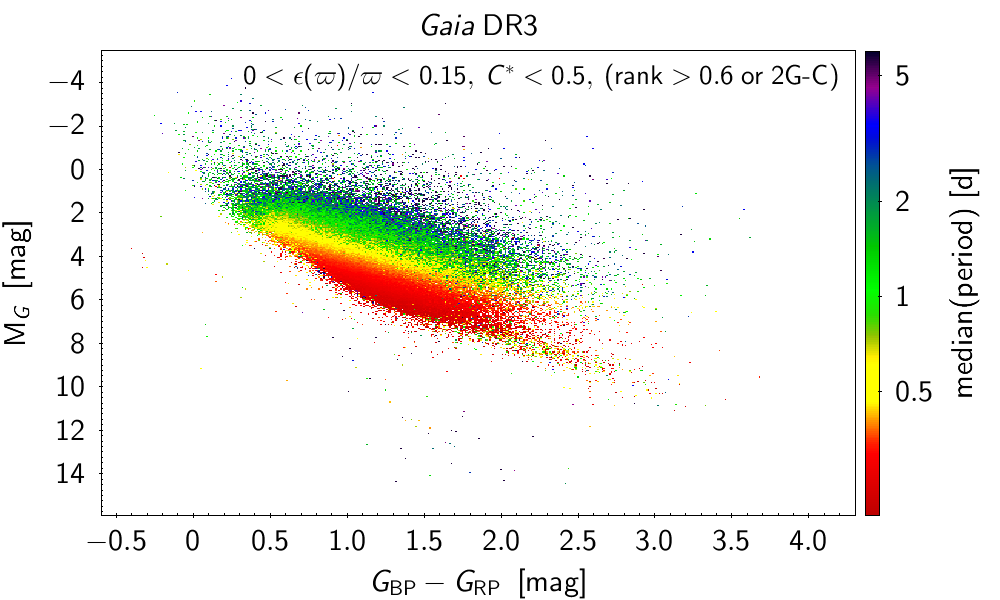}
  \caption{Same as third panel of Fig.~\ref{Fig:obsHR_15perc}, but for the subset with global rankings larger then 0.6 or belonging to the group 2G-C. 
           %\textbf{Top panel:} Density map similar to the second panel of Fig.~\ref{Fig:obsHR_15perc}.
           %\textbf{Bottom panel:} Colour-coded with the median value of the orbital period in each bin similar to the third panel of Fig.~\ref{Fig:obsHR_15perc}.
           %The axes ranges are kept similar to those in Fig.~\ref{Fig:obsHR_15perc} for easy comparison. 
          }
\label{Fig:obsHR_15perc_rank0p6}
\end{figure}
%-----------

%-----------
\begin{figure}
  \centering
  \includegraphics[trim={0 77 0 42},clip,width=\linewidth]{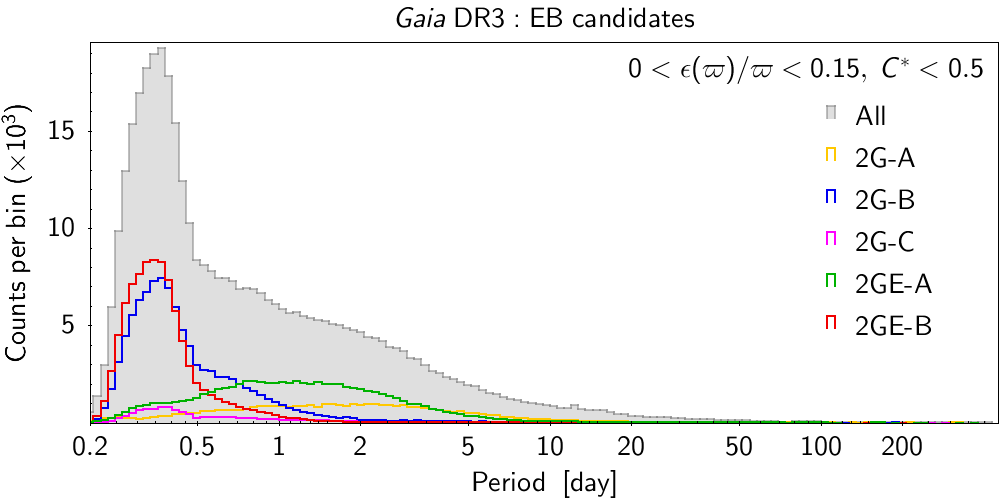}
  \vskip -0.25mm
  \includegraphics[trim={0 0 0 42},clip,width=\linewidth]{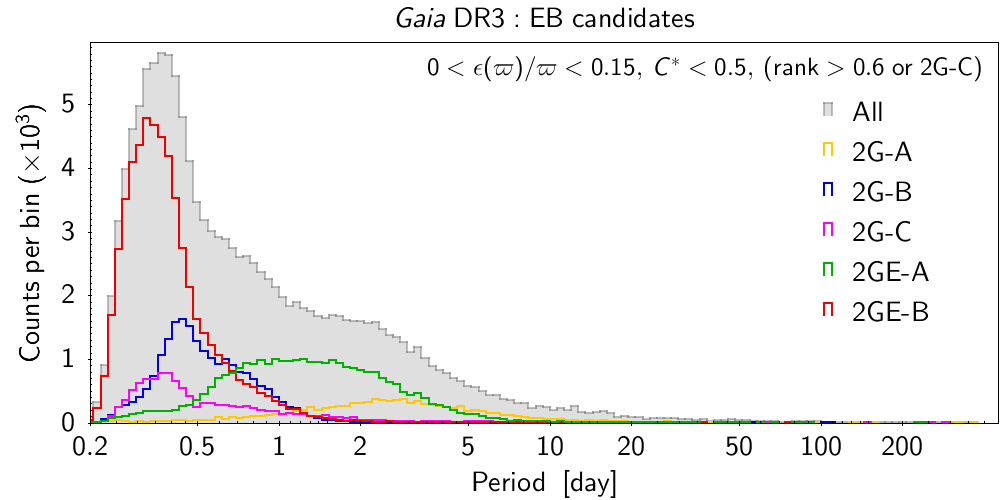}
  \caption{Period distributions of the good parallax sample.
           The filled grey histogram represents the full distribution, while the orange, blue, magenta, red and green histograms represent the 2G-A, 2G-B, 2G-C, 2GE-A and 2GE-B subsamples, respectively.
           \textbf{Top panel:} All candidates in the good parallax sample with $C^*<0.5$.
           \textbf{Bottom panel:} Same as top panel, but restricted to candidates with global rankings larger than 0.6 except for group 2G-C.
          }
    \label{Fig:histo_period_15perc}
\end{figure}
%-----------

%topcat -stilts plot2plane \
%   xpix=1450 ypix=590 \
%   xlog=true xlabel='(median_mag_g_fov<13)? 1/frequency : NULL' ylabel='max(e_fromModel, 1.e-3)' \
%   xmin=0.2 xmax=220 ymin=-0.02 ymax=1.1 \
%   auxmap=rainbow auxclip=0.089,1 auxquant=10 auxmin=10 auxmax=12 \
%   auxvisible=true auxlabel='median_mag_g_fov / mag' \
%   legend=true legborder=false legopaque=false legpos=0.0,1.0 \
%   in='Copy of 2' icmd='select sample_2G_A' x='(median_mag_g_fov<13)? 1/frequency : NULL' y='max(e_fromModel, 1.e-3)' leglabel='4: sample_2G_A' \
%   layer_1=XYError \
%      xerrhi_1=0 xerrlo_1=0 yerrhi_1=0 yerrlo_1=e_fromModel_error \
%      shading_1=auto colour_1=light_grey \
%   layer_2=Size \
%      size_2='0.5+derived_secondary_ecl_depth / derived_primary_ecl_depth' aux_2=median_mag_g_fov \
%      shading_2=aux scale_2=0.32 \
%   layer_3=Function \
%      fexpr_3='0.98 - 3.25 * exp(-pow(6.3*x,0.23))' colour_3=blue thick_3=2 dash_3=8,4 \
%   legseq=_1,_2 

%-----------
\begin{figure*}
  \centering
  \includegraphics[trim={0 0 0 50},clip,width=\linewidth]{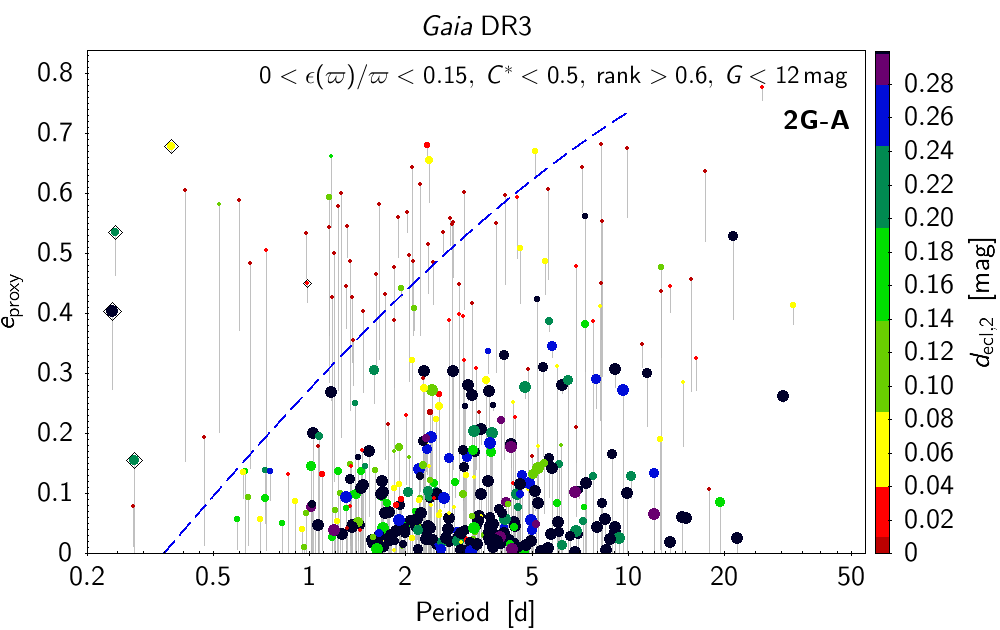}
  \caption{Eccentricity proxy versus orbital period of Sample~2G-A eclipsing binaries (well-detached and without ellipsoidal component) that are brighter than 12~mag in \gmag and that have parallax uncertainties better than 15\%, corrected \BPRPexcess smaller than 0.5, and global rankings larger than 0.6.
           The colour of each marker is related to the depth of the secondary eclipse (in \gmag magnitude) according to the colour-scale drawn on the right of the figure, with ratios larger than 0.45 rendered in black.
           The size of each marker is proportional to the secondary over primary depth ratio.
           The vertical line segments indicate the 1-$\sigma$ uncertainty of the eccentricity proxy.
           For clarity, it has only been drawn on the small eccentricity side.
           The blue dashed line is Eq.~(4.4) from \citet{Mazeh08} with $E=0.98$, $A=3.25$, $B=6.3$ and $C=0.23$.
           The five sources above this line that are highlighted with open diamonds have their light curves displayed in Fig.~\ref{Fig:lcs_tooLargeEcc}.
          }
\label{Fig:period_ecc_brightStars}
\end{figure*}
%-----------

%-----------
%\begin{figure*}
%  \centering
%  \includegraphics[width=\linewidth]{Figures/fig_period_ecc_brightStars_pLit.png}
%  \caption{Same as Fig.~\ref{Fig:period_ecc_brightStars}, but for $\eccProxy-\epsilon(\eccProxy)$ versus orbital period, and limited to  eclipsing binary candidates with secondary over primary eclipse depth ratios larger than 0.2.
%           Sources that have non-\textit{Gaia} periods published in the literature (79\% of the sample) are indicated with a white dot in the marker. 
%           In that case, the value of non-\textit{Gaia} period is shown by a small cross connected to the \Gaia period by a horizontal grey line.
%          }
%\label{Fig:period_ecc_brightStars_pLit}
%\end{figure*}
%-----------

%-----------
\begin{figure}
  \centering
  \includegraphics[trim={0 80 0 50},clip,width=\linewidth]{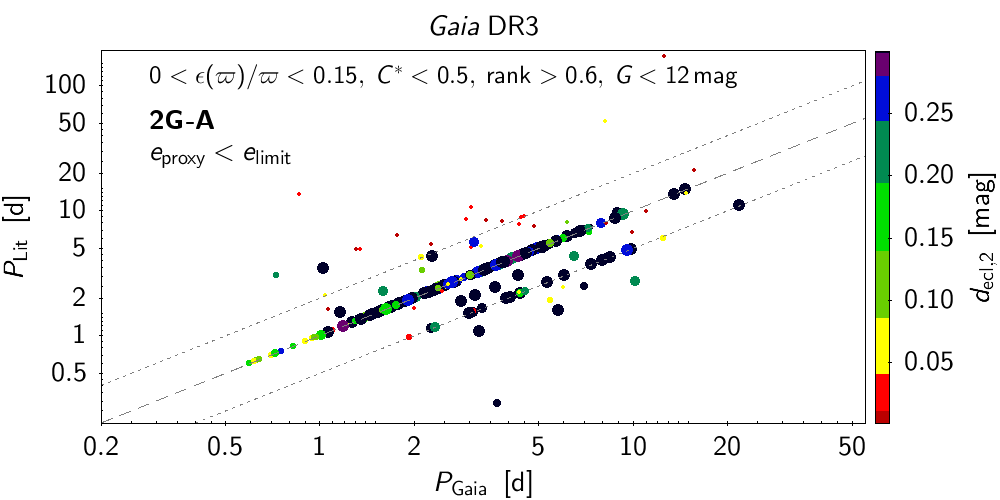}
  \vskip -0.5mm
  \includegraphics[trim={0 0 0 50},clip,width=\linewidth]{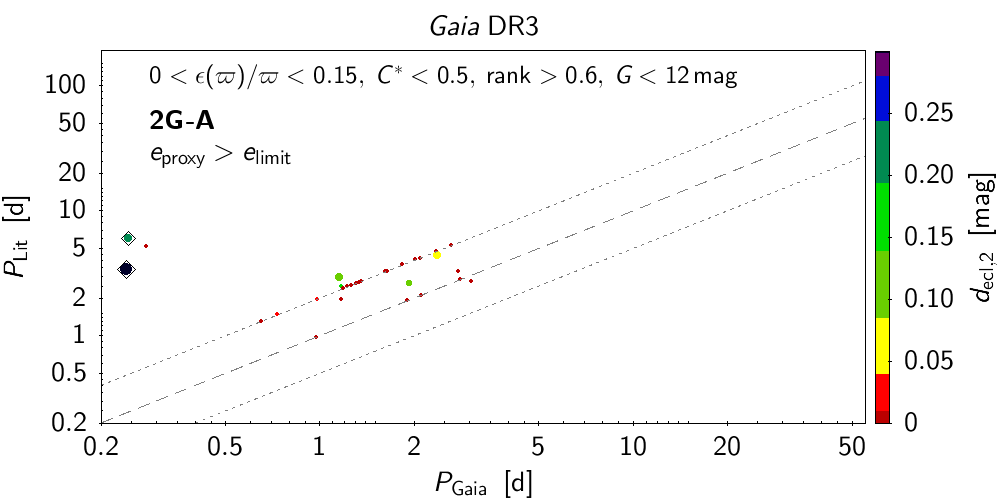}
  \caption{Same as Fig.~\ref{Fig:period_ecc_brightStars}, but for non-\Gaia literature period versus \Gaia period for those sources that have a period published in the literature.
           The dashed line is $P_\mathrm{Lit} = P_\mathrm{Gaia}$, and the dotted lines are $P_\mathrm{Lit} = 2\,P_\mathrm{Gaia}$ and $P_\mathrm{Lit} = 0.5\,P_\mathrm{Gaia}$ relations. 
           \textbf{Top panel:} sources that have eccentricity proxies smaller than the eccentricity limit shown in Fig.~\ref{Fig:period_ecc_brightStars}.
           \textbf{Bottom panel:} sources with eccentricity proxies larger than this limit.
          }
\label{Fig:periodVsPeriodLit_brightStars}
\end{figure}
%-----------

%-----------
\begin{figure}
  \centering
  \includegraphics[trim={40 150 0 70},clip,width=\linewidth]{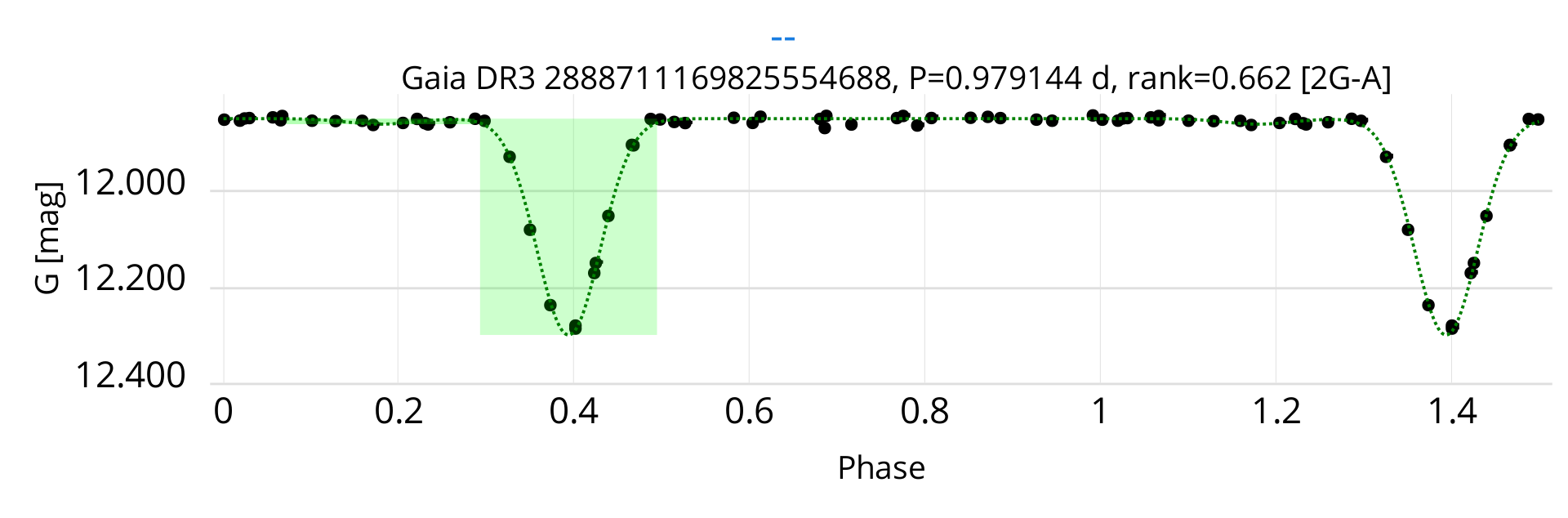} % 0.979 d
  \vskip -0.5mm
  \includegraphics[trim={40 145 0 80},clip,width=\linewidth]{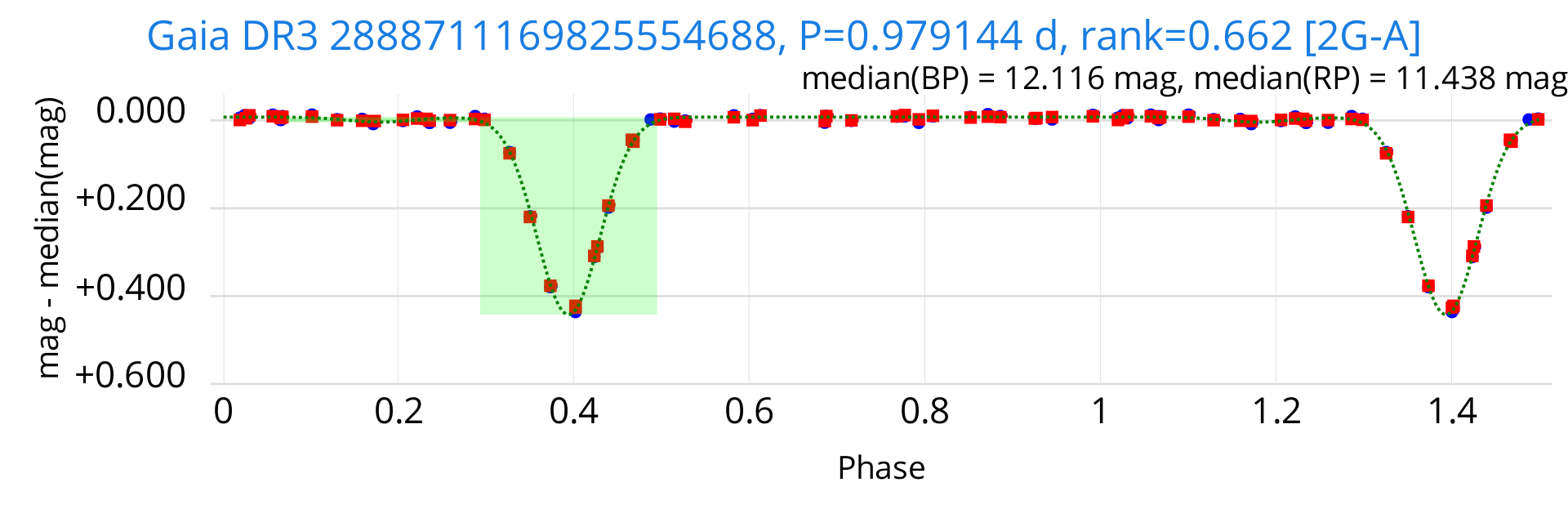} % 0.979 d
  \vskip -0.5mm
  \includegraphics[trim={40 150 0 70},clip,width=\linewidth]{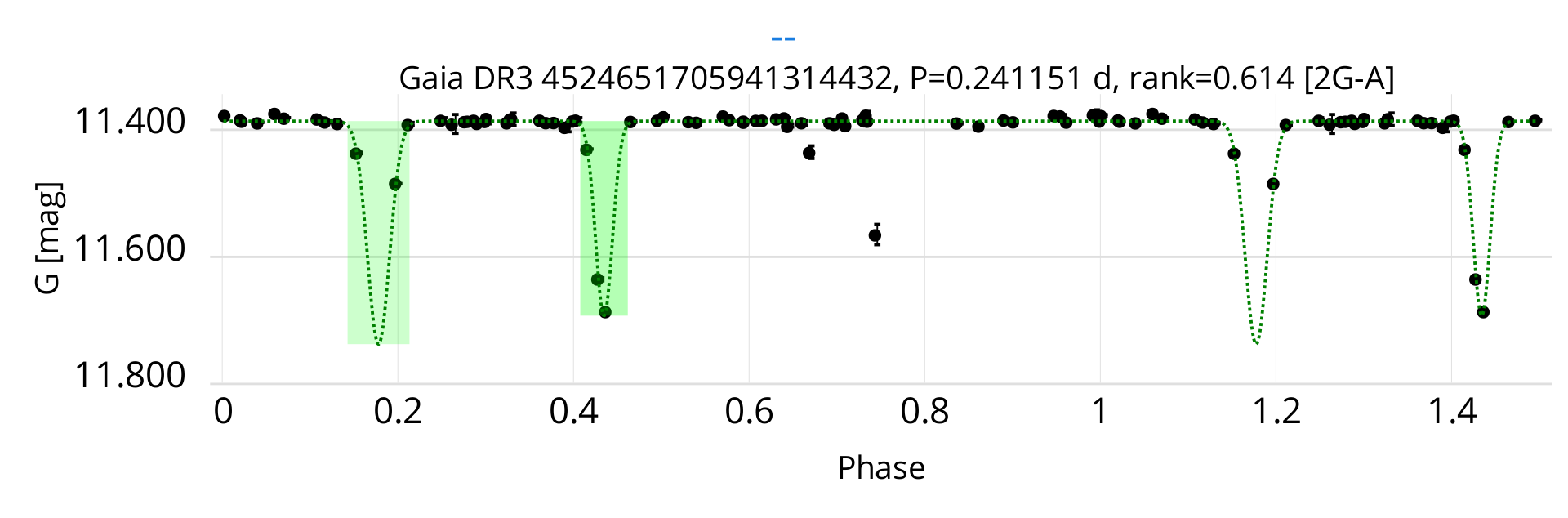} % 0.241 d
  \vskip -0.5mm
  \includegraphics[trim={40 145 0 80},clip,width=\linewidth]{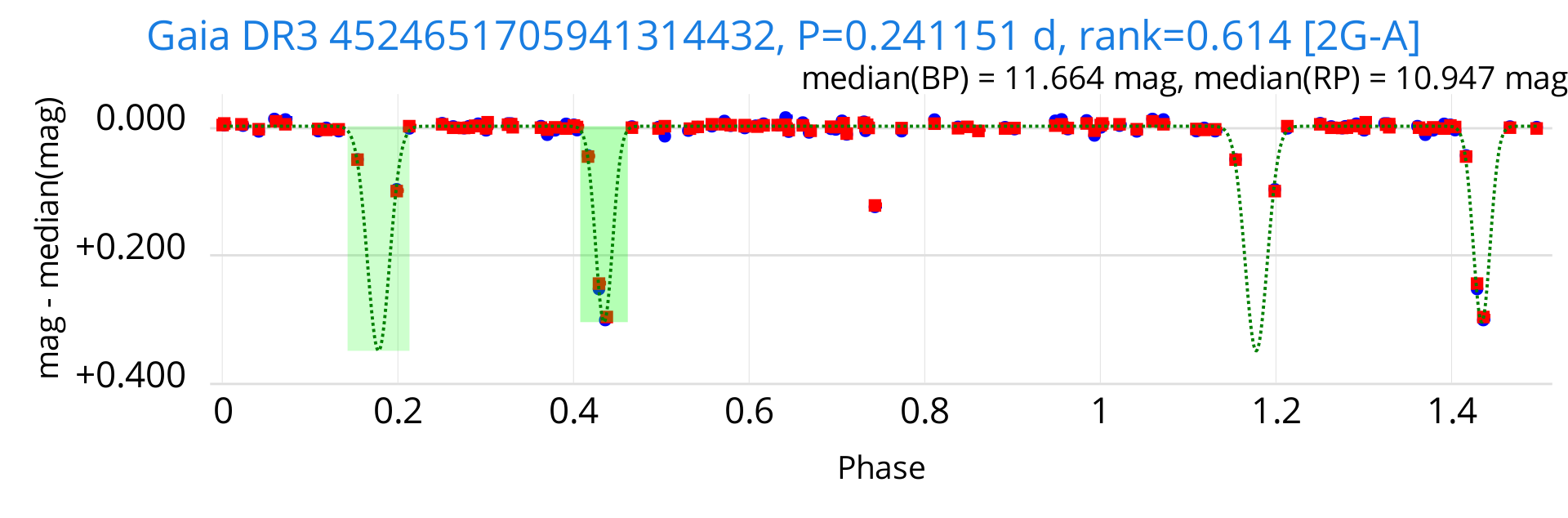} % 0.241 d
  \vskip -0.5mm
  \includegraphics[trim={40 150 0 70},clip,width=\linewidth]{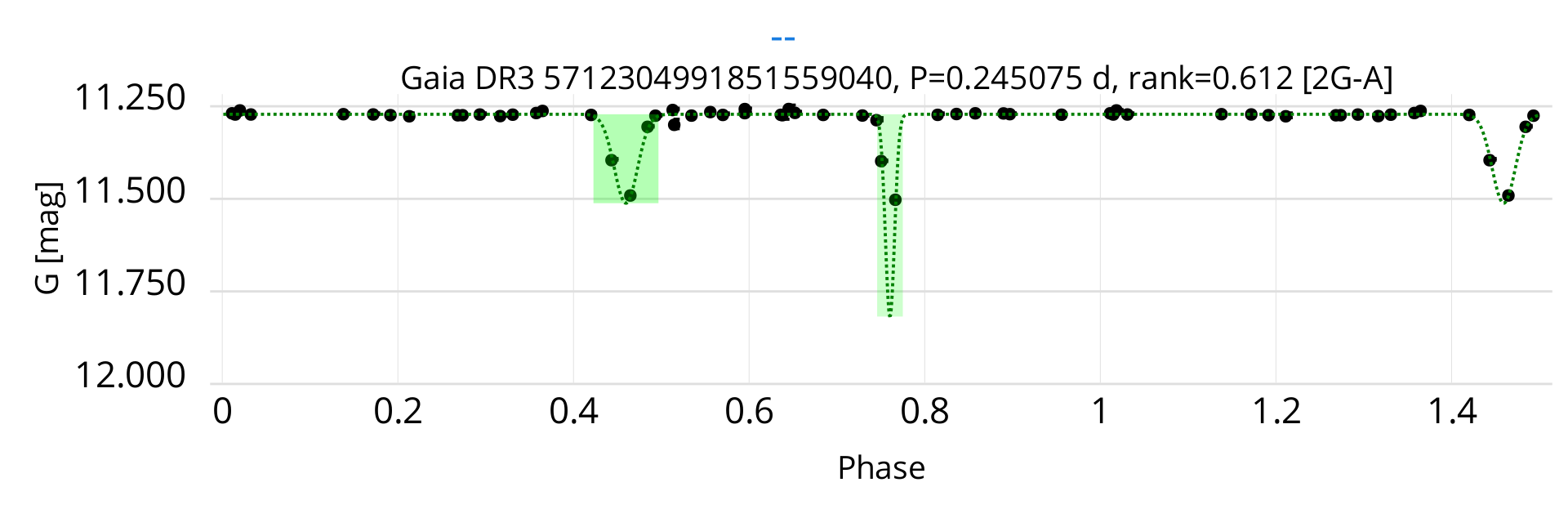} % 0.245 d
  \vskip -0.5mm
  \includegraphics[trim={40 145 0 80},clip,width=\linewidth]{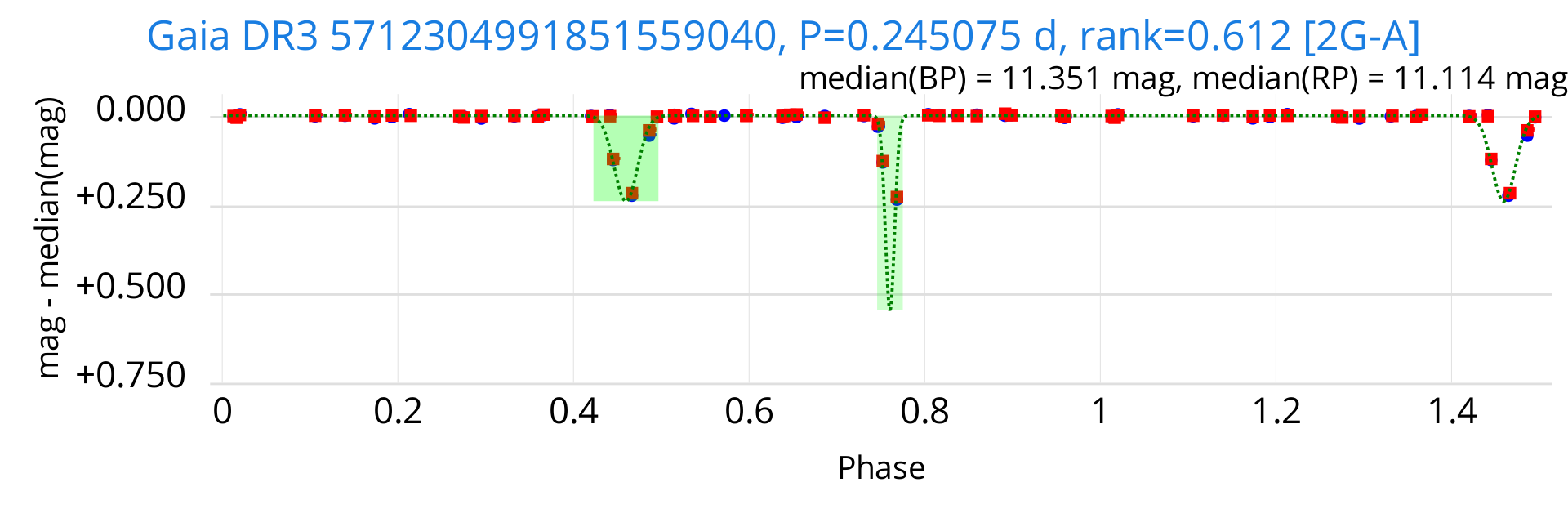} % 0.245 d
  \vskip -0.5mm
  \includegraphics[trim={40 150 0 70},clip,width=\linewidth]{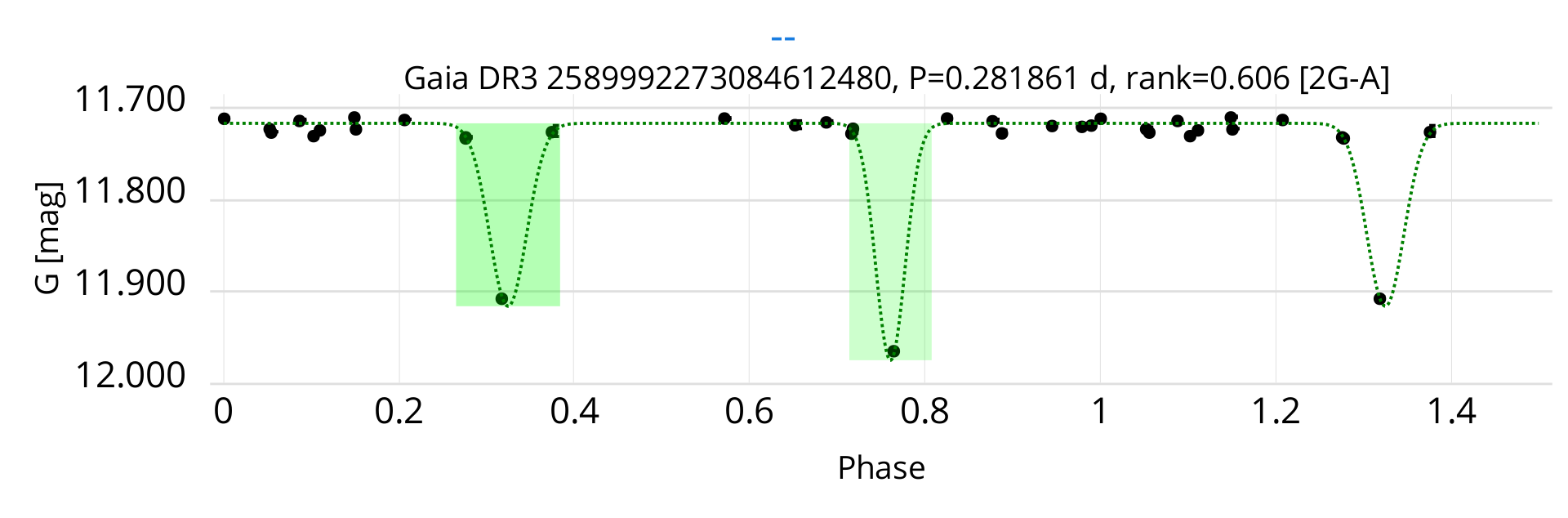} % 0.282 d
  \vskip -0.5mm
  \includegraphics[trim={40 145 0 80},clip,width=\linewidth]{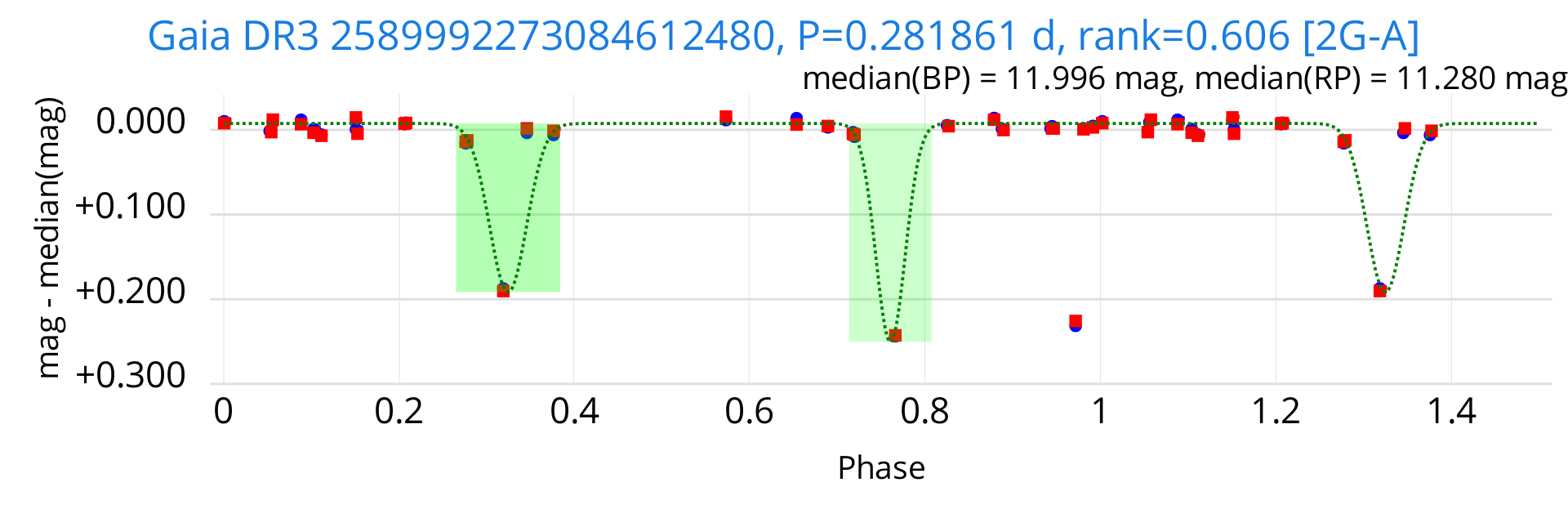} % 0.282 d
  \vskip -0.5mm
  \includegraphics[trim={40 150 0 70},clip,width=\linewidth]{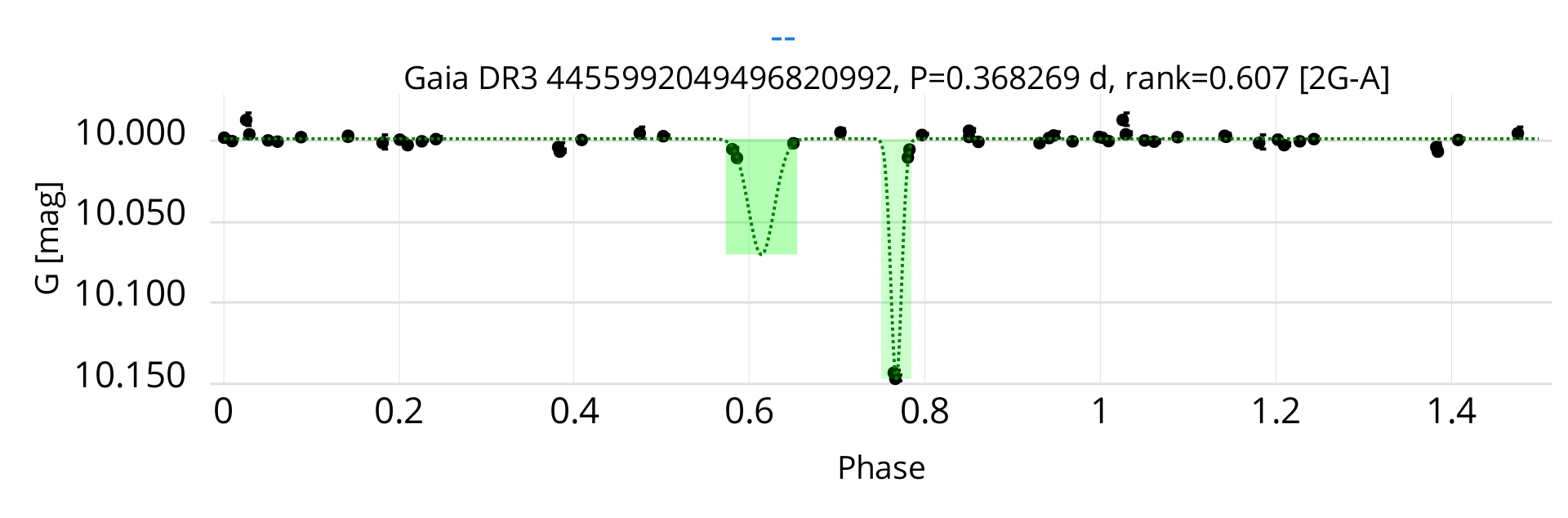} % 0.368 d
  \vskip -0.5mm
  \includegraphics[trim={40 45 0 80},clip,width=\linewidth]{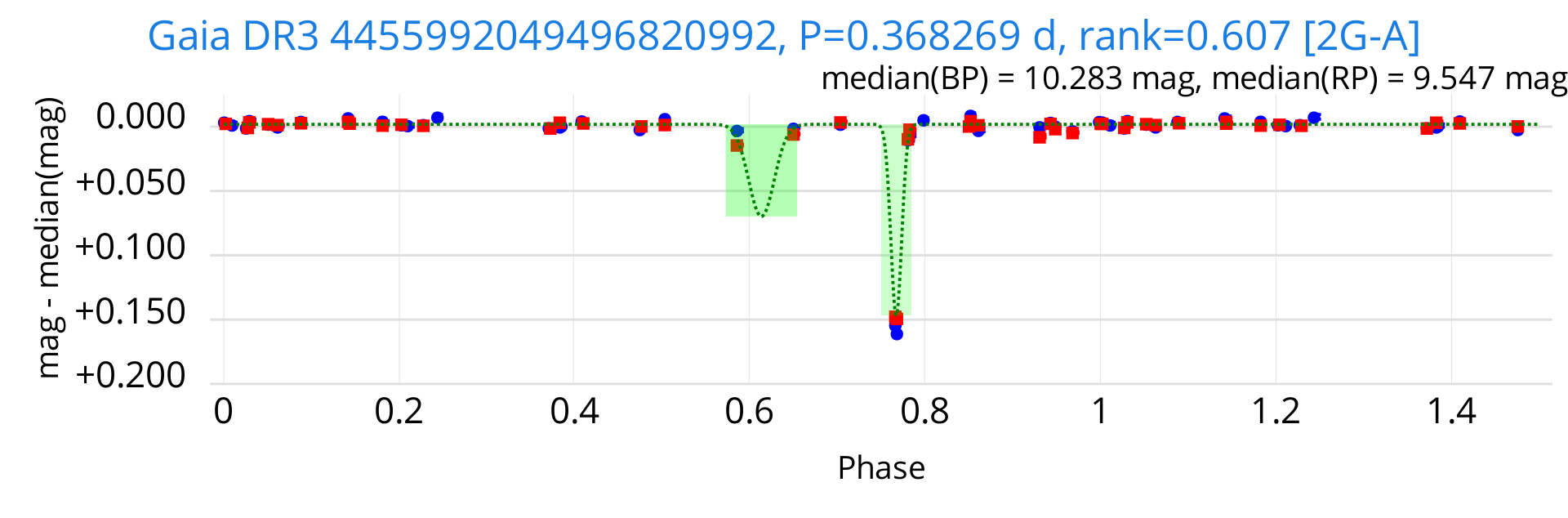} % 0.368 d
    \caption{Same as Fig.~\ref{Fig:lcs_example_2GA}, but for five bright candidates with eccentricity proxies larger than the expected limit shown in Fig.~\ref{Fig:periodVsPeriodLit_brightStars}.
             From top to bottom:
             \GaiaSrcIdInCaption{2888711169825554688, 4524651705941314432, 5712304991851559040, 2589992273084612480 and 4455992049496820992}.
            }
\label{Fig:lcs_tooLargeEcc}
\end{figure}
% -- out of bar:
% source_id==4655269942119876864L || source_id==4651511051157939968L || source_id==4651581458546544000L || source_id==4651853274236480256L || source_id==4651018852223135104L
% -- in bar:
% source_id==4657899076482979328L || source_id==4658068847785992064L || source_id==4657993978084740992L || source_id==4657143097661657728L || source_id==4657909281189517056L || source_id==4657914336358760448L
%-----------
% 2888711169825554688, 4524651705941314432, 5712304991851559040, 2589992273084612480, 4455992049496820992

Samples of the \Gaia DR3 eclipsing binary candidates towards the LMC and the Galactic Bulge have been briefly discussed in Sects.~\ref{Sect:quality_completeness} and \ref{Sect:quality_newGaia}.
In this section, we illustrate the catalogue with samples of candidates with good parallaxes.
We consider positive relative parallax uncertainties better than 15\% (409\,437 sources), and restrict to sources with good {\BPRPexcess}s to exclude obviously wrong \BPminusRP colours in colour-magnitude diagrams.
We use for this the corrected \BPRPexcess $C^*$ proposed by \citet{Riello_etal21}, using the \gbp and \grp median values in their Eq.~(6).
The distribution of this quantity for the sample with good parallaxes is shown in Fig.~\ref{Fig:bprpExcessCorrected_par15perc}.
We limit our sample to $C^*<0.5$.
This removes 2\% of the initial sample, leading to 400\,996 sources.
The resulting parallax distribution versus \gmag magnitude is shown in Fig.~\ref{Fig:parallaxVsG}.
Most sources are brighter than $\sim$18~mag.
Sources fainter than this value lie within 2~kpc of the Sun.

The absolute \absMeanG magnitude versus \BPminusRP colour diagram, hereafter called the observational Hertzsprung-Russell (HR) diagram, is shown in Fig.~\ref{Fig:obsHR_15perc}.
In the top panel, we show for reference the distribution of a sample of ten million sources extracted randomly from \Gaia DR3 under the conditions of positive relative parallax uncertainty better than 5\%, with at least 135 CCD measurements in \gmag (field \texttt{\small{phot\_g\_n\_obs}} in the \Gaia archive) and at least 15 measurements in \gbp and \grp (\texttt{\small{phot\_rp\_n\_obs}} and \texttt{\small{phot\_bp\_n\_obs}}, respectively), and with no multiple source detection (\texttt{\small{ipd\_frac\_multi\_peak}}=0).
The distribution of the eclipsing binaries with good parallaxes is shown in the second panel.
For comparison, contour lines of the ten million sample distribution are added in grey.
The sample of eclipsing binaries is seen to have a lower envelope of the main sequence shifted by $\sim$0.75~mag to the bright side compared to that of the ten million sample (a convincing representation is shown in Fig.~\ref{Fig:obsHR_15perc_Gplus0p75mag} of Appendix~\ref{Appendix:additionalFigures}), consistent with the magnitude shift expected for binary systems containing two main sequence stars of similar luminosities.

Almost half of the eclipsing binary candidates in this good parallax sample have an ellipsoidal component in their light curve model, with 18\% of the sample belonging to group 2GE-A and 22\% to 2GE-B).
An additional 24\% are tight systems with the light curves described by two wide Gaussians (group 2G-B), and 10\% belong to group 2G-A.

The orbital period distribution across the observational HR diagram is shown in the third panel from top of Fig.~\ref{Fig:obsHR_15perc}, with the per-bin median period colour-coded.
The period is seen to be well correlated, on the mean, with the stellar radius, as expected for tight systems.
This, however, is not observed for intrinsically faint main-sequence candidates with absolute M$_\gmag \gtrsim 7$~mag, neither for candidates below the main sequence where cataclysmic variables are found.
The global rankings of these faint candidates are generally also very low, as can be seen from the bottom panel of Fig.~\ref{Fig:obsHR_15perc}.
The observational HR diagram is much cleaner if we restrict the sample to candidates with high global rankings.
This is illustrated in Fig.~\ref{Fig:obsHR_15perc_rank0p6} with the sample of good parallaxes restricted to sources with a global ranking larger than 0.6, to which all 24'081 tight binaries from Sample 2G-C are also added\footnote{
The tight binaries in Sample 2G-C have, on the mean, lower global rankings than the candidates in Samples 2G-A, 2G-B, 2GE-A and 2GE-B, despite their light curves being generally very good (see 
Fig.~\ref{Fig:histo_globalRanking_2G} in Appendix~\ref{Sect:catalogue_usage_model} and Fig.~\ref{Fig:rank_magG_samples} in Appendix~\ref{Appendix:additionalFigures}).
}.
More than half of this new sample of 144\,994 sources have an ellipsoidal component (among which 21\% in Sample 2GE-A and 38\% in Sample 2GE-B), and 15\% are in Sample 2G-B.
The period distributions in the two samples, the full sample with good parallaxes and the one restricted to high global rankings, are very similar, but systems with a strong ellipsoidal component (2GE-B) become predominant at short periods in the sample restricted to high global rankings, as shown in Fig.~\ref{Fig:histo_period_15perc}.

%- - - - - - - - - - - - - - - - - - - - -
\paragraph{Bright candidates without detected ellipsoidal variability --}

About ten percent of the sample with good parallaxes have two Gaussians in their light curve and no detected ellipsoidal variability (group 2G-A), pointing to detached systems (and some semi-detached ones, see Sect.~\ref{Sect:catalogue_usage_model}).
As expected, they have, on the mean, longer periods than the tighter systems (see Fig.~\ref{Fig:histo_period_15perc}).
A key question concerns the circularization of these systems at short periods.
While we do not model the binary systems, and hence do not know their eccentricity, an eccentricity proxy \eccProxy can be derived from the relative eclipse locations and durations provided by the two-Gaussian model.
The relevant equations are recalled in Appendix~\ref{Appendix:eccentricty}.
It is shown in that appendix that most candidates in groups with potential tidal interactions (groups 2G-B, 2G-C, 2GE-A and 2GE-B) have eccentricity proxies compatible with circular systems, while large eccentricity proxies are found in Sample 2G-A of systems without detected tidal interaction.
However, the analysis also reveals the unexpected presence of short-period systems with large eccentricity proxies (top panel of Fig.~\ref{Fig:eccentricity_period_samples} in the appendix), while these systems are expected to have been circularised.  

In order to check the status of these short-period systems with large eccentricity proxies, we focus on the subset of bright candidates with \gmag\,<\,12~mag.
There are 401 such candidates in the good parallax sample with large global rankings defined above.
Their $P$--\eccProxy diagram is shown in Fig.~\ref{Fig:period_ecc_brightStars}.
The eccentricity limit above which all systems at a given period are expected to be circular is shown by the dashed blue line, following Eq.~(4.4) of \citet{Mazeh08} with $E=0.98$, $A=3.25$, $B=6.3$ and $C=0.23$.
The figure shows 48 systems that are unexpectedly above this limit, that we call outlier sources.
The uncertainty on their eccentricity proxy, a quantity that can be computed from the two-Gaussian model parameter uncertainties (see Eq.~\ref{Eq:eError} in Appendix~\ref{Appendix:eccentricty}), does not resolve the issue.
If we take the $1\sigma$ lower values of the eccentricity proxies considering their \eccProxyError uncertainties, 27 candidates would remain above the limit (the $1\sigma$ downward corrections are shown as vertical grey line segments in Fig.~\ref{Fig:period_ecc_brightStars}), and still 14 with a $2\sigma$ downward correction.
The depth of the secondary eclipse is colour-coded in Fig.~\ref{Fig:period_ecc_brightStars}, red corresponding to shallow dips and blue to deep dips.
Most of the outliers are seen to have very shallow secondary eclipses, less than 10~mmag deep (red markers in the figure).
The figure also encodes the secondary over primary depths ratios, with the size of the markers proportional to this ratio such that smaller circles correspond to smaller depth ratios.
The large majority of the outliers are seen to have their secondary eclipse much shallower than their primary eclipse (small marker size).
These features invite for a careful check of the reliability of the secondary eclipse.

A comparison of the \Gaia periods with literature periods, when available, provides additional insight.
The top panel of Fig.~\ref{Fig:periodVsPeriodLit_brightStars} shows this comparison for candidates that have acceptable eccentricity proxies, i.e. below the expected upper limit.
A very good match between $P_\mathrm{Gaia}$ and $P_\mathrm{lit}$ is seen for a large majority of the sources, with most of them having a 1:1 ratio and some a 1:2 ratio ($P_\mathrm{Gaia} \simeq 2\,P_\mathrm{lit}$).
In contrast, only very few outlier sources have $P_\mathrm{Gaia} \simeq P_\mathrm{lit}$, as shown in the bottom panel of Fig.~\ref{Fig:periodVsPeriodLit_brightStars}.
But the majority of them still have a \Gaia period `compatible' with the literature period within a factor of two, with $P_\mathrm{Gaia} \simeq 0.5\,P_\mathrm{lit}$.
This suggests that the \Gaia pipeline considered the two distinct eclipses as one unique eclipse, and detected a (usually very shallow) artefact in the folded light curve to represent an imaginary secondary eclipse.
A typical example of such a case is shown in the top panel of Fig.~\ref{Fig:lcs_tooLargeEcc}, where the secondary has a very shallow secondary dip identified at phase around 0.1, barely visible in the figure. 
If we increase the \Gaia period of the outlier sources by a factor of two, the majority of them become compatible with the expected maximum eccentricity given their period in Fig.~\ref{Fig:period_ecc_brightStars}.
Note that in that case, the eccentricity proxy would also take another value as the two-Gaussian model will be different.

The above analyses suggest a quick way to clean the $P$--\eccProxy diagram in Fig.~\ref{Fig:period_ecc_brightStars} by removing all small red points.
If we do this, a few outlier sources would still remain, that have significant secondary eclipse depths both in terms of absolute depth (non-red colour) and of secondary over primary depth ratio (large marker sizes).
We investigate here the four systems with the shortest periods.
They are, in increasing order of orbital period,
\GaiaSrcId{4524651705941314432} (0.2411507~d), % ($P=0.2411507018$~d)
\GaiaSrcIdWithoutDR3{5712304991851559040} (0.2450746~d), % ($P=0.2450745924$~d)
\GaiaSrcIdWithoutDR3{2589992273084612480} (0.2818606~d),
and
\GaiaSrcIdWithoutDR3{4455992049496820992} (0.3682685~d).
They are identified with open diamonds in Fig.~\ref{Fig:period_ecc_brightStars}, and their \gmag, \gbp and \grp light curves are shown in Fig.~\ref{Fig:lcs_tooLargeEcc}.
All two-Gaussian models to the \gmag light curves (dashed lines in the figure) look acceptable.
A closer look, however, reveals some features that put in question the periods extracted from the \gmag light curves, which have only a small number of measurements in the eclipses.
%The first source, for example, while having several points in eclipse, has an out-of-eclipse measurement at phase 0.75 in Fig.~\ref{Fig:period_ecc_brightStars} that is much fainter than the out-of-eclipse base magnitude.
%While its large \gmag uncertainty tends to give a low weight to this measurement, the simultaneous \gbp and \grp measurements confirm with good precision the faint luminosity of the system at this epoch.
Investigation of the \gbp and \grp light curves, which where not used in the DR3 processing of eclipsing binaries, suggest that the periods derived from the \gmag light curves may be incorrect, at least for the second and third sources.
%Finally, the fourth source has only one point in its primary eclipse and lacks measurement in the core of the secondary eclipse.
Therefore, the periods of these outlier sources should be double checked.

Two of these four outliers are identified in the ASAS-SN survey as eclipsing binaries \citep{JayasingheStanekKochanek_etal19}, but with different periods than the \Gaia period.
\GaiaSrcId{4524651705941314432} is mentioned with a period of 3.3972697~d (\texttt{\small{ASASSN-V J075432.26-211826.4}}),
and \GaiaSrcId{5712304991851559040} with a period of 5.9965616~d (\texttt{\small{ASASSN-V J184156.16+192755.8}}).
Their $P_\mathrm{lit}$ versus $P_\mathrm{Gaia}$ values are highlighted by the diamonds in the bottom panel of Fig.~\ref{Fig:periodVsPeriodLit_brightStars}.
The \GaiaSrcId{4524651705941314432} light curve is compatible with the ASAS-SN period (see Fig.~\ref{Fig:lcs_tooLargeEcc_ASASSN_periods} in Appendix~\ref{Appendix:additionalFigures} which displays the \Gaia light curve folded with the ASAS-SN period).
This period, however, was not selected by the automated \Gaia pipeline due to the scarcity of points that would result in the secondary eclipse.
Regarding \GaiaSrcId{5712304991851559040}, the period proposed by ASAS-SN is not compatible with the \Gaia light curve.
A check in the ASAS-SN database of light curves\footnote{\url{https://asas-sn.osu.edu/variables}} actually reveals that the ASAS-SN period is not very good for the ASAS-SN light curve neither.
However, twice this period (11.9931232~d) would be compatible with both ASAS-SN and \Gaia light curves.
But here too, the scarcity of \Gaia measurements that would result in the secondary eclipse prevented the \Gaia pipeline to chose this period.

% For varidashboard: 4524651705941314432, 5712304991851559040

In summary, \Gaia results of eclipsing binaries are generally very good given the available \Gaia time series.
Further investigations must, however, be performed for specific cases.
This is especially true for well-detached eclipsing binaries with long periods, for which an incorrect (very) short period may instead be chosen as the best solution by the DR3 pipeline.
%The analysis presented in this section proposes hints to proceed with such additional investigations on the \Gaia DR3 data.

%==================================================================
\section{Summary and conclusions}
\label{Sect:conclusions}

This paper presents the first \Gaia catalogue of eclipsing binaries made available in June 2022 within \Gaia DR3.
It contains more than two million candidates, filtered from a larger set of eclipsing binary candidates identified by the general classification pipeline of the variability processing modules within the \Gaia DPAC \citep{DR3-DPACP-165}.
The orbital periods are determined based on the cleaned \gmag light curves.
A two-Gaussian model is used to characterise the morphology of their light curves, containing up to two Gaussian and one sine functions, and a global ranking is provided that quantifies the quality of the model fit to the \gmag light curves.
The model adequately identifies the eclipses and ellipsoidal variability when they are clearly detectable in the light curves, and several groups of eclipsing binaries, from wide to tight systems, are identified in Sect.~\ref{Sect:catalogue_usage_model} based on the model parameters.
The two-Gaussian model, however, can contain components that are not relevant to describe real eclipses or ellipsoidal variability, though they reliably describe the geometry of the light curves.
The \gbp and \grp light curves were not considered in the processing of the eclipsing binaries in DR3.

About 600\,000 of the \Gaia candidates have a crossmatch in the literature, of which 88\% are also identified as eclipsing binaries in the literature.
The \Gaia and literature periods are similar for 86\% of the sources identified as eclipsing binaries in both \Gaia and the literature.
This number increases to 93\% when also considering period ratios of two.
The overall completeness of the catalogue is estimated to lie between 25\% and 50\% depending on the sky region, based on a comparison with OGLE4 catalogues of eclipsing binaries towards the Magellanic Clouds and the Galactic Bulge.
More than half of the catalogue consist of new candidates, with larger percentages of new candidates in dense regions of the sky.

%Tips for the exploitation of the catalogue are given in Sect.~\ref{Sect:catalogue_usage}.
%Various samples are presented in Sect.~\ref{Sect:catalogue_usage_model} based on the two-Gaussian model parameters. 
%They are defined in Table~\ref{Tab:sample_definition}, and the types of binaries they contain are summarised in Table~\ref{Tab:sample_summary}.
%Sources whose light curves are modelled with two and only two Gaussian functions, for example, are categorized based on the widths of the two Gaussians, leading to samples 2G-A, 2G-B, 2G-C, 2G-D, 2G-X and 2G-Y (see Fig.~\ref{Fig:sigmaGaussians_2G_samples}).
%When an additional ellipsoidal component is present in the model, samples 2GE-A, 2GE-B and 2GE-Z are constructed based on the amplitude of the cosine function and the phase separation of the locations of the two dips in the model. 
%We remind that the two-Gaussian model can contain components that are not relevant to describe real eclipses or ellipsoidal variability, though they reliably describe the geometry of the light curves. 
%This will be the case, for example, when the eclipse significances are low, when they don't contain enough measurements, when the period is wrongly determined (even to a factor of two), or when artefacts in the light curves lead to erroneous physical component identification in the light curves.
%Samples 2G-X, 2G-Y and 2GE-Z contain a majority of such cases.
%But such cases can also occur in the other samples.
%The sample definitions in Table~\ref{Tab:sample_summary} are given as examples, and should be study-specific tailored with a proper analysis.

Illustrative samples with good parallaxes, containing in majority tight systems, confirm that their properties such as distribution in the observational HR diagram and periods, are overall as expected.
The analysis of the period-eccentricity diagram for a subset of detached systems, on the other hand, highlights the challenges in dealing with these systems, and illustrates the usage of the catalogue.

This release represents the largest catalogue of eclipsing binary candidates available so far in the literature.
Looking to the future, the next \Gaia Data Release 4 (DR4) will provide an even larger catalogue of candidates, with improved characterisation due not only to the larger time baseline and increased number of measurements (DR4 will be based on 66 months of data instead of the 34 months in DR3), but also due to further improvements in our processing pipeline by, for example, taking into account \gbp and \grp time series.

%==================================================================
\begin{acknowledgements}

Specific acknowledgements related to the \Gaia data on which this work is based are given in Appendix~\ref{Appendix:acknowledgements}.
This research has made use of the free Starlink Tables Infrastructure Library \citep[STILTS and Topcat,][]{Taylor_2006}.
This research has made use of NASA's Astrophysics Data System Bibliographic Services.
\end{acknowledgements}

%==================================================================
\bibliographystyle{aa} % style aa.bst
% bib file: ~/Library/texmf/bibtex/bib/bibTex.bib
% Managed with BibDesk
\bibliography{GaiaDR3_EB}

\begin{thebibliography}{49}
\expandafter\ifx\csname natexlab\endcsname\relax\def\natexlab#1{#1}\fi

\bibitem[{{Alcock} {et~al.}(1997){Alcock}, {Allsman}, {Alves}, {Axelrod},
  {Becker}, {Bennett}, {Cook}, {Freeman}, {Griest}, {Guern}, {Lehner},
  {Marshall}, {Peterson}, {Pratt}, {Quinn}, {Rodgers}, {Stubbs}, {Sutherland},
  {Welch}, \& {MACHO Collaboration}}]{AlcockAllsmanAlves97}
{Alcock}, C., {Allsman}, R.~A., {Alves}, D., {et~al.} 1997, \apj, 486, 697

\bibitem[{{Andersen}(1991)}]{Andersen_1991}
{Andersen}, J. 1991, \aapr, 3, 91

\bibitem[{{Avvakumova} {et~al.}(2013){Avvakumova}, {Malkov}, \&
  {Kniazev}}]{AvvakumovaMalkovKniazev13}
{Avvakumova}, E.~A., {Malkov}, O.~Y., \& {Kniazev}, A.~Y. 2013, Astronomische
  Nachrichten, 334, 860

\bibitem[{{Burke} {et~al.}(1970){Burke}, {Rolland}, \&
  {Boy}}]{BurkeRollandBoy70}
{Burke}, Edward~W., J., {Rolland}, W.~W., \& {Boy}, W.~R. 1970, \jrasc, 64, 353

\bibitem[{{Chen} {et~al.}(2020){Chen}, {Wang}, {Deng}, {de Grijs}, {Yang}, \&
  {Tian}}]{ChenWangDeng_etal20}
{Chen}, X., {Wang}, S., {Deng}, L., {et~al.} 2020, \apjs, 249, 18

\bibitem[{{Cumming} {et~al.}(1999){Cumming}, {Marcy}, \&
  {Butler}}]{CummingMarcyBulter99}
{Cumming}, A., {Marcy}, G.~W., \& {Butler}, R.~P. 1999, \apj, 526, 890

\bibitem[{{Devor} {et~al.}(2008){Devor}, {Charbonneau}, {O'Donovan},
  {Mandushev}, \& {Torres}}]{TRES_ECL_DEVOR_2008}
{Devor}, J., {Charbonneau}, D., {O'Donovan}, F.~T., {Mandushev}, G., \&
  {Torres}, G. 2008, \aj, 135, 850

\bibitem[{{Drake} {et~al.}(2017){Drake}, {Djorgovski}, {Catelan}, {Graham},
  {Mahabal}, {Larson}, {Christensen}, {Torrealba}, {Beshore}, {McNaught},
  {Garradd}, {Belokurov}, \& {Koposov}}]{CATALINA_VAR_DRAKE_2017}
{Drake}, A.~J., {Djorgovski}, S.~G., {Catelan}, M., {et~al.} 2017, \mnras, 469,
  3688

\bibitem[{{Ekstr{\"o}m} {et~al.}(2012){Ekstr{\"o}m}, {Georgy}, {Eggenberger},
  {Meynet}, {Mowlavi}, {Wyttenbach}, {Granada}, {Decressin}, {Hirschi},
  {Frischknecht}, {Charbonnel}, \& {Maeder}}]{EkstromGeorgyEggenberger_etal12}
{Ekstr{\"o}m}, S., {Georgy}, C., {Eggenberger}, P., {et~al.} 2012, \aap, 537,
  A146

\bibitem[{ESA(1997)}]{Hipparcos_1997}
ESA, ed. 1997, ESA Special Publication, Vol. 1200, {The HIPPARCOS and TYCHO
  catalogues. Astrometric and photometric star catalogues derived from the ESA
  HIPPARCOS Space Astrometry Mission}, ed. ESA

\bibitem[{{Eyer} {et~al.}(2022){Eyer}, {Audard}, {Holl}, {Rimoldini},
  {Carnerero}, {Clementini}, {De Ridder}, {Distefano}, {Evans}, {Gavras},
  {Gomel}, {Lebzelter}, {Marton}, {Mowlavi}, {Panahi}, {Ripepi}, {Wyrzykowski},
  {Nienartowicz}, {Jevardat de Fombelle}, {Lecoeur-Taibi}, {Rohrbasser},
  {Riello}, {Garcia-Lario}, {Lanzafame}, {Mazeh}, {Raiteri}, {Zucker},
  {Abraham}, {Aerts}, {Aguado}, {Anderson}, {Bashi}, {Binnenfeld}, {Faigler},
  {Garofalo}, {Karbevska}, {Kospal}, {Kruszynska}, {Kun}, {Lanza}, {Leccia},
  {Marconi}, {Messina}, {Molinaro}, {Molnar}, {Muraveva}, {Musella}, {Nagy},
  {Pagano}, {Palaversa}, {Plachy}, {Rybicki}, {Shahaf}, {Szabados},
  {Szegedi-Elek}, {Trabucchi}, {Barblan}, \& {Roelens}}]{2022arXiv220606416E}
{Eyer}, L., {Audard}, M., {Holl}, B., {et~al.} 2022, arXiv e-prints,
  arXiv:2206.06416

\bibitem[{{Eyer} {et~al.}(2017){Eyer}, {Mowlavi}, {Evans}, {Nienartowicz},
  {Ordonez}, {Holl}, {Lecoeur-Taibi}, {Riello}, {Clementini}, {Cuypers}, {De
  Ridder}, {Lanzafame}, {Sarro}, {Charnas}, {Guy}, {Jevardat de Fombelle},
  {Rimoldini}, {S{\"u}veges}, {Mignard}, {Busso}, {De Angeli}, {van Leeuwen},
  {Dubath}, {Beck}, {Aguado}, {Debosscher}, {Distefano}, {Fuchs}, {Koubsky},
  {Lebzelter}, {Leccia}, {Lopez}, {Moitinho}, {Regibo}, {Ripepi}, {Roelens},
  {Szabados}, {Tingley}, {Votruba}, {Zucker}, {Aerts}, {Barblan},
  {Blanco-Cuaresma}, {Grenon}, {Jan}, {Lorenz}, {Miranda}, {Morgenthaler},
  {Ordenovic}, {Palaversa}, {Prsa}, {Ruiz-Fuertes}, {Anderson}, {Delgado},
  {Dzigan}, {Hudec}, {Jonckheere}, {Klagyivik}, {Kutka}, {Moniez}, {Nicoletti},
  {Park}, {Van Hemelryck}, {Varadi}, {Kochoska}, {Lanza}, {Marconi},
  {Marschalko}, {Messina}, {Musella}, {Pagano}, {Sadowski}, \&
  {Schultheis}}]{2017arXiv170203295E}
{Eyer}, L., {Mowlavi}, N., {Evans}, D.~W., {et~al.} 2017, arXiv e-prints,
  arXiv:1702.03295

\bibitem[{{Feigelson} \& {Babu}(2012)}]{FeigelsonBabu12}
{Feigelson}, E.~D. \& {Babu}, G.~J. 2012, {Modern Statistical Methods for
  Astronomy} (Cambridge University Press)

\bibitem[{{Gaia Collaboration} {et~al.}(2022{\natexlab{a}}){Gaia
  Collaboration}, {Arenou}, {Babusiaux}, {Barstow}, {Faigler}, {Jorissen},
  {Kervella}, {Mazeh}, {Mowlavi}, {Panuzzo}, \&
  et~al.}]{GaiaArenouBabusiaux_etal22}
{Gaia Collaboration}, {Arenou}, F., {Babusiaux}, C., {et~al.}
  2022{\natexlab{a}}, arXiv e-prints, arXiv:2206.05595

\bibitem[{{Gaia Collaboration} {et~al.}(2021){Gaia Collaboration}, {Brown},
  {Vallenari}, {Prusti}, {de Bruijne}, {Babusiaux}, {Biermann}, {Creevey},
  {Evans}, {Eyer}, {Hutton}, {Jansen}, {Jordi}, {Klioner}, {Lammers},
  {Lindegren}, {Luri}, {Mignard}, {Panem}, {Pourbaix}, {Randich}, {Sartoretti},
  {Soubiran}, {Walton}, {Arenou}, {Bailer-Jones}, {Bastian}, {Cropper},
  {Drimmel}, {Katz}, {Lattanzi}, {van Leeuwen}, {Bakker}, {Cacciari},
  {Casta{\~n}eda}, {De Angeli}, {Ducourant}, {Fabricius}, {Fouesneau},
  {Fr{\'e}mat}, {Guerra}, {Guerrier}, {Guiraud}, {Jean-Antoine Piccolo},
  {Masana}, {Messineo}, {Mowlavi}, {Nicolas}, {Nienartowicz}, {Pailler},
  {Panuzzo}, {Riclet}, {Roux}, {Seabroke}, {Sordo}, {Tanga}, {Th{\'e}venin},
  {Gracia-Abril}, {Portell}, {Teyssier}, {Altmann}, {Andrae}, {Bellas-Velidis},
  {Benson}, {Berthier}, {Blomme}, {Brugaletta}, {Burgess}, {Busso}, {Carry},
  {Cellino}, {Cheek}, {Clementini}, {Damerdji}, {Davidson}, {Delchambre},
  {Dell'Oro}, {Fern{\'a}ndez-Hern{\'a}ndez}, {Galluccio}, {Garc{\'\i}a-Lario},
  {Garcia-Reinaldos}, {Gonz{\'a}lez-N{\'u}{\~n}ez}, {Gosset}, {Haigron},
  {Halbwachs}, {Hambly}, {Harrison}, {Hatzidimitriou}, {Heiter},
  {Hern{\'a}ndez}, {Hestroffer}, {Hodgkin}, {Holl}, {Jan{\ss}en}, {Jevardat de
  Fombelle}, {Jordan}, {Krone-Martins}, {Lanzafame}, {L{\"o}ffler}, {Lorca},
  {Manteiga}, {Marchal}, {Marrese}, {Moitinho}, {Mora}, {Muinonen}, {Osborne},
  {Pancino}, {Pauwels}, {Petit}, {Recio-Blanco}, {Richards}, {Riello},
  {Rimoldini}, {Robin}, {Roegiers}, {Rybizki}, {Sarro}, {Siopis}, {Smith},
  {Sozzetti}, {Ulla}, {Utrilla}, {van Leeuwen}, {van Reeven}, {Abbas}, {Abreu
  Aramburu}, {Accart}, {Aerts}, {Aguado}, {Ajaj}, {Altavilla}, {{\'A}lvarez},
  {{\'A}lvarez Cid-Fuentes}, {Alves}, {Anderson}, {Anglada Varela}, {Antoja},
  {Audard}, {Baines}, {Baker}, {Balaguer-N{\'u}{\~n}ez}, {Balbinot}, {Balog},
  {Barache}, {Barbato}, {Barros}, {Barstow}, {Bartolom{\'e}}, {Bassilana},
  {Bauchet}, {Baudesson-Stella}, {Becciani}, {Bellazzini}, {Bernet}, {Bertone},
  {Bianchi}, {Blanco-Cuaresma}, {Boch}, {Bombrun}, {Bossini}, {Bouquillon},
  {Bragaglia}, {Bramante}, {Breedt}, {Bressan}, {Brouillet}, {Bucciarelli},
  {Burlacu}, {Busonero}, {Butkevich}, {Buzzi}, {Caffau}, {Cancelliere},
  {C{\'a}novas}, {Cantat-Gaudin}, {Carballo}, {Carlucci}, {Carnerero},
  {Carrasco}, {Casamiquela}, {Castellani}, {Castro-Ginard}, {Castro Sampol},
  {Chaoul}, {Charlot}, {Chemin}, {Chiavassa}, {Cioni}, {Comoretto}, {Cooper},
  {Cornez}, {Cowell}, {Crifo}, {Crosta}, {Crowley}, {Dafonte}, {Dapergolas},
  {David}, {David}, {de Laverny}, {De Luise}, {De March}, {De Ridder}, {de
  Souza}, {de Teodoro}, {de Torres}, {del Peloso}, {del Pozo}, {Delbo},
  {Delgado}, {Delgado}, {Delisle}, {Di Matteo}, {Diakite}, {Diener},
  {Distefano}, {Dolding}, {Eappachen}, {Edvardsson}, {Enke}, {Esquej}, {Fabre},
  {Fabrizio}, {Faigler}, {Fedorets}, {Fernique}, {Fienga}, {Figueras},
  {Fouron}, {Fragkoudi}, {Fraile}, {Franke}, {Gai}, {Garabato},
  {Garcia-Gutierrez}, {Garc{\'\i}a-Torres}, {Garofalo}, {Gavras}, {Gerlach},
  {Geyer}, {Giacobbe}, {Gilmore}, {Girona}, {Giuffrida}, {Gomel}, {Gomez},
  {Gonzalez-Santamaria}, {Gonz{\'a}lez-Vidal}, {Granvik},
  {Guti{\'e}rrez-S{\'a}nchez}, {Guy}, {Hauser}, {Haywood}, {Helmi}, {Hidalgo},
  {Hilger}, {H{\l}adczuk}, {Hobbs}, {Holland}, {Huckle}, {Jasniewicz},
  {Jonker}, {Juaristi Campillo}, {Julbe}, {Karbevska}, {Kervella}, {Khanna},
  {Kochoska}, {Kontizas}, {Kordopatis}, {Korn}, {Kostrzewa-Rutkowska},
  {Kruszy{\'n}ska}, {Lambert}, {Lanza}, {Lasne}, {Le Campion}, {Le Fustec},
  {Lebreton}, {Lebzelter}, {Leccia}, {Leclerc}, {Lecoeur-Taibi}, {Liao},
  {Licata}, {Lindstr{\o}m}, {Lister}, {Livanou}, {Lobel}, {Madrero Pardo},
  {Managau}, {Mann}, {Marchant}, {Marconi}, {Marcos Santos}, {Marinoni},
  {Marocco}, {Marshall}, {Martin Polo}, {Mart{\'\i}n-Fleitas}, {Masip},
  {Massari}, {Mastrobuono-Battisti}, {Mazeh}, {McMillan}, {Messina},
  {Michalik}, {Millar}, {Mints}, {Molina}, {Molinaro}, {Moln{\'a}r},
  {Montegriffo}, {Mor}, {Morbidelli}, {Morel}, {Morris}, {Mulone}, {Munoz},
  {Muraveva}, {Murphy}, {Musella}, {Noval}, {Ord{\'e}novic}, {Orr{\`u}},
  {Osinde}, {Pagani}, {Pagano}, {Palaversa}, {Palicio}, {Panahi}, {Pawlak},
  {Pe{\~n}alosa Esteller}, {Penttil{\"a}}, {Piersimoni}, {Pineau}, {Plachy},
  {Plum}, {Poggio}, {Poretti}, {Poujoulet}, {Pr{\v{s}}a}, {Pulone}, {Racero},
  {Ragaini}, {Rainer}, {Raiteri}, {Rambaux}, {Ramos}, {Ramos-Lerate}, {Re
  Fiorentin}, {Regibo}, {Reyl{\'e}}, {Ripepi}, {Riva}, {Rixon}, {Robichon},
  {Robin}, {Roelens}, {Rohrbasser}, {Romero-G{\'o}mez}, {Rowell}, {Royer},
  {Rybicki}, {Sadowski}, {Sagrist{\`a} Sell{\'e}s}, {Sahlmann}, {Salgado},
  {Salguero}, {Samaras}, {Sanchez Gimenez}, {Sanna}, {Santove{\~n}a},
  {Sarasso}, {Schultheis}, {Sciacca}, {Segol}, {Segovia}, {S{\'e}gransan},
  {Semeux}, {Shahaf}, {Siddiqui}, {Siebert}, {Siltala}, {Slezak}, {Smart},
  {Solano}, {Solitro}, {Souami}, {Souchay}, {Spagna}, {Spoto}, {Steele},
  {Steidelm{\"u}ller}, {Stephenson}, {S{\"u}veges}, {Szabados}, {Szegedi-Elek},
  {Taris}, {Tauran}, {Taylor}, {Teixeira}, {Thuillot}, {Tonello}, {Torra},
  {Torra}, {Turon}, {Unger}, {Vaillant}, {van Dillen}, {Vanel}, {Vecchiato},
  {Viala}, {Vicente}, {Voutsinas}, {Weiler}, {Wevers}, {Wyrzykowski}, {Yoldas},
  {Yvard}, {Zhao}, {Zorec}, {Zucker}, {Zurbach}, \&
  {Zwitter}}]{Gaia_BrownVallenariPrusti_etal21}
{Gaia Collaboration}, {Brown}, A.~G.~A., {Vallenari}, A., {et~al.} 2021, \aap,
  649, A1

\bibitem[{{Gaia Collaboration} {et~al.}(2016){Gaia Collaboration}, {Prusti},
  {de Bruijne}, {Brown}, {Vallenari}, {Babusiaux}, {Bailer-Jones}, {Bastian},
  {Biermann}, {Evans}, \& et~al.}]{Gaia_PrustiDeBruijneBrown_etal16}
{Gaia Collaboration}, {Prusti}, T., {de Bruijne}, J.~H.~J., {et~al.} 2016,
  \aap, 595, A1

\bibitem[{{Gaia Collaboration} {et~al.}(2022{\natexlab{b}}){Gaia
  Collaboration}, {Vallenari}, \& {et al.}}]{DR3-DPACP-185}
{Gaia Collaboration}, {Vallenari}, A., \& {et al.} 2022{\natexlab{b}}, \aap, in
  prep.

\bibitem[{{Gavras et al.}(2022)}]{DR3-DPACP-177}
{Gavras et al.} 2022, \aap\ submitted

\bibitem[{{Heck} {et~al.}(1985){Heck}, {Manfroid}, \&
  {Mersch}}]{HeckManfroidMersch_85}
{Heck}, A., {Manfroid}, J., \& {Mersch}, G. 1985, \aaps, 59, 63

\bibitem[{{Heinze} {et~al.}(2018){Heinze}, {Tonry}, {Denneau}, {Flewelling},
  {Stalder}, {Rest}, {Smith}, {Smartt}, \&
  {Weiland}}]{HeinzeTonryDenneau_etal18}
{Heinze}, A.~N., {Tonry}, J.~L., {Denneau}, L., {et~al.} 2018, \aj, 156, 241

\bibitem[{{Holl et al.}(2022)}]{DR3-DPACP-164}
{Holl et al.} 2022, \aap\ in prep.

\bibitem[{{Jayasinghe} {et~al.}(2019){Jayasinghe}, {Stanek}, {Kochanek},
  {Shappee}, {Holoien}, {Thompson}, {Prieto}, {Dong}, {Pawlak}, {Pejcha},
  {Shields}, {Pojmanski}, {Otero}, {Britt}, \&
  {Will}}]{JayasingheStanekKochanek_etal19}
{Jayasinghe}, T., {Stanek}, K.~Z., {Kochanek}, C.~S., {et~al.} 2019, \mnras,
  486, 1907

\bibitem[{{Jurkevich}(1971)}]{Jurkevich71}
{Jurkevich}, I. 1971, \apss, 13, 154

\bibitem[{{Kim} {et~al.}(2014){Kim}, {Protopapas}, {Bailer-Jones}, {Byun},
  {Chang}, {Marquette}, \& {Shin}}]{KimProtopapasBailerJones_etal14}
{Kim}, D.-W., {Protopapas}, P., {Bailer-Jones}, C. A.~L., {et~al.} 2014, \aap,
  566, A43

\bibitem[{{Kirk} {et~al.}(2016){Kirk}, {Conroy}, {Pr{\v{s}}a}, {Abdul-Masih},
  {Kochoska}, {Matijevi{\v{c}}}, {Hambleton}, {Barclay}, {Bloemen}, {Boyajian},
  {Doyle}, {Fulton}, {Hoekstra}, {Jek}, {Kane}, {Kostov}, {Latham}, {Mazeh},
  {Orosz}, {Pepper}, {Quarles}, {Ragozzine}, {Shporer}, {Southworth},
  {Stassun}, {Thompson}, {Welsh}, {Agol}, {Derekas}, {Devor}, {Fischer},
  {Green}, {Gropp}, {Jacobs}, {Johnston}, {LaCourse}, {Saetre}, {Schwengeler},
  {Toczyski}, {Werner}, {Garrett}, {Gore}, {Martinez}, {Spitzer}, {Stevick},
  {Thomadis}, {Vrijmoet}, {Yenawine}, {Batalha}, \&
  {Borucki}}]{KEPLER_ECL_KIRK_2016}
{Kirk}, B., {Conroy}, K., {Pr{\v{s}}a}, A., {et~al.} 2016, \aj, 151, 68

\bibitem[{{Lafler} \& {Kinman}(1965)}]{LaflerKinman65}
{Lafler}, J. \& {Kinman}, T.~D. 1965, \apjs, 11, 216

\bibitem[{{Malkov}(2020)}]{Malkov20}
{Malkov}, O.~Y. 2020, \mnras, 491, 5489

\bibitem[{{Mazeh}(2008)}]{Mazeh08}
{Mazeh}, T. 2008, in EAS Publications Series, Vol.~29, EAS Publications Series,
  ed. M.~J. {Goupil} \& J.~P. {Zahn}, 1--65

\bibitem[{{Mowlavi} {et~al.}(2017){Mowlavi}, {Lecoeur-Ta{\"\i}bi}, {Holl},
  {Rimoldini}, {Barblan}, {Prsa}, {Kochoska}, {S{\"u}veges}, {Eyer},
  {Nienartowicz}, {Jevardat}, {Charnas}, {Guy}, \&
  {Audard}}]{MowlaviLecoeurHoll_etal17}
{Mowlavi}, N., {Lecoeur-Ta{\"\i}bi}, I., {Holl}, B., {et~al.} 2017, \aap, in
  press (arXiv 1703.10597) [\eprint[arXiv]{1703.10597}]

\bibitem[{{Paczy{\'n}ski} {et~al.}(2006){Paczy{\'n}ski}, {Szczygie{\l}},
  {Pilecki}, \& {Pojma{\'n}ski}}]{PaczynskiSzczygielPilecki_etal06}
{Paczy{\'n}ski}, B., {Szczygie{\l}}, D.~M., {Pilecki}, B., \& {Pojma{\'n}ski},
  G. 2006, \mnras, 368, 1311

\bibitem[{{Palaversa} {et~al.}(2013){Palaversa}, {Ivezi{\'c}}, {Eyer},
  {Ru{\v{z}}djak}, {Sudar}, {Galin}, {Kroflin}, {Mesari{\'c}}, {Munk},
  {Vrbanec}, {Bo{\v{z}}i{\'c}}, {Loebman}, {Sesar}, {Rimoldini}, {Hunt-Walker},
  {VanderPlas}, {Westman}, {Stuart}, {Becker}, {Srdo{\v{c}}}, {Wozniak}, \&
  {Oluseyi}}]{LINEAR_VAR_PALAVERSA_2013}
{Palaversa}, L., {Ivezi{\'c}}, {\v{Z}}., {Eyer}, L., {et~al.} 2013, \aj, 146,
  101

\bibitem[{{Pawlak} {et~al.}(2016){Pawlak}, {Soszy{\'n}ski}, {Udalski},
  {Szyma{\'n}ski}, {Wyrzykowski}, {Ulaczyk}, {Poleski}, {Pietrukowicz},
  {Koz{\l}owski}, {Skowron}, {Skowron}, {Mr{\'o}z}, \&
  {Hamanowicz}}]{PawlakSoszynskiUdalski_etal16}
{Pawlak}, M., {Soszy{\'n}ski}, I., {Udalski}, A., {et~al.} 2016, \actaa, 66,
  421

\bibitem[{{Pigulski} {et~al.}(2009){Pigulski}, {Pojma{\'n}ski}, {Pilecki}, \&
  {Szczygie{\l}}}]{ASAS_KEPLER_VAR_PIGULSKI_2009}
{Pigulski}, A., {Pojma{\'n}ski}, G., {Pilecki}, B., \& {Szczygie{\l}}, D.~M.
  2009, \actaa, 59, 33

\bibitem[{{Pojmanski}(2002)}]{Pojmanski_2002}
{Pojmanski}, G. 2002, \actaa, 52, 397

\bibitem[{{Pr{\v{s}}a} {et~al.}(2022){Pr{\v{s}}a}, {Kochoska}, {Conroy},
  {Eisner}, {Hey}, {IJspeert}, {Kruse}, {Fleming}, {Johnston}, {Kristiansen},
  {LaCourse}, {Mortensen}, {Pepper}, {Stassun}, {Torres}, {Abdul-Masih},
  {Chakraborty}, {Gagliano}, {Guo}, {Hambleton}, {Hong}, {Jacobs}, {Jones},
  {Kostov}, {Lee}, {Omohundro}, {Orosz}, {Page}, {Powell}, {Rappaport}, {Reed},
  {Schnittman}, {Schwengeler}, {Shporer}, {Terentev}, {Vanderburg}, {Welsh},
  {Caldwell}, {Doty}, {Jenkins}, {Latham}, {Ricker}, {Seager}, {Schlieder},
  {Shiao}, {Vanderspek}, \& {Winn}}]{PrsaKochoskaConroa_etal22}
{Pr{\v{s}}a}, A., {Kochoska}, A., {Conroy}, K.~E., {et~al.} 2022, \apjs, 258,
  16

\bibitem[{{Renault} {et~al.}(1998){Renault}, {Aubourg}, {Bareyre}, {Brehin},
  {Gros}, {Lachieze-Rey}, {Laurent}, {Lesquoy}, {Magneville}, {Milsztajn},
  {Moscoso}, {Palanque-Delabrouille}, {Queinnec}, {Rich}, {Spiro}, {Vigroux},
  {Zylberajch}, {Ansari}, {Cavalier}, {Moniez}, {Beaulieu}, {Ferlet}, {Grison},
  {Vidal-Madjar}, {Guibert}, {Moreau}, {Maurice}, {Prevot}, {Gry}, {Char},
  {Fernandez}, \& {EROS Collaboration}}]{RenaultAubourgBareyre98}
{Renault}, C., {Aubourg}, E., {Bareyre}, P., {et~al.} 1998, \aap, 329, 522

\bibitem[{{Riello} {et~al.}(2021){Riello}, {De Angeli}, {Evans}, {Montegriffo},
  {Carrasco}, {Busso}, {Palaversa}, {Burgess}, {Diener}, {Davidson}, {Rowell},
  {Fabricius}, {Jordi}, {Bellazzini}, {Pancino}, {Harrison}, {Cacciari}, {van
  Leeuwen}, {Hambly}, {Hodgkin}, {Osborne}, {Altavilla}, {Barstow}, {Brown},
  {Castellani}, {Cowell}, {De Luise}, {Gilmore}, {Giuffrida}, {Hidalgo},
  {Holland}, {Marinoni}, {Pagani}, {Piersimoni}, {Pulone}, {Ragaini}, {Rainer},
  {Richards}, {Sanna}, {Walton}, {Weiler}, \& {Yoldas}}]{Riello_etal21}
{Riello}, M., {De Angeli}, F., {Evans}, D.~W., {et~al.} 2021, \aap, 649, A3

\bibitem[{{Rimoldini et al.}(2022)}]{DR3-DPACP-165}
{Rimoldini et al.} 2022, \aap\ in prep.

\bibitem[{{Schwarzenberg-Czerny}(1997)}]{SchwarzenbergCzerny97}
{Schwarzenberg-Czerny}, A. 1997, in Astrophysics and Space Science Library,
  Vol. 218, Astronomical Time Series, ed. D.~{Maoz}, A.~{Sternberg}, \& E.~M.
  {Leibowitz}, 183

\bibitem[{{Sesar} {et~al.}(2017){Sesar}, {Hernitschek}, {Mitrovi{\'c}},
  {Ivezi{\'c}}, {Rix}, {Cohen}, {Bernard}, {Grebel}, {Martin}, {Schlafly},
  {Burgett}, {Draper}, {Flewelling}, {Kaiser}, {Kudritzki}, {Magnier},
  {Metcalfe}, {Tonry}, \& {Waters}}]{SesarBranimirHernitschek_etal17}
{Sesar}, B., {Hernitschek}, N., {Mitrovi{\'c}}, S., {et~al.} 2017, \aj, 153,
  204

\bibitem[{{Siopis} {et~al.}(2022){Siopis}, {Sadowski}, \&
  {Pourbaix}}]{DR3-DPACP-179}
{Siopis}, C., {Sadowski}, D., \& {Pourbaix}, D. 2022, \aap, in prep.

\bibitem[{{Soszy{\'n}ski} {et~al.}(2016){Soszy{\'n}ski}, {Pawlak},
  {Pietrukowicz}, {Udalski}, {Szyma{\'n}ski}, {Wyrzykowski}, {Ulaczyk},
  {Poleski}, {Koz{\l}owski}, {Skowron}, {Skowron}, {Mr{\'o}z}, \&
  {Hamanowicz}}]{SoszynskiPawlakPietrukowicz_etal16}
{Soszy{\'n}ski}, I., {Pawlak}, M., {Pietrukowicz}, P., {et~al.} 2016, \actaa,
  66, 405

\bibitem[{{Southworth}(2015)}]{Southworth_2015}
{Southworth}, J. 2015, in Astronomical Society of the Pacific Conference
  Series, Vol. 496, Living Together: Planets, Host Stars and Binaries, ed.
  S.~M. {Rucinski}, G.~{Torres}, \& M.~{Zejda}, 164

\bibitem[{{Stellingwerf}(1978)}]{Stellingwerf78}
{Stellingwerf}, R.~F. 1978, \apj, 224, 953

\bibitem[{{Taylor}(2006)}]{Taylor_2006}
{Taylor}, M.~B. 2006, in Astronomical Society of the Pacific Conference Series,
  Vol. 351, Astronomical Data Analysis Software and Systems XV, ed.
  C.~{Gabriel}, C.~{Arviset}, D.~{Ponz}, \& S.~{Enrique}, 666

\bibitem[{{Udalski} {et~al.}(1992){Udalski}, {Szymanski}, {Kaluzny}, {Kubiak},
  \& {Mateo}}]{UdalskiSzymanskiKaluzny92}
{Udalski}, A., {Szymanski}, M., {Kaluzny}, J., {Kubiak}, M., \& {Mateo}, M.
  1992, \actaa, 42, 253

\bibitem[{{Wall} \& {Jenkins}(2003)}]{WallJenkins03}
{Wall}, J.~V. \& {Jenkins}, C.~R. 2003, {Practical Statistics for Astronomers},
  Vol.~3 (?)

\bibitem[{{Watson} {et~al.}(2006){Watson}, {Henden}, \&
  {Price}}]{COMP_VAR_VSX_2019}
{Watson}, C.~L., {Henden}, A.~A., \& {Price}, A. 2006, Society for Astronomical
  Sciences Annual Symposium, 25, 47

\bibitem[{{Zechmeister} \& {K{\"u}rster}(2009)}]{ZechmeisterKurster09}
{Zechmeister}, M. \& {K{\"u}rster}, M. 2009, \aap, 496, 577

\end{thebibliography}

%\bibliographystyle{aa}
%\bibliography{gaiaLPV_DR2}

%==================================================================
%==================================================================
\begin{appendix}

%==================================================================
\section{Analysis of the two-Gaussian model parameters}
\label{Appendix:twoGaussianModel}

\begin{figure}
  \centering
  \includegraphics[trim={0 75 0 42},clip,width=0.815\linewidth]{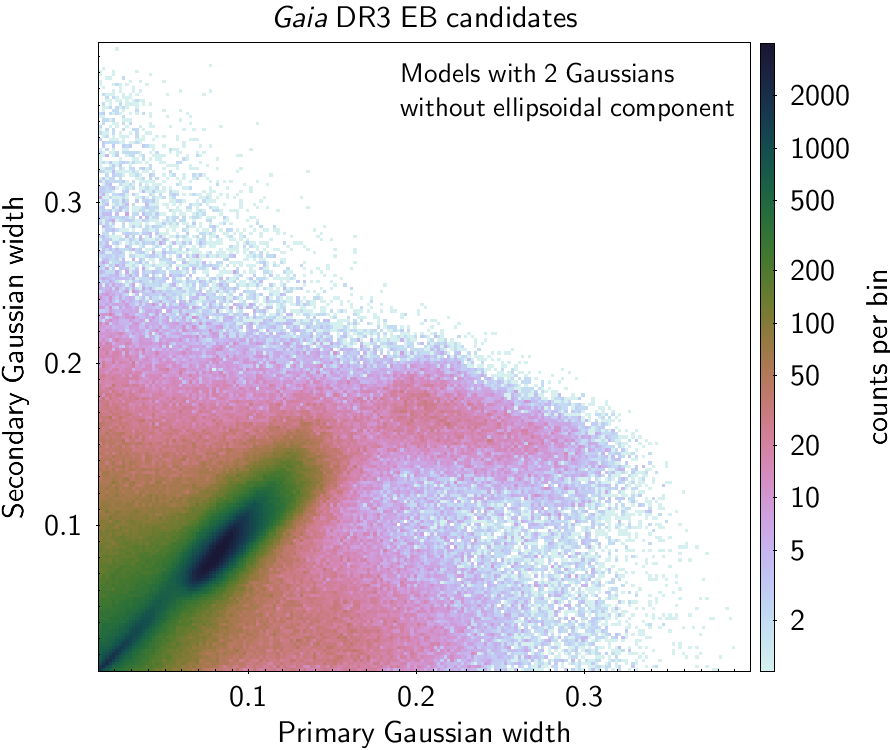}
  \vskip -0.5mm
  \includegraphics[trim={0 75 0 42},clip,width=0.815\linewidth]{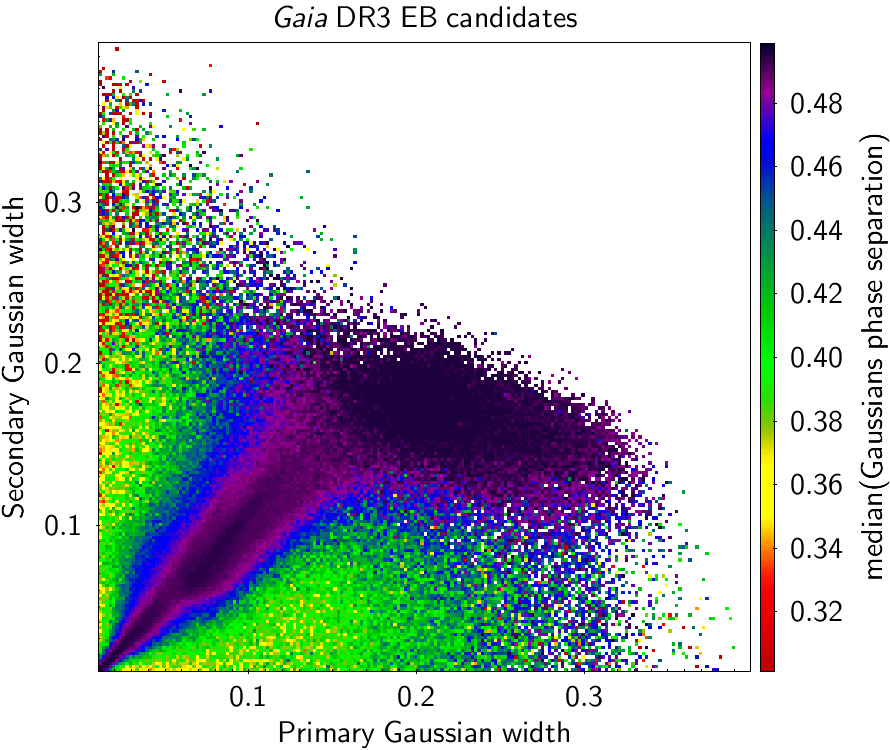}
  \vskip -0.5mm
  \includegraphics[trim={0 75 0 42},clip,width=0.815\linewidth]{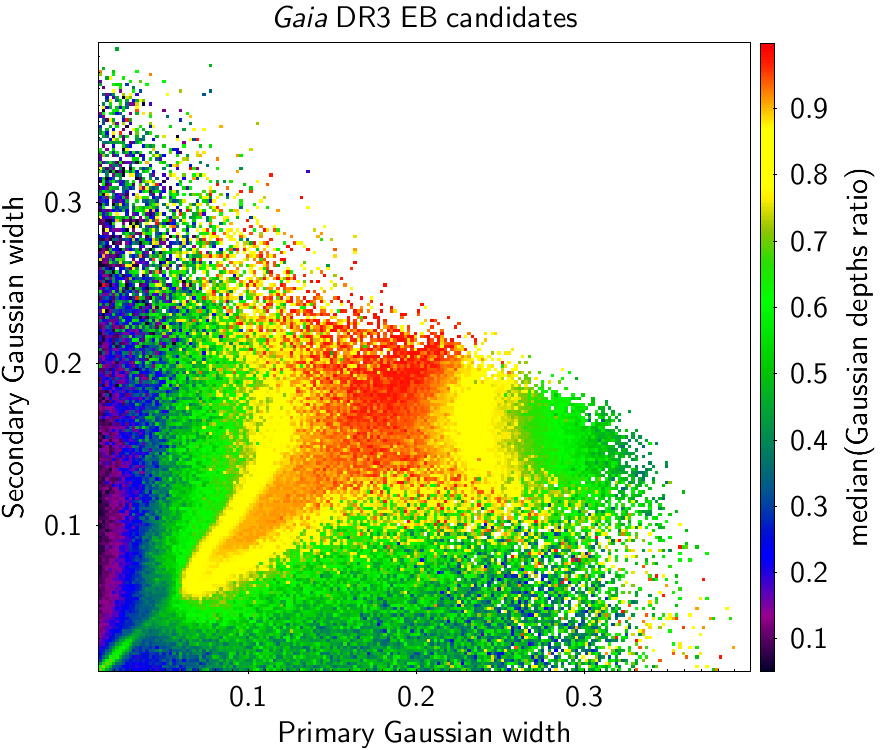}
  \vskip -0.5mm  
  \includegraphics[trim={0 0 0 42},clip,width=0.815\linewidth]{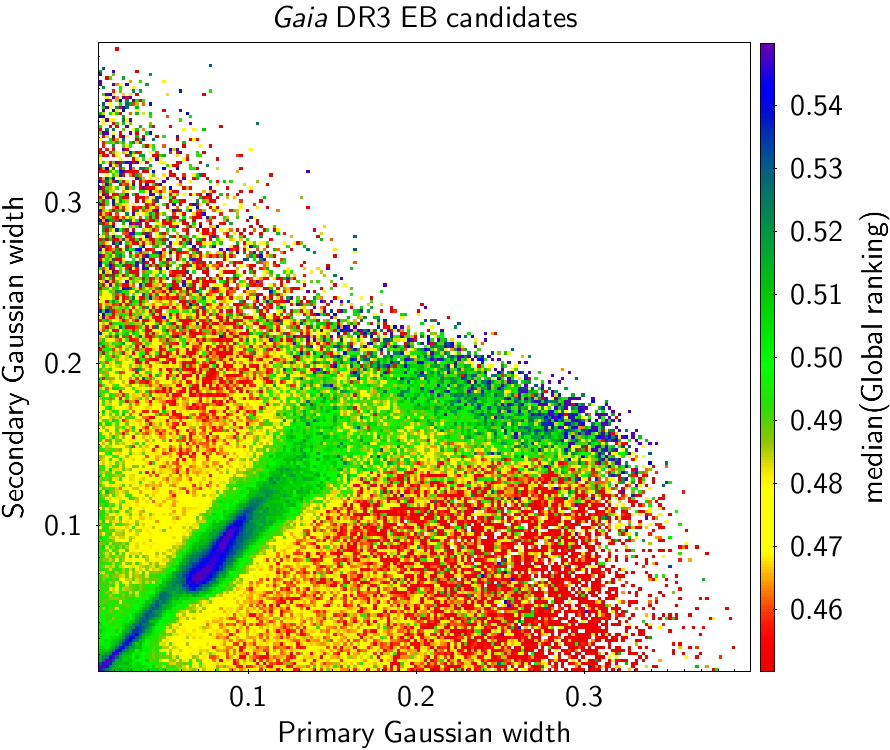}
  \caption{Secondary versus primary Gaussian widths for models without an ellipsoidal component.
           The widths are expressed in phase units.
           \textbf{Top panel}: Density map of the sample, colour-coded according to the colour-scale shown on the right of the panel.
           \textbf{Second panel}: Bin-median value of the phase separation of the two Gaussians, shifted between 0 and 0.5.
           \textbf{Third panel}: Bin-median value of secondary to primary Gaussian depths ratio.
           \textbf{Bottom panel}: Bin-median value of global ranking.
           The values are colour-coded according to the colour-scales on the right of each panel.
           The axes ranges have been restricted for better visibility.
          }
\label{Fig:sigmaGaussians_2G}
\end{figure}

\begin{figure}
  \centering
  \includegraphics[trim={0 0 0 42},clip,width=0.815\linewidth]{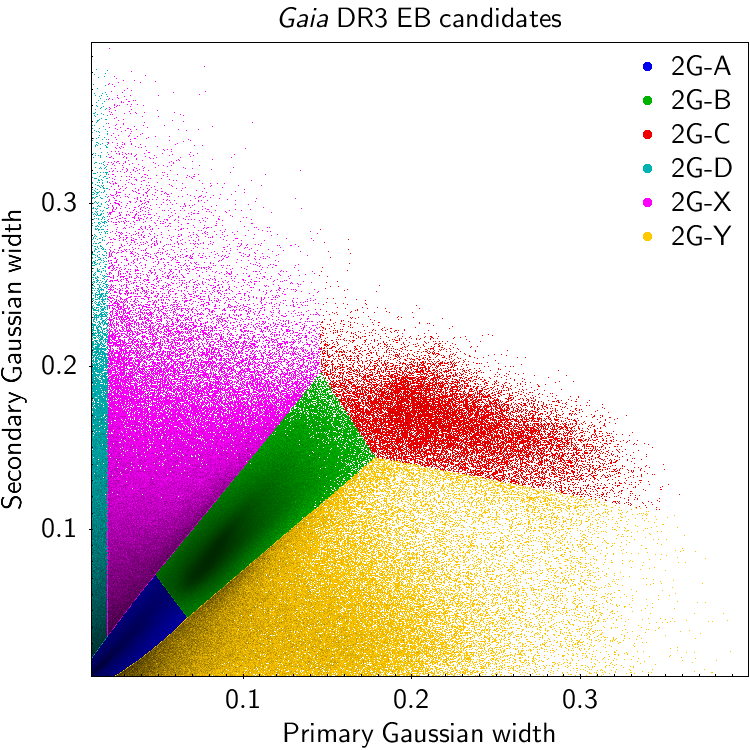}
  \caption{Distribution in the primary versus secondary Gaussian widths plane of the samples defined in Table~\ref{Tab:sample_definition} for the models without an ellipsoidal component.
           %The background grey points represent the full sample of eclipsing binary candidates whose light curves are modelled with two Gaussians and without an ellipsoidal component.
          }
\label{Fig:sigmaGaussians_2G_samples}
\end{figure}

\begin{table*}
\caption{Sample definition.
        }
\centering
\begin{tabular}{l l r}
\hline\hline
Sample &\!\!Definition & \!\!\!\!\!\!\!\!\!\!\!\!\!Nbr. sources\\
\hline
\hline
%--------------------------
\multicolumn{3}{l}{\rule{-8pt}{3.0ex} ----- Two Gaussians only (\small{\texttt{num\_model\_parameters}=8})} \\
2G-A
  &\!\!$-0.007406+0.7\,\sigma_\mathrm{p}+0.4(4.3\,\sigma_\mathrm{p}-0.1)^2 < \sigma_\mathrm{s} < \min(0.0119+1.26\,\sigma_\mathrm{p}-0.6(2\,\sigma_\mathrm{p}-0.1)^2,0.14-1.4\,\sigma_\mathrm{p})$
  & 285\,320 \\
2G-B
  &\!\!$\max(0.14-1.4\,\sigma_\mathrm{p},-0.013972+0.88624\,\sigma_\mathrm{p}) < \sigma_\mathrm{s} < \min(0.012+1.26\,\sigma_\mathrm{p}, 0.43-1.6\,\sigma_\mathrm{p})$
  & 834\,093 \\
2G-C
  &\!\!$\sigma_\mathrm{s} > \max(0.18-0.2\,\sigma_\mathrm{p}, 0.43-1.6\,\sigma_\mathrm{p}, 0.204-51(\sigma_\mathrm{p}-0.146)$
  & 24\,081 \\
2G-D
  &\!\!$\sigma_\mathrm{s}>\sigma_\mathrm{p}$, \; $\sigma_\mathrm{p}<0.02$, \; not in Sample~2G-A
  & 111\,820 \\
2G-X
  &\!\!$\sigma_\mathrm{s}>\sigma_\mathrm{p}$, \; not in samples~2G-[A,B,C,D]
  & 182\,280 \\
2G-Y
  &\!\!$\sigma_\mathrm{s}<\sigma_\mathrm{p}$, \; not in samples~2G-[A,B,C,D]
  & 150\,332 \\
%--------------------------
%\hline
\multicolumn{3}{l}{\rule{-8pt}{3.0ex} ----- Two Gaussians $+$ cosine (\small{\texttt{num\_model\_parameters}=9})} \\
2GE-A
  &\!\!$2\,\Aell < 0.11$, \; $\deltaPhi<0.07$
  & 162\,630 \\
2GE-B
  &\!\!$2\,\Aell \ge 0.11$, \; $\deltaPhi<0.07$
  & 265\,276 \\
2GE-Z
  &\!\!$\deltaPhi \ge 0.07$
  & 47\,219 \\
%--------------------------
%\hline
\multicolumn{3}{l}{\rule{-8pt}{3.0ex} ----- One Gaussian only (\small{\texttt{num\_model\_parameters}=5})} \\
1G
  &\!\!\textit{All}
  & 36\,984 \\
%--------------------------
%\hline
\multicolumn{3}{l}{\rule{-8pt}{3.0ex} ----- One Gaussian $+$ cosine (\small{\texttt{num\_model\_parameters}=6})} \\
1GE
  &\!\!\textit{All}
  & 48\,215 \\
%--------------------------
%\hline
\multicolumn{3}{l}{\rule{-8pt}{3.0ex} ----- One cosine only (\small{\texttt{num\_model\_parameters}=4})} \\
0GE
  &\!\!\textit{All}
  & 36\,227 \\
\hline
\hline
\end{tabular}
\label{Tab:sample_definition}
\end{table*}

This appendix provides some insight on the two-Gaussian model properties of the \Gaia DR3 catalogue of eclipsing binary candidates.
In Sects.~\ref{Sect:catalogue_usage_model_2G} and \ref{Sect:catalogue_usage_model_2GE} we discuss the cases of models including two Gaussians, which constitute the overwhelming majority of the catalogue (94\% of the candidates, see Table~\ref{Tab:modelTypes} in the main body of the article).
Models containing one Gaussian (4\% of the catalogue) are then considered in Sects.~\ref{Sect:catalogue_usage_model_1G} and \ref{Sect:catalogue_usage_model_1GE}.
The remaining 2\% of candidates, whose light curves are modelled with only an ellipsoidal component, are discussed in Sect.~\ref{Sect:catalogue_usage_model_ellipsoidal}.
A summary is provided in Sect.~\ref{Sect:catalogue_usage_summary}.

%- - - - - - - - - - - - - - - - - - - -
\subsection{Models with only two Gaussians}
\label{Sect:catalogue_usage_model_2G}

Almost three quarters (73\%) of the sources in the catalogue have their light curves modelled by two Gaussians and without an ellipsoidal component.
They are identified in the \Gaia table with {\small\texttt{model\_type=TWOGAUSSIANS}} or {\small\texttt{num\_model\_parameters}=8}.

The distributions of the phase widths $\sigma_p$ and $\sigma_s$ of their primary and secondary Gaussian functions, respectively, are shown in Fig.~\ref{Fig:sigmaGaussians_2G} (top panel).
The bulk of the sources lies along the diagonal line of equal widths, where two distinct samples can be discerned.
We define samples 2G-A and 2G-B as the ones at smaller and larger widths, respectively.
A third distinct sample is visible at $\sigma_\mathrm{p} \gtrsim 0.17$, which we label Sample~2G-C.
The exact definitions of each sample is given in  Table~\ref{Tab:sample_definition}, and their locations in the $\sigma_p$ -- $\sigma_s$ plane is summarised in Fig.~\ref{Fig:sigmaGaussians_2G_samples}.
We briefly describe their properties in the next paragraphs.

%- - - - - - - - - - - - - - - - - - - -
\paragraph{Sample 2G-A.}

\begin{figure}
  \centering
  \includegraphics[trim={40 150 0 70},clip,width=\linewidth]{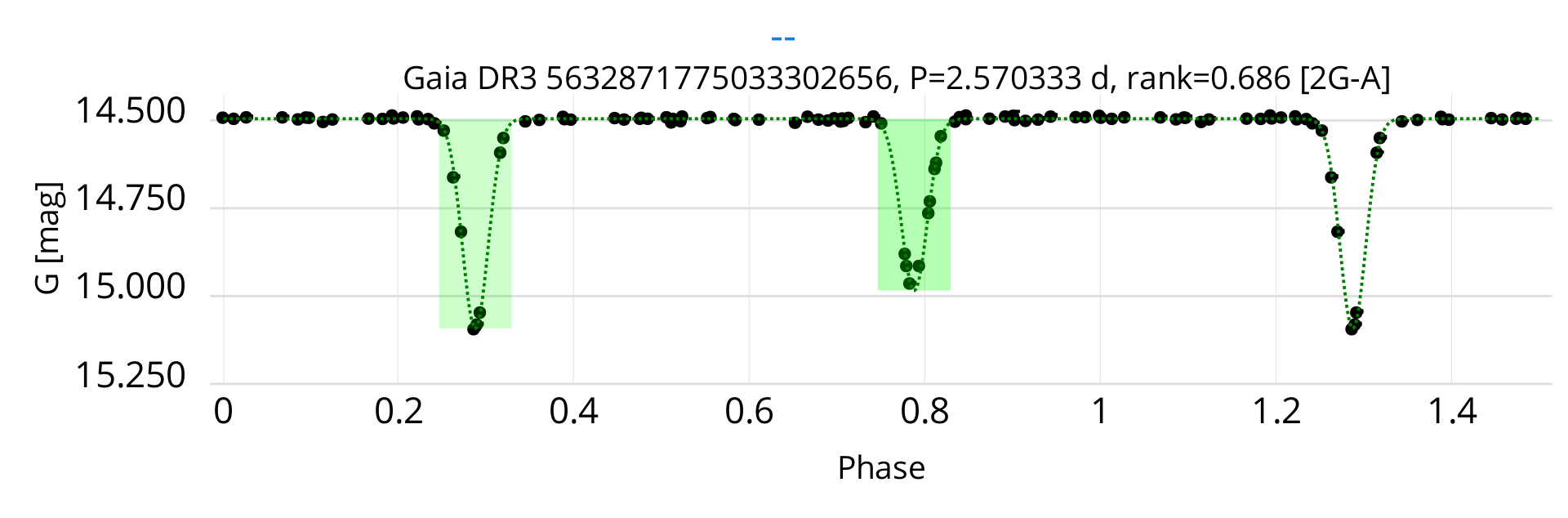}
  \vskip -0.5mm
  \includegraphics[trim={40 145 0 80},clip,width=\linewidth]{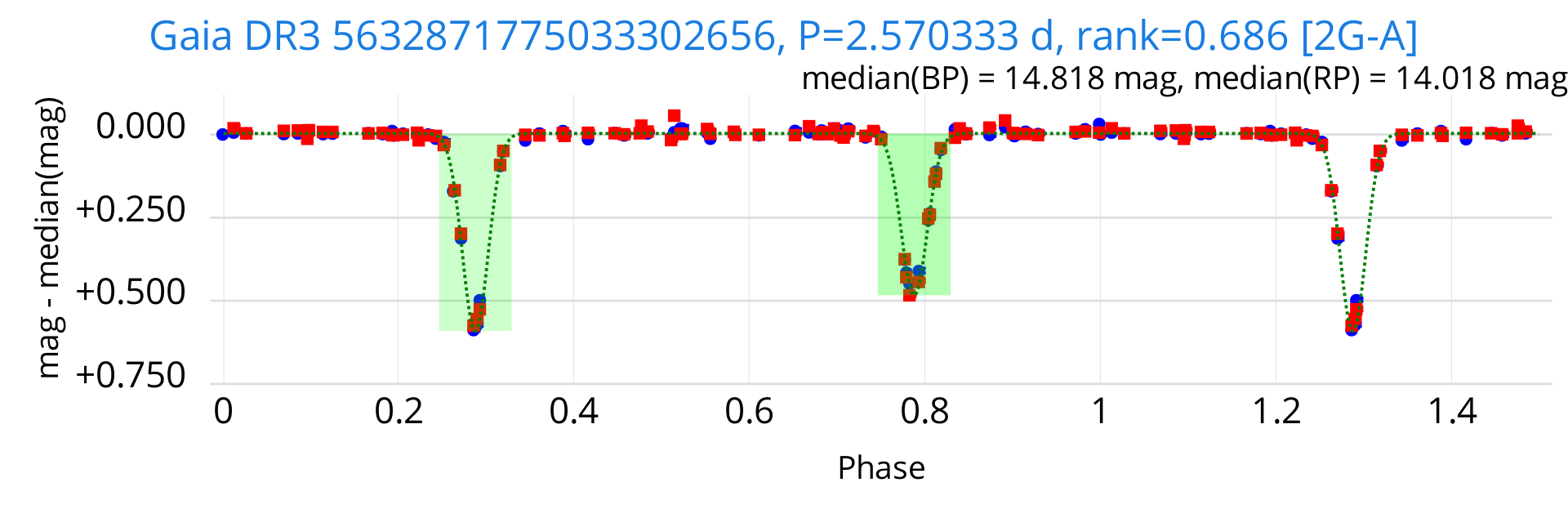}
  \vskip -0.5mm
  \includegraphics[trim={40 150 0 70},clip,width=\linewidth]{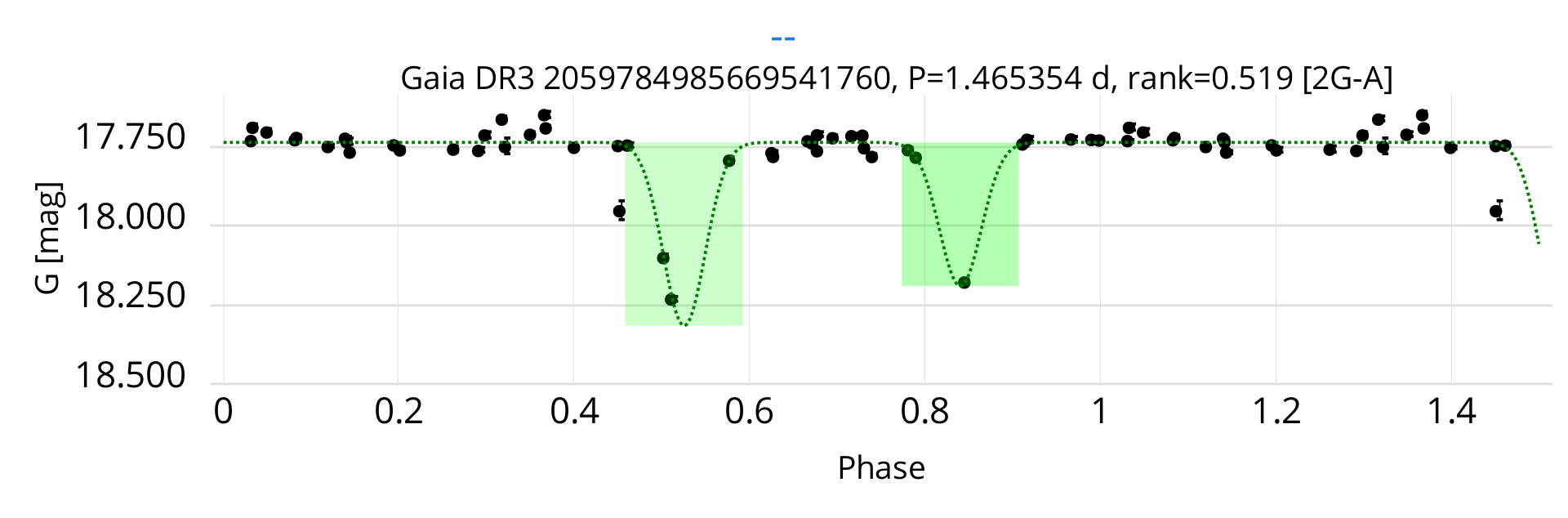}
  \vskip -0.5mm
  \includegraphics[trim={40 45 0 80},clip,width=\linewidth]{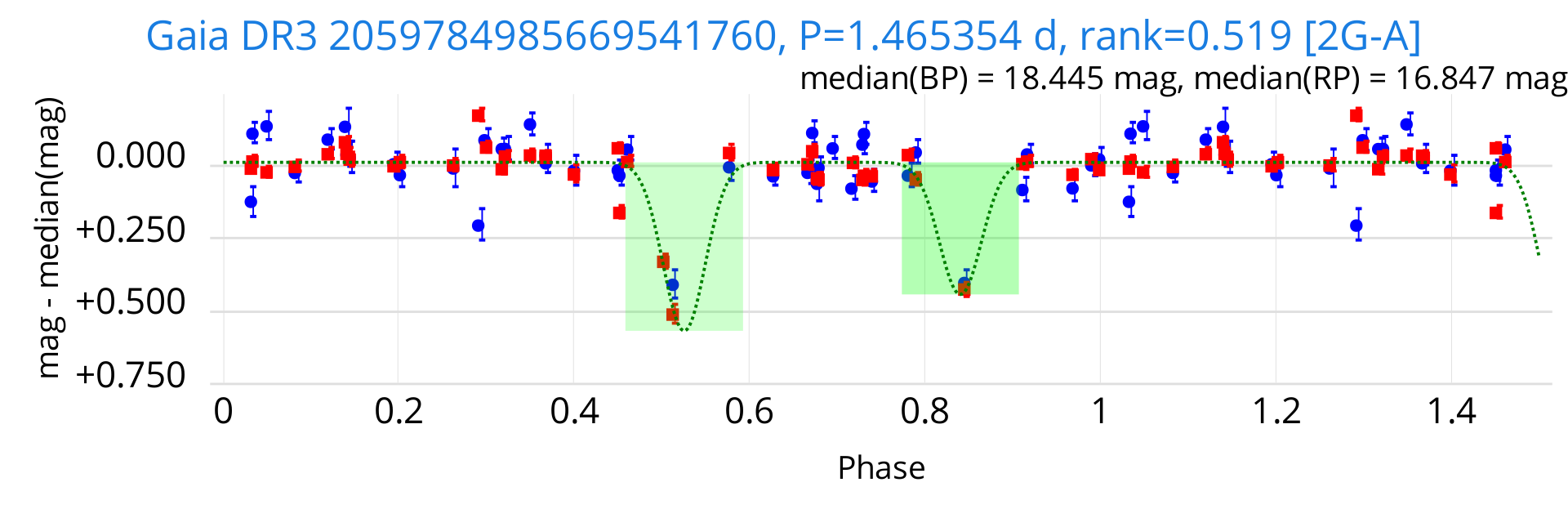}
  \caption{Example \gmag light curves  (top panels in each set) and \gbp and \grp light curves (bottom panels in each set) of sources in Sample~2G-A modelled with only two Gaussians.
           The \gbp and \grp magnitudes are shifted by a value equal to their median magnitudes, the values of which are given in the panel.
           The two-Gaussian model obtained from the \gmag light curve is superposed in dotted line in both panels of each set.
           The green areas indicate the derived eclipse durations.
           The top set is for the circular candidate \GaiaSrcIdInCaption{5632871775033302656} and the bottom set is for the eccentric candidate \GaiaSrcIdInCaption{2059784985669541760}.
            }
\label{Fig:lcs_2G_A}
\end{figure}

%-----------
\begin{figure}
  \centering
  \includegraphics[trim={40 150 0 70},clip,width=\linewidth]{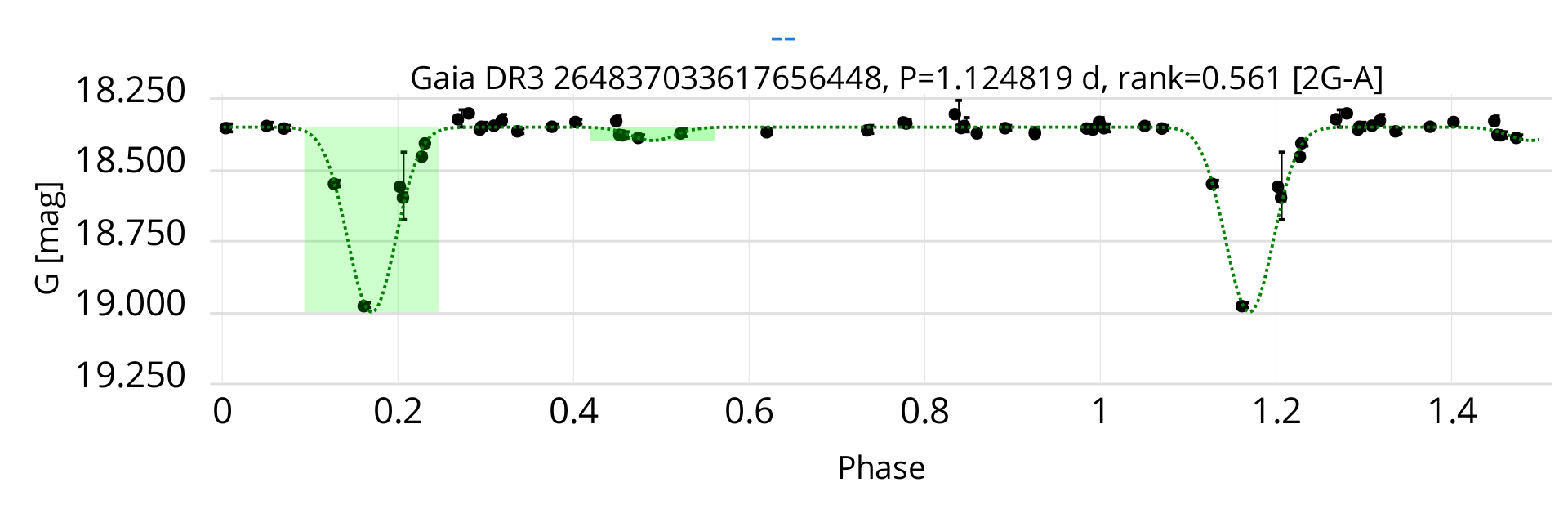}
  \vskip -0.5mm
  \includegraphics[trim={40 145 0 80},clip,width=\linewidth]{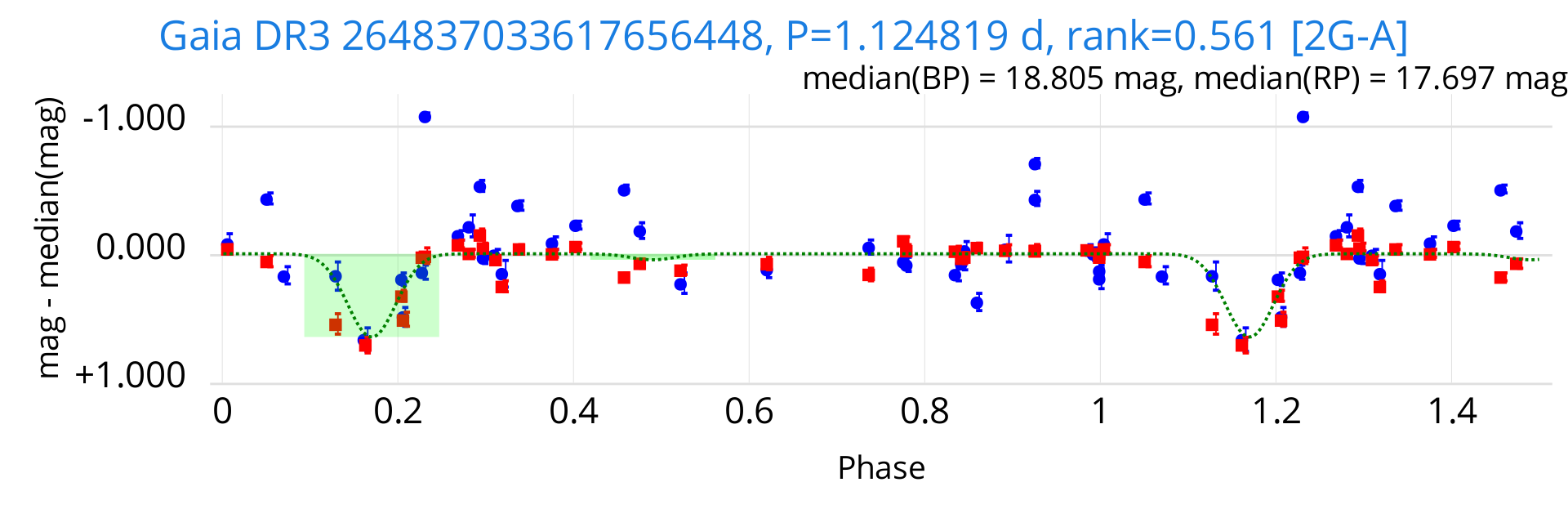}
  \vskip -0.5mm
  \includegraphics[trim={40 150 0 70},clip,width=\linewidth]{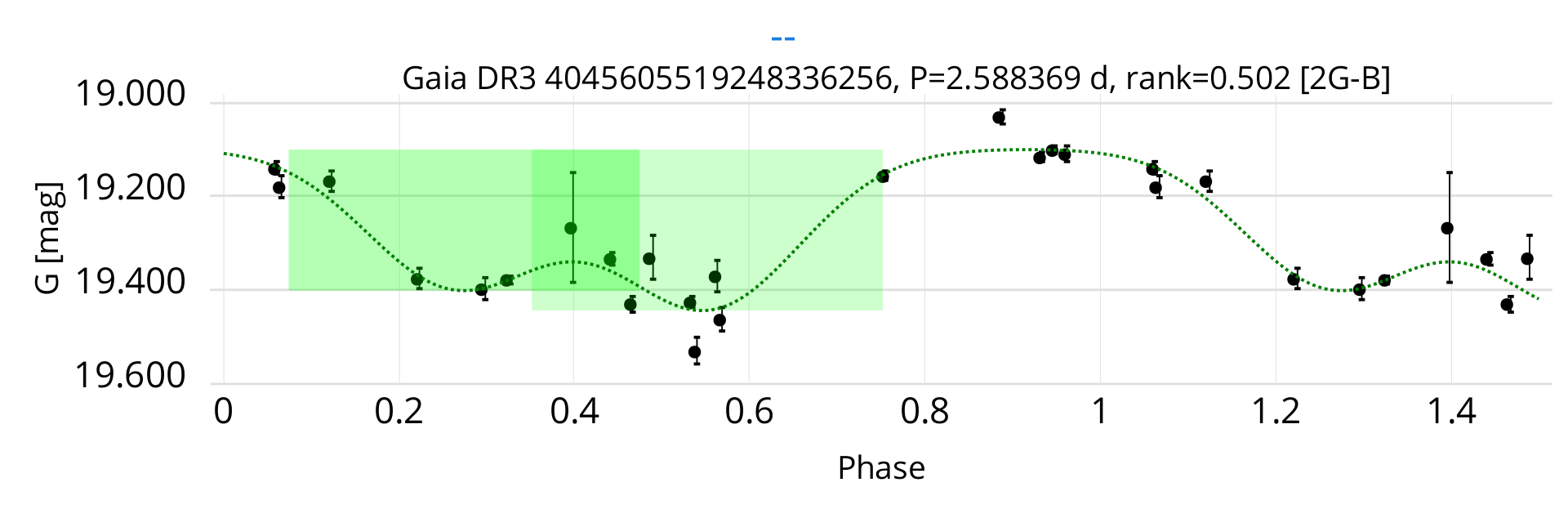}
  \vskip -0.5mm
  \includegraphics[trim={40 145 0 80},clip,width=\linewidth]{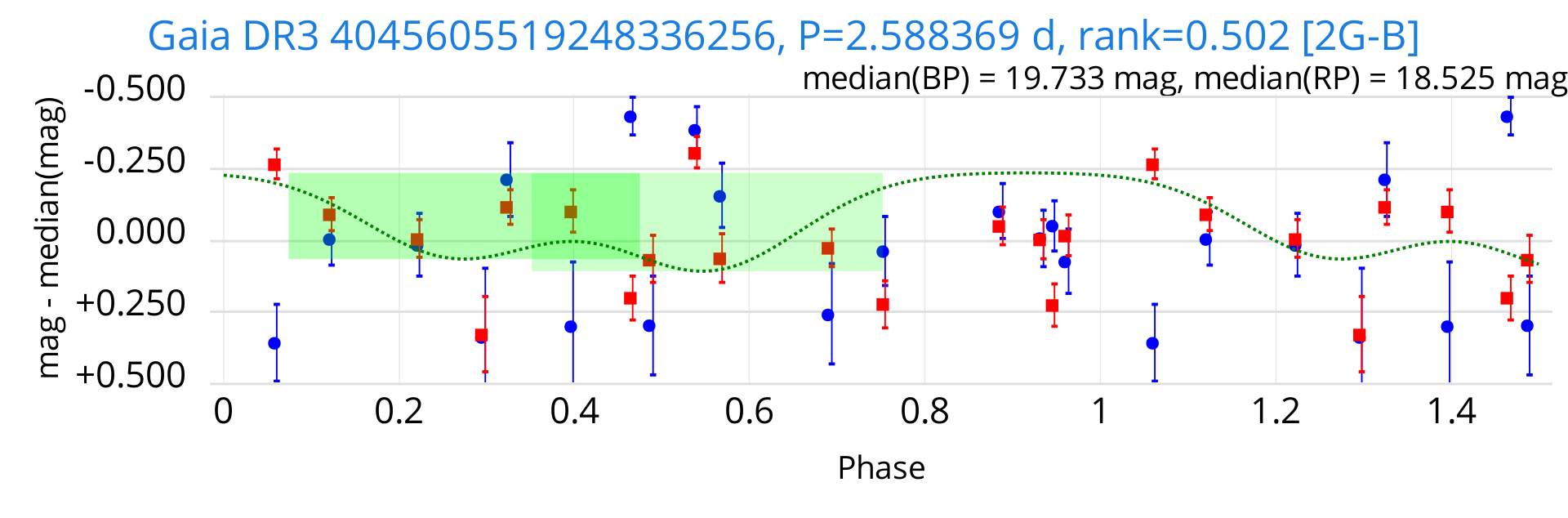}
  \vskip -0.5mm
  \includegraphics[trim={40 150 0 70},clip,width=\linewidth]{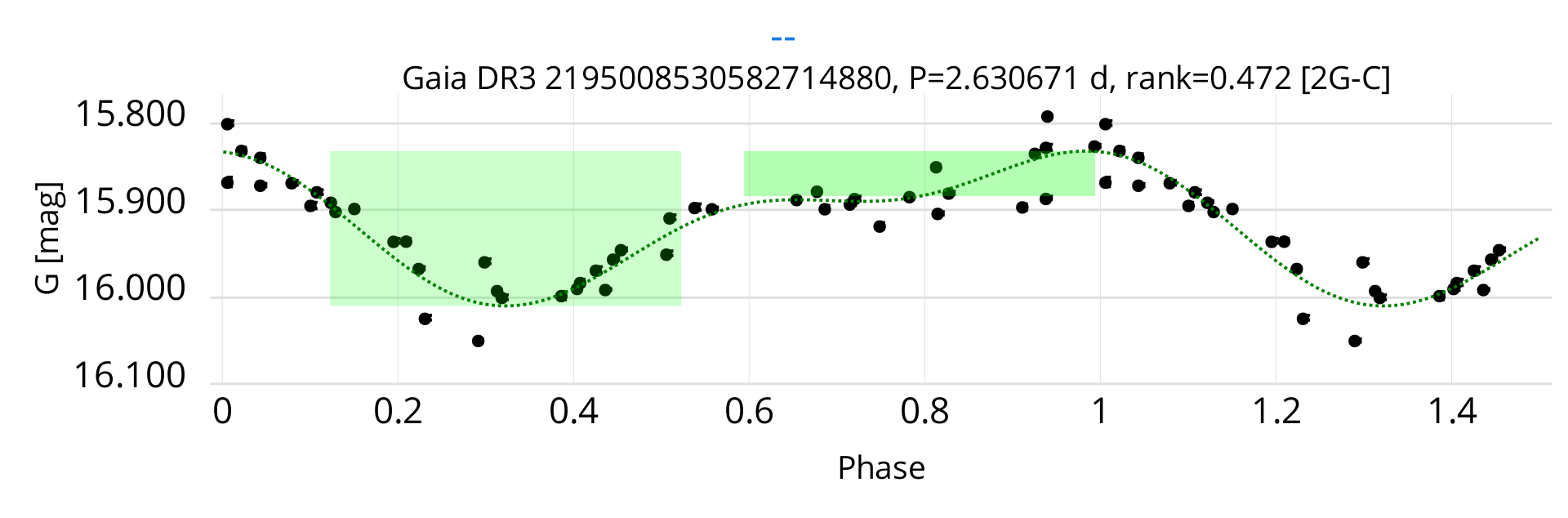}
  \vskip -0.5mm
  \includegraphics[trim={40 45 0 80},clip,width=\linewidth]{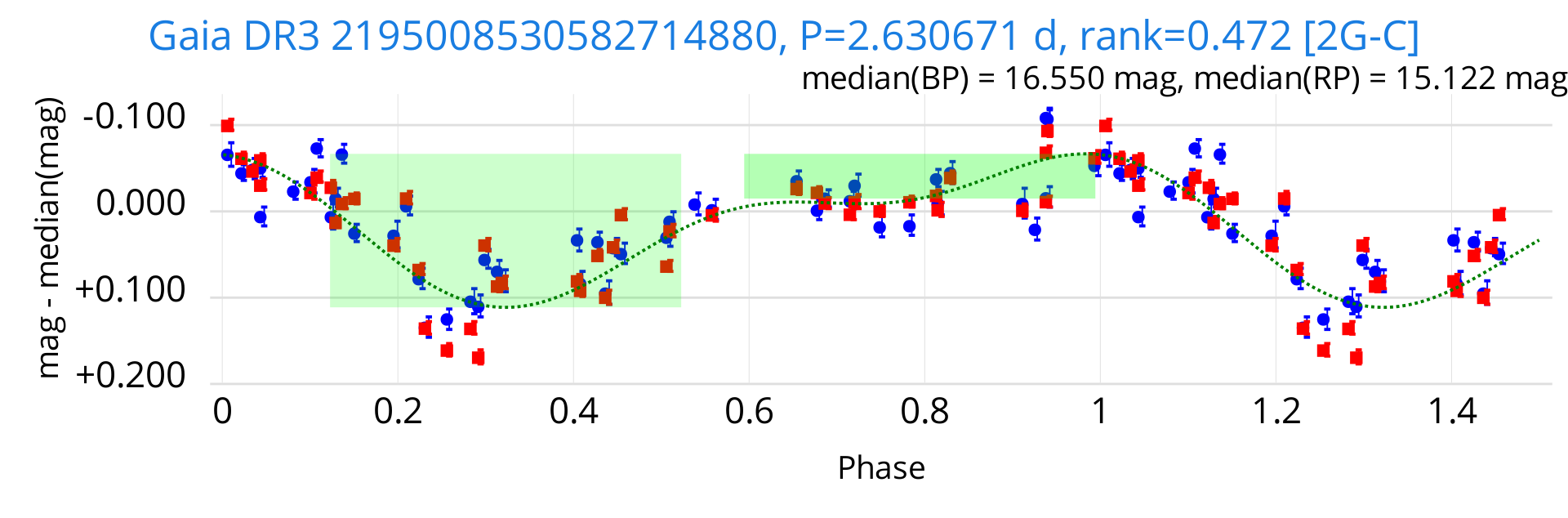}
  \caption{Same as Fig.~\ref{Fig:lcs_2G_A}, but illustrating cases of wrong model identifications.
           From top to bottom sets: spurious secondary eclipse identification in Sample~2G-A (\GaiaSrcIdInCaption{264837033617656448}), wrong period estimate in Sample~2G-B (\GaiaSrcIdInCaption{4045605519248336256}), spurious ellipsoidal variable in Sample~2G-C (\GaiaSrcIdInCaption{2195008530582714880}).
          }
\label{Fig:lcs_2G_spurious}
\end{figure}
%-----------

\begin{figure}
  \centering
  \includegraphics[trim={40 150 0 70},clip,width=\linewidth]{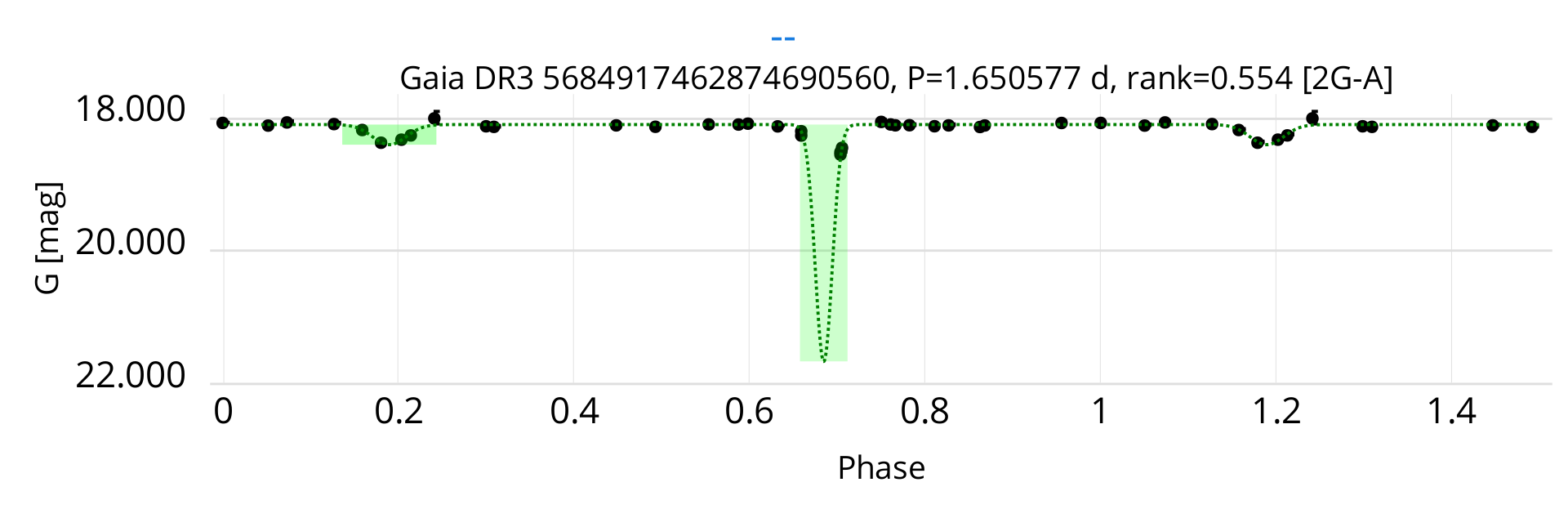}
  \vskip -0.5mm
  \includegraphics[trim={40 45 0 80},clip,width=\linewidth]{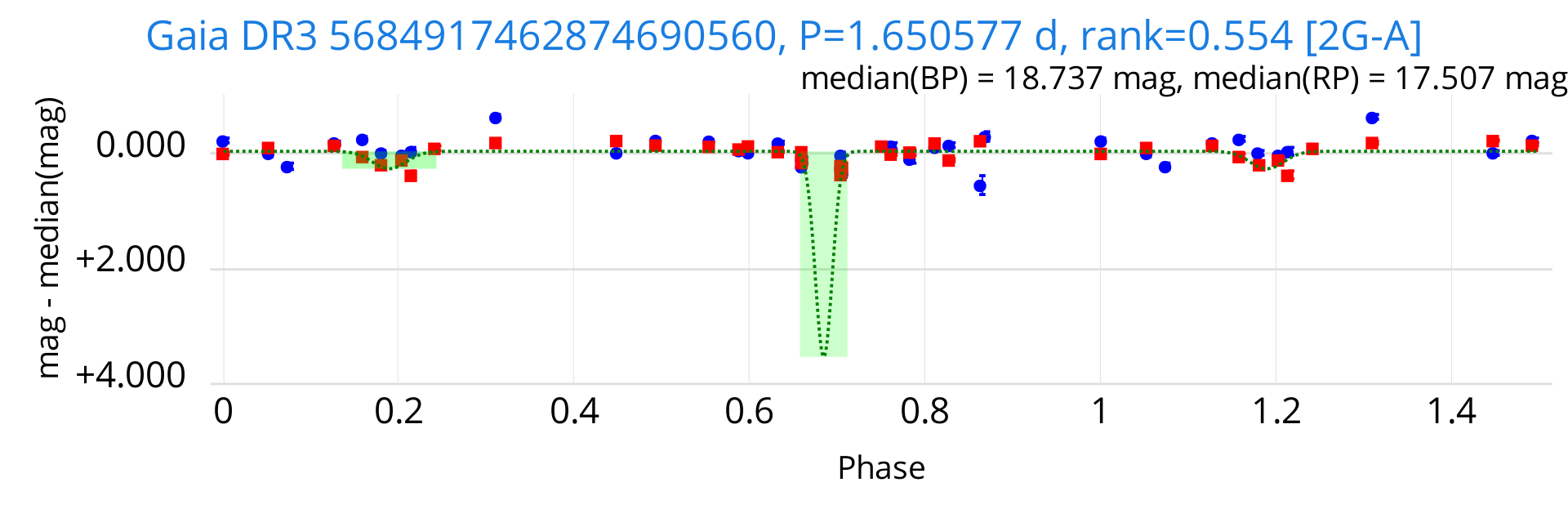}
  \caption{Same as Fig.~\ref{Fig:lcs_2G_A}, but for a case (\GaiaSrcIdInCaption{5684917462874690560}) with an insufficient coverage of the primary eclipse that leads to a poor constrain of its depth.
            }
\label{Fig:lcs_2G_A_unconstrainedDepth}
\end{figure}

\begin{figure}
  \centering
  \includegraphics[trim={0 80 0 42},clip,width=\linewidth]{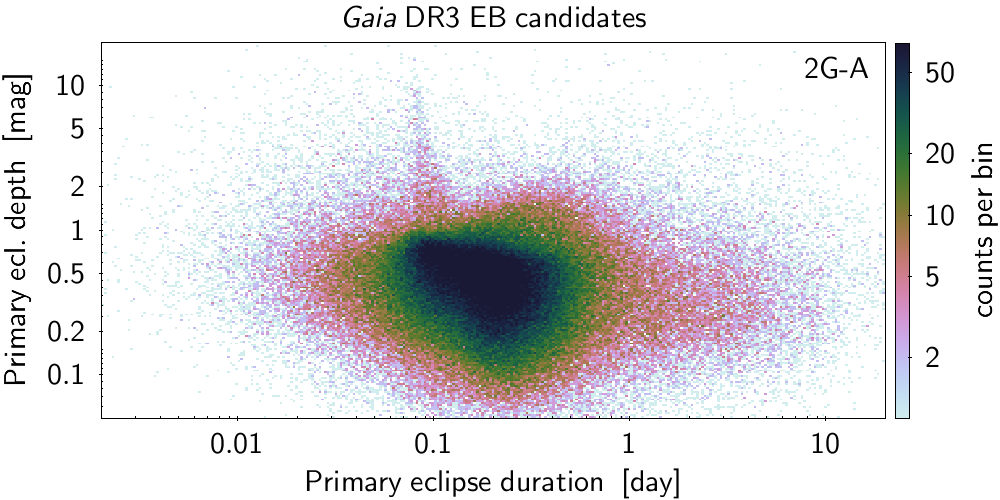}
  \vskip -0.5mm
  \includegraphics[trim={0 80 0 42},clip,width=\linewidth]{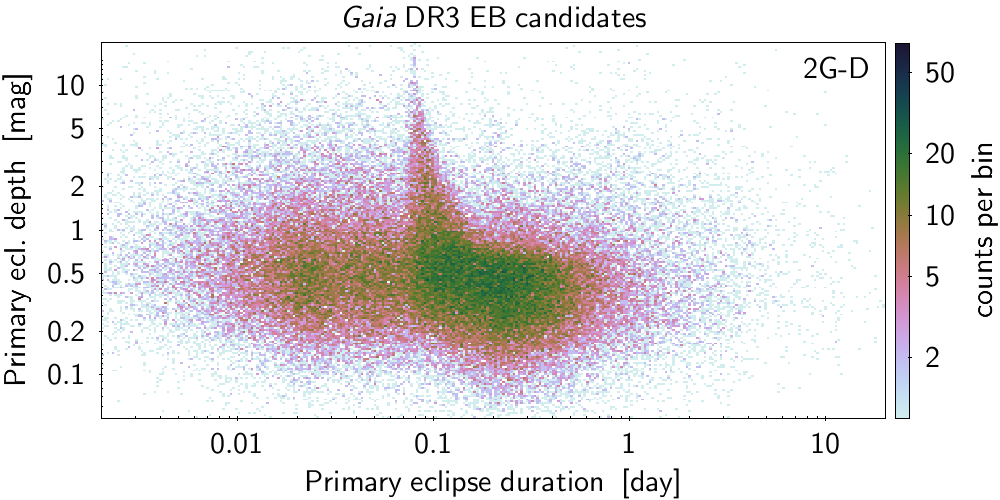}
  \vskip -0.5mm
  \includegraphics[trim={0 0 0 42},clip,width=\linewidth]{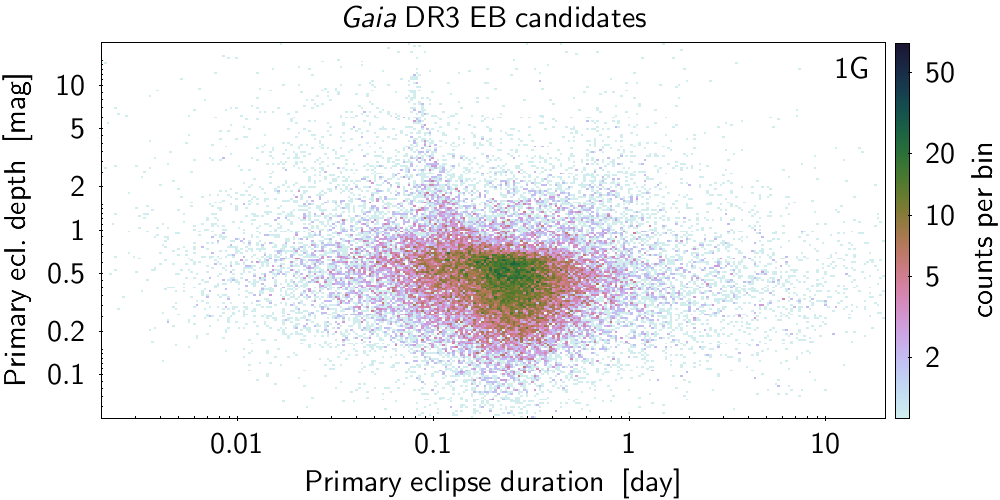}
  \caption{Density map of the primary eclipse depth versus duration (in days) for the samples 2G-A (top panel), 2G-D (middle panel) and 1G (bottom panel).
           The primary eclipse duration is taken equal to $\durationEclPrimary\,P_\mathrm{orb}$.
           The larger occurrence of eclipse depths larger than about one magnitude for eclipse durations between $\sim$0.07 and $\sim$0.17 days is linked to the equivalent time intervals between successive observations in the two \Gaia fields of view (see text, Sect.~\ref{Sect:catalogue_usage_model_2G}).
           The axes ranges have been limited for better visibility.
          }
\label{Fig:eclDepthVsEclDration}
\end{figure}

The first sample identified along the diagonal line of equal widths at phase widths smaller than $\sim$0.06 (blue region in Fig.~\ref{Fig:sigmaGaussians_2G_samples}) contains well detached eclipsing binaries.
It can also contain semi-detached systems if the star that fills its Roche Lobe is much fainter than the primary star such that the ellipsoidal variability induced by the pear-shaped faint companion remains undetected in the light curve.
In this case, the secondary eclipse is expected to be much shallower than the primary eclipse.
The two Gaussians in Sample~2G-A are separated by 0.5 in phase in the majority of cases, as shown in the second panel from top in Fig.~\ref{Fig:sigmaGaussians_2G}, suggesting circular orbits for the majority of them.
Two example light curves\footnote{
The \gbp and \grp light curves are shown together with the \gmag light curves in all the examples shown in this Appendix for information purpose only.
These \gbp and \grp light curves were not used in the processing pipeline that led to the results published in \Gaia DR3.
}
are shown in Fig.~\ref{Fig:lcs_2G_A}, one with a circular orbit and one with an eccentric orbit.

The depth ratio between the two eclipses takes all values between (close-to) zero and one (thin blue histogram in the top panel of Fig.~\ref{Fig:histo_depthRatios_2G} in the main body of the text), suggesting, for close-to-circular systems, a wide range of luminosity ratios between the two binary stars.
The median value of this ratio is about 0.6 in all bins in the $\sigma_p$ -- $\sigma_s$ diagram where $\sigma_p \simeq \sigma_s$, as seen in the third panel of Fig.~\ref{Fig:sigmaGaussians_2G}.
The per-bin median depth ratio, however, drops quickly when moving away from the equal-width area, with ratios reaching below 0.3 at the edges of the 2G-A region.
This may indicate the presence of sources for which the two Gaussians are not catching the real eclipse dips.
Two-gaussian models that realistically describe the eclipse properties of well-detached eclipsing binaries are expected to have similar Gaussian widths.

Detection of spurious eclipses by the automated algorithm is unavoidable.
An example of such a case is shown in Fig.~\ref{Fig:lcs_2G_spurious} (top example), where a spurious shallow dip detected close to the primary eclipse is identified as the secondary eclipse, probably due to the lack of measurements at a phase distant by 0.5 from the primary, or due to a period too small by a factor of two.
We note that the primary eclipse seems still correct in that example.
A visual check of a random set of 100 sources in Sample~2G-A leads to an estimated 5-10\% sources that could have spurious secondary eclipse identifications.

Finally, we must caution that an eclipse may lack enough phase coverage to properly constrain its depth.
An example of such a case is shown in Fig.~\ref{Fig:lcs_2G_A_unconstrainedDepth}.
The Gaussian depths and derived eclipse depths are consequently much larger than the magnitude range of the observations, as expected from the lack of observations in the faint parts of the eclipse.
Additional observations are needed to confirm the true eclipse depth of these cases.
Figure~\ref{Fig:eclDepthVsEclDration} (top panel) reveals that this mainly occurs for eclipse durations between $\sim$0.07 and $\sim$0.17 days.
These durations correspond to the time intervals between successive passages in the \Gaia field of views, which are 1.775 hours from FOV1 to FOV2 and 4.225 hours from FOV2 to FOV1 given the 6-hour rotation period of the spacecraft and the 106.5 degree basic angle between the two fields of view.
For eclipse durations in this time interval, observations of a particular source may be lacking in the core of its eclipse depending on the observation time distribution over the 34 months covered in DR3.
Let us consider such a source where only very few observations fall in its eclipse time window during these 34 months.
% Visual inspection of the light curves suggests that these cases have consecutive observations in FOV1 and FOV2 at magnitudes that are convincingly deeper than the rest of the light curve. The way we sele
For durations larger than 4.2~h, the probability to have measurements inside the eclipse is large as the source will be observed during a minimum of two successive FOV passages.
For durations smaller than 4.2~h (but larger than 1.8~h), the probability to have enough observations in the middle of the eclipse decreases with decreasing eclipse duration.
If measurements are still available at the edge of the eclipse, the eclipse will be caught by the pipeline, but with rather unconstrained eclipse depth.
For durations below 1.8~h, the probability to have observations only at the edges of the eclipse decreases considerably as this duration is shorter than the shortest time interval between two successive FOVs.
These aspects explain the excess of sources with (too) large primary eclipse depths in Fig.~\ref{Fig:eclDepthVsEclDration} for eclipse durations between $\sim$0.07 and $\sim$0.17 days.

%- - - - - - - - - - - - - - - - - - - -
\paragraph{Sample 2G-B.}

\begin{figure}
  \centering
  \includegraphics[trim={40 150 0 70},clip,width=\linewidth]{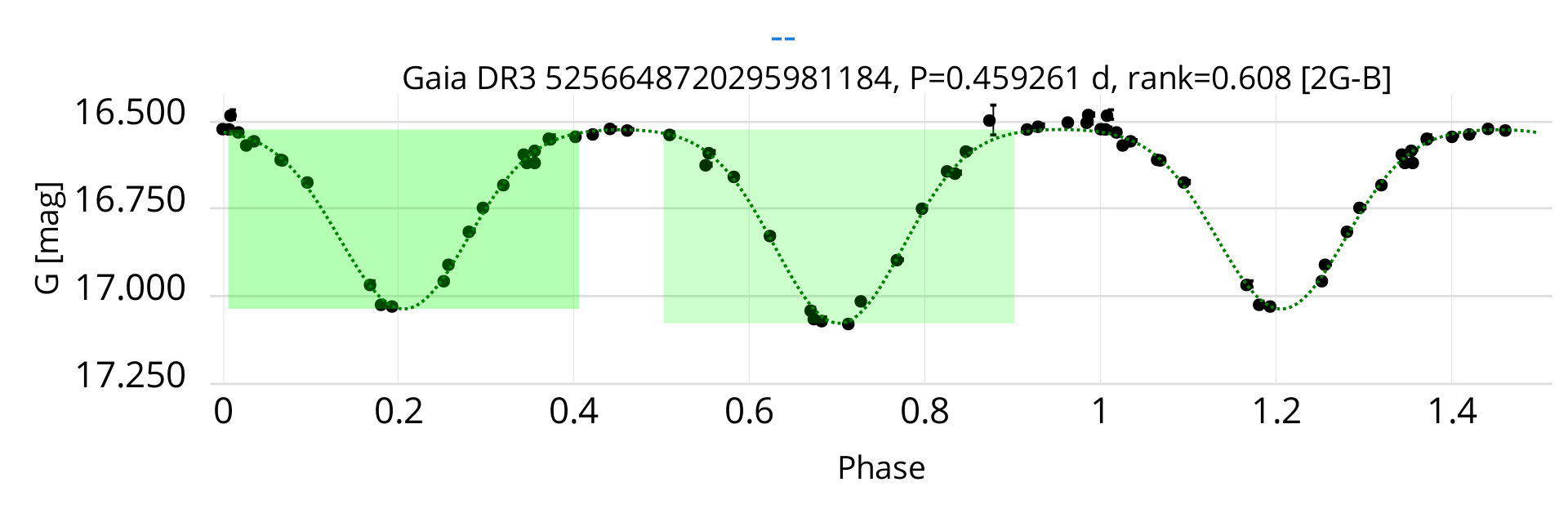}
  \vskip -0.5mm
  \includegraphics[trim={40 145 0 80},clip,width=\linewidth]{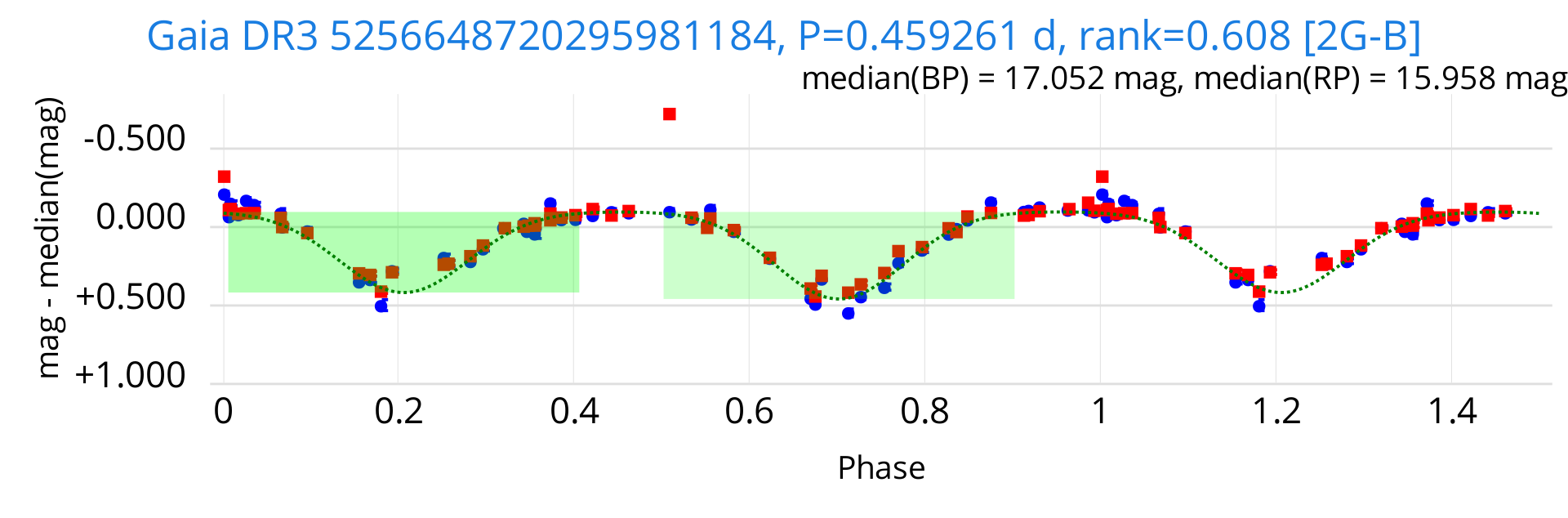}
  \vskip -0.5mm
  \includegraphics[trim={40 150 0 70},clip,width=\linewidth]{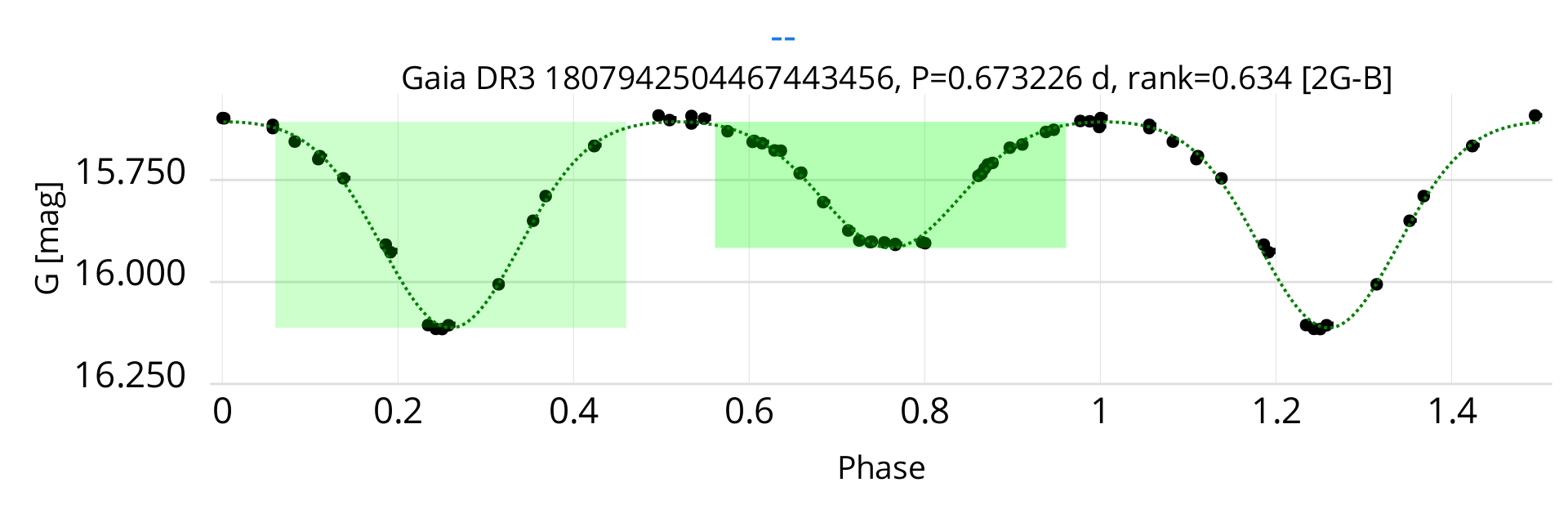}
  \vskip -0.5mm
  \includegraphics[trim={40 45 0 80},clip,width=\linewidth]{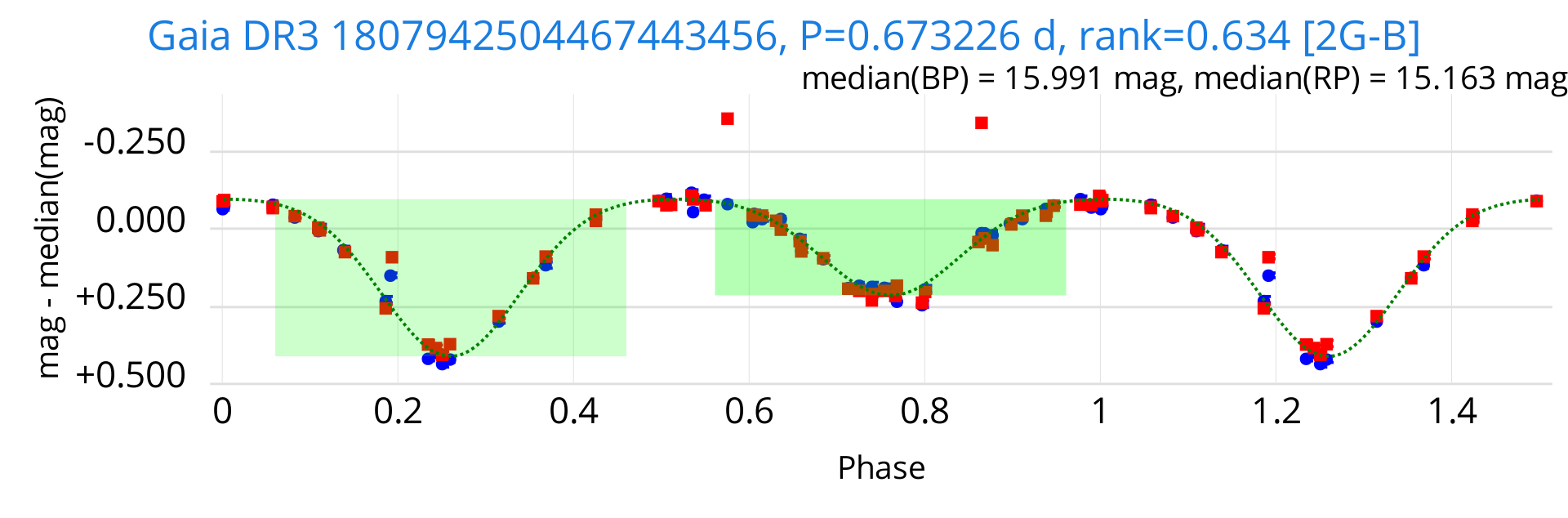}
    \caption{Same as Fig.~\ref{Fig:lcs_2G_A}, but for two sources in Sample~2G-B of light curves modelled with only two Gaussians.
             The top set is for EW-type eclipsing binary candidate \GaiaSrcIdInCaption{5256648720295981184} and the bottom set is for EB-type eclipsing binary candidate \GaiaSrcIdInCaption{1807942504467443456}.
            }
\label{Fig:lcs_2G_B}
\end{figure}
 
The second sample identified along the line of equal-Gaussian widths in the top panel of Fig.~\ref{Fig:sigmaGaussians_2G} lies at phase widths between 0.06 and 0.15.
It is the most populated region in the diagram.
Their larger Gaussian widths lead to the absence of flat inter-eclipse phases.
These are tighter binaries than candidates in Sample~2G-A.
The phase separation between the two eclipses is close to 0.5 (see second panel from top of Fig.~\ref{Fig:sigmaGaussians_2G}), as expected for these types of eclipsing binaires.
The distribution of the eclipse depth ratio (green thick histogram in the top panel of Fig.~\ref{Fig:histo_depthRatios_2G}) peaks at one with a tail extending down to below 0.4.
Example light curves are shown in Fig.~\ref{Fig:lcs_2G_B}.

In this sample~2G-B, spurious cases can happen when a potentially wrong period is obtained.
An example of such a case is shown in the second source from top in Fig.~\ref{Fig:lcs_2G_spurious}.
Visual inspection of a random set of 100 sources in Sample~2G-B reveals only one such case, suggesting a very low fraction of spurious cases on the order of the percent.
%\NM{It may be possible to identify these spurious cases using the extent of overlap of the eclipses. This technique can further be explored and documented, if time permits...}

%- - - - - - - - - - - - - - - - - - - -
\paragraph{Sample 2G-C.}

\begin{figure}
  \centering
  \includegraphics[trim={40 150 0 70},clip,width=\linewidth]{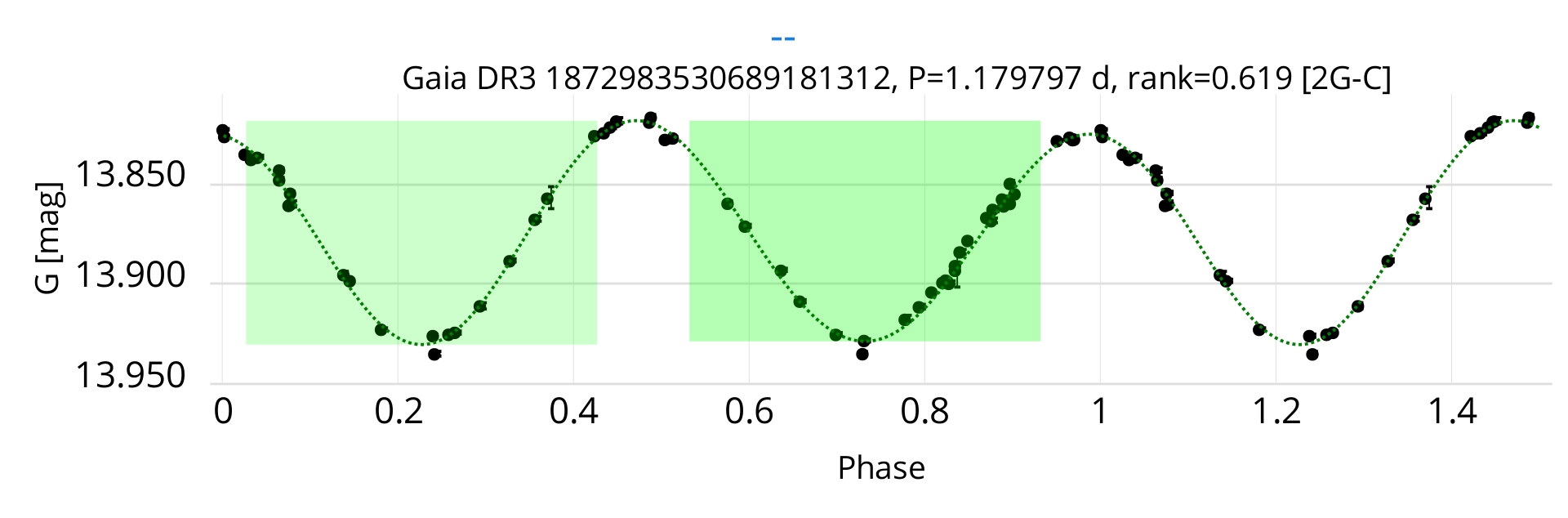}
  \vskip -0.5mm
  \includegraphics[trim={40 145 0 80},clip,width=\linewidth]{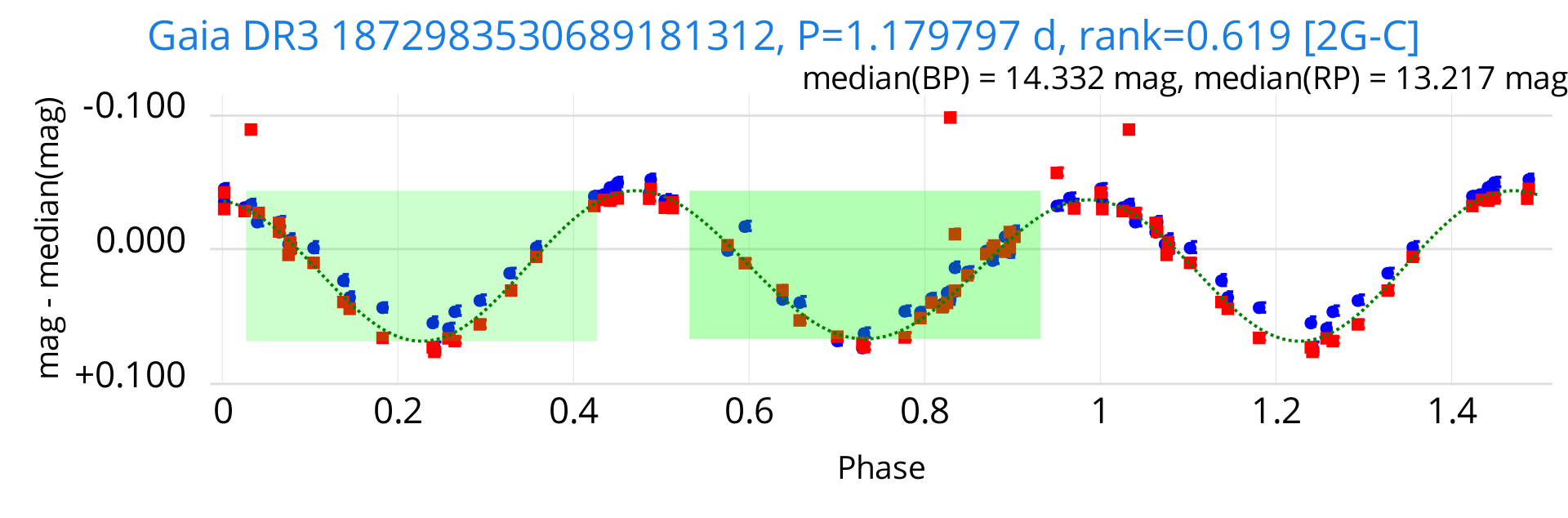}
  \vskip -0.5mm
  \includegraphics[trim={40 150 0 70},clip,width=\linewidth]{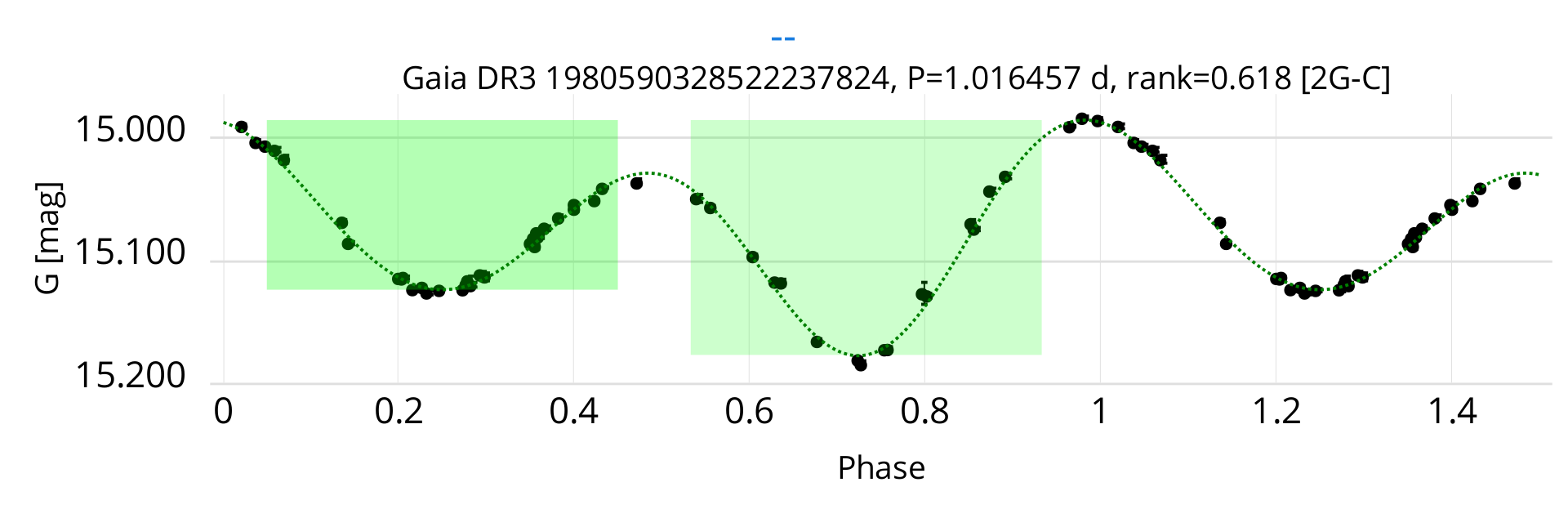}
  \vskip -0.5mm
  \includegraphics[trim={40 45 0 80},clip,width=\linewidth]{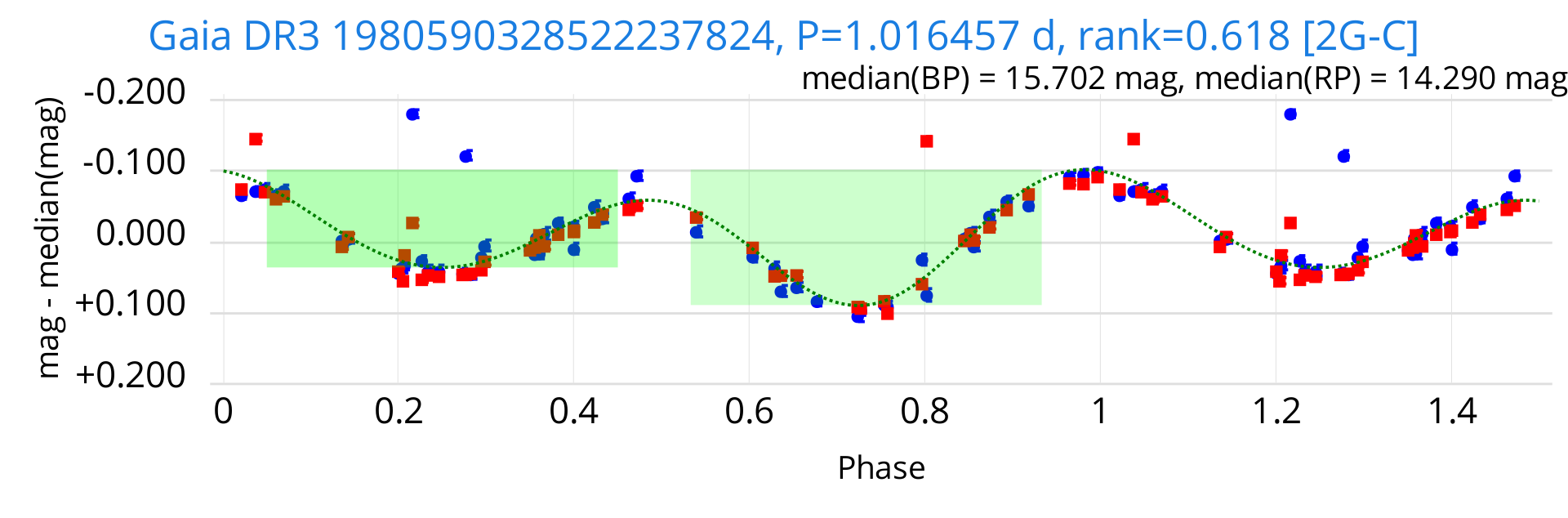}
    \caption{Same as Fig.~\ref{Fig:lcs_2G_A}, but for two ellipsoidal variable candidates from Sample~2G-C of light curves modelled with only two Gaussians.
             The top set is for a typical ellipsoidal variable (\GaiaSrcIdInCaption{1872983530689181312}) and the bottom panel for an ellipsoidal variable with light amplitude modulation (\GaiaSrcIdInCaption{1980590328522237824}).
            }
\label{Fig:lcs_2G_C}
\end{figure}

\begin{figure}
  \centering
  \includegraphics[trim={0 80 0 42},clip,width=\linewidth]{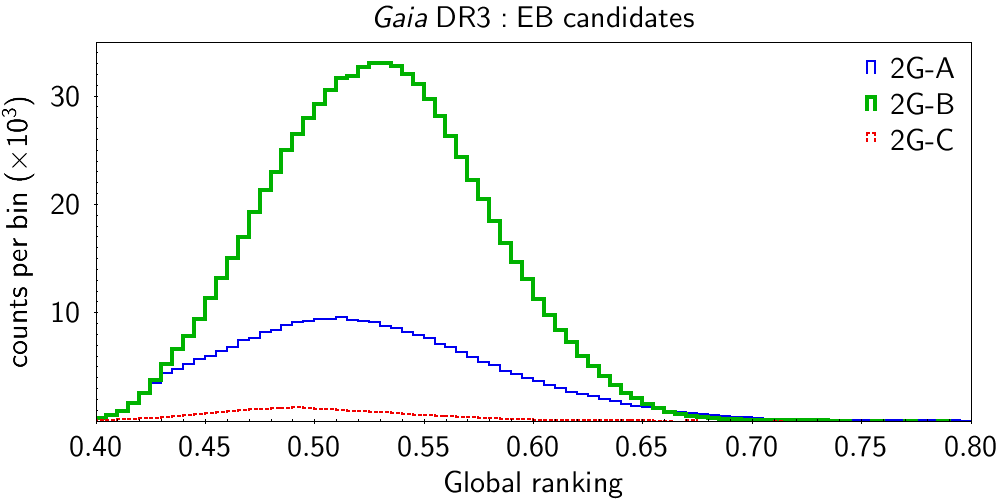}
  \vskip -0.5mm
  \includegraphics[trim={0 0 0 42},clip,width=\linewidth]{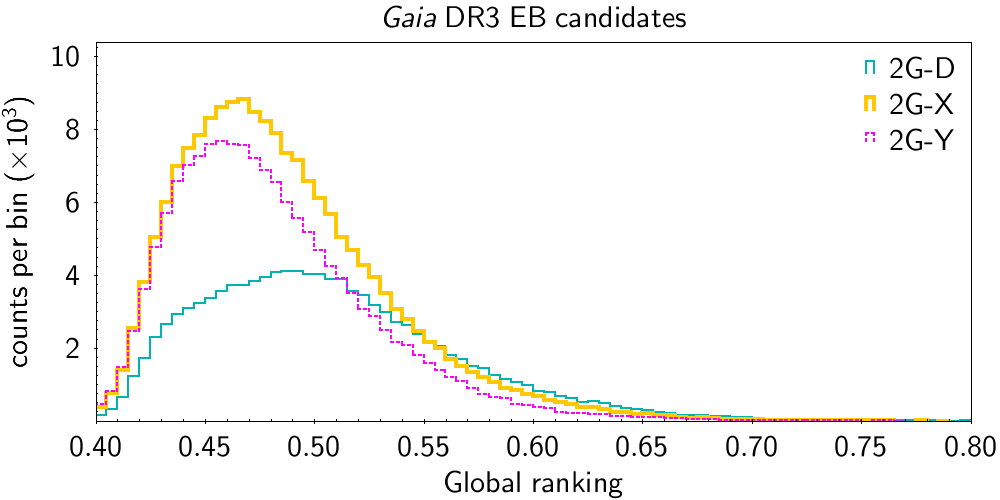}
  \caption{Same as Fig.~\ref{Fig:histo_depthRatios_2G}, but for the global ranking.
           The histograms are not area-normalized.
          }
    \label{Fig:histo_globalRanking_2G}
\end{figure}

\begin{figure}
  \centering
  \includegraphics[trim={0 74 0 42},clip,width=\linewidth]{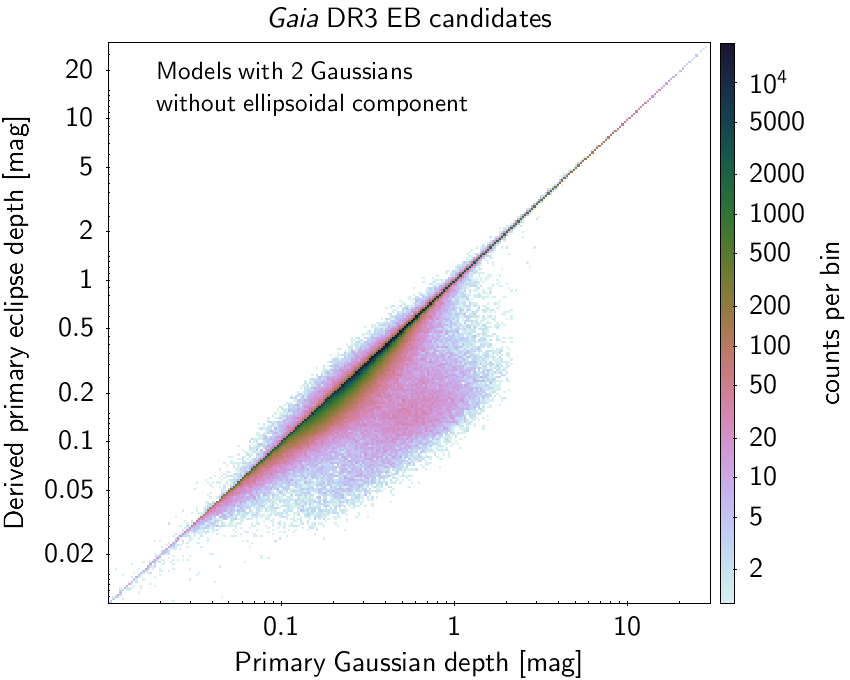}
  \vskip -0.5mm
  \includegraphics[trim={0 0 0 44},clip,width=\linewidth]{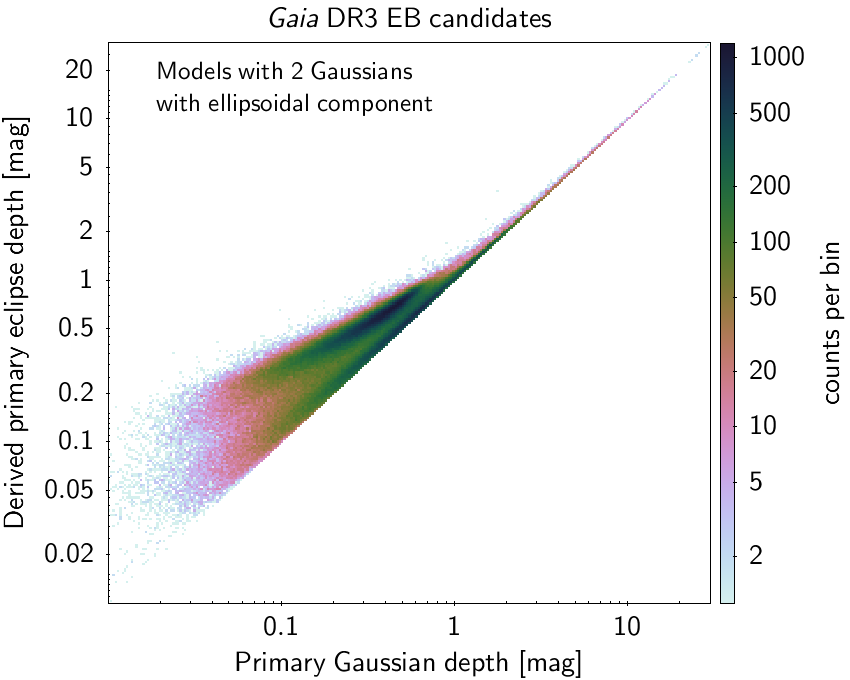}
  \caption{Density map of the depth of the primary eclipse versus the depth of the deepest Gaussian.
           The sample of sources whose light curves are modelled with two Gaussians and without an ellipsoidal component is shown in the top panel, while those with two Gaussians and a cosine function are shown in the bottom panel.
           The density in the maps is colour coded according to the colour scales shown on the right of each panel.
           The axes ranges have been restricted for better visibility.
          }
\label{Fig:derivedVsGaussianPrimaryDepth}
\end{figure}

A subset of the sources shown in the top panel of Fig.~\ref{Fig:sigmaGaussians_2G} forms a distinct group from the mainstream of sources, at primary Gaussian widths larger than $\sim$0.17.
We define it as Sample~2G-C.
It represents less than 2\% of the {\small\texttt{TWOGAUSSIANS}} models.
The light curves of these sources are modelled with two wide and deep overlapping Gaussians.
When the Gaussians have similar depths and widths, which is the case at $\sigma_p \simeq 0.2$, the light curve shape is close to sinusoidal.
An example is shown in Fig.~\ref{Fig:lcs_2G_C}, top case.
For $\sigma_p$ values above 0.2, $\sigma_s$ decreases with increasing $\sigma_p$.
These sources are ellipsoidal binaries with light curve modulations.
An example is given by the second case in Fig.~\ref{Fig:lcs_2G_C}.
The change of light curve shape with increasing $\sigma_p$ of these Sample~2G-C sources is illustrated in the third panel of Fig.~\ref{Fig:sigmaGaussians_2G}, where the per-bin median depth ratio decreases from above 0.9 at $\sigma_p \simeq 0.2$ (red colours) to about 0.6 at $\sigma_p \simeq 0.35$ (green colours).

The phase separation between the locations of the two Gaussians is expected to be $\sim$0.5 for these tight systems.
This is indeed confirmed for the majority of the sources in the sample (see second panel from top in Fig.~\ref{Fig:sigmaGaussians_2G}), with 90\% of them having a deviation from a 0.5 separation of less than 0.015 in phase.
The ones with larger deviations are spurious cases.
An example light curve of such an apparently spurious case is shown by the third case in Fig.~\ref{Fig:lcs_2G_spurious}.

The global ranking of the candidates in Sample~2G-C is generally lower than the ones of samples 2G-A and 2G-B, with only very few cases above 0.6.
This is shown in the top panel of Fig.~\ref{Fig:histo_globalRanking_2G}.
Nevertheless, the light curves are very good in the majority of cases.

Finally, one must mention a note on the value of \texttt{derived\_primary\_ecl\_depth} reported in the catalogue for these 2G-C binaries (a similar note applies to  \texttt{derived\_secondary\_ecl\_depth}).
This value represents the depth of the faintest point of the primary dip in the modelled light curve.
For most of the candidates in the catalogue, the value of \texttt{derived\_primary\_ecl\_depth} is similar to the depth of the deepest Gaussian.
When there is a significant overlap between the two Gaussian components, as is the case for candidates in Sample 2G-C, a significant difference exists between the depths of the Gaussians and the derived depths in the modelled light curve. 
This is illustrated in the top panel of Fig.~\ref{Fig:derivedVsGaussianPrimaryDepth}, which compares these two values for the primary Gaussian of models containing only two Gaussians.
Sample~2G-C is identified as the distinctive subsample below (and somehow parallel to) the diagonal line, with primary Gaussian depths (on the abscissa) that are two to ten times larger than the actual depth in their light curves (on the ordinate).

%- - - - - - - - - - - - - - - - - - - -
\paragraph{Sample 2G-D.}

\begin{figure}
  \centering
  \includegraphics[trim={40 150 0 70},clip,width=\linewidth]{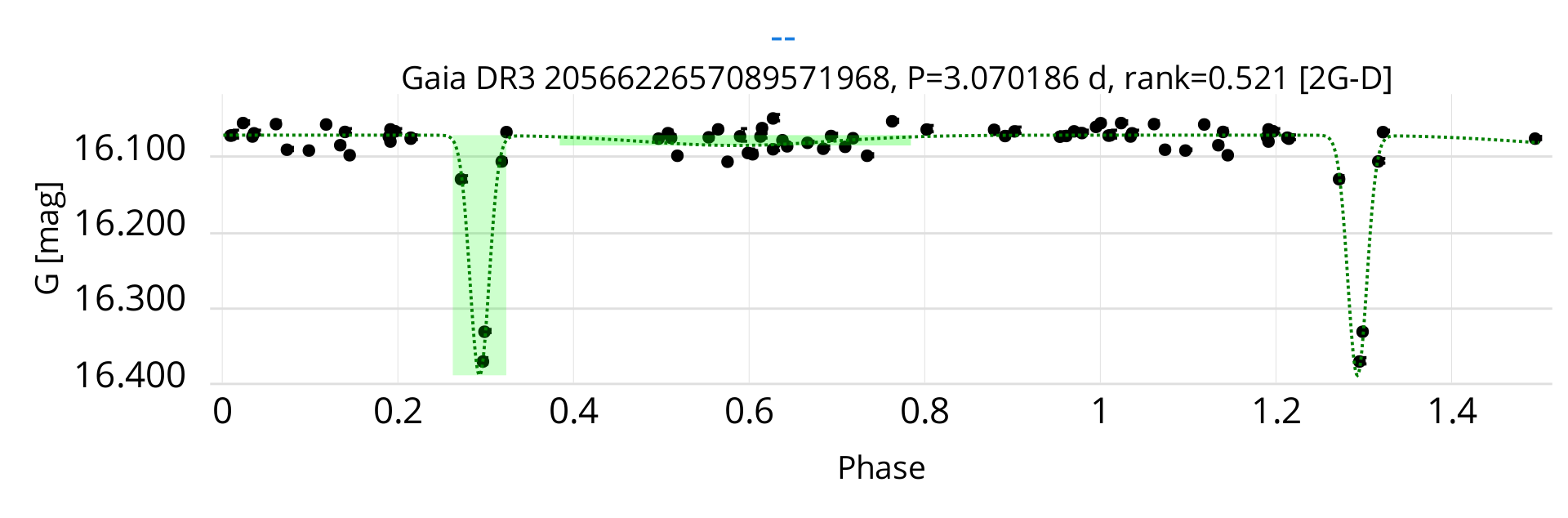}
  \vskip -0.5mm
  \includegraphics[trim={40 145 0 80},clip,width=\linewidth]{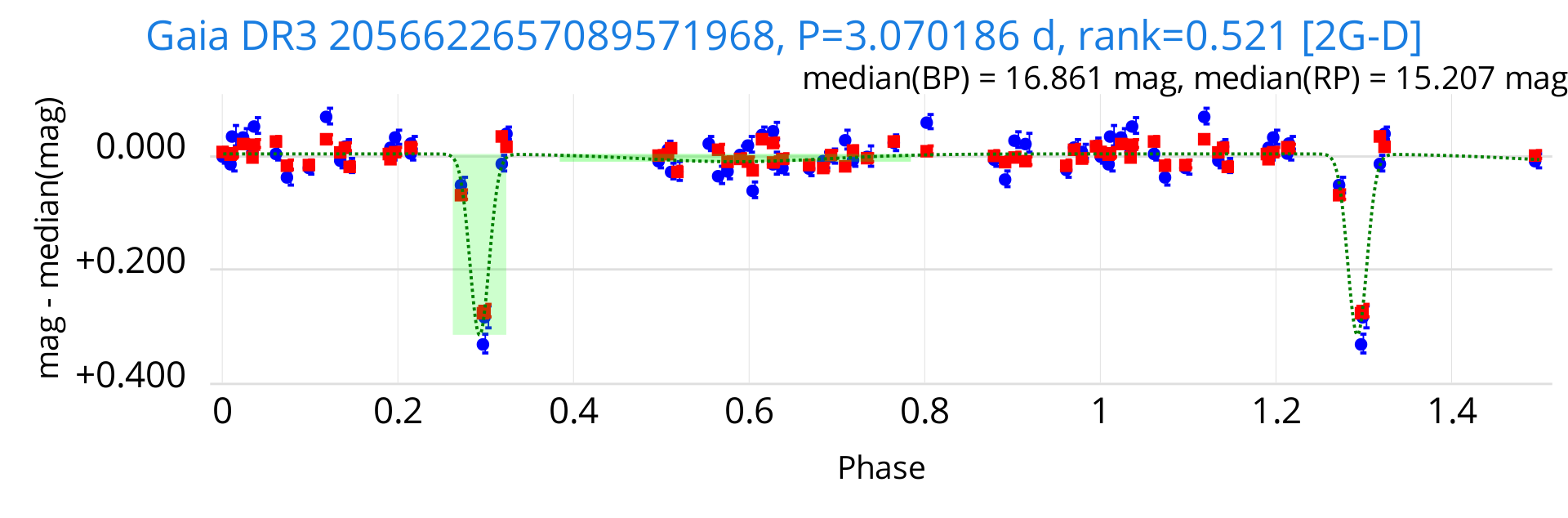}
  \vskip -0.5mm
  \includegraphics[trim={40 150 0 70},clip,width=\linewidth]{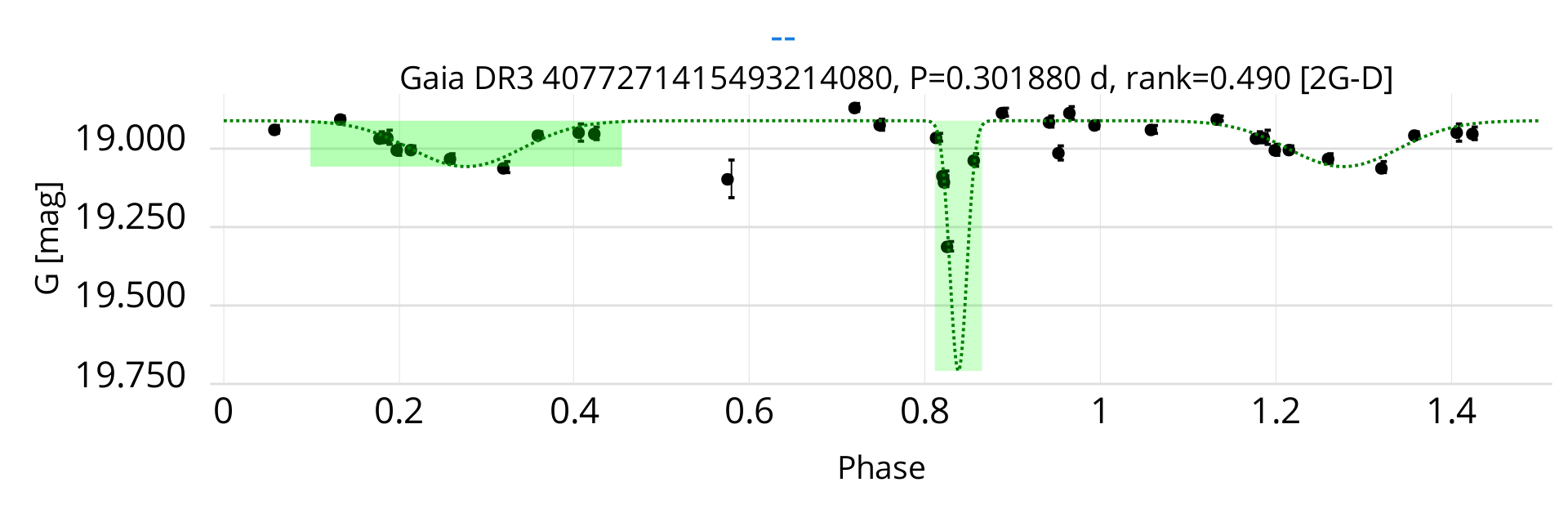}
  \vskip -0.5mm
  \includegraphics[trim={40 45 0 80},clip,width=\linewidth]{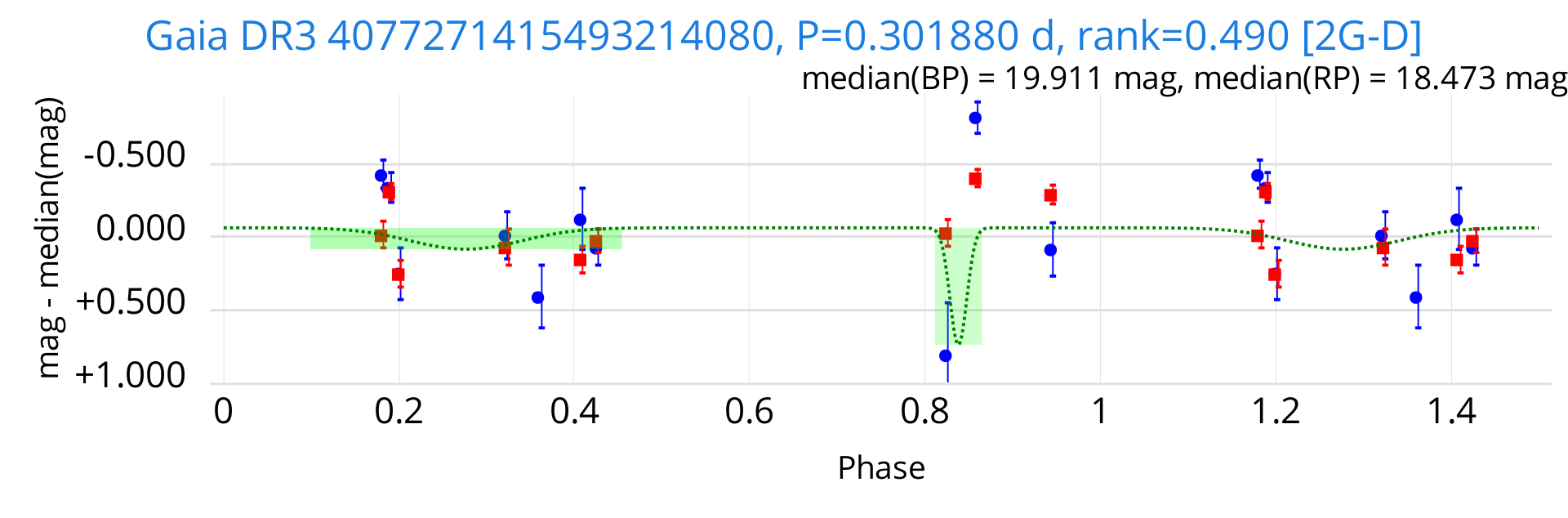}
    \caption{Same as Fig.~\ref{Fig:lcs_2G_A}, but for two candidates in Sample~2G-D of light curves modelled with only two Gaussians.
             The top set shows a case with a convincing primary eclipse detection while the secondary eclipse is spurious (\GaiaSrcIdInCaption{2056622657089571968}).
             The bottom set shows a seemingly good case from the \gmag light curve (\GaiaSrcIdInCaption{4077271415493214080}).
            }
\label{Fig:lcs_2G_D}
\end{figure}

The remaining areas in Fig.~\ref{Fig:sigmaGaussians_2G} (top panel) other than the ones defined by samples 2G-A, 2G-B and 2G-C contain a variety of light curve geometries.
Among them, the ones having narrow primary and much wider secondary Gaussians stand out, with secondary Gaussian depths much smaller, in general, than the primary Gaussian depths, as seen in the third panel of Fig.~\ref{Fig:sigmaGaussians_2G}.
We therefore define Sample 2G-D with $\sigma_\mathrm{p} \lesssim 0.02$ and $\sigma_\mathrm{s} \gtrsim 3\,\sigma_\mathrm{p}$.
The histogram of their depth ratio is shown by the cyan dotted histogram in Fig.~\ref{Fig:histo_depthRatios_2G} (bottom panel).
Most of them have depth ratios smaller than 0.2.
It contains about 5\% of the full catalogue.

While their primary Gaussians correctly identify the presence of a detached eclipse in most cases, caution must be taken on the reality of the second Gaussian identification.
The automated algorithm can indeed fail to detect a narrow secondary eclipse in the case of inadequate phase coverage and/or too shallow secondary, and pick up, instead, a wide and shallow feature in the light curve geometry for the secondary eclipse.
This can also result from an incorrect orbital period, possibly by a factor of two.
An example of such a light curve is shown in the top panel of Fig.~\ref{Fig:lcs_2G_D}, where the failure to detect a secondary eclipse may be due to either a lack of observations in the phase range distant by 0.5 from the primary eclipse (if circular orbit), or to a period determination (3.07~d) twice too short of the true value.
Instead, a physically improbable wide and shallow secondary is picked up by the automated pipeline at a phase 0.3 apart from the primary.
The \gbp and \grp light curves confirm these conclusions drawn from the \gmag light curve.
The second example in Fig.~\ref{Fig:lcs_2G_D} illustrates a case where the \gbp and \grp light curves would not be useful, were they to be considered in the analysis, as they lack reliable observations due to the source being faint and lying in a dense region on the sky close to the Galactic bulge.

It must also be mentioned that a fraction of sources in this 2G-D sample lack enough observations to properly constrain their eclipse depth.
This was already noted to happen in Sample~2G-A at eclipse durations between 0.07 and 1.5 days (top panel of Fig.~\ref{Fig:eclDepthVsEclDration}).
The same is true in this sample 2G-D, as clearly seen in the middle panel of Fig.~\ref{Fig:eclDepthVsEclDration}, affecting an even larger fraction of sources in this sample than in Sample 2G-A due to the shorter eclipse durations of the candidates in Sample 2G-D. 
Sources in this sample deserve additional investigations, using \Gaia data themselves and/or complementary observations.

%- - - - - - - - - - - - - - - - - - - -
\paragraph{Samples 2G-X and 2G-Y.}

\begin{figure}
  \centering
  \includegraphics[trim={0 0 0 42},clip,width=0.815\linewidth]{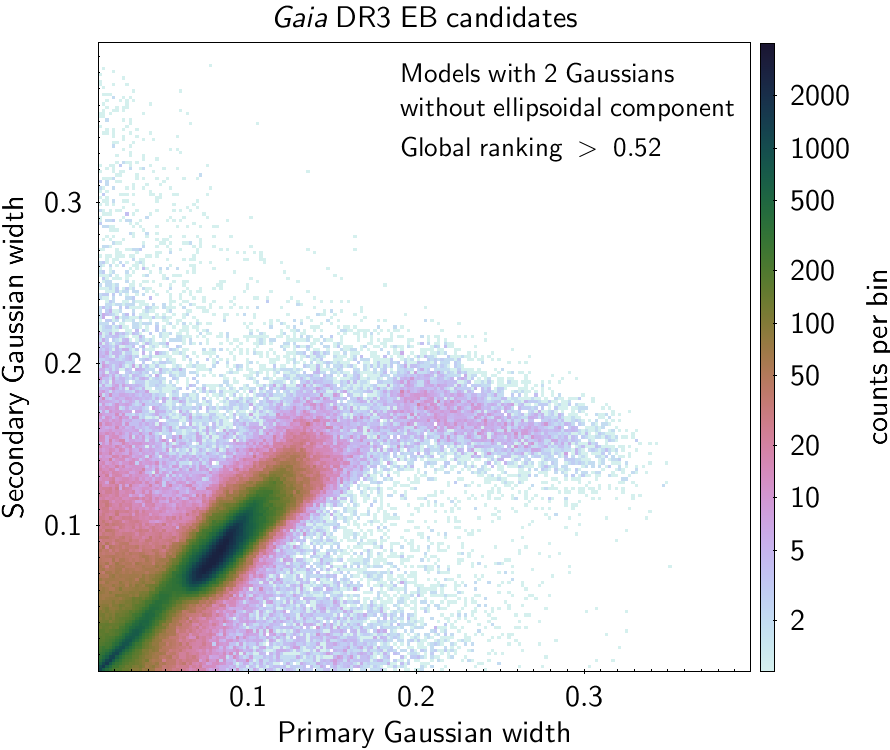}
  \caption{Same as the top panel of Fig.~\ref{Fig:sigmaGaussians_2G}, but for the subset of sources with global rankings larger than 0.52.
           The colour-scale is kept identical to that of the top panel of Fig.~\ref{Fig:sigmaGaussians_2G} to allow a direct comparison.
          }
\label{Fig:sigmaGaussians_2G_rankGT0p52}
\end{figure}

The two remaining areas in Fig.~\ref{Fig:sigmaGaussians_2G}, the one above the equal-widths diagonal area (which defines Sample~2G-X), and the one below that area (defining Sample~2G-Y), contain sources with various light curve model geometries.
Their Gaussian depths ratios span all values from almost zero up to one (yellow and pink histograms in the bottom panel of Fig.~\ref{Fig:histo_depthRatios_2G}, respectively).
The phase separation between the two Gaussians has also a wider distribution than those of Samples~2G-A to C, with median values between 0.4 and 0.45 (greenish regions in the second panel of Fig.~\ref{Fig:sigmaGaussians_2G}).
Both these features ask for a careful investigation of the light curves, as the models may be unreliable to represent physical features of the binary system.
This is also suggested from their generally low global rankings shown in the bottom panel of Fig.~\ref{Fig:sigmaGaussians_2G}, with median per-bin values below 0.48 for a majority of them (yellowish and reddish colours in the panel).
The histograms of their global rankings shown in the bottom panel of Fig.~\ref{Fig:histo_globalRanking_2G} confirm this.
Restricting the samples to sources with global rankings greater than 0.52 removes much of the sources in samples~2G-X and 2G-Y, as shown in Fig.~\ref{Fig:sigmaGaussians_2G_rankGT0p52} (to be compared with the top panel of Fig.~\ref{Fig:sigmaGaussians_2G}). 
The total number of sources in each of these two samples amounts to about 10\% of the full catalogue.

%- - - - - - - - - - - - - - - - - - - -
\subsection{Models with two Gaussians and an ellipsoidal component}
\label{Sect:catalogue_usage_model_2GE}

\begin{figure}
  \centering
  \includegraphics[trim={0 0 0 42},clip,width=0.815\linewidth]{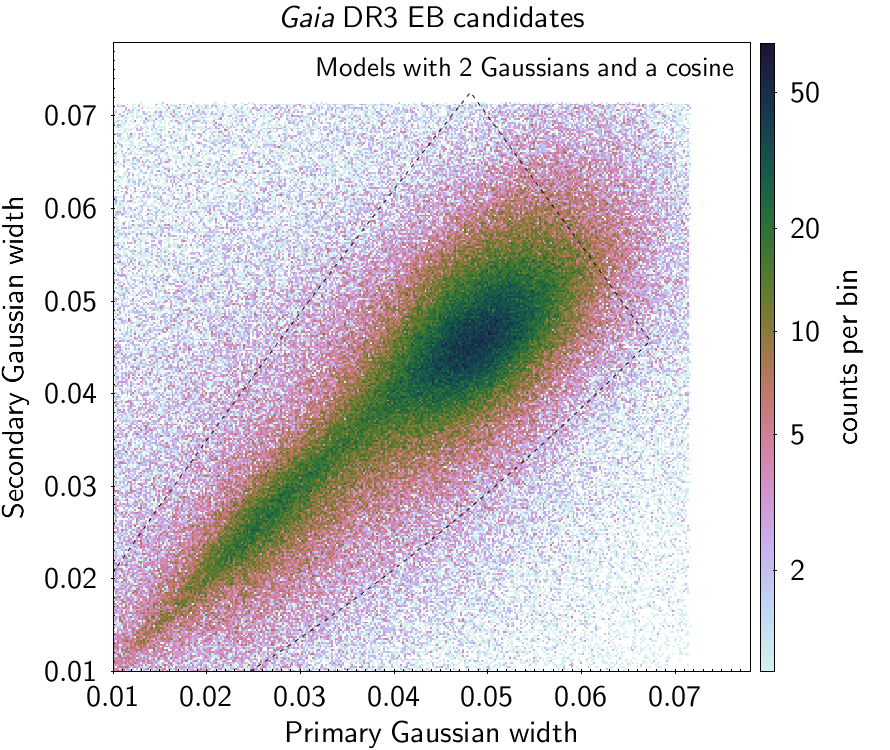}
  \caption{Same as the top panel of Fig.~\ref{Fig:sigmaGaussians_2G}, but for the sample of sources whose light curves are modelled with two Gaussians and an ellipsoidal component (samples 2GE-A, 2GE-B and 2GE-Z).
           The dashed line delineates the region defined for Sample~2G-A eclipsing binaries of the models containing two Gaussians but no ellipsoidal component (Table~\ref{Tab:sample_definition} and Fig.~\ref{Fig:sigmaGaussians_2G_samples}).
          }
\label{Fig:sigmaGaussians_2GE}
\end{figure}

\begin{figure}
  \centering
  \includegraphics[trim={0 0 0 42},clip,width=\linewidth]{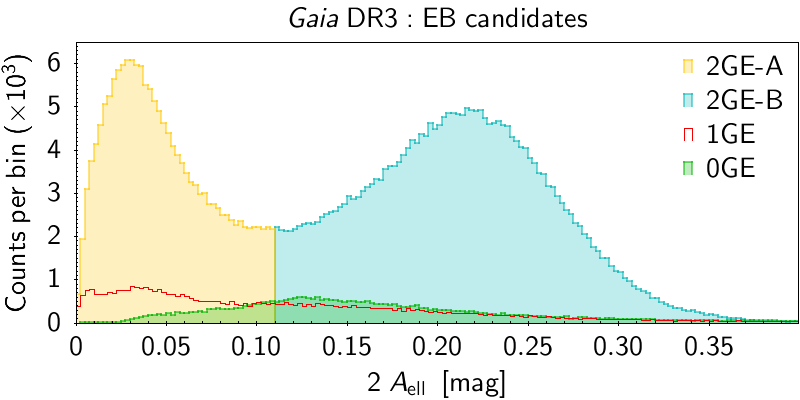}
  \caption{Peak-to-peak amplitude distributions of the ellipsoidal component (cosine) of the two-Gaussian models in the various samples containing the cosine component as labeled in the figure.
           The abscissa scale has been limited for better visibility.
          }
\label{Fig:histo_Aell}
\end{figure}

\begin{figure}
  \centering
  \includegraphics[trim={0 0 0 42},clip,width=\linewidth]{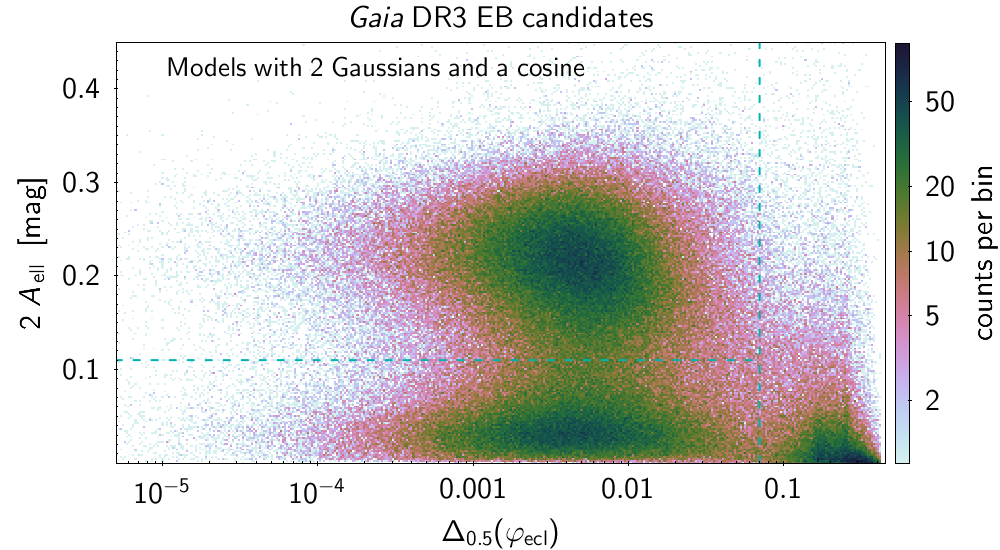}
  \caption{Density map of the ellipsoidal varation amplitude (peak-to-peak) versus deviation from 0.5 of the phase separation between primary and secondary eclipse locations for the sample of sources having two Gaussians and an ellipsoidal component.
           The expression of \deltaPhi is given by Eq.~(\ref{Eq:deltaphi}).
           The dashed lines delineate the three samples 2GE-A (lower-left region), 2GE-B (upper-left region) and 2GE-Z (right region) defined in the text.
           The axes scales have been limited for better visibility.
          }
\label{Fig:depthRatioVsEccentricity_2GE}
\end{figure}

About a quarter (23\%) of the sources in the catalogue have their \gmag light curves modelled with two Gaussians and an additional ellipsoidal (cosine) component.
The two Gaussians have similar widths%
\footnote{
When the model includes an ellipsoidal (cosine) term, the Gaussian widths are, by construction, smaller than 0.4/5.6=0.0714 to avoid a degenerative competition between the wide Gaussian and the cosine functions (see Sect.~\ref{Sect:catalogue_method} in the main body of the text).
} %
(see Fig.~\ref{Fig:sigmaGaussians_2GE}), reminiscent of Sample 2G-A of wide binaries identified in Sect.~\ref{Sect:catalogue_usage_model_2G}.
Contrary to Sample 2G-A, however, Sample 2GE-A candidates are all tight systems with a visible ellipsoidal component.

The amplitude distribution of the ellipsoidal component is shown in Fig.~\ref{Fig:histo_Aell}.
It reveals two main sub-samples.
The one at small amplitudes ($2\,\Aell<0.11$~mag; pink filled histogram) defines Sample~2GE-A and the one with larger amplitudes ($2\,\Aell\ge 0.11$~mag; blue filled histogram in the figure) defines Sample~2GE-B.
We further restrict the sources in these two samples to the ones which have eclipse separations close to 0.5 in phase, because our choice to use a cosine function to describe the ellipsoidal variability implies, in principle, a circular orbit.
To do this, we define the separation \deltaPhi between the locations of the primary (\phaseEclPrimary) and secondary (\phaseEclSecondary) eclipses in the models measured relative to a separation of 0.5 as
\begin{equation}
  \deltaPhi = | \; |\phaseEclPrimary - \phaseEclSecondary|-0.5 \; |\;\;.
\label{Eq:deltaphi}
\end{equation}
The distribution of \deltaPhi versus the ellipsoidal peak-to-peak amplitude is shown in Fig.~\ref{Fig:depthRatioVsEccentricity_2GE}.
It confirms that the majority of sources with two Gaussians and an ellipsoidal component are nearly circular, with $\deltaPhi \lesssim 0.07$ (i.e. a phase separation of $0.5 \pm 0.07$ between the eclipses) for 90\% of them.
A clear separation is actually seen between samples with $\deltaPhi<0.07$ and $\deltaPhi>0.07$. 
We therefore restrict Samples~2GE-A and 2GE-B to $\deltaPhi<0.07$, and put candidates with $\deltaPhi \ge 0.07$ in Sample~2GE-Z.
The definition of the three samples are summarised in Table~\ref{Tab:sample_definition}.
They are successively discussed in more details in the next paragraphs.

%- - - - - - - - - - - - - - - - - - - -
\paragraph{Sample 2GE-A.}

\begin{figure}
  \centering
  \includegraphics[trim={3 78 0 42},clip,width=\linewidth]{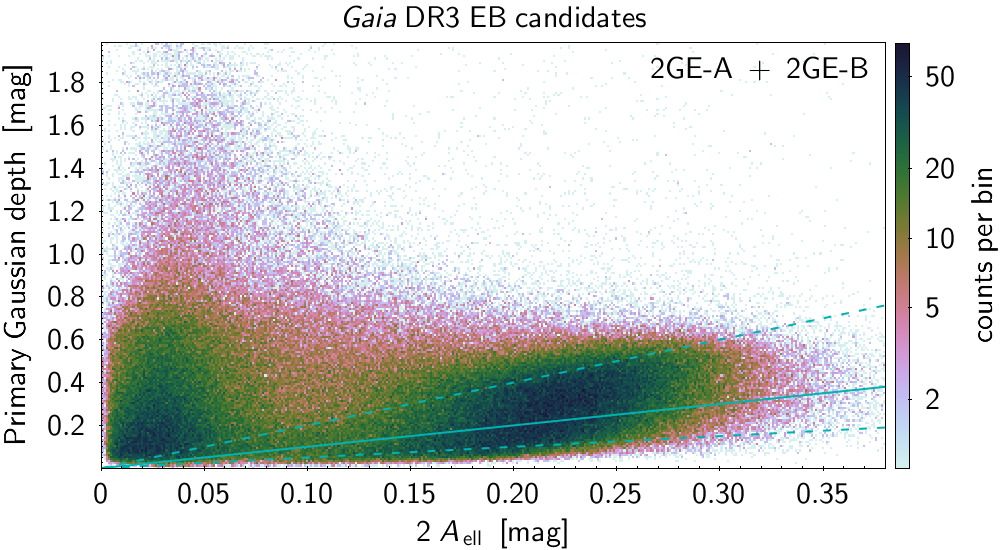}
  \vskip -0.5mm
  \includegraphics[trim={0 0 0 42},clip,width=\linewidth]{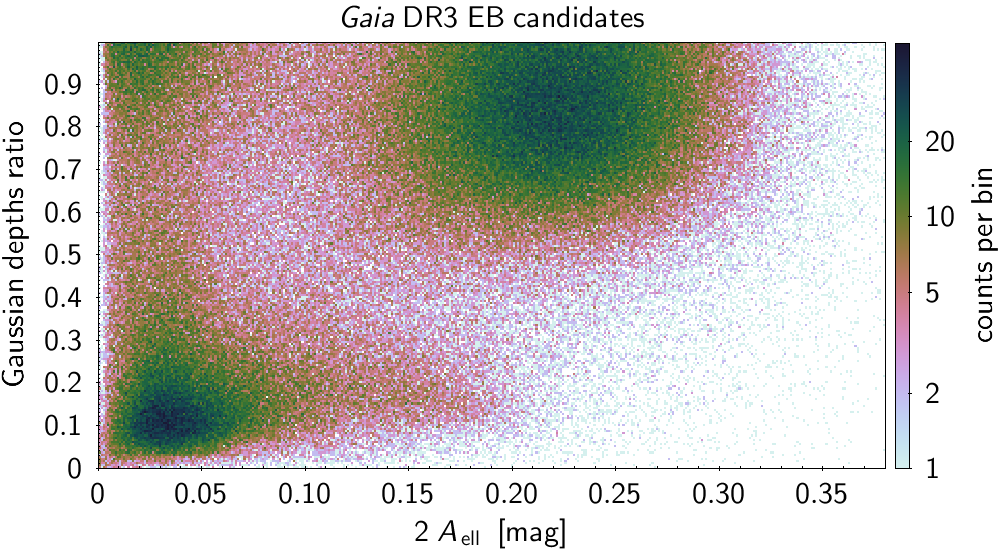}
  \caption{Density maps of two-Gaussian related quantities versus ellipsoidal amplitude (peak-to-peak) of the samples 2GE-A ($2\,\Aell< 0.11$) and 2GE-B ($2\,\Aell\ge 0.11$) of sources having two Gaussians and an ellipsoidal component.
           \textbf{Top panel:} Primary Gaussian depth, with 1:1 (solid), 2:1 (upper dashed) and 1:2 (lower dashed) lines to guide the eyes.
           \textbf{Bottom panel:} Secondary to primary Gaussian depths ratio.
           The axes scales are truncated for better visibility
          }
\label{Fig:AellVsGaussDepth_E1E2}
\end{figure}

\begin{figure}
  \centering
  \includegraphics[trim={40 150 0 70},clip,width=\linewidth]{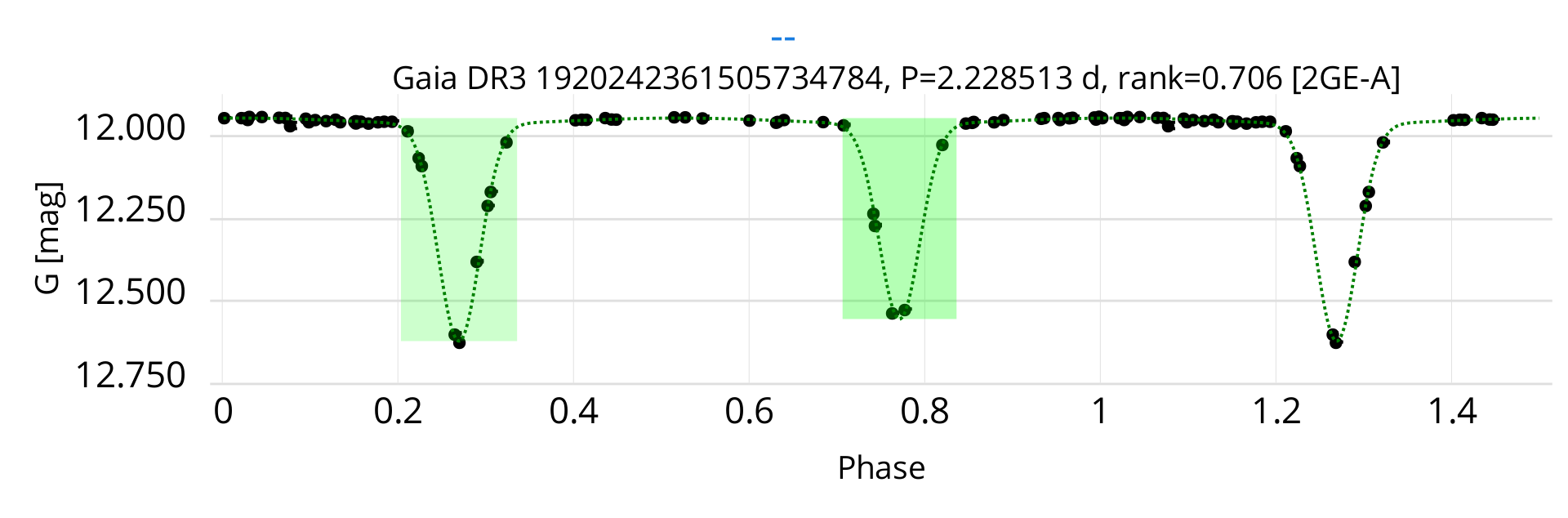}
  \vskip -0.5mm
  \includegraphics[trim={40 145 0 80},clip,width=\linewidth]{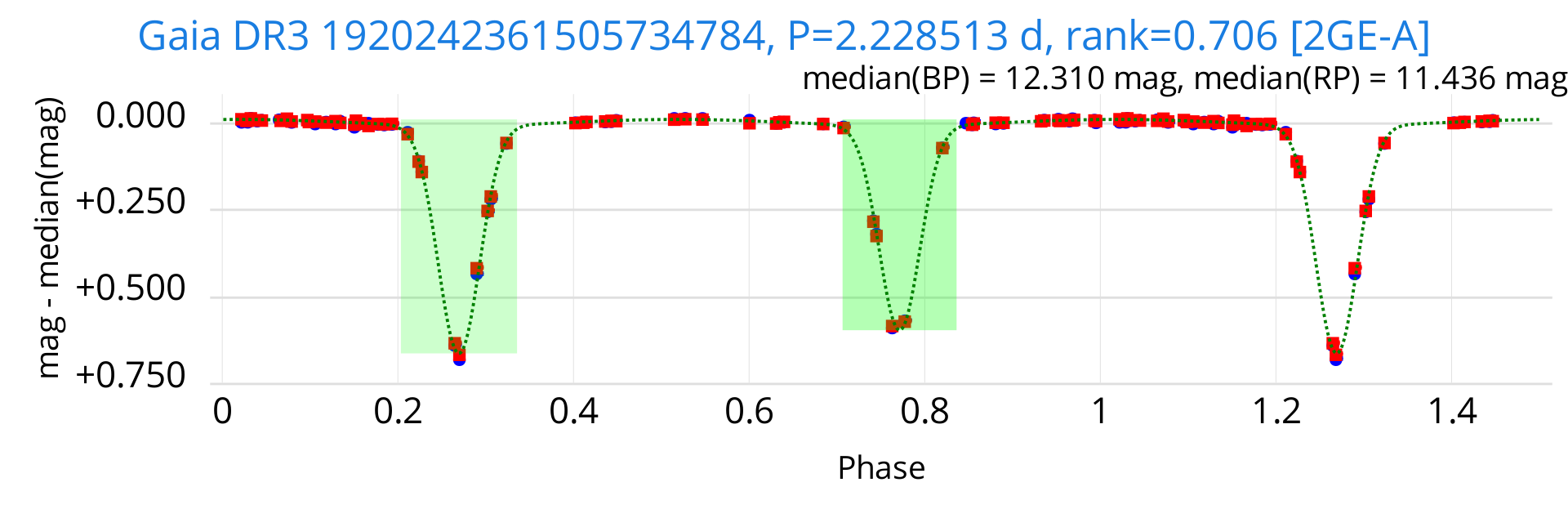}
  \vskip -0.5mm
  \includegraphics[trim={40 150 0 70},clip,width=\linewidth]{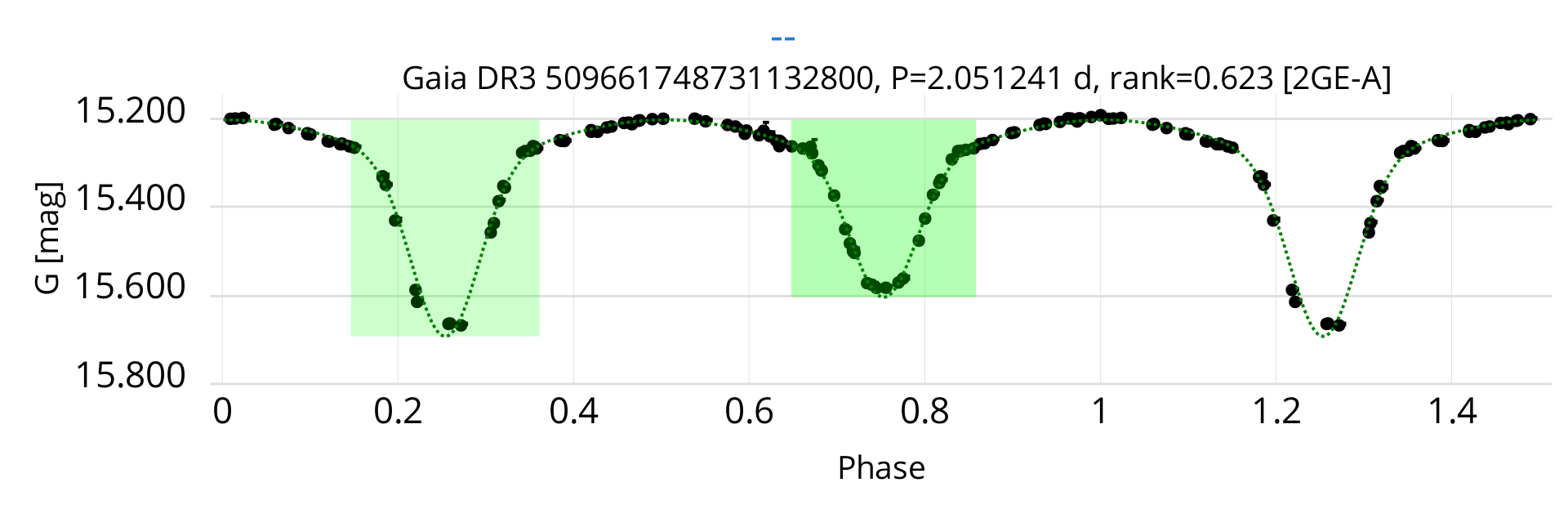}
  \vskip -0.5mm
  \includegraphics[trim={40 145 0 80},clip,width=\linewidth]{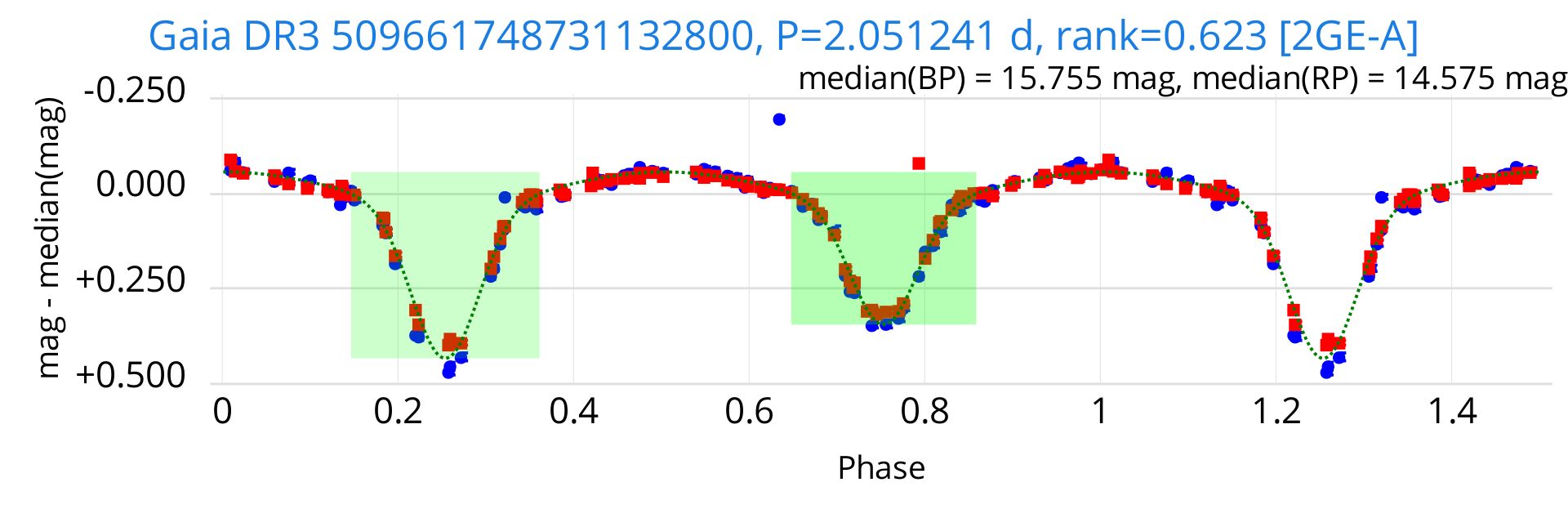}
  \vskip -0.5mm
  \includegraphics[trim={40 150 0 70},clip,width=\linewidth]{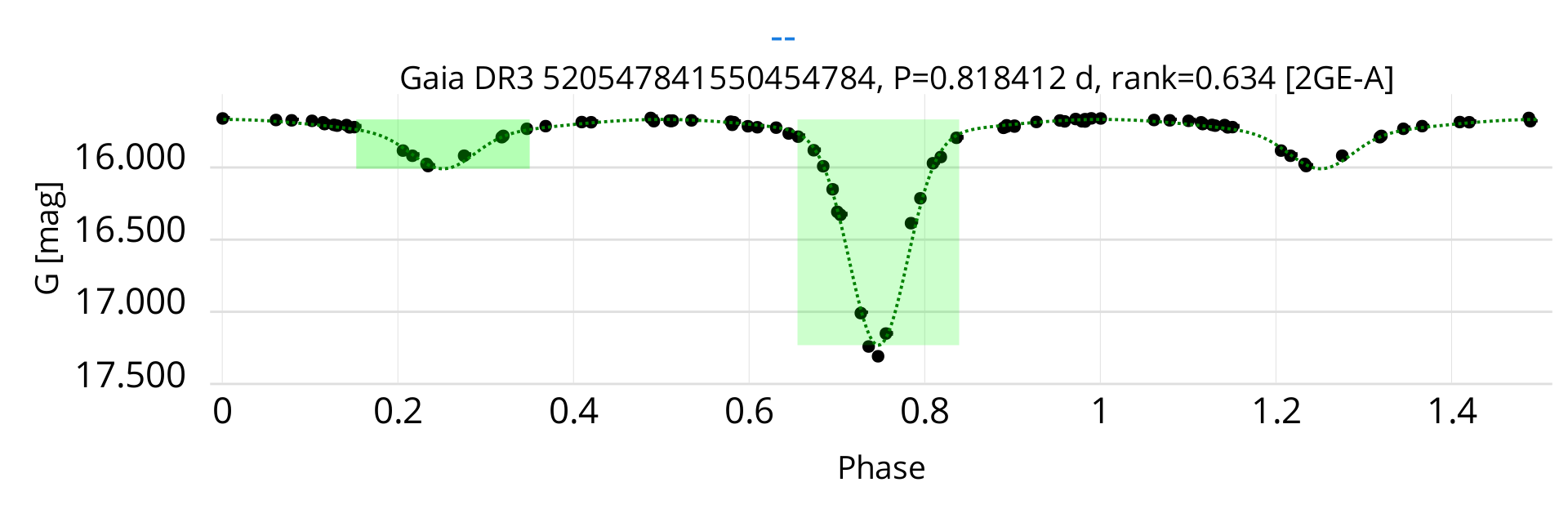}
  \vskip -0.5mm
  \includegraphics[trim={40 45 0 80},clip,width=\linewidth]{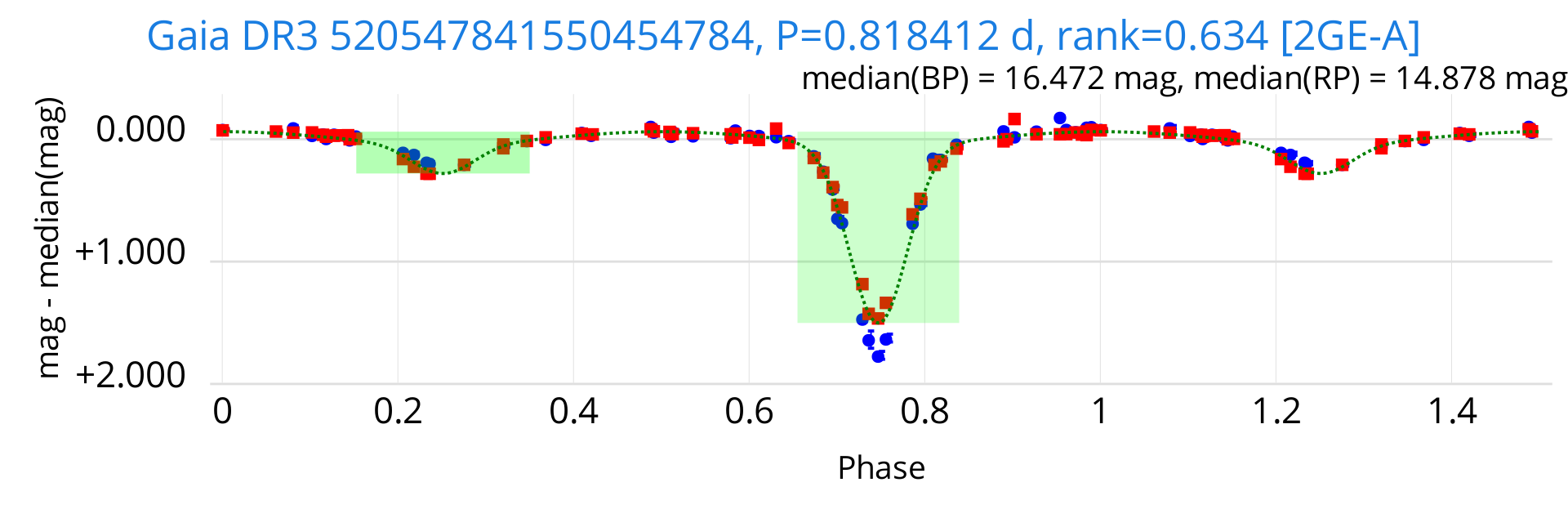}
    \caption{Same as Fig.~\ref{Fig:lcs_2G_A}, but for three candidates in Sample~2GE-A of light curves modelled with two Gaussians and an ellipsoidal component.
             The top set shows a case with a weak ellipsoidal variability (\GaiaSrcIdInCaption{1920242361505734784}),
             the middle set with a mild ellipsoidal component (\GaiaSrcIdInCaption{509661748731132800}),
             and the bottom set with very different eclipse depths (\GaiaSrcIdInCaption{520547841550454784}).
            }
\label{Fig:lcs_2GE_A}
\end{figure}

Sample~2GE-A mainly consists of eclipsing binaries with inter-eclipse brightness variations of small to moderate amplitudes due to ellipsoidal variability.
It contains one third of the binaries whose light curves are modelled with two Gaussians and a cosine (Table~\ref{Tab:sample_definition}).
The top panel of Fig~\ref{Fig:AellVsGaussDepth_E1E2} displays the primary Gaussian depth versus (peak-to-peak) amplitude of the ellipsoidal component for the combined 2GE-A $+$ 2GE-B samples.
The 2GE-A sample lies, by definition, at $2\,\Aell\le 0.11$~mag.
The depth of the primary Gaussian goes up to $\sim$0.8~mag for the bulk of the sample, and up to more than two magnitudes at the tail of the distribution.
Samples 2GE-A and 2GE-B are relatively well separated in the figure.

Three example light curves are shown in Fig.~\ref{Fig:lcs_2GE_A}.
The two first examples show cases with about equal eclipse depths, the top one with a mild ellipsoidal variability and the middle one with a stronger ellipsoidal variabiltiy.
The bottom example illustrates a case with significantly unequal eclipse depths.
This last case represents the majority of candidates in Sample~2GE-A, as seen from the bottom panel of Fig~\ref{Fig:AellVsGaussDepth_E1E2}.
The bulk distribution of the 2GE-A sample is seen in the figure to have depths ratios smaller than 0.3, distinct from the second concentration at close-to-equal depth ratios.

%- - - - - - - - - - - - - - - - - - - -
\paragraph{Sample 2GE-B.}

\begin{figure}
  \centering
  \includegraphics[trim={40 150 0 70},clip,width=\linewidth]{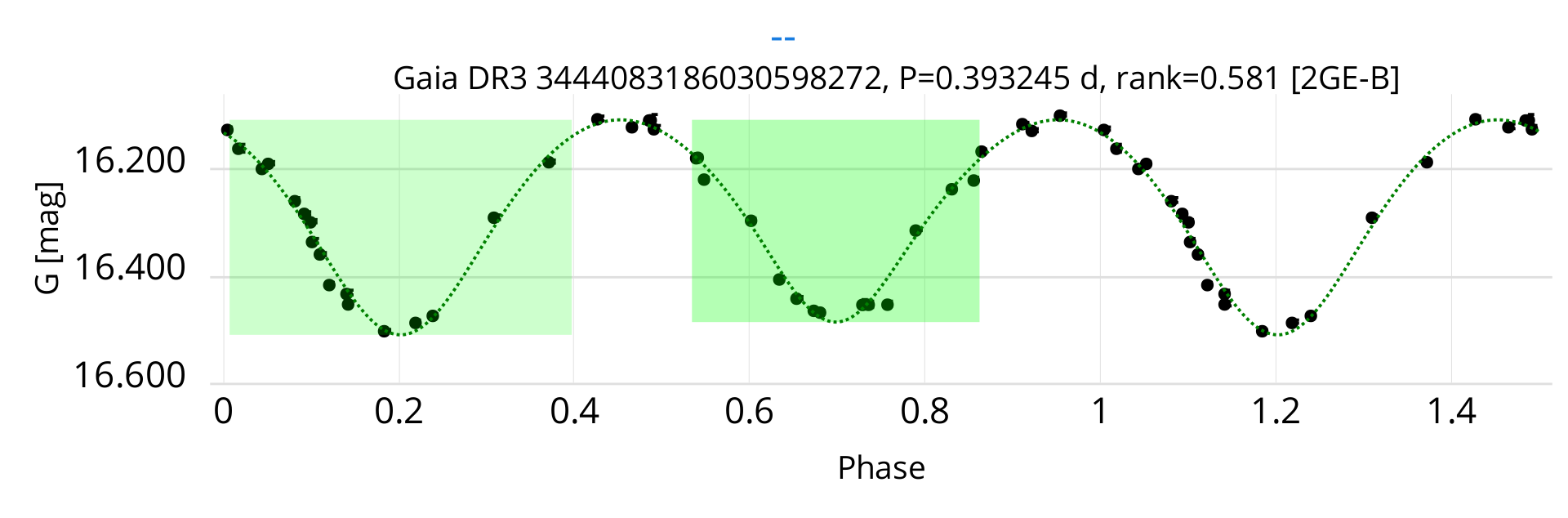}
  \vskip -0.5mm
  \includegraphics[trim={40 145 0 80},clip,width=\linewidth]{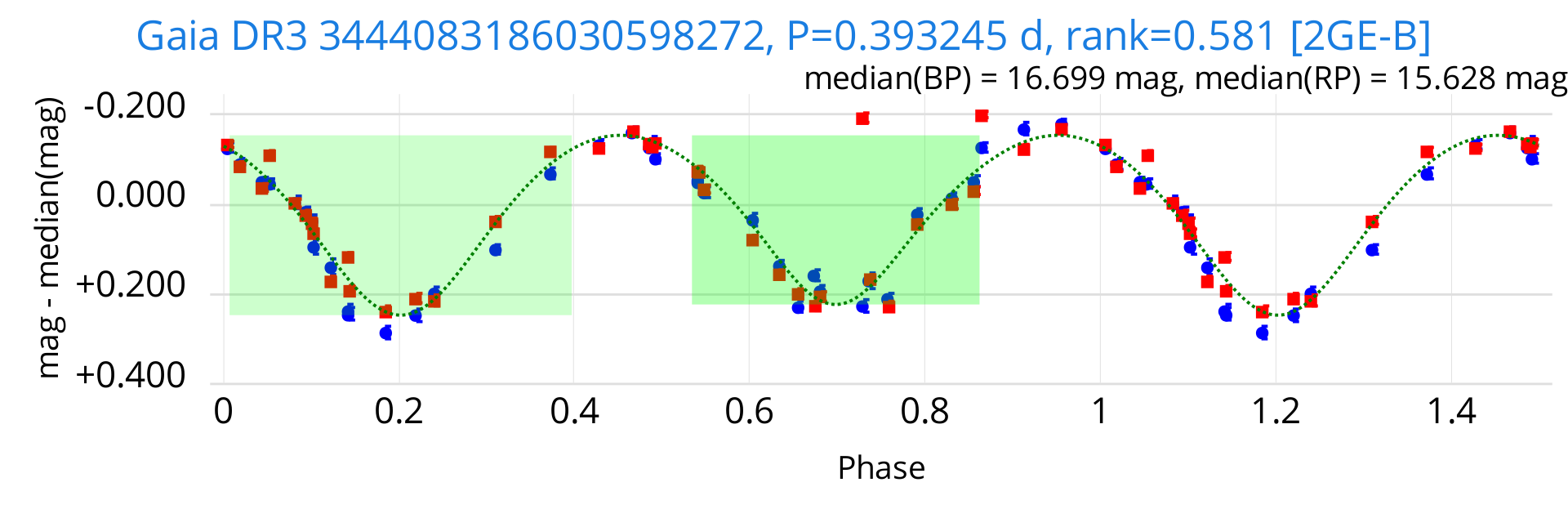}
  \vskip -0.5mm
  \includegraphics[trim={40 150 0 70},clip,width=\linewidth]{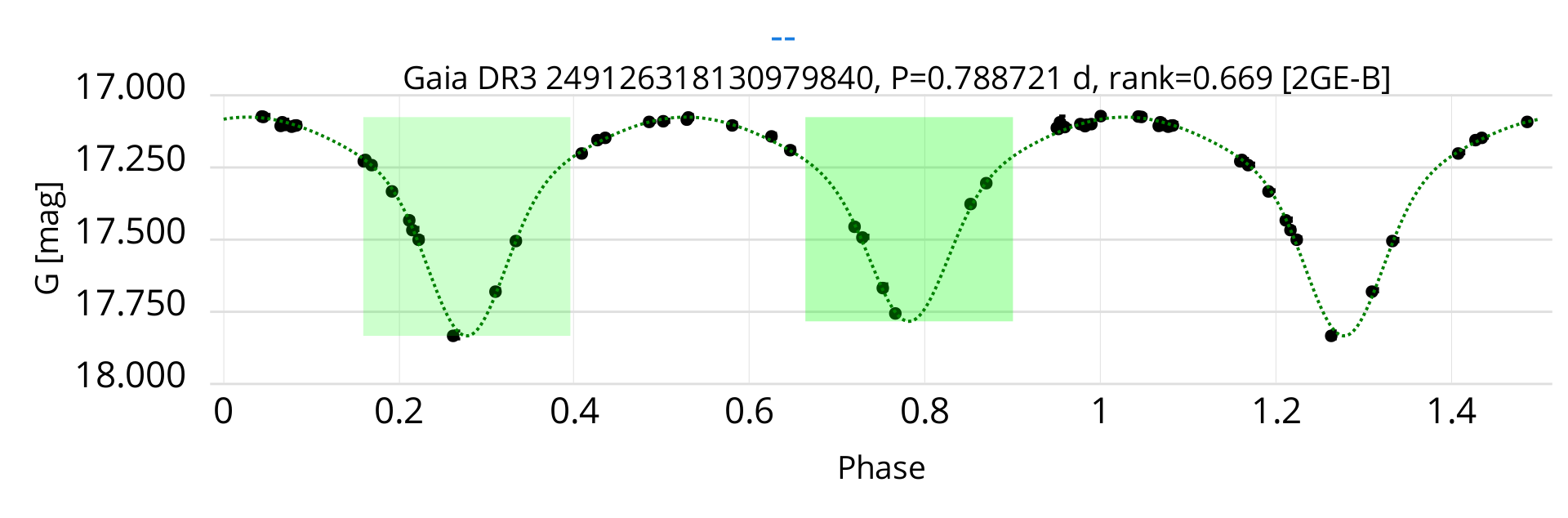}
  \vskip -0.5mm
  \includegraphics[trim={40 145 0 80},clip,width=\linewidth]{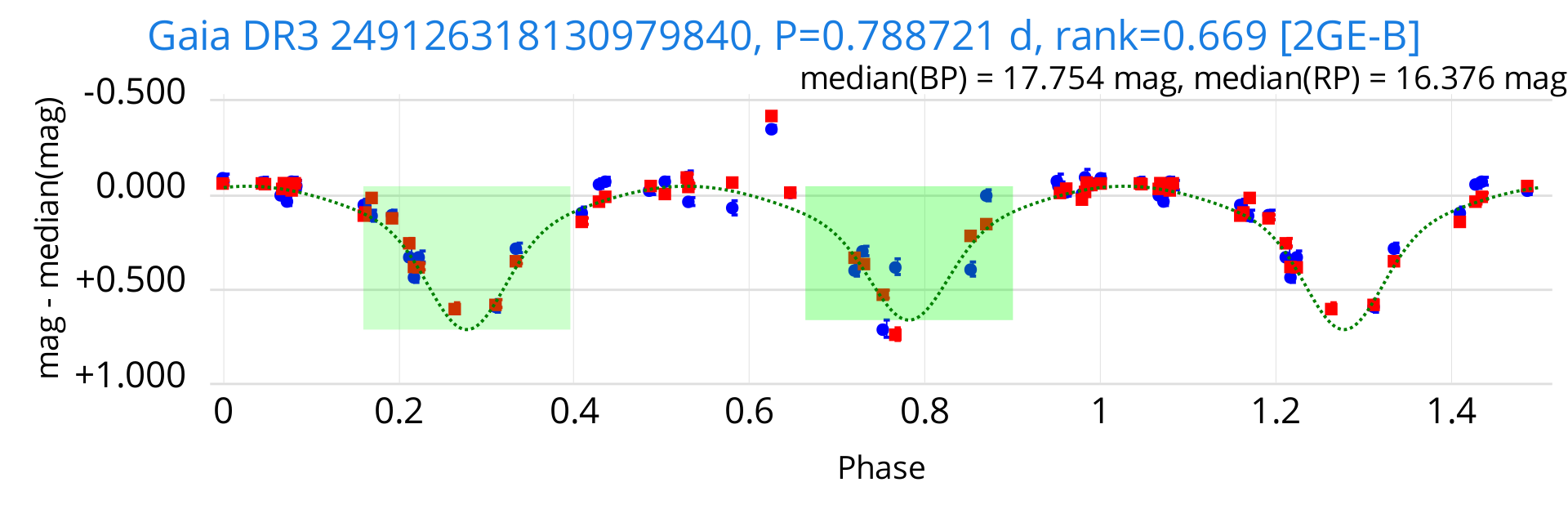}
  \vskip -0.5mm
  \includegraphics[trim={40 150 0 70},clip,width=\linewidth]{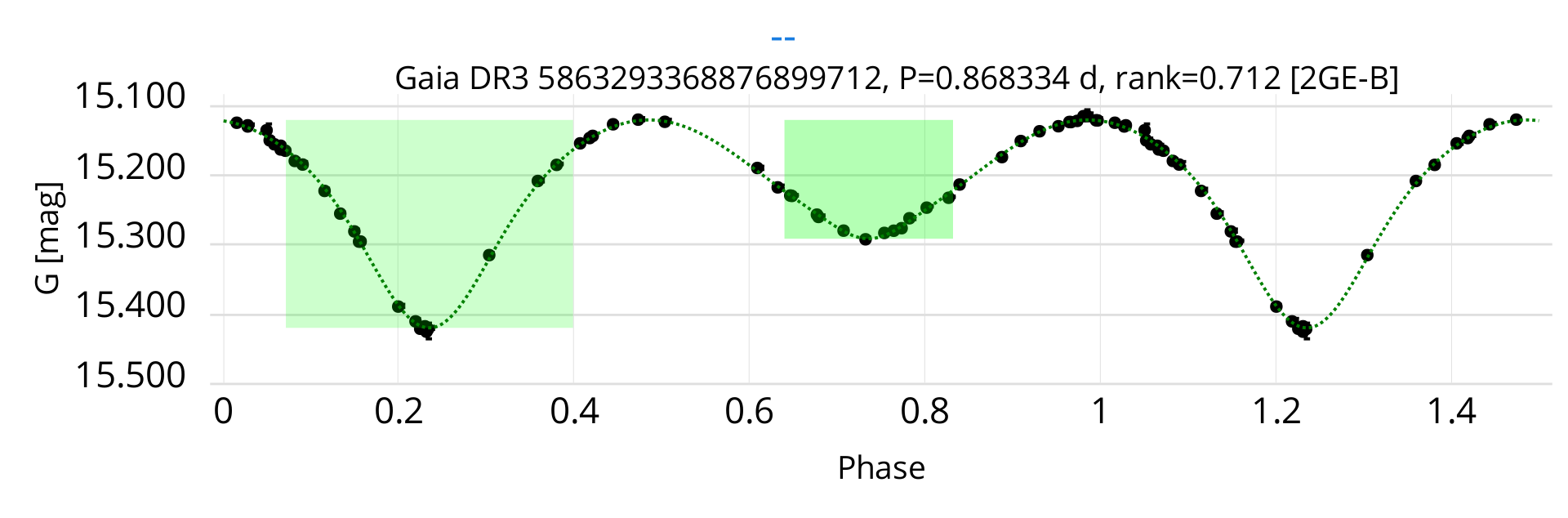}
  \vskip -0.5mm
  \includegraphics[trim={40 45 0 80},clip,width=\linewidth]{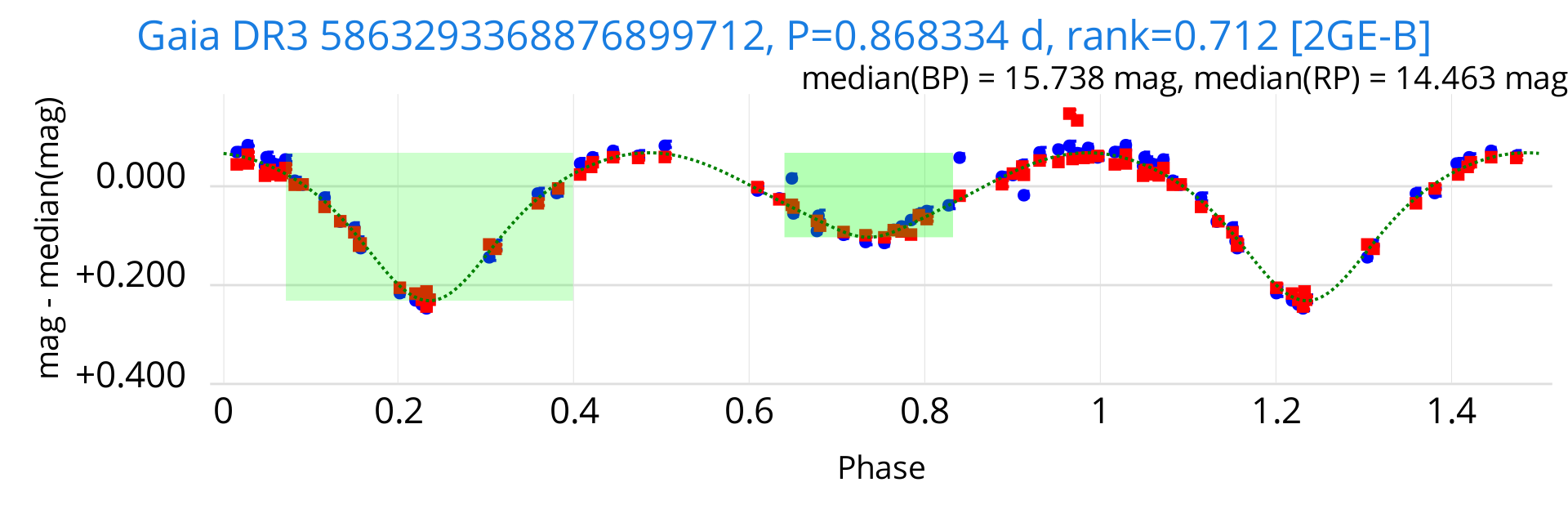}
    \caption{Same as Fig.~\ref{Fig:lcs_2G_A}, but for three candidates in Sample~2GE-B of light curves modelled with two Gaussians and an ellipsoidal component.
             The top case shows an example of light curve where the ellipsoidal component is the major contributor to the light shape (\GaiaSrcIdInCaption{3444083186030598272}), while the second case shows an example with a larger Gaussian depth than the ellipsoidal variability amplitude (\GaiaSrcIdInCaption{249126318130979840}).
             The bottom case exemplifies a Sample~2GE-B source with a small depth ratio (of 0.14) between the secondary and primary Gaussian depths.
            }
\label{Fig:lcs_2GE_B}
\end{figure}

In this sample, the large amplitudes of the ellipsoidal component in the two-Gaussian models dictate the overall morphology of the light curves.
The peak-to-peak amplitudes range from 0.11~mag (by definition) to above 0.35~mag (see Fig.~\ref{Fig:AellVsGaussDepth_E1E2}, top panel).
The Gaussian components, on the other hand, determine the sharpness of the eclipses in the light curve.

The depth of the Gaussian component is typically between half and twice the peak-to-peak amplitude of the ellipsoidal amplitude.
Three example light curves are shown in Fig.~\ref{Fig:lcs_2GE_B}.
In the top example, the Gaussian (depth of 0.15~mag) is less prominent than the ellipsoidal component (peak-to-peak amplitude of $0.25$~mag).
The middle example shows a case with a stronger Gaussian component (depth of 0.51~mag) than the ellipsoidal amplitude (peak-to-peak amplitude of $0.25$~mag).
The impact of the Gaussian component on the otherwize sine-like shape of the light curve is clearly visible.
The bottom example illustrates a case with a secondary Gaussian much shallower than the primary Gaussian.
This last case characterises a small fraction of the candidates in Sample 2GE-B that have Gaussian depth ratios smaller than about 0.3.
In Fig.~\ref{Fig:AellVsGaussDepth_E1E2} (bottom panel), they are seen to be an extension of the distribution of Sample 2GE-A towards larger ellipsoidal amplitudes.

%\NM{I am tempted to define three samples from the bottom panel of Fig.~\ref{Fig:AellVsGaussDepth_E1E2}, instead of the two 2GE-B and 2GE-A samples, with Sample 2GE-A further splitted in two sub-sampless based on the bimodal distribution of their Gaussian depths ratios, but it would need more time which is lacking...}

%3444083186030598272: primary Gaussian depth: 0.1528539~mag, $\Aell=0.124187686$~mag, derived primary depth = 0.40098023
%249126318130979840: primary Gaussian depth: 0.51222354~mag, $\Aell=0.1236117$~mag, derived primary depth = 0.7594467
%source_id==3444083186030598272L ||  source_id==249126318130979840L

%- - - - - - - - - - - - - - - - - - - -
\paragraph{Sample 2GE-Z}

\begin{figure}
  \centering
  \includegraphics[trim={0 83 0 42},clip,width=\linewidth]{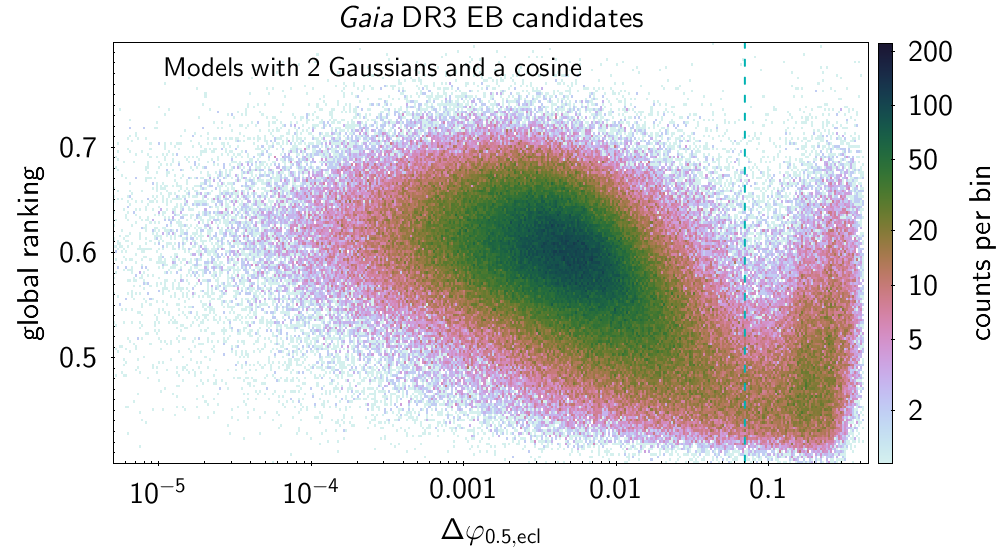}
  \vskip -0.5mm
  \includegraphics[trim={0 0 0 42},clip,width=\linewidth]{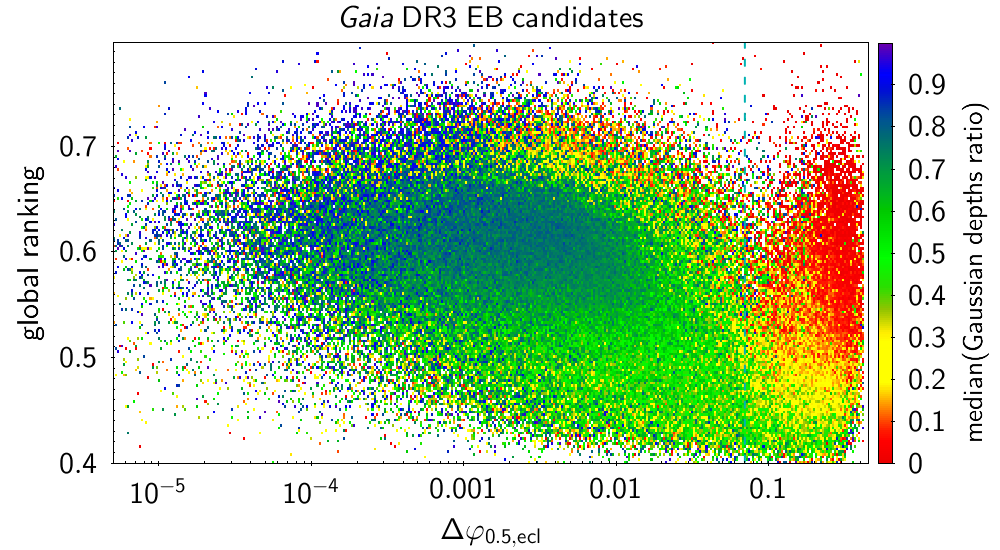}
  \caption{Same as Fig.~\ref{Fig:depthRatioVsEccentricity_2GE}, but for the global ranking versus eclipse phase separation relative to 0.5.
           \textbf{Top panel:} Density map.
           \textbf{Bottom panel:} Primary to secondary Gaussian depth ratio colour-coded according to the colour scale shown on the right of the panel.
          }
\label{Fig:rankVsEccentricity_2GE}
\end{figure}

Only 10\% of the models with two Gaussians and a cosine fall in Sample~2GE-Z, characterised by eclipse separations deviating from 0.5 by $\deltaPhi>0.07$.
The majority of them have small ellipsoidal variability amplitudes (Fig.~\ref{Fig:depthRatioVsEccentricity_2GE}, lower right sample).
An analysis based on their Gaussian widths, similarly to what is done in Sect.~\ref{Sect:catalogue_usage_model_2G} for models containing only two Gaussians, suggests that the model components in Sample 2GE-Z may not reflect physical features of binary systems and that further investigation is required before drawing conclusions.
The Gaussian widths are indeed widely distributed in the $\sigma_1$ -- $\sigma_2$ plane, contrary to the distributions of the 2GE-A and 2GE-B samples.
Sample~2GE-Z is reminiscent of Samples 2G-X and 2G-Y in Sect.~\ref{Sect:catalogue_usage_model_2G}.

Another indication that the Gaussian models of Sample 2GE-Z should be taken with caution comes from their global rankings.
The distribution of the rankings of all sources with two Gaussians and an cosine is shown in Fig.~\ref{Fig:rankVsEccentricity_2GE} against \deltaPhi. 
Sample 2GE-Z, defined by $\deltaPhi>0.07$, has a distribution peaked towards low global rankings, while samples 2GE-A and 2GE-B, located at $\deltaPhi<0.07$, are predominantly found at larger rankings.
Moreover, the candidates in Sample 2GE-Z that do have large global rankings, have, on average, very shallow secondary Gaussians, as shown in the bottom panel of Fig.~\ref{Fig:rankVsEccentricity_2GE}.
The second eclipse of these models, despite their good global rankings, may thus be spurious.
We note that samples 2GE-A and 2GE-B have, on average, good global rankings.
Sample 2GE-Z, therefore, harbors a variety of cases that require additional investigations before using their two-Gaussian model parameters.
Sample 2GE-Z represents only 2\% of the full catalogue.

%- - - - - - - - - - - - - - - - - - - -
\subsection{Models with only one Gaussian}
\label{Sect:catalogue_usage_model_1G}

\begin{figure}
  \centering
  \includegraphics[trim={40 150 0 70},clip,width=\linewidth]{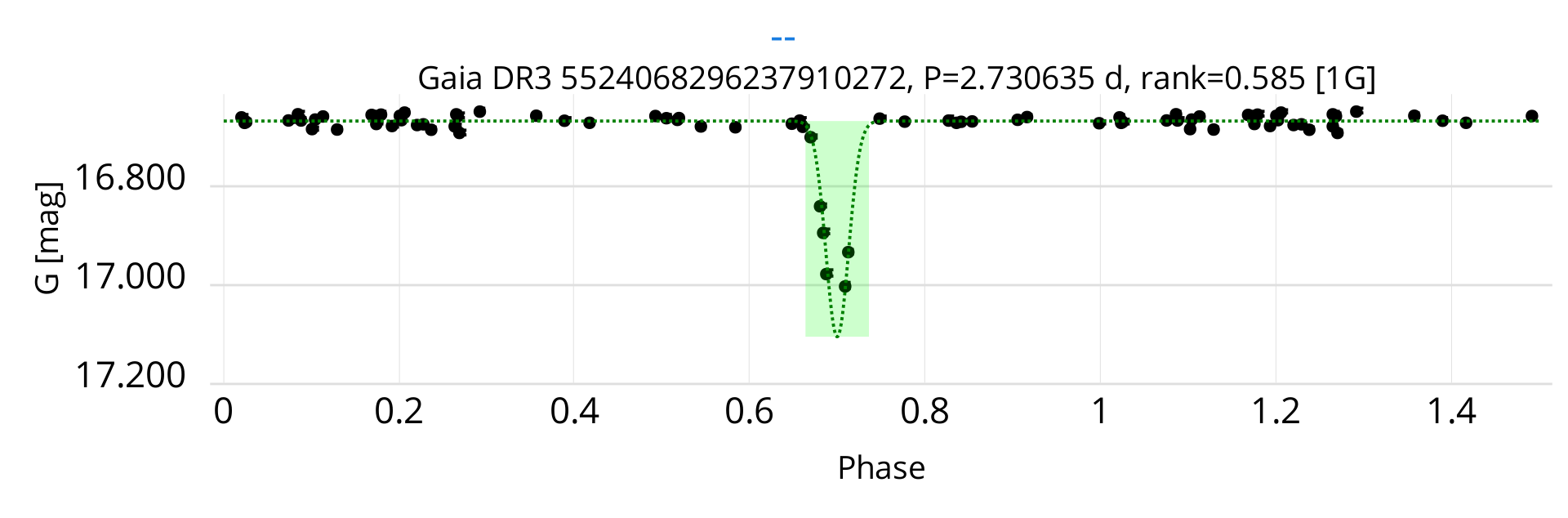}
  \vskip -0.5mm
  \includegraphics[trim={40 145 0 80},clip,width=\linewidth]{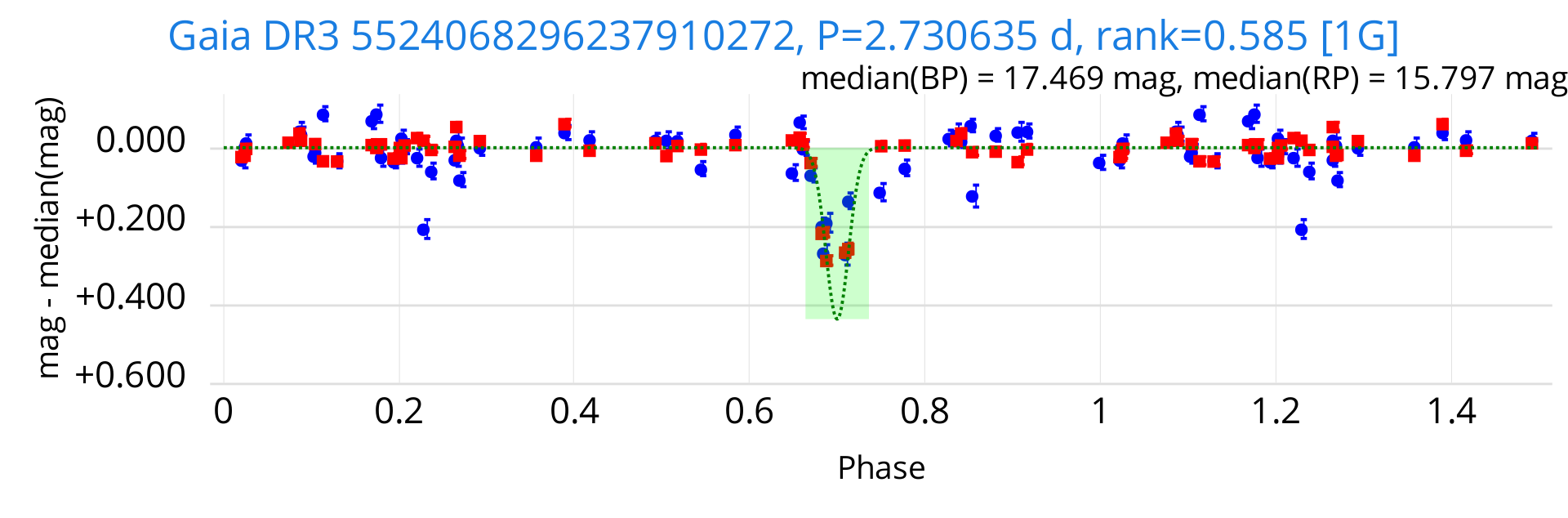}
  \vskip -0.5mm
  \includegraphics[trim={40 150 0 70},clip,width=\linewidth]{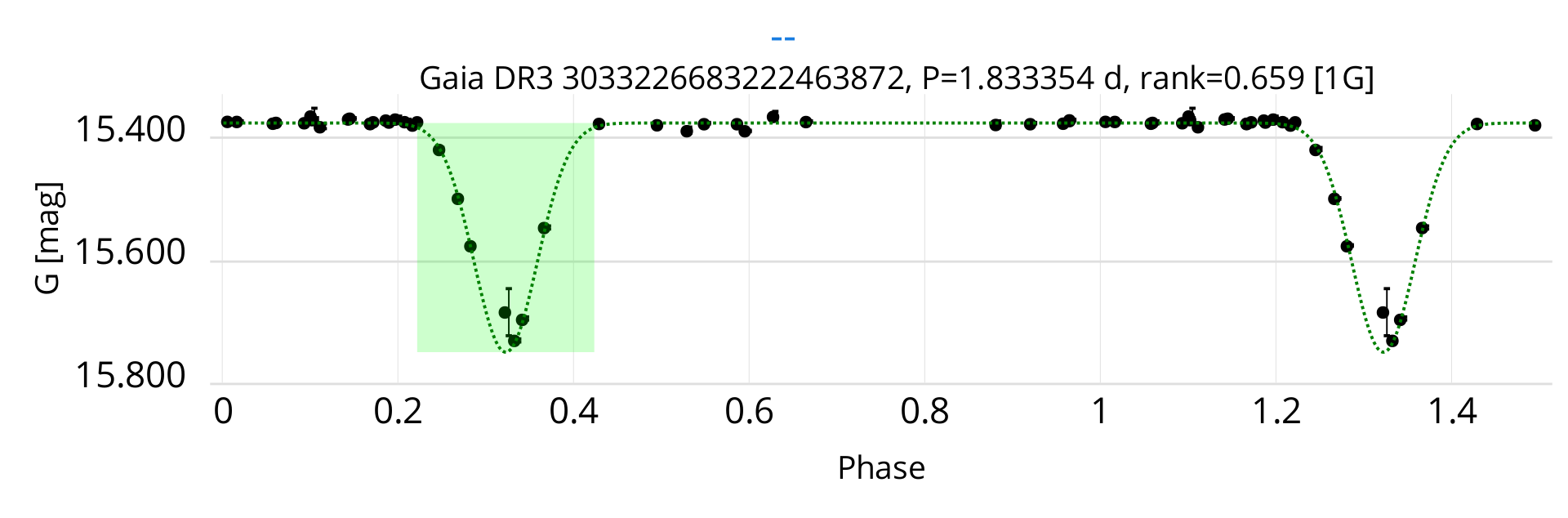}
  \vskip -0.5mm
  \includegraphics[trim={40 145 0 80},clip,width=\linewidth]{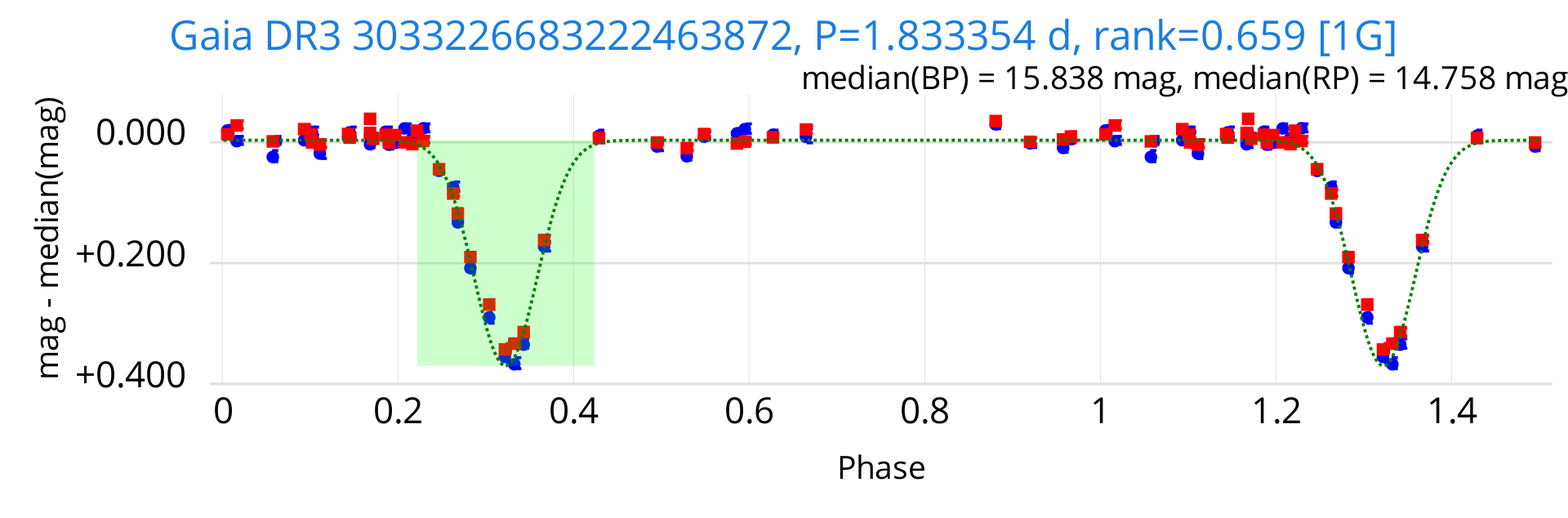}
  \vskip -0.5mm
  \includegraphics[trim={40 150 0 70},clip,width=\linewidth]{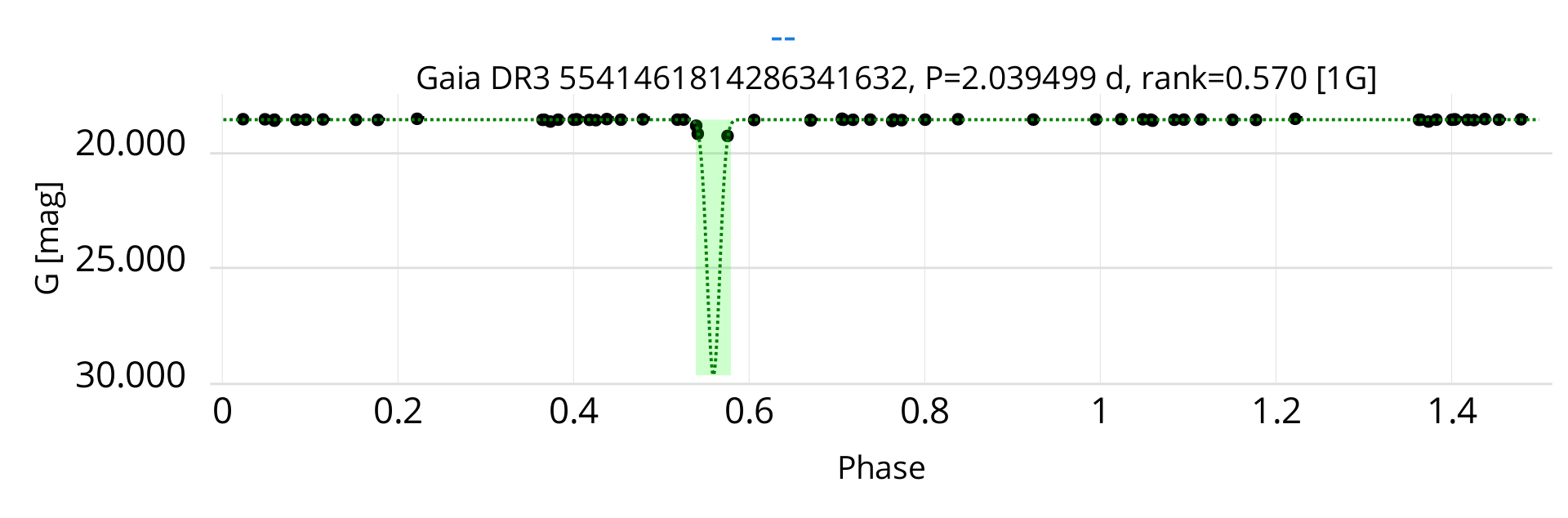}
  \vskip -0.5mm
  \includegraphics[trim={40 45 0 80},clip,width=\linewidth]{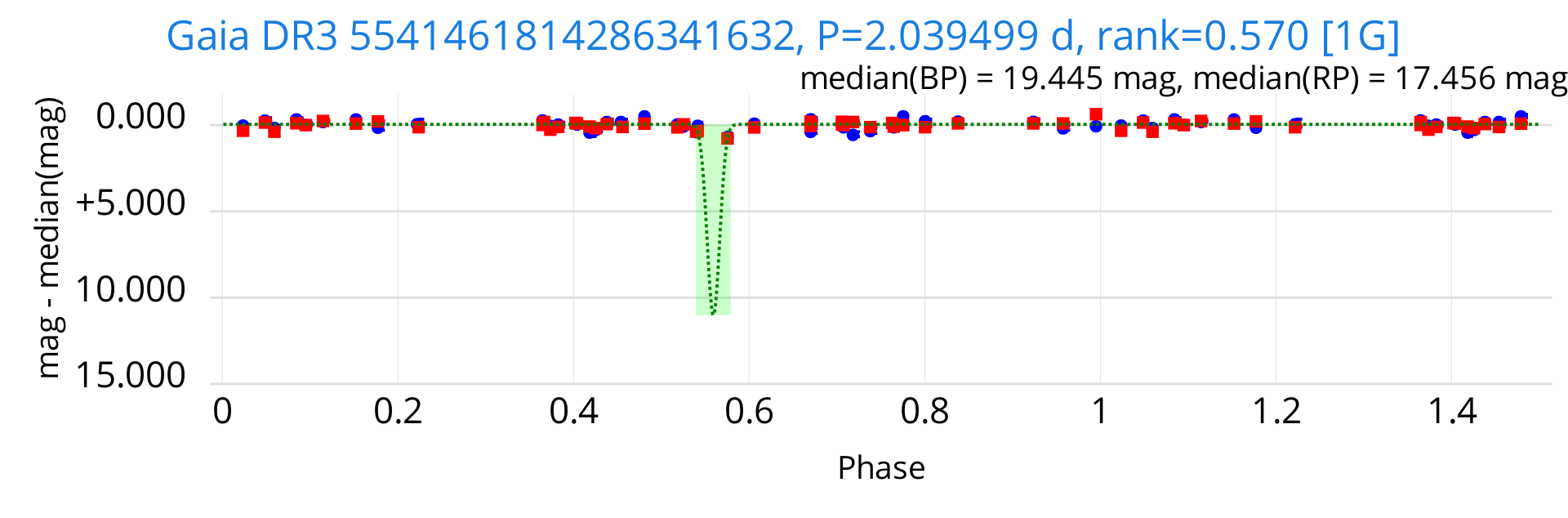}
    \caption{Same as Fig.~\ref{Fig:lcs_2G_A}, but for three candidates in Sample~1G of light curves modelled with only one Gaussian.
             The top set (\GaiaSrcIdInCaption{5524068296237910272}) shows a case where the period is probably a factor of two too small,
             the second set (\GaiaSrcIdInCaption{3033226683222463872}) shows a case where observation may be missing at the phase window of the second eclipse,
             and the third case (\GaiaSrcIdInCaption{5541461814286341632}) illustrates a case where the lack of enough phase coverage within an eclipse leads to an overestimate of the eclipse depth.
            }
\label{Fig:lcs_1G}
\end{figure}

%- - - - - - - - - - - - - - - - - - - -
\paragraph{Sample 1G.}

The number of sources whose light curve model contains only a Gaussian, defining Sample~1G, amounts to 1.7\% of the full catalogue. %\NM{36'984 / 2'184'477 = 1.69\%}
Eccentric systems can result in the occurence of only one eclipse.
The detection of only one eclipse can also happen if the light emitted by the companion in the considered photometric band is below instrument detection limit.
But such a situation can also happen if the period is wrong (typically by a factor of two), or if there are not enough observations in the eclipse(s).
Three example light curves are shown in Fig.~\ref{Fig:lcs_1G}.
In the top example, the period may be a factor of two too small.
In the second example, the secondary eclipse may lack enough observations.
In the third example, the lack of sufficient eclipse coverage leads to an unconstrained eclipse depth, resulting in a too large depth.
This last case is similar to what was noticed in Sect.~\ref{Sect:catalogue_usage_model_2G} for some sources in samples 2G-A and 2G-D.
The distribution of the primary eclipse depth versus eclipse duration is shown in the bottom panel of Fig.~\ref{Fig:eclDepthVsEclDration} for Sample 1G.
In summary, the eclipsing binary candidates in this sample 1G require additional investigations to confirm their properties.

%- - - - - - - - - - - - - - - - - - - -
\subsection{Models with one Gaussian and an ellipsoidal component}
\label{Sect:catalogue_usage_model_1GE}

\begin{figure}
  \centering
  \includegraphics[trim={0 0 0 42},clip,width=\linewidth]{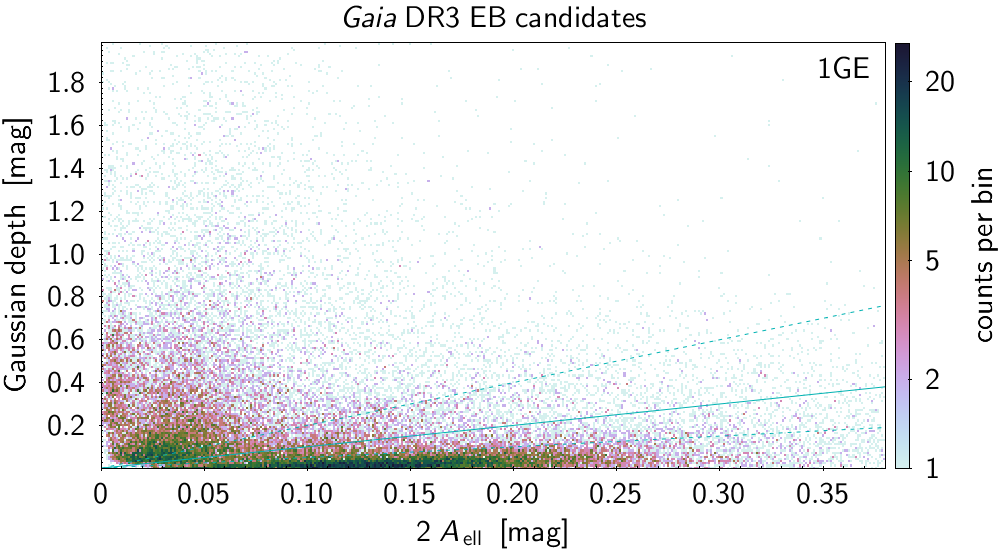}
  \caption{Same as the top panel of Fig.~\ref{Fig:AellVsGaussDepth_E1E2}, but for the sample 1GE with one Gaussian and an ellipsoidal component.
           The axes scales are kept identical to those in Fig.~\ref{Fig:AellVsGaussDepth_E1E2} for easy comparison.
          }
\label{Fig:AellVsGaussDepth_E3}
\end{figure}

\begin{figure}
  \centering
  \includegraphics[trim={40 150 0 70},clip,width=\linewidth]{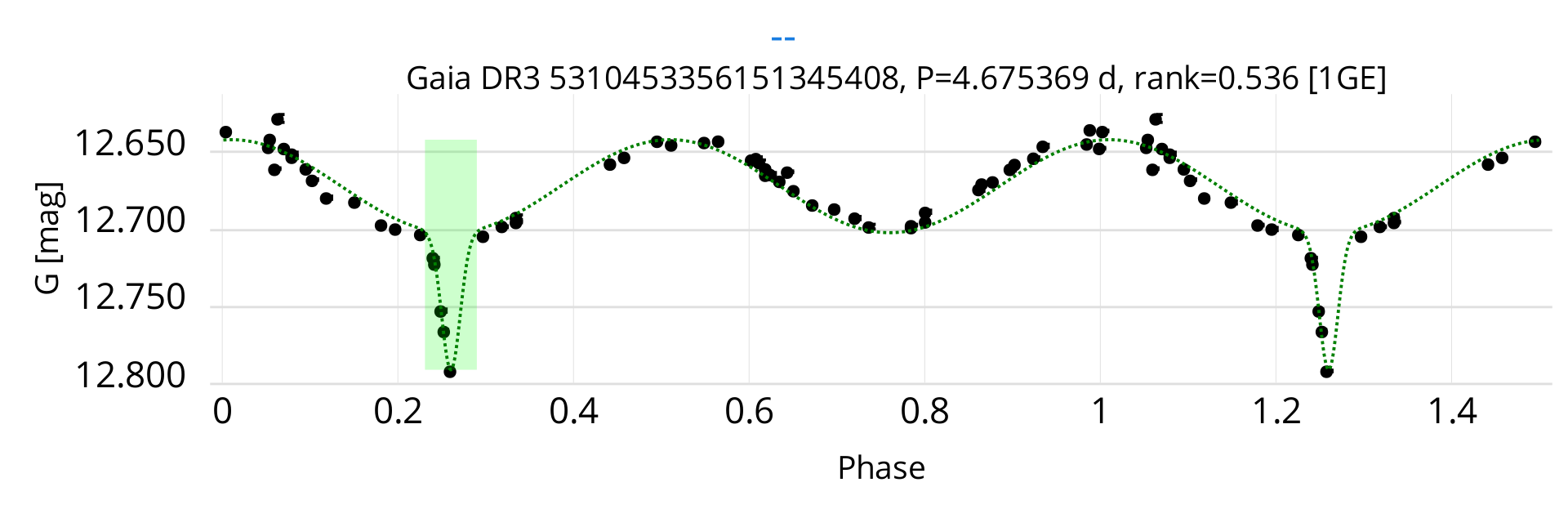}
  \vskip -0.5mm
  \includegraphics[trim={40 145 0 80},clip,width=\linewidth]{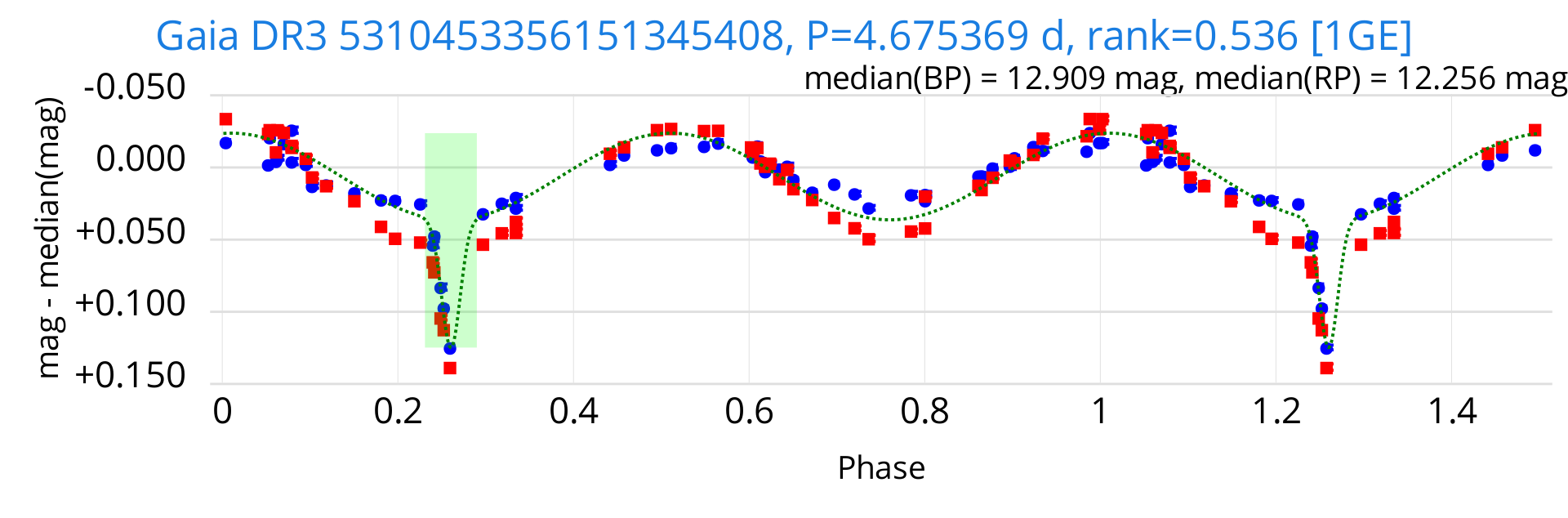}
  \vskip -0.5mm
  \includegraphics[trim={40 150 0 70},clip,width=\linewidth]{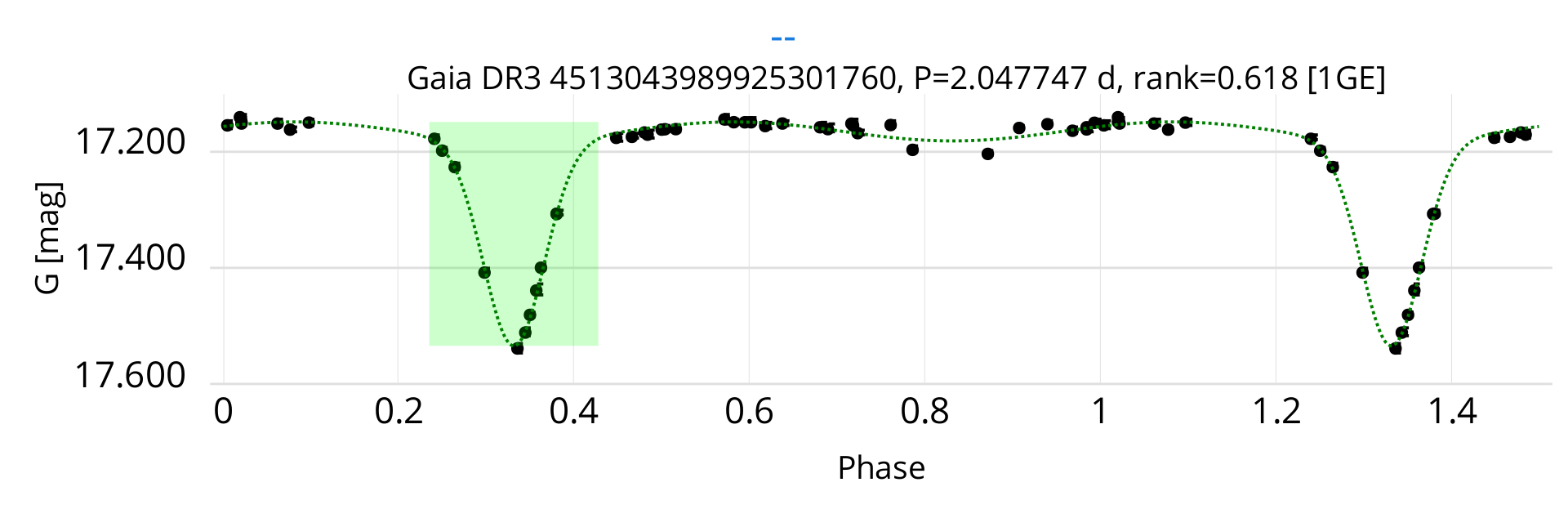}
  \vskip -0.5mm
  \includegraphics[trim={40 145 0 80},clip,width=\linewidth]{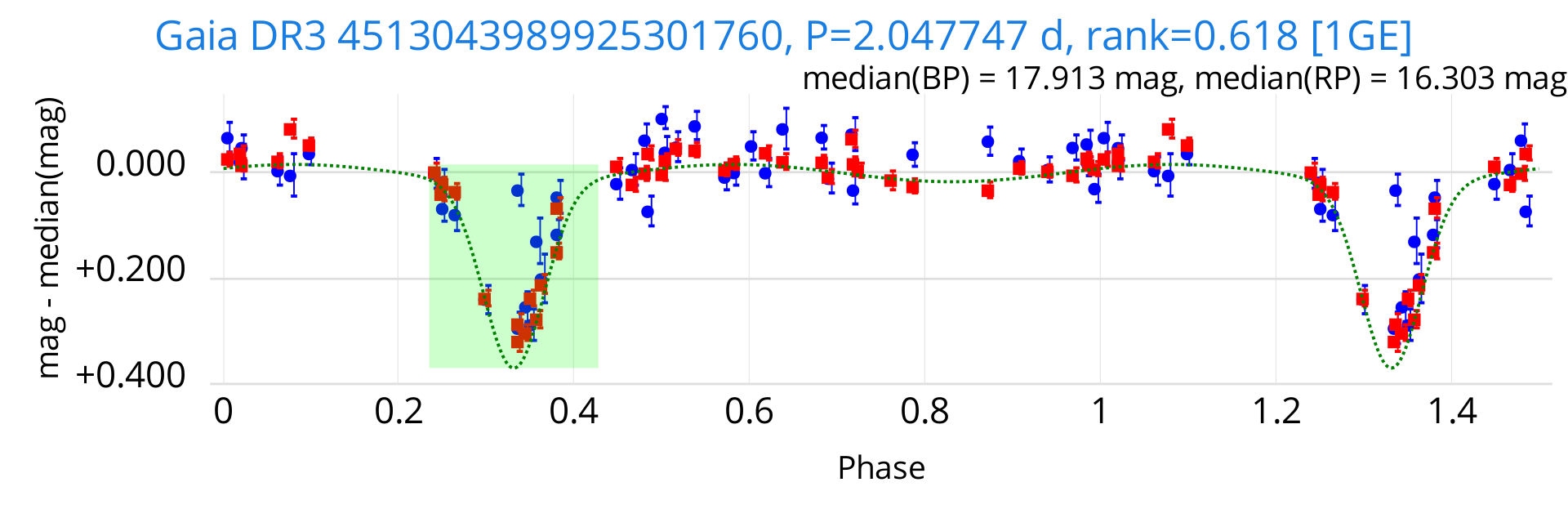}
  \vskip -0.5mm
  \includegraphics[trim={40 150 0 70},clip,width=\linewidth]{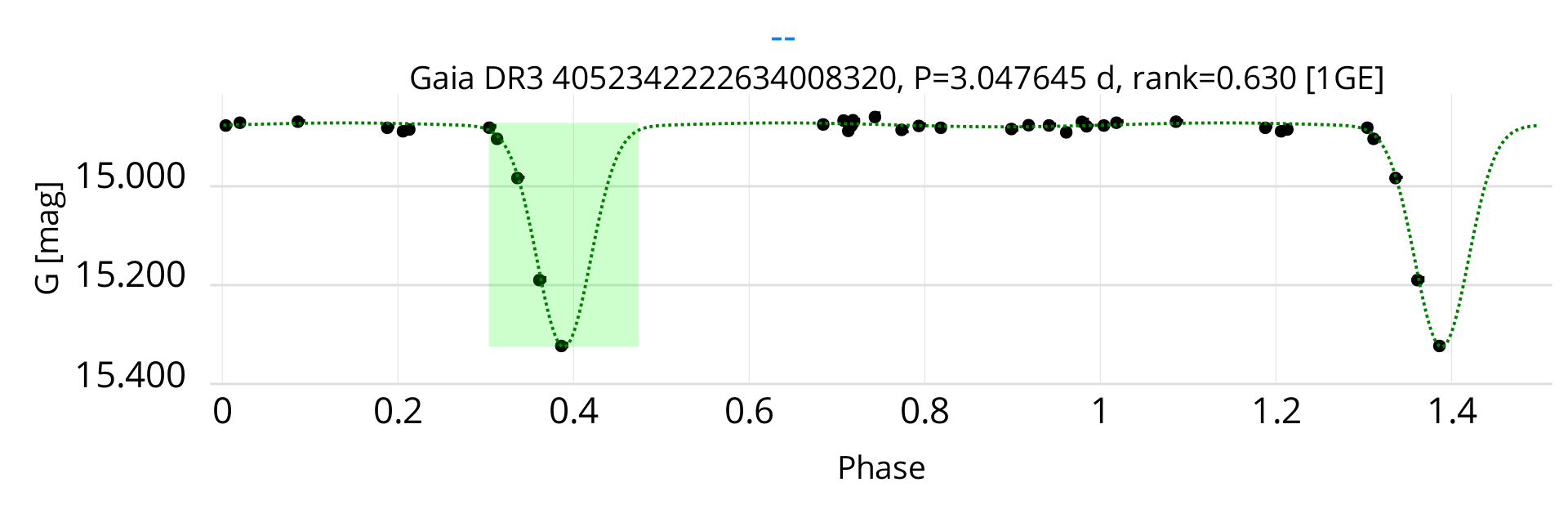}
  \vskip -0.5mm
  \includegraphics[trim={40 45 0 80},clip,width=\linewidth]{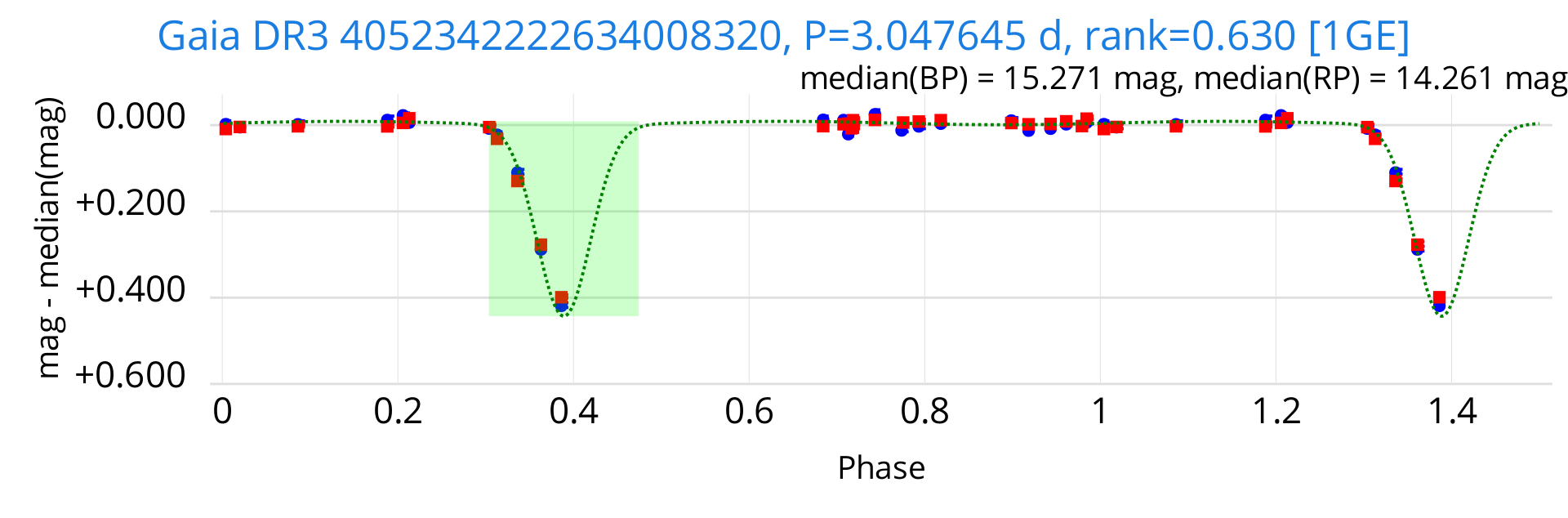}
    \caption{Same as Fig.~\ref{Fig:lcs_2G_A}, but for three candidates in Sample~1GE of light curves modelled with one Gaussian and an ellipsoidal component.
             The top, middle and bottom sets show cases with large (\GaiaSrcIdInCaption{5310453356151345408}), mild (\GaiaSrcIdInCaption{4513043989925301760}), and small (\GaiaSrcIdInCaption{4052342222634008320}) ellipsoidal component relative to the Gaussian depth.
             The middle case may have its secondary eclipse gone undetected due to a lack of \Gaia measurement at a phase of 0.5 apart from the primary eclipse.
            }
\label{Fig:lcs_1GE}
\end{figure}

%- - - - - - - - - - - - - - - - - - - -
\paragraph{Sample 1GE.}

For 2.2\% of the sources in the catalogue, the \gmag light curve is modelled with one Gaussian and an ellipsoidal component.
They define Sample~1GE.
They are comparable to sources modelled with two Gaussians and an ellipsoidal component, but with a secondary eclipse (on top of the ellipsoidal variability) too faint to be detected, leading to the absence of a second Gaussian in the model.
The Gaussian depth is shown in Fig.~\ref{Fig:AellVsGaussDepth_E3} versus peak-to-peak amplitude of the ellipsoidal variability.
A comparison of this figure with the similar figure for models containing two Gaussians (Fig.~\ref{Fig:AellVsGaussDepth_E1E2}, top panel) confirms that Sample~1GE can be considered to be an extension at small primary Gaussian depths of samples 2GE-A and 2GE-B.
The peak-to-peak amplitude distribution of the ellipsoidal component in Sample~1GE is shown by the red histogram in Fig.~\ref{Fig:histo_Aell}.
Example light curves of sources in this sample are displayed in Fig.~\ref{Fig:lcs_1GE}.

%- - - - - - - - - - - - - - - - - - - -
\subsection{Models with only an ellipsoidal component}
\label{Sect:catalogue_usage_model_ellipsoidal}

\begin{figure}
  \centering
  \includegraphics[trim={0 0 0 42},clip,width=\linewidth]{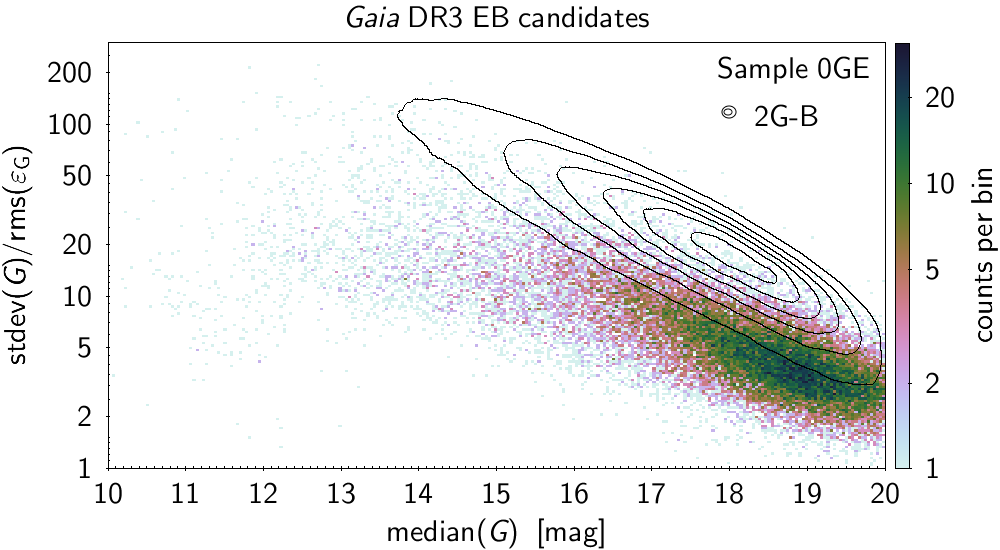}
  \caption{Density map of the signal-to-noise (computed as the ratio of the standard deviation over the root-mean-square of the \gmag magnitude uncertainties) versus \gmag magnitude for the sample of candidates modelled with only a cosine function (Sample~0GE).
           The contours delineate the density of sources in the sample 2GE-B (modelled with two Gaussians and an ellipsoidal component with large amplitude (see text).
           Six contours are shown on a linear scale of the density of sources on the map.
          }
    \label{Fig:snVsG_E0E1}
\end{figure}

\begin{figure}
  \centering
  \includegraphics[trim={0 0 0 42},clip,width=\linewidth]{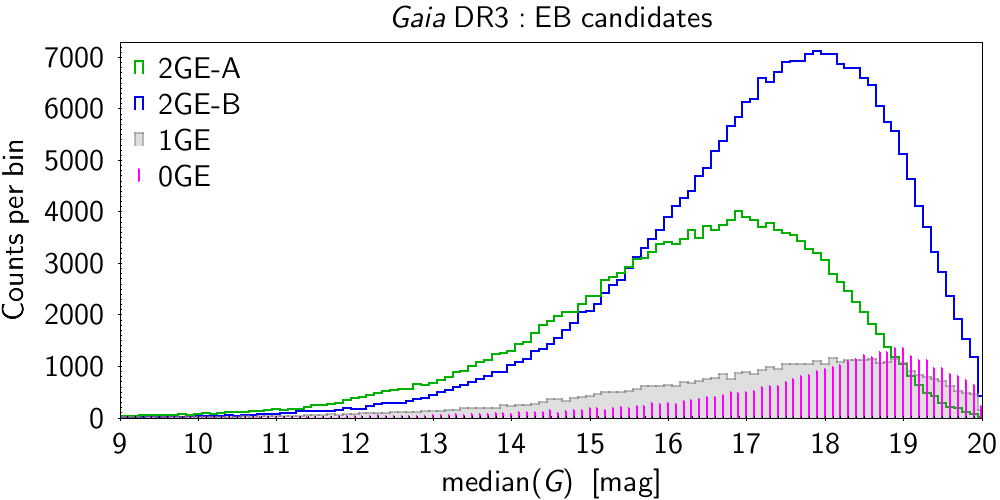}
  \caption{G magnitude distribution of the various samples containing an ellipsoidal component in their light curve models, as labeled in the figure 
           The abscissa has been truncated at the bright side for better visibility.
          }
    \label{Fig:figHisto_Gmag_withEll}
\end{figure}

\begin{figure}
  \centering
  \includegraphics[trim={40 150 0 70},clip,width=\linewidth]{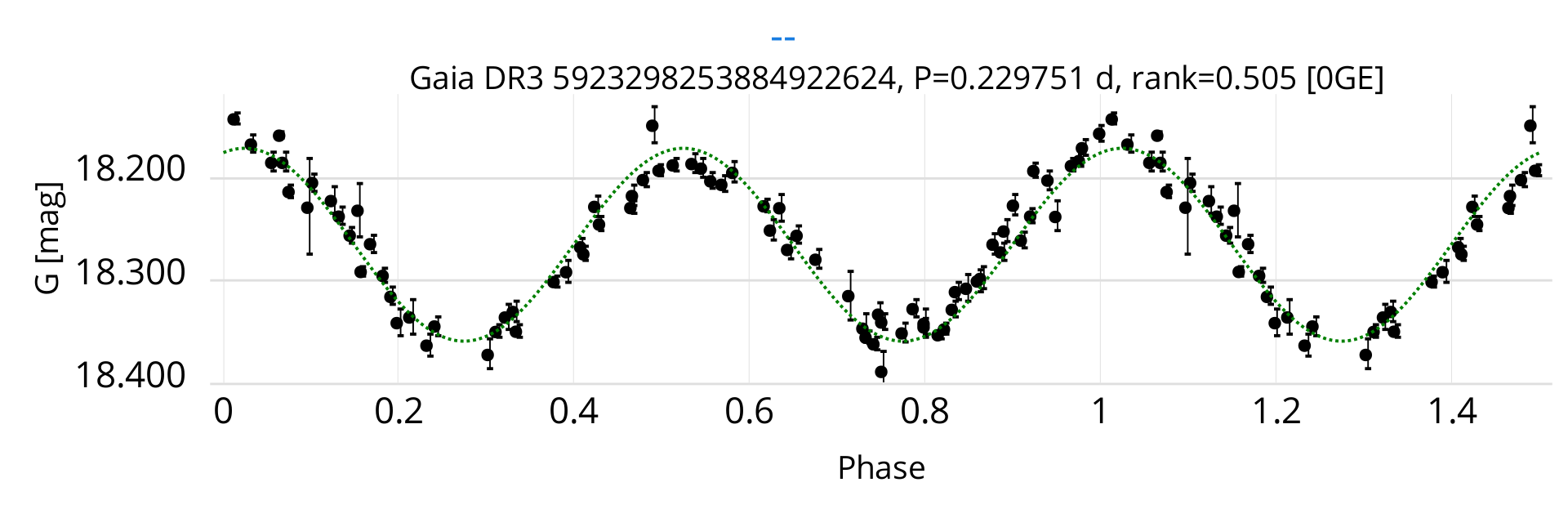}
  \vskip -0.5mm
  \includegraphics[trim={40 145 0 80},clip,width=\linewidth]{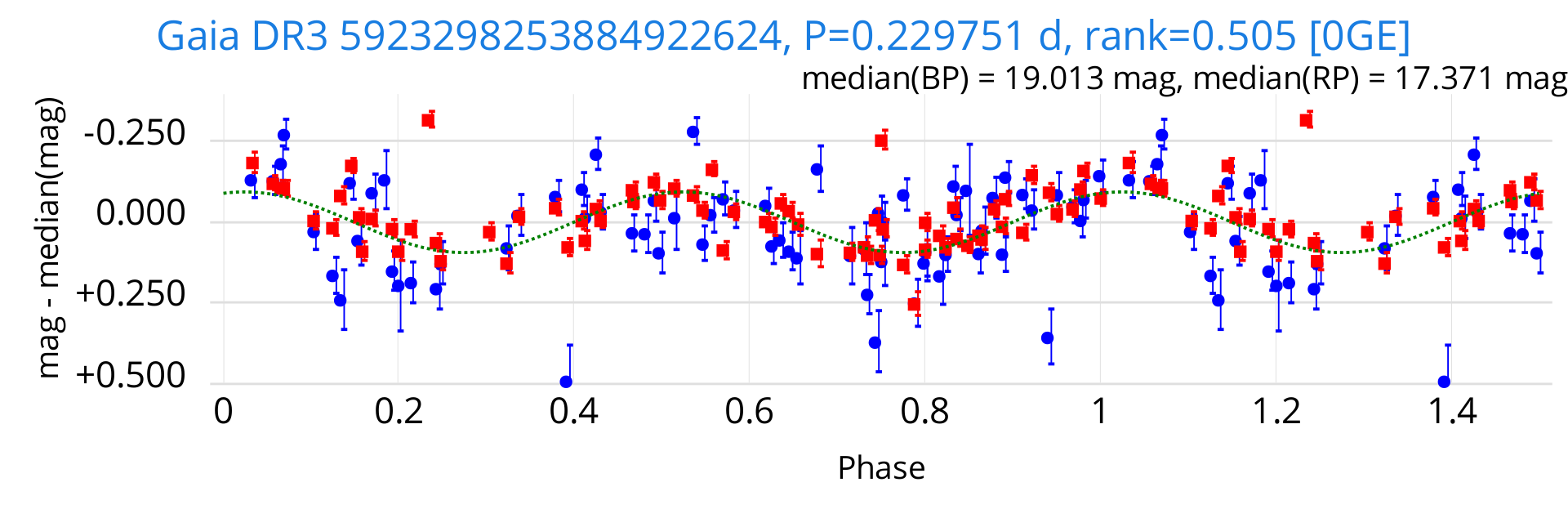}
  \vskip -0.5mm
  \includegraphics[trim={40 150 0 70},clip,width=\linewidth]{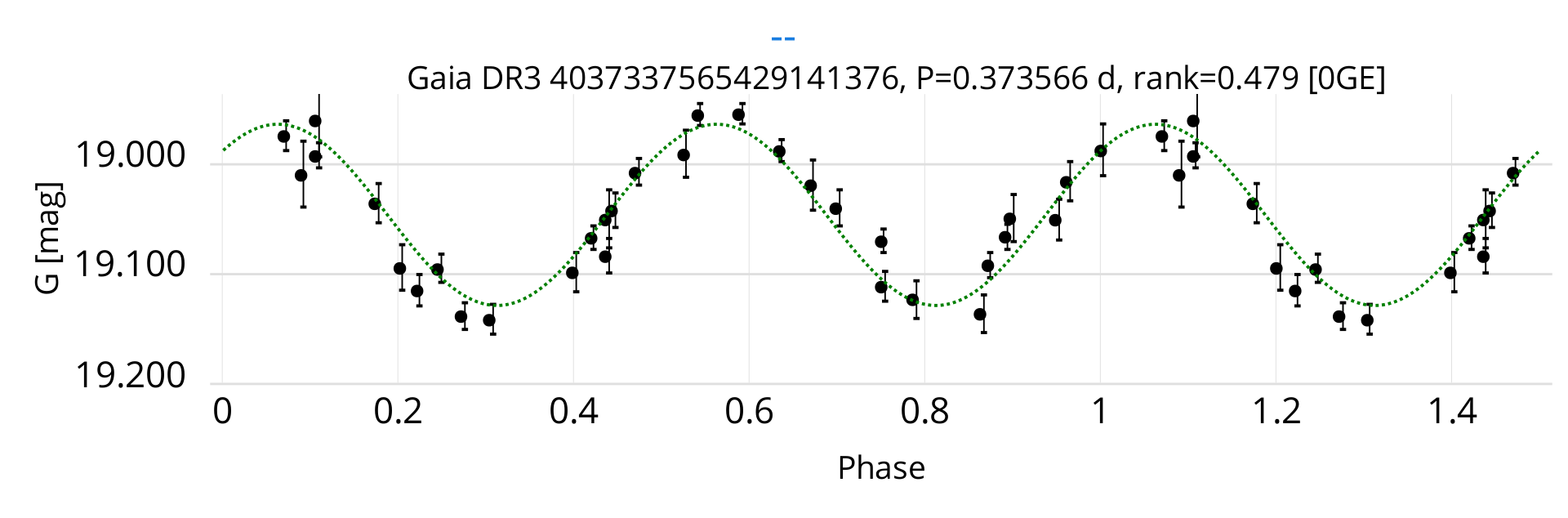}
  \vskip -0.5mm
  \includegraphics[trim={40 145 0 80},clip,width=\linewidth]{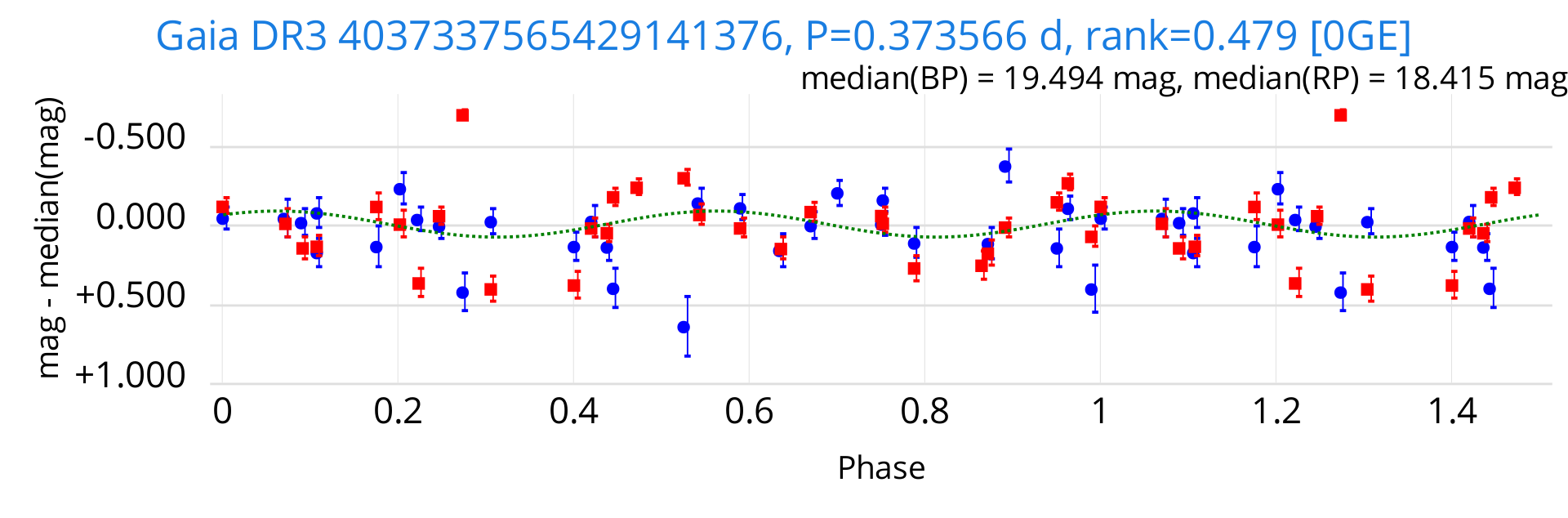}
  \vskip -0.5mm
  \includegraphics[trim={40 150 0 70},clip,width=\linewidth]{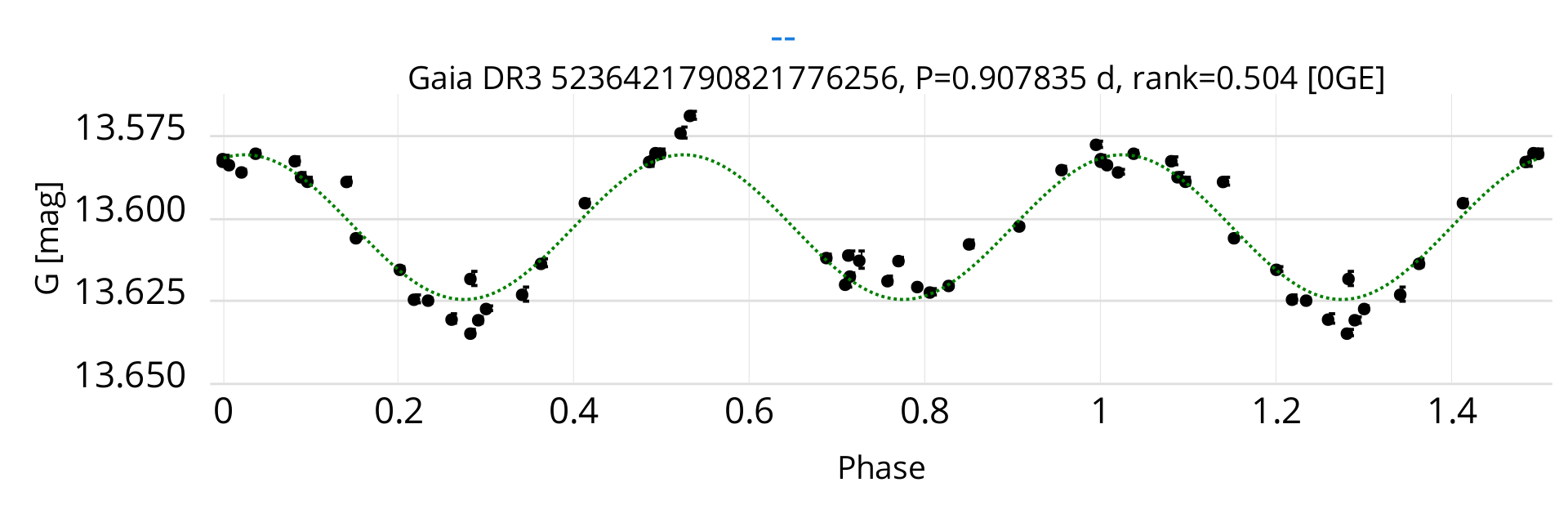}
  \vskip -0.5mm
  \includegraphics[trim={40 45 0 80},clip,width=\linewidth]{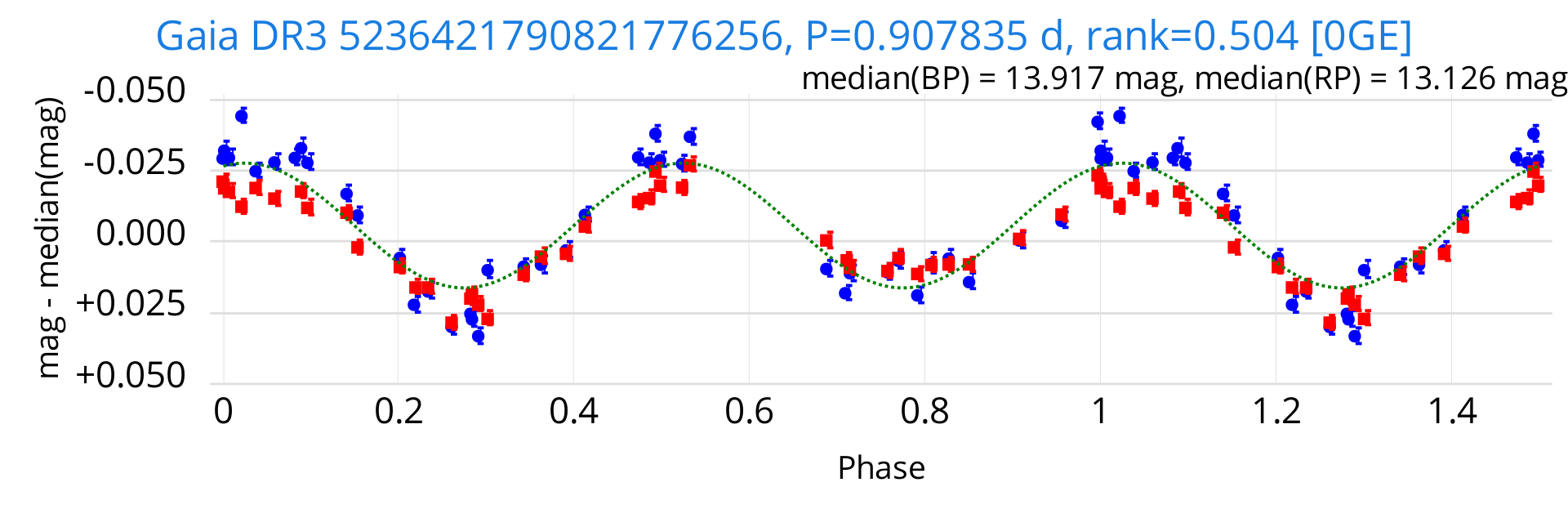}
    \caption{Same as Fig.~\ref{Fig:lcs_2G_A}, but for three candidates in Sample~0GE (light curves modelled with only a cosine).
             The first two examples show the most common cases of faint candidates, with (top case, \GaiaSrcIdInCaption{5923298253884922624} and without (middle case, \GaiaSrcIdInCaption{4037337565429141376}) a clear confirmation of the variability in the \gbp and \grp light curves.
             The bottom example shows a relatively rarer case in this sample of a bright candidate (\GaiaSrcIdInCaption{5236421790821776256}).
            }
\label{Fig:lcs_0GE}
\end{figure}

%- - - - - - - - - - - - - - - - - - - -
\paragraph{Sample 0GE.}

\begin{table*}
\caption{Main binary system types expected in the samples defined in Table~\ref{Tab:sample_summary}.
        }
\centering
\begin{tabular}{l l c}
\hline\hline
Samples                  & Main binary types    & \% in catalogue\\
\hline
\hline
%--------------------------
2G-A, 2GE-A, 2G-D, 1G       & Wide systems (detached or, under some conditions, semi-detached) &  27\%\\
2G-B, 2G-C, 2GE-B, 1GE, 0GE & Tight systems    &  56\%\\
2G-X, 2G-Y, 2GE-Z           & To be investigated  &  17\% \\
\hline
\hline
\end{tabular}
\label{Tab:sample_summary}
\end{table*}

The remaining candidates in the catalogue are modelled with only a cosine and no Gaussian function.
We label them Sample~0GE.
It contains only 2\% of the full catalogue.
%They may be compared with Sample~2G-C having two very wide Gaussians (Fig.~\ref{Fig:lcs_2G_C}, top example) or with Sample~2GE-B having a dominant ellipsoidal component and a Gaussian (Fig.~\ref{Fig:lcs_2GE_B}, top example).
In most cases, a purely cosine model is favored by the automated procedure over a solution involving Gaussians when photometric uncertainties are large.
This is shown in Fig.~\ref{Fig:snVsG_E0E1}, where the signal-to-noise ratio is plotted versus \gmag magnitude.
For comparison, the distribution of the 2GE-B sample is also shown in the figure by six contour lines equally distant on a linear scale.
As a result, since photometric uncertainty increases with increasing magnitude, the magnitudes of the 0GE sample are, on the mean, fainter than any of the other samples containing an ellipsoidal component (see Fig.~\ref{Fig:figHisto_Gmag_withEll}).
The presence of intrinsic scatter in the light curve may also lead to the automated procedure to favor a purely cosine model over models containing Gaussian functions.

An example light curve of a faint ($\sim$18.25~mag in \gmag) 0GE source is shown in Fig.~\ref{Fig:lcs_0GE} (top case).
The \gbp and \grp light curves confirm the ellipsoidal-like variability, though with much larger uncertainties on the measurements.
The second example displays another 0GE faint source ($\sim$19.05~mag in \gmag) with a clear sinusoidal-like variability in \gmag, but with no such clear variability in \gbp and \grp.
Sources like this one require further confirmation of their variability.
The last example in Fig.~\ref{Fig:lcs_0GE} shows a (rarer) bright case, with \gmag around 13.3~mag.

%The ellipsoidal amplitudes of the sources Sammple~0GE are distributed between those of samples 2GE-A and 2GE-B, as shown by the green filled histogram in Fig.~\ref{Fig:histo_Aell}.
%This supports the suggestion that Sample~0GE complements samples 2GE-A and 2GE-B.

%------------------------------------------------------------------
\subsection{Summary}
\label{Sect:catalogue_usage_summary}

% All sources with LCs in the appendix (for VariDashboard):
% 5632871775033302656, 5684917462874690560, 1872983530689181312, 1807942504467443456, 1980590328522237824, 2059784985669541760, 520547841550454784, 5310453356151345408, 4052342222634008320, 4037337565429141376, 2195008530582714880, 5863293368876899712, 1920242361505734784
% 5256648720295981184, 264837033617656448, 509661748731132800, 4513043989925301760, 2056622657089571968, 3033226683222463872, 249126318130979840, 5524068296237910272, 5541461814286341632, 4077271415493214080, 3444083186030598272, 5923298253884922624, 5236421790821776256

% 4045605519248336256

%-----------
\begin{figure}
  \centering
  \includegraphics[trim={0 77 0 44},clip,width=\linewidth,height=1.95cm]{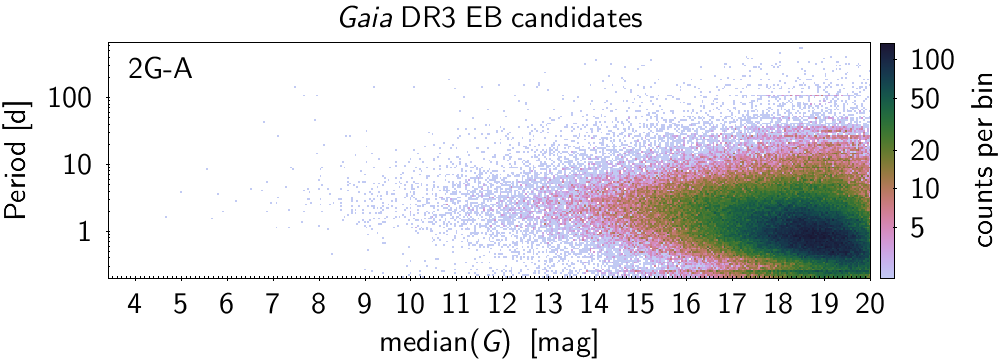}
  \vskip -0.5mm
  \includegraphics[trim={0 77 0 44},clip,width=\linewidth,height=1.95cm]{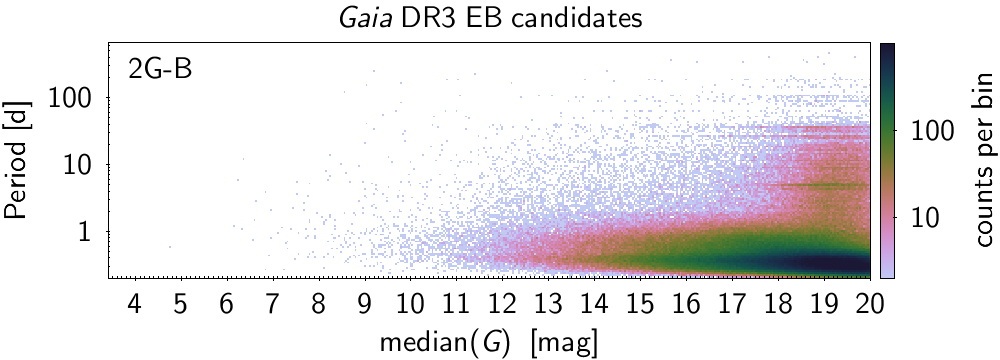}
  \vskip -0.5mm
  \includegraphics[trim={0 77 0 44},clip,width=\linewidth,height=1.95cm]{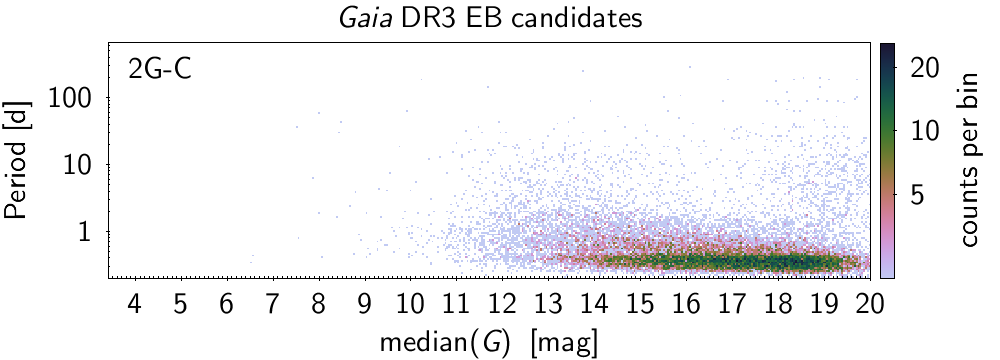}
  \vskip -0.5mm
  \includegraphics[trim={0 77 0 44},clip,width=\linewidth,height=1.95cm]{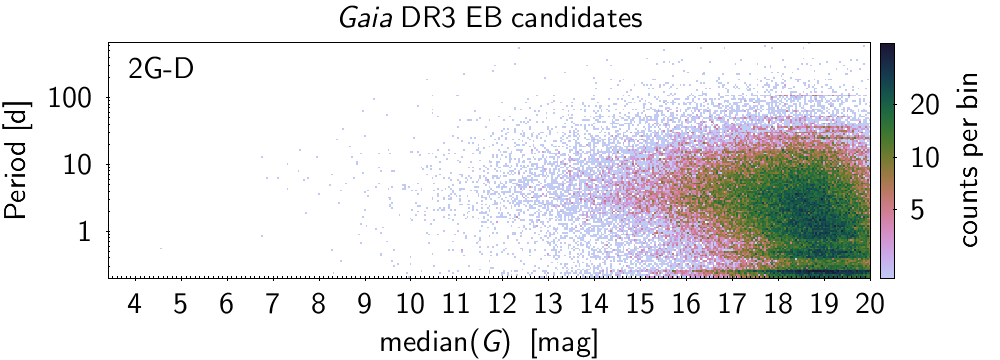}
  \vskip -0.5mm
  \includegraphics[trim={0 77 0 44},clip,width=\linewidth,height=1.95cm]{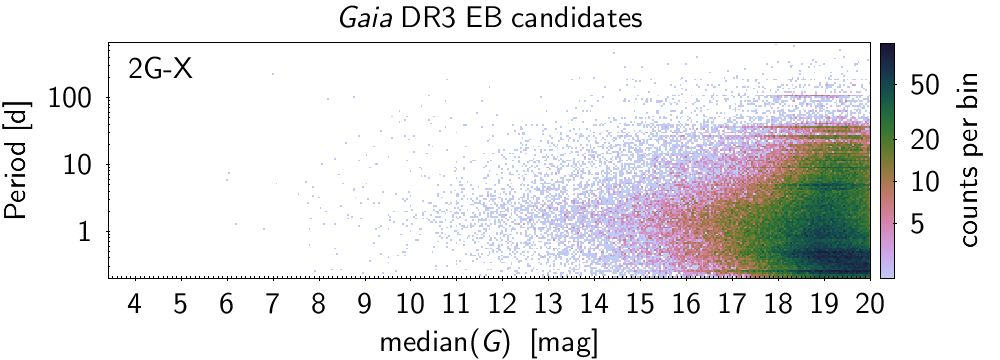}
  \vskip -0.5mm
  \includegraphics[trim={0 77 0 44},clip,width=\linewidth,height=1.95cm]{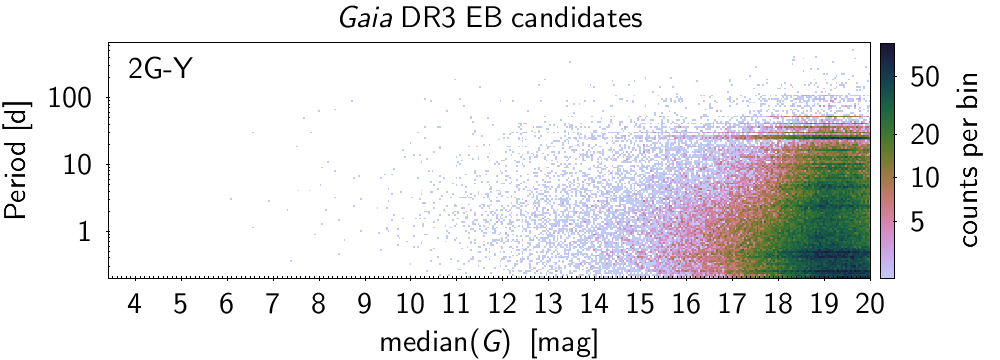}
  \vskip -0.5mm
  \includegraphics[trim={0 77 0 44},clip,width=\linewidth,height=1.95cm]{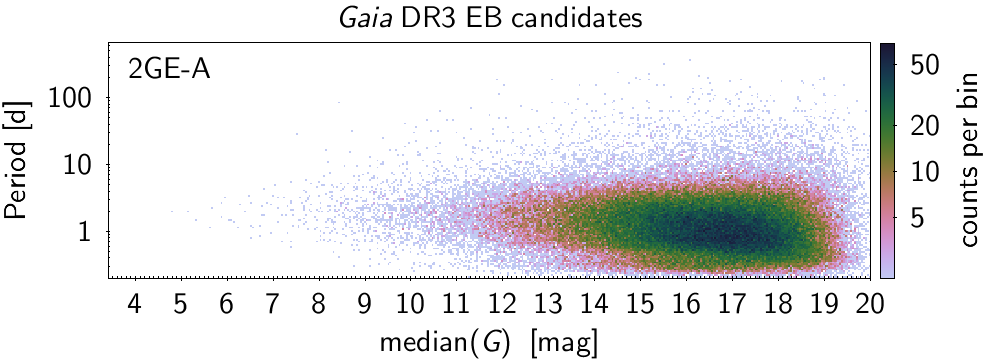}
  \vskip -0.5mm
  \includegraphics[trim={0 77 0 44},clip,width=\linewidth,height=1.95cm]{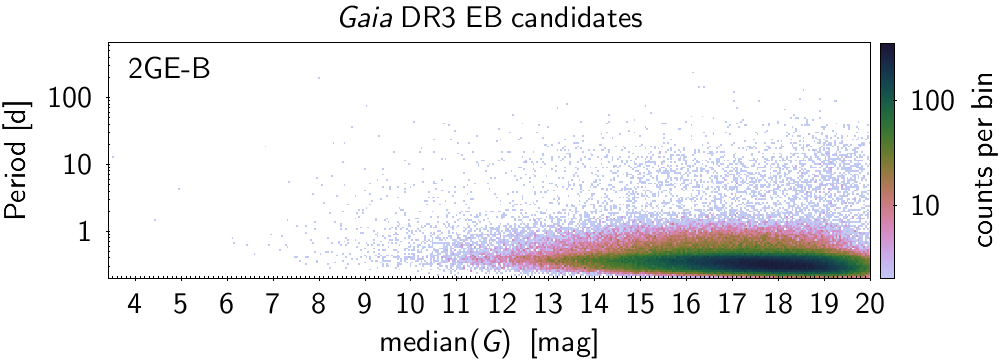}
  \vskip -0.5mm
  \includegraphics[trim={0 77 0 44},clip,width=\linewidth,height=1.95cm]{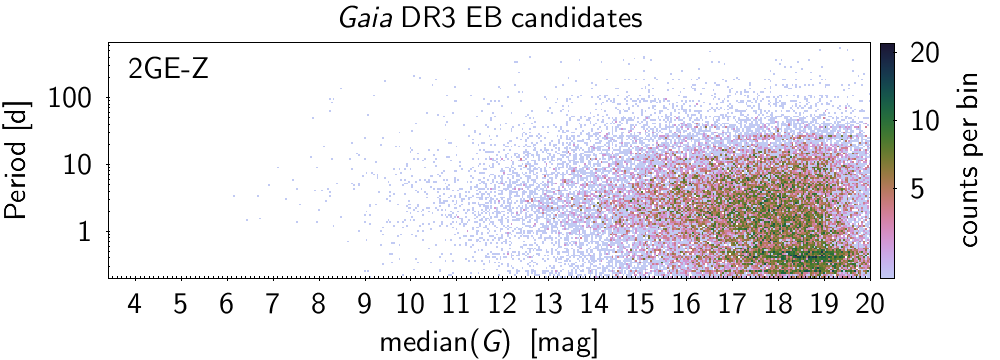}
  \vskip -0.5mm
  \includegraphics[trim={0 77 0 44},clip,width=\linewidth,height=1.95cm]{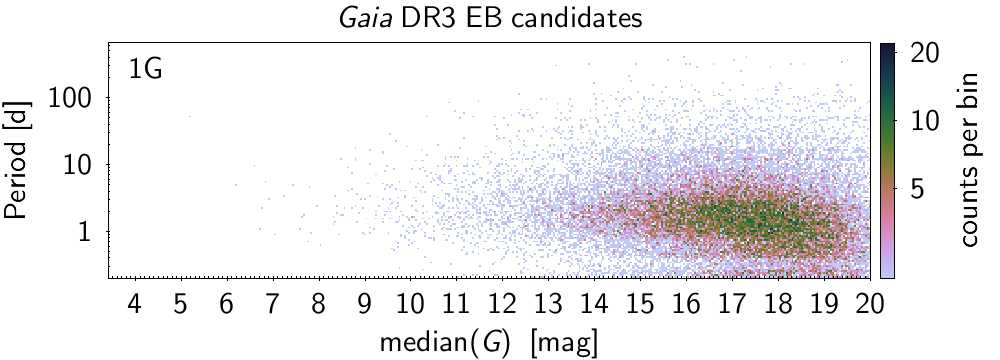}
  \vskip -0.5mm
  \includegraphics[trim={0 77 0 44},clip,width=\linewidth,height=1.95cm]{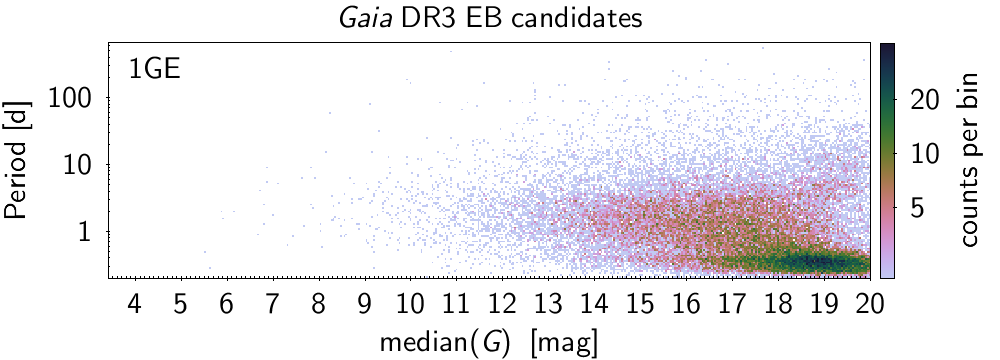}
  \vskip -0.5mm
  \includegraphics[trim={0 0 0 44},clip,width=\linewidth,height=2.5cm]{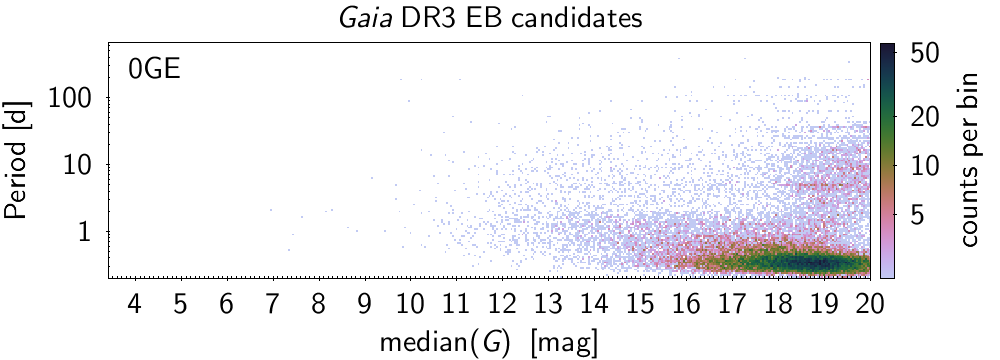}
  \caption{Density maps of period versus \gmag magnitude for the various samples defined in Table~\ref{Tab:sample_definition}.
          }
    \label{Fig:period_magG_samples}
\end{figure}
%-----------

%-----------
\begin{figure}
  \centering
  \includegraphics[trim={0 77 0 44},clip,width=\linewidth,height=1.95cm]{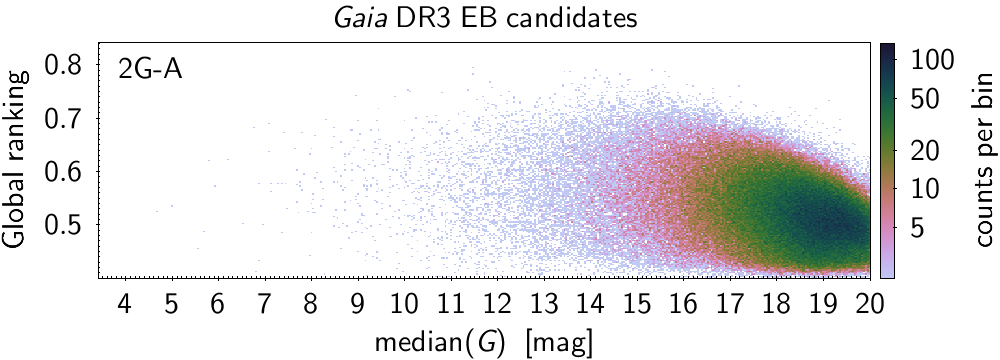}
  \vskip -0.5mm
  \includegraphics[trim={0 77 0 44},clip,width=\linewidth,height=1.95cm]{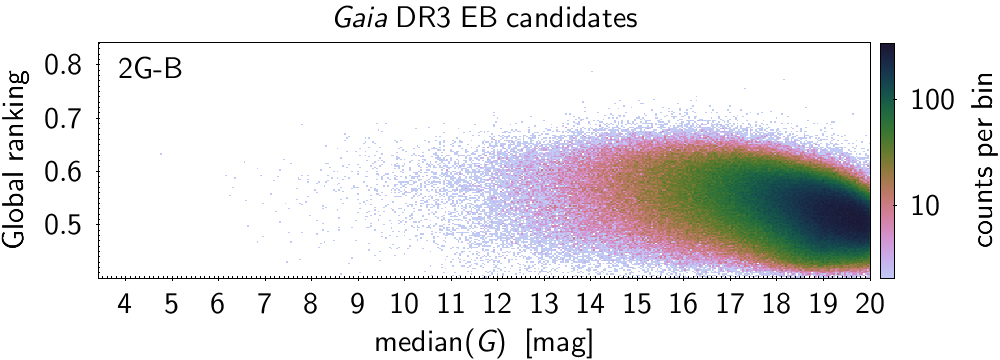}
  \vskip -0.5mm
  \includegraphics[trim={0 77 0 44},clip,width=\linewidth,height=1.95cm]{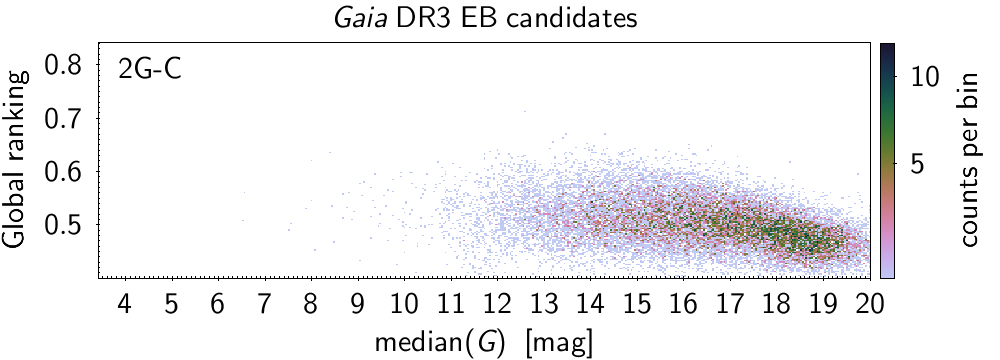}
  \vskip -0.5mm
  \includegraphics[trim={0 77 0 44},clip,width=\linewidth,height=1.95cm]{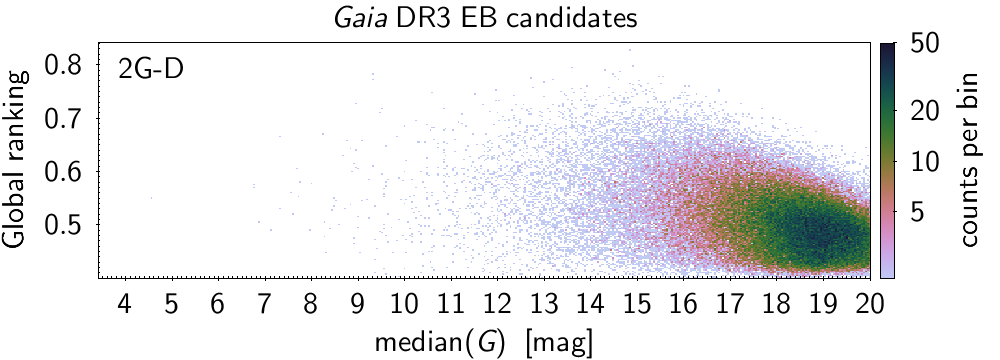}
  \vskip -0.5mm
  \includegraphics[trim={0 77 0 44},clip,width=\linewidth,height=1.95cm]{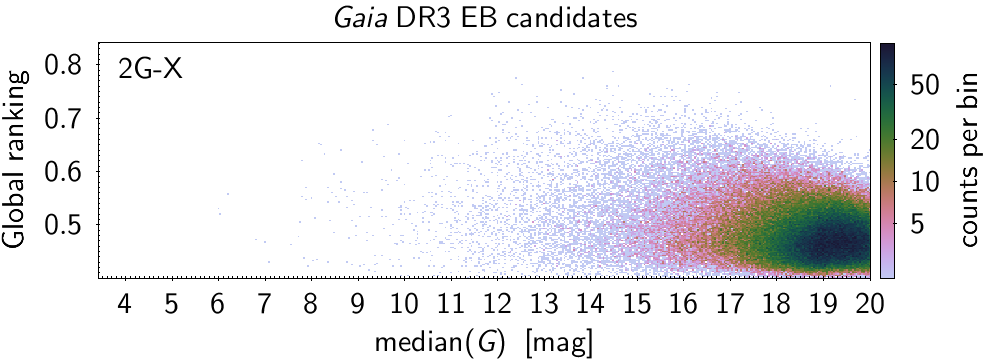}
  \vskip -0.5mm
  \includegraphics[trim={0 77 0 44},clip,width=\linewidth,height=1.95cm]{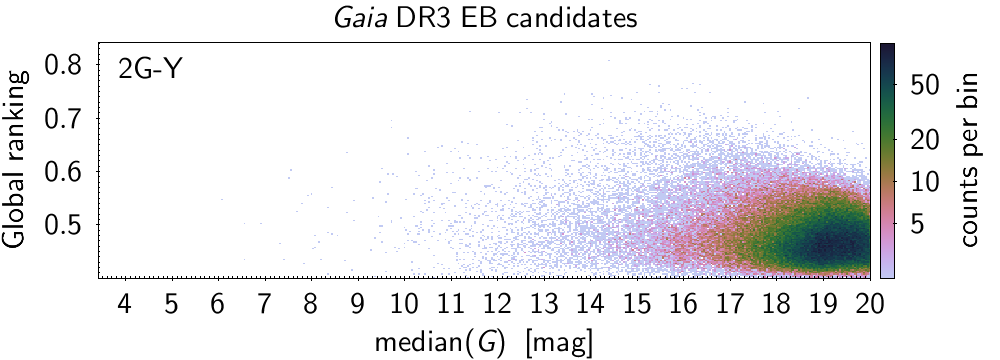}
  \vskip -0.5mm
  \includegraphics[trim={0 77 0 44},clip,width=\linewidth,height=1.95cm]{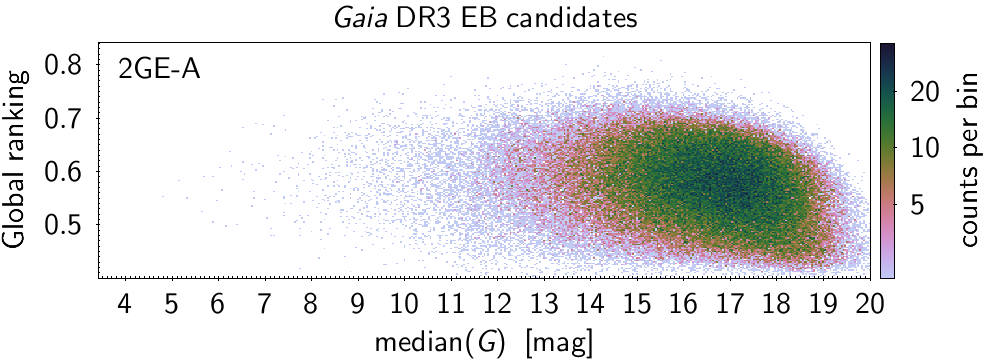}
  \vskip -0.5mm
  \includegraphics[trim={0 77 0 44},clip,width=\linewidth,height=1.95cm]{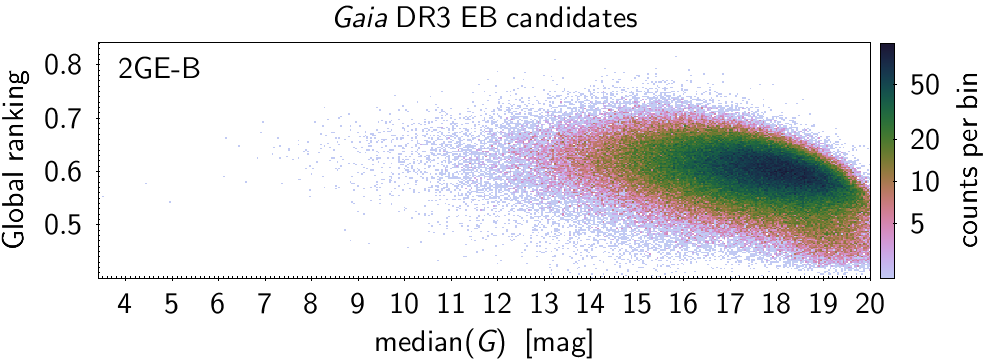}
  \vskip -0.5mm
  \includegraphics[trim={0 77 0 44},clip,width=\linewidth,height=1.95cm]{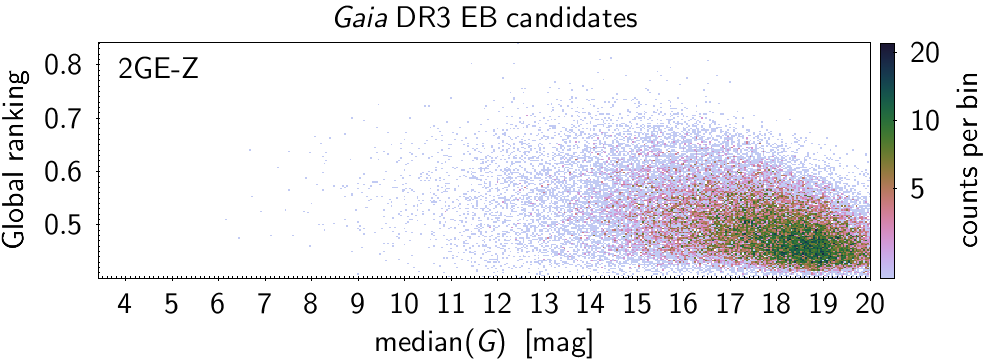}
  \vskip -0.5mm
  \includegraphics[trim={0 77 0 44},clip,width=\linewidth,height=1.95cm]{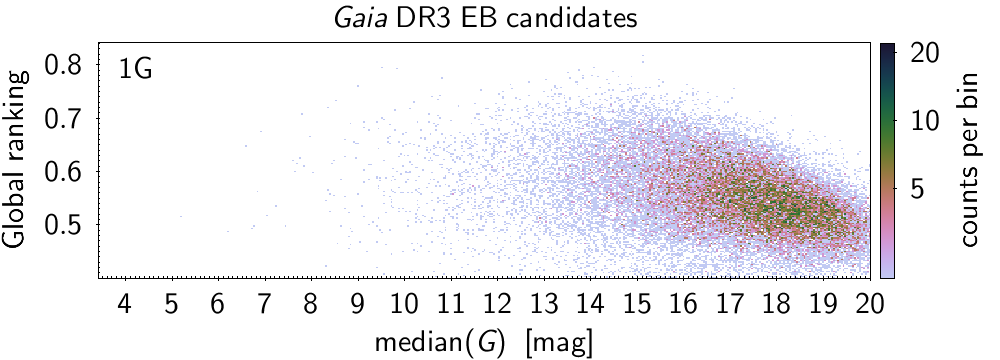}
  \vskip -0.5mm
  \includegraphics[trim={0 77 0 44},clip,width=\linewidth,height=1.95cm]{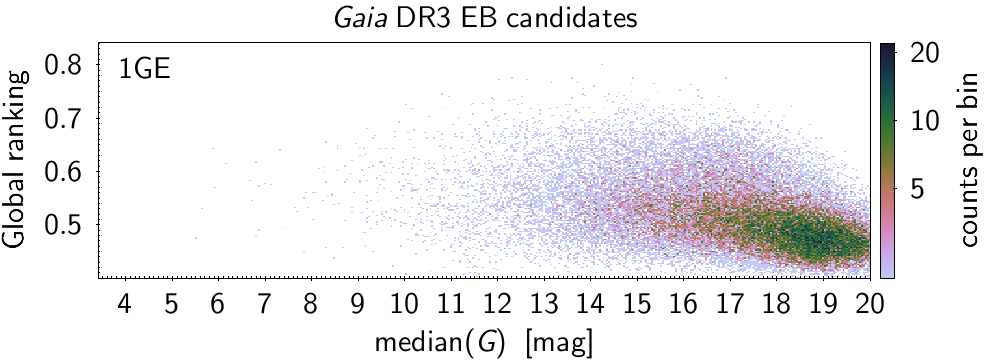}
  \vskip -0.5mm
  \includegraphics[trim={0 0 0 44},clip,width=\linewidth,height=2.5cm]{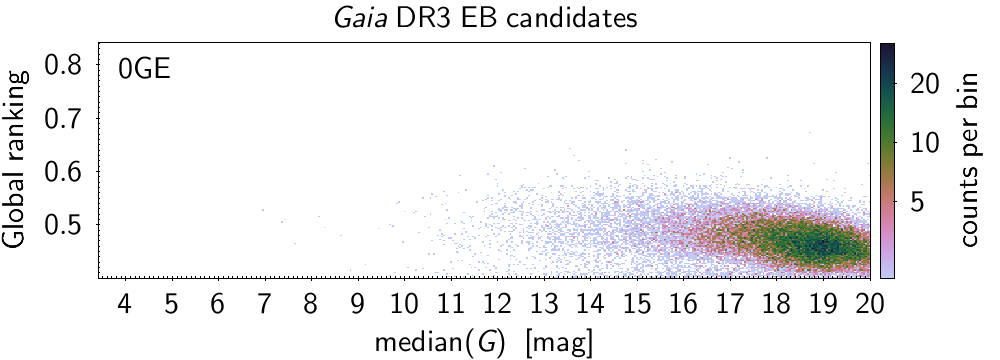}
  \caption{Same as Fig.~\ref{Fig:period_magG_samples}, but for global ranking versus \gmag magnitude.
          }
    \label{Fig:rank_magG_samples}
\end{figure}
%-----------

Table~\ref{Tab:sample_summary} summarises the various samples identified in in this Appendix, categorized according to the type of binary system expected for the majority of candidates in each sample.
Light curve alone does not always allow to uniquely identify the type of binary system.
The light curve of an Algol-type eclipsing binary, for example, in which a star that fills its Roche lobe is much fainter than its companion, can resemble that of a fully detached system.
We therefore consider wide versus tight systems without further sub-classification.
Wide detached systems lead to light curves that are either devoid of ellipsoidal variability (samples 2G-A, 2G-D, 1G) or that have a mild ellipsoidal component (2GE-A). 
Algol-type systems where the secondary is much fainter than the primary would preferentially be classified in this category as well.
This first category gathers about one fourth of the catalogue.
The second category contains tighter systems, displaying light curves that have either a predominant ellipsoidal component (samples 2GE-B, 1GE, 0GE), or are described by two wide overlapping Gaussians (2G-B, 2G-C).
They represent more than half of the catalogue.
We note that the two famous tight systems $\beta$~Lyr and W~Uma belong to Sample 2GE-B (see Fig.~\ref{Fig:lcs_examples_2GE_B} in the main body of the article).
The third category, containing less than one fifth of the catalogue, gathers samples 2G-X, 2G-Y and 2GE-Z.
The nature of the majority of candidates therein require further investigations.
The grouping of the samples in these three categories is supported by their period distributions shown in Fig.~\ref{Fig:histo_period_samples}.
The samples in the first category (2G-A, 2GE-A, 2G-D, 1G) all display wide period distributions reaching values above 10~days (top panel in the figure), as expected for wide systems, while the samples in the second category have period distributions peaking below 0.5~days (second and third panels in the figure), as expected for tight systems.
The distributions of the period versus magnitude for all samples are shown in Fig.~\ref{Fig:period_magG_samples}.
Likewise, the distributions of their global ranking versus magnitude are shown in Fig.~\ref{Fig:rank_magG_samples}.
The abscissa and ordinate scales are kept identical in all panels of each figure for easier comparison. 

The categorization presented in Table~\ref{Tab:sample_summary} is not intended to provide a thorough classification of the two million eclipsing binary candidates, a task that would require additional analysis.
Rather, it offers a convenient quick analysis and overview of the catalogue content.
It must also be stressed that the definition of the samples as given in Table~\ref{Tab:sample_definition} is based on well-defined cuts on $\sigma_\mathrm{p}$, $\sigma_\mathrm{s}$, \Aell and \deltaPhi, which introduces an additional source of uncertainty in the classification.

Figures~\ref{Fig:period_magG_samples} and \ref{Fig:rank_magG_samples} summarise the periods and global rankings versus \gmag magnitude for each sample.

%==================================================================
\section{Eccentricity proxy}
\label{Appendix:eccentricty}

%-----------
\begin{figure}
  \centering
  \includegraphics[trim={0 68 0 42},clip,width=\linewidth,height=2.5cm]{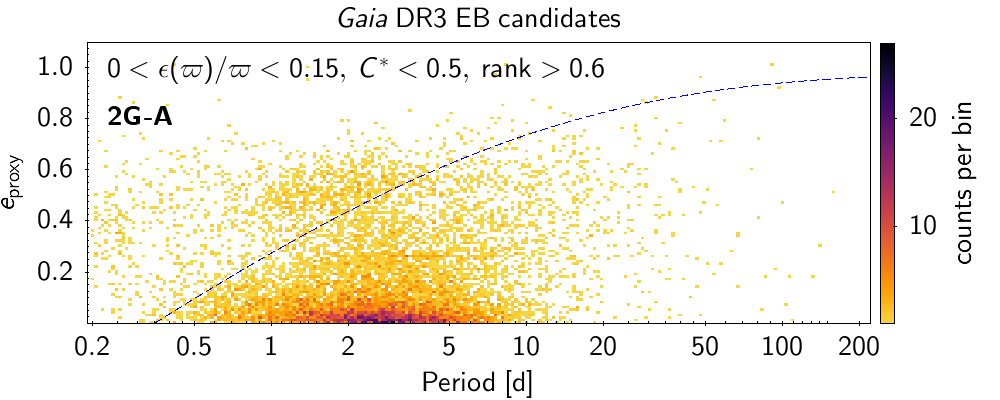}
  \vskip -0.2mm
  \includegraphics[trim={0 68 0 42},clip,width=\linewidth,height=2.5cm]{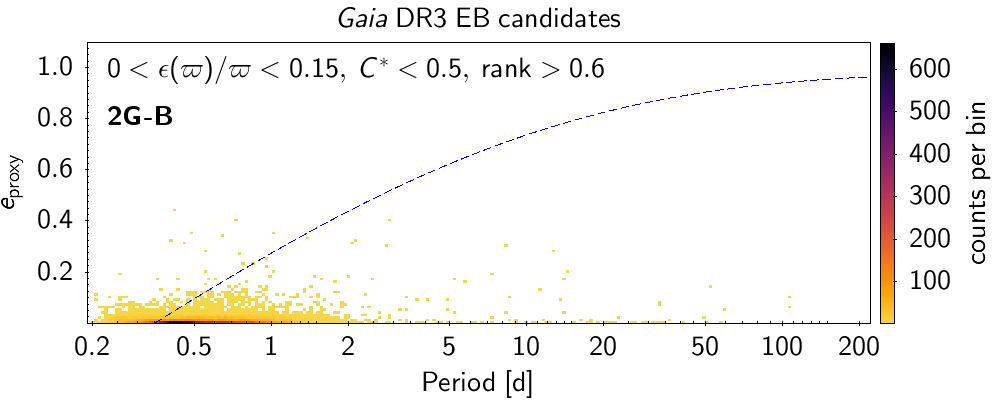}
  \vskip -0.2mm
  \includegraphics[trim={0 68 0 42},clip,width=\linewidth,height=2.5cm]{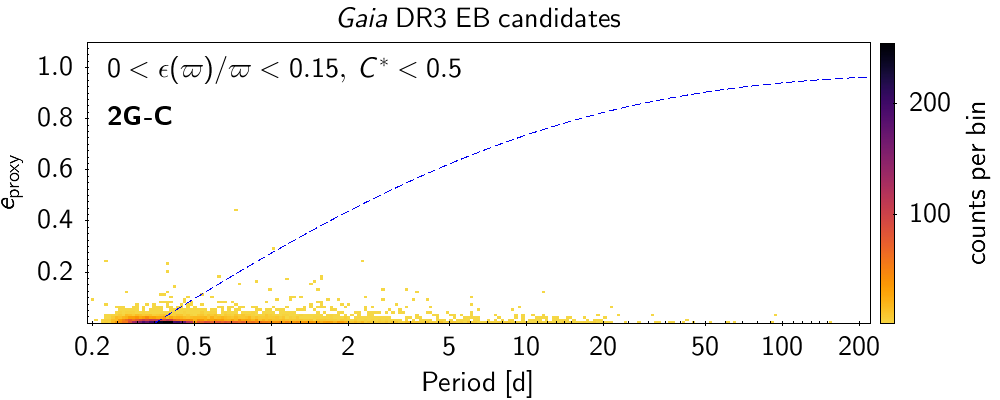}
  \vskip -0.2mm
  \includegraphics[trim={0 68 0 42},clip,width=\linewidth,height=2.5cm]{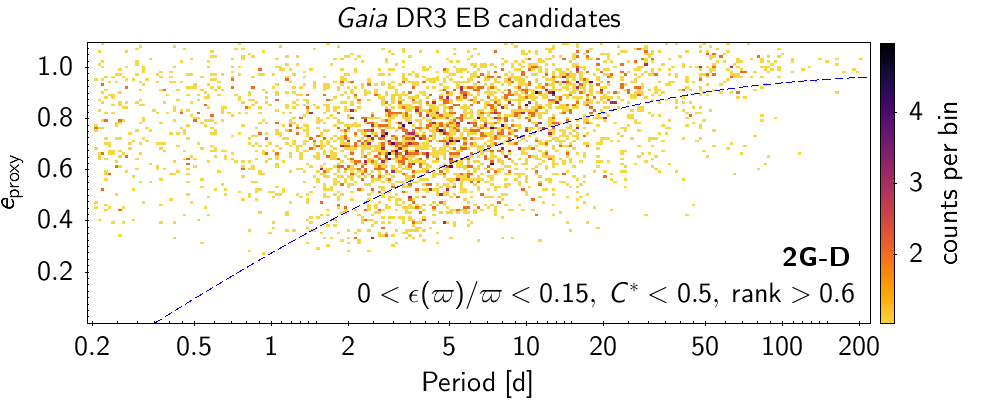}
  \vskip -0.2mm
  \includegraphics[trim={0 68 0 42},clip,width=\linewidth,height=2.5cm]{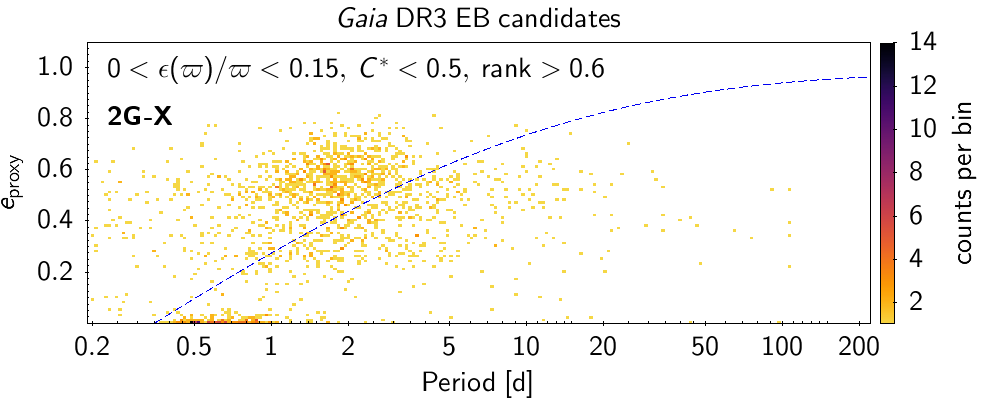}
  \vskip -0.2mm
  \includegraphics[trim={0 68 0 42},clip,width=\linewidth,height=2.5cm]{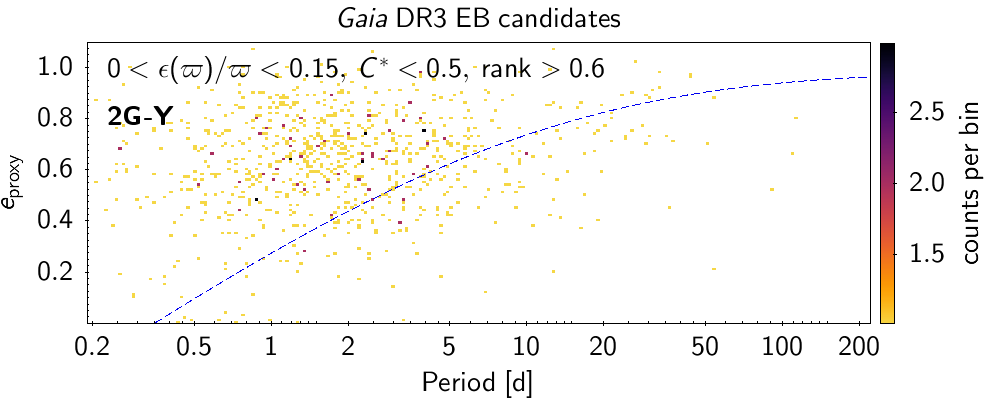}
  \vskip -0.2mm
  \includegraphics[trim={0 68 0 42},clip,width=\linewidth,height=2.5cm]{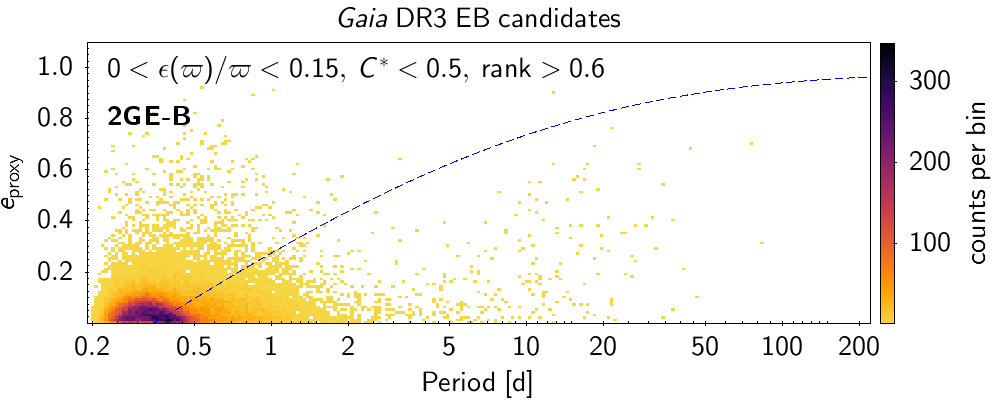}
  \vskip -0.2mm
  \includegraphics[trim={0 68 0 42},clip,width=\linewidth,height=2.5cm]{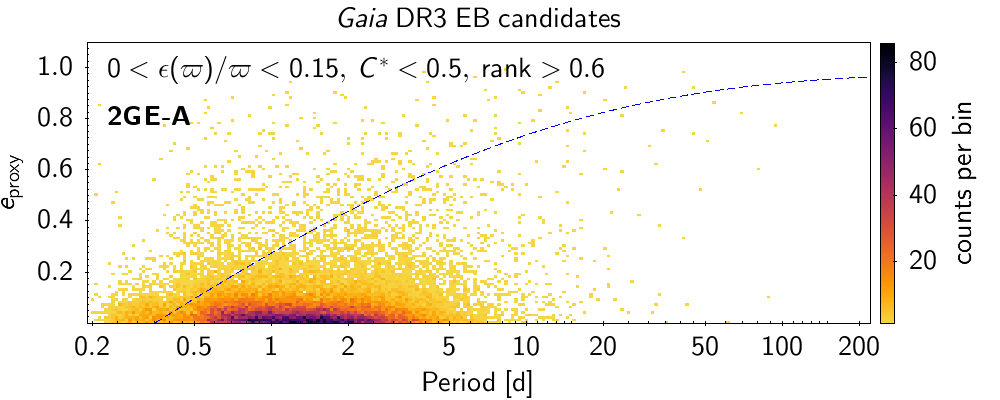}
  \vskip -0.2mm
  \includegraphics[trim={0  0 0 42},clip,width=\linewidth,height=3.1cm]{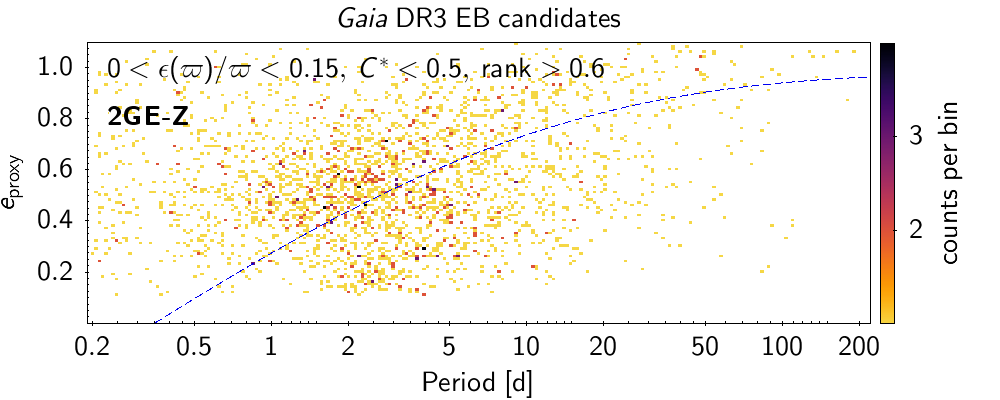}
  \caption{Density maps of the eccentricity proxy versus orbital period for samples with good parallaxes and global rankings as defined in the text.
           %The panels show the maps for groups that include two Gaussians in their light curve model (see Table~\ref{Tab:sample_definition}).
           The dashed lines are Eq.~(4.4) from \citet{Mazeh08} with $E=0.98$, $A=3.25$, $B=6.3$ and $C=0.23$.
           The axes ranges are truncated for better visibility.
          }
    \label{Fig:eccentricity_period_samples}
\end{figure}
%-----------

%-----------
\begin{figure}
  \centering
  \includegraphics[trim={0 68 0 42},clip,width=\linewidth,height=2.5cm]{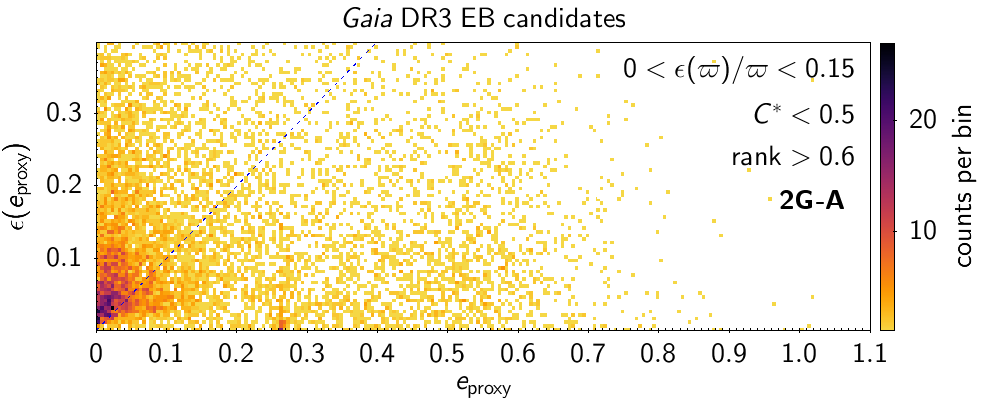}
  \vskip -0.2mm
  \includegraphics[trim={0 68 0 42},clip,width=\linewidth,height=2.5cm]{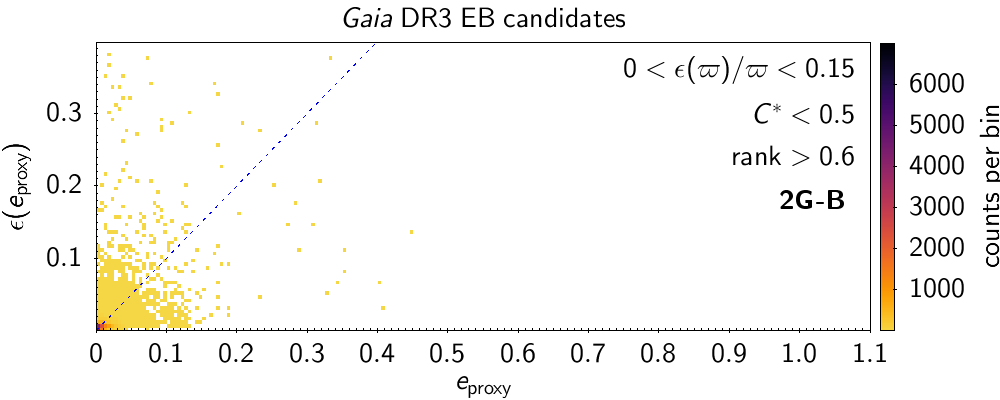}
  \vskip -0.2mm
  \includegraphics[trim={0 68 0 42},clip,width=\linewidth,height=2.5cm]{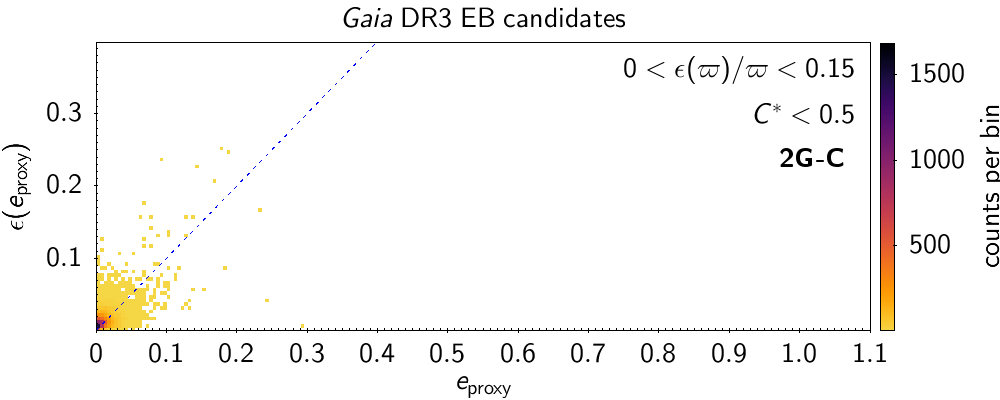}
  \vskip -0.2mm
  \includegraphics[trim={0 68 0 42},clip,width=\linewidth,height=2.5cm]{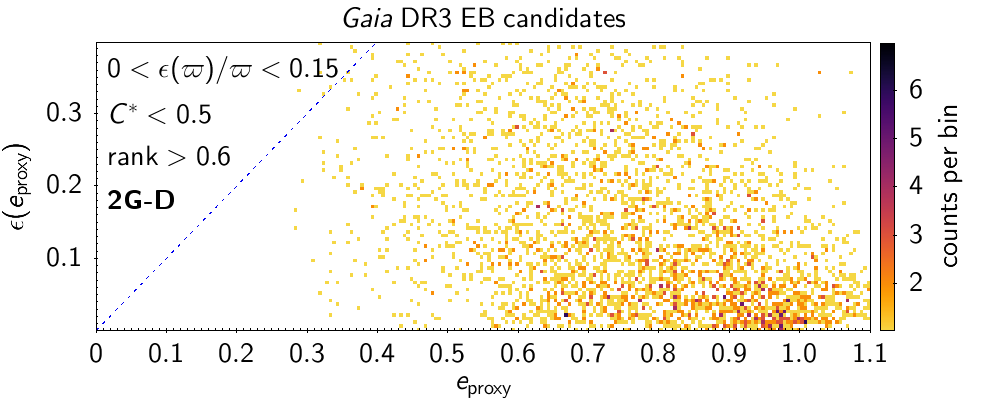}
  \vskip -0.2mm
  \includegraphics[trim={0 68 0 42},clip,width=\linewidth,height=2.5cm]{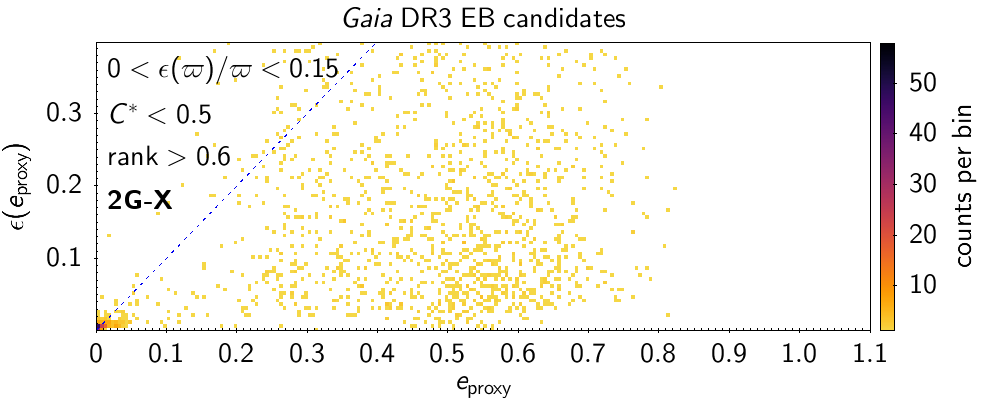}
  \vskip -0.2mm
  \includegraphics[trim={0 68 0 42},clip,width=\linewidth,height=2.5cm]{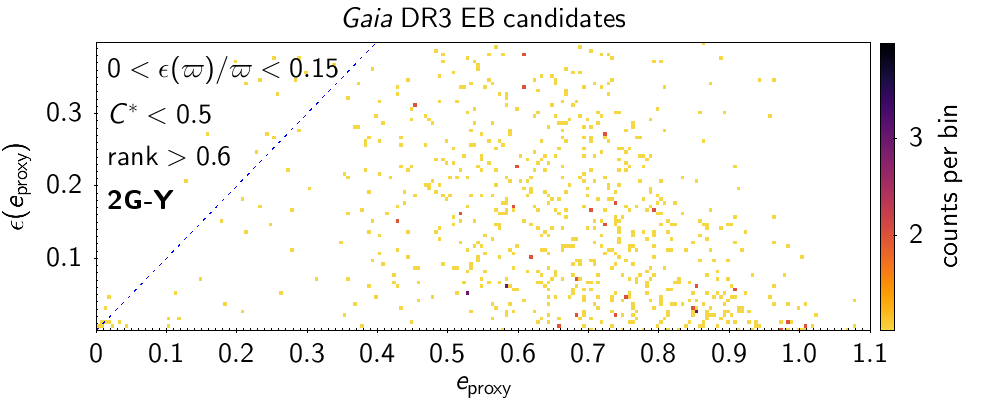}
  \vskip -0.2mm
  \includegraphics[trim={0 68 0 42},clip,width=\linewidth,height=2.5cm]{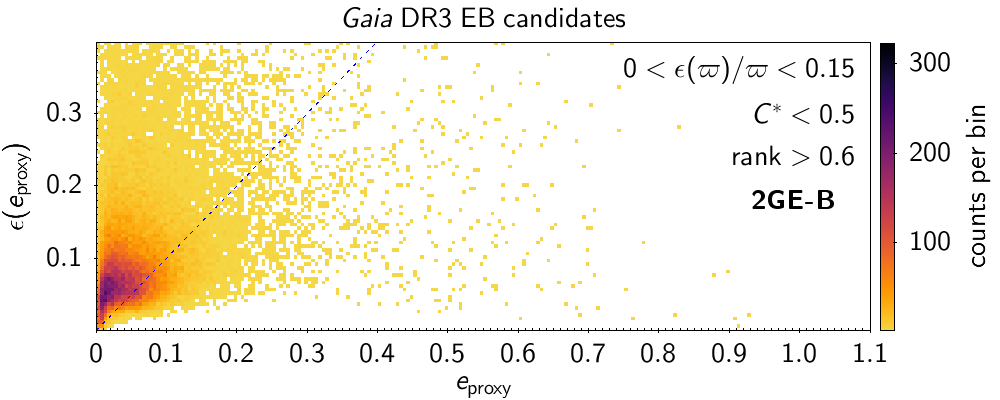}
  \vskip -0.2mm
  \includegraphics[trim={0 68 0 42},clip,width=\linewidth,height=2.5cm]{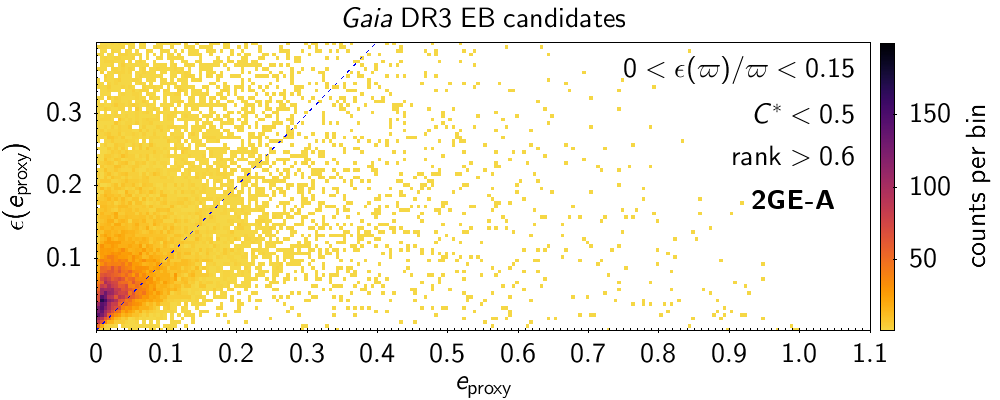}
  \vskip -0.2mm
  \includegraphics[trim={0  0 0 42},clip,width=\linewidth,height=3.1cm]{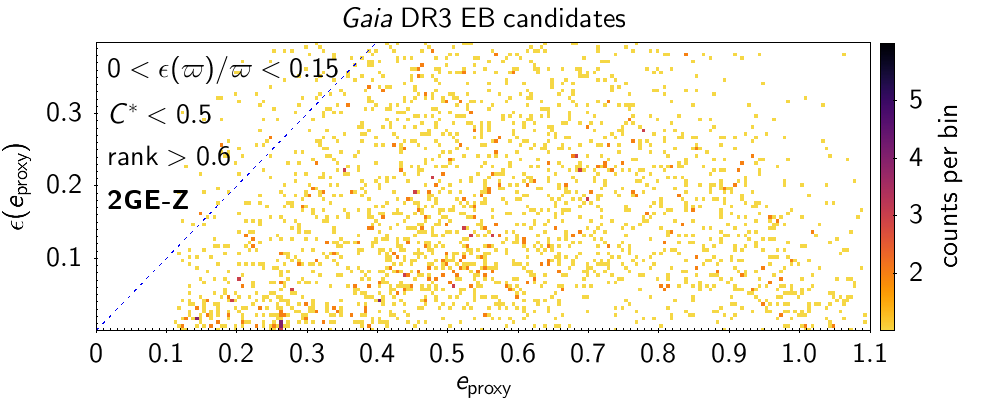}
  \caption{Same as Fig.~\ref{Fig:eccentricity_period_samples}, but for the uncertainty on the eccentricity proxy versus the eccentricity proxy.
           The dashed lines are one-to-one lines.
           %The axes ranges have been truncated for better visibility.
           %~{\color{white}.}
           %\hspace{6cm}{\color{white}.}
           %\hspace{6cm}{\color{white}.}
           %\hspace{6cm}{\color{white}.}
           %\hspace{6cm}{\color{white}.}
           %\hspace{6cm}{\color{white}.}
          }
    \label{Fig:error_eccentricity_samples}
\end{figure}
%-----------

%-----------
\begin{figure}
  \centering
  \includegraphics[trim={0 0 0 42},clip,width=\linewidth]{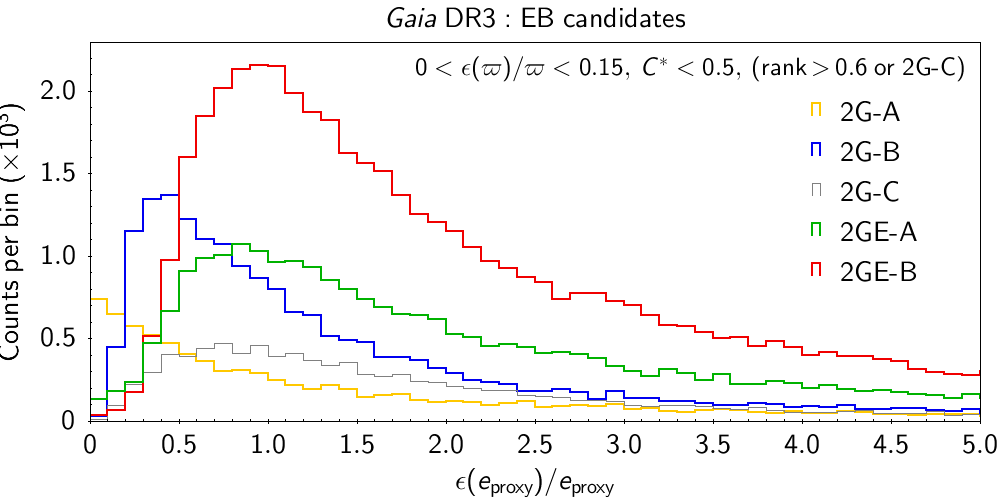}
  \caption{Histograms of the relative uncertainty on the eccentricity proxy for various samples with conditions on the parallax, corrected \BPRPexcess and global ranking as written in the figure.
          }
    \label{Fig:histo_eccRelErr}
\end{figure}
%-----------

A proxy for the eccentricity can be derived using the two-Gaussian model results based on the derived relative locations and durations of the eclipses.
At small eccentricities, the projected eccentricity $\eccProxy \cos\,\omega$ can be approximated from the phase separation of the eclipses with
\begin{equation}
  \eccProxy \cos\,\omega \simeq \frac{\pi}{2} (|\phaseEclSecondary - \phaseEclPrimary| - 0.5) \;,
\label{Eq:ecosw}
\end{equation}
where $\eccProxy$ is the eccentricity proxy and $\omega$ is the periastron argument.
Equation~\ref{Eq:ecosw} is readily computable from the derived model parameters.
In addition to eclipse locations, the models also provide the durations \durationEclPrimary and \durationEclSecondary of the primary and secondary eclipses, respectively.
From these parameters, $\eccProxy \sin\,\omega$ can be computed using
\begin{equation}
  \eccProxy \sin\,\omega = \frac{\durationEclSecondary - \durationEclPrimary}{\durationEclSecondary + \durationEclPrimary} \; .
\label{Eq:esinw}
\end{equation}
The eccentricity is then easily derived from Eqs.~\ref{Eq:ecosw} and \ref{Eq:esinw}, and writes
\begin{equation}
  \eccProxy \simeq \left[ \frac{\pi^2}{4} (|\phaseEclSecondary - \phaseEclPrimary| - 0.5)^2 + \left( \frac{\durationEclSecondary - \durationEclPrimary}{\durationEclSecondary + \durationEclPrimary}\right)^2 \right]^{1/2} \; .
\label{Eq:e}
\end{equation}
The uncertainty \eccProxyError on the eccentricity proxy can be computed from Eq.~\ref{Eq:e} by propagation of the uncertainties $\varepsilon(\phaseEclPrimary)$, $\varepsilon(\phaseEclSecondary)$, $\varepsilon(\durationEclPrimary)$ and $\varepsilon(\durationEclSecondary)$ of \phaseEclPrimary, \phaseEclSecondary, \durationEclPrimary and \durationEclSecondary, respectively.
Assuming these uncertainties are uncorrelated, we obtain
\begin{eqnarray}
  \eccProxyError \!\!& \!\!=\!\!\!\! & \frac{1}{2\,\eccProxy}  \; \times \nonumber\\
                 \!\!&  \!\!\!\!     & \hspace{-14mm} \left\{ \frac{\pi^2}{2} \;\; |(|\phaseEclSecondary - \phaseEclPrimary| - 0.5)| \; [\varepsilon(\phaseEclPrimary) + \varepsilon(\phaseEclSecondary)] \right. \nonumber\\
                 \!\!&  \!\!\!\!     & \hspace{-11mm} \left. + \; \frac{4\,|\durationEclSecondary - \durationEclPrimary|}{(\durationEclSecondary + \durationEclPrimary)^3} \; \left[ \durationEclSecondary \, \varepsilon(\durationEclPrimary) + \durationEclPrimary \, \varepsilon(\durationEclSecondary) \right] \right\} .
\label{Eq:eError}
\end{eqnarray}

%
% e_fromModel_error for topcat:
%e_fromModel_error_A = 0.5*PI*PI * abs(abs(derived_secondary_ecl_phase-derived_primary_ecl_phase)-0.5) * (derived_primary_ecl_phase_error+derived_secondary_ecl_phase_error)
%e_fromModel_error_B = (4*abs(derived_secondary_ecl_duration-derived_primary_ecl_duration)/pow(derived_secondary_ecl_duration+derived_primary_ecl_duration,3)) * (derived_secondary_ecl_duration*derived_primary_ecl_duration_error+derived_primary_ecl_duration*derived_secondary_ecl_duration_error)
%e_fromModel_error_new = (e_fromModel_error_A+e_fromModel_error_B) / (2*e_fromModel)
%

The eccentricities are shown versus period in Fig.~\ref{Fig:eccentricity_period_samples} for the various samples defined in Sect.~\ref{Sect:catalogue_usage_model} that contain two Gaussians in their light curve models.
The samples displayed in the figure are limited to the candidates analyzed in Sect.~\ref{Sect:overview}, that have good parallaxes (uncertainties better than 15\%), corrected \BPRPexcess $C^*$ smaller than 0.5, and that either have global rankings larger than 0.6 or belong to group 2G-C.
Figure~\ref{Fig:eccentricity_period_samples} shows that most candidates in groups with potential tidal interactions (groups 2G-B, 2G-C, 2GE-A and 2GE-B) have eccentricity proxies smaller than 0.1.
The uncertainties on these values, shown versus eccentricity in Fig.~\ref{Fig:error_eccentricity_samples}, can be as large as 0.15 even at these small eccentricity proxies.
These small eccentricities are therefore compatible with circularised systems.

In contrast, large eccentricity proxies ($\eccProxy \gtrsim 0.3$) are found in group 2G-A (top panel in Fig.~\ref{Fig:eccentricity_period_samples}), which contains well detached systems with no tidal effect.
In particular, many short-period systems are seen to have large eccentricity proxies, contrary to the expectation of them being circularised.
This can be seen in the figure, where the eccentricity limit, at any given orbital period, above which systems are expected to be circularised is shown by the dashed blue line  based on Eq.~(4.4) of \citet{Mazeh08} (with $E=0.98$, $A=3.25$, $B=6.3$ and $C=0.23$).
However, a careful analysis of the systems with eccentricities larger than this limit, performed in Sect.~\ref{Sect:overview} of the main body of this article, concludes that the eccentricity and/or orbital period of these systems are incorrect.
Caution must thus be taken in the interpretation of the two-gaussian model results when analyzing specific cases.

Large eccentricity proxies are also derived for candidates in groups 2G-D, 2G-X, 2G-Y and 2GE-Z (Fig.~\ref{Fig:eccentricity_period_samples}).
These groups, however, have been shown in Sect.~\ref{Sect:catalogue_usage_model} to have unreliable light curve models.
Their large eccentricities are thus mostly spurious, and need confirmation on a case-by-case study.

%==================================================================
\section{Additional figures}
\label{Appendix:additionalFigures}

%----
\begin{figure}
  \centering
  \includegraphics[trim={40 145 0 0},clip,width=\linewidth]{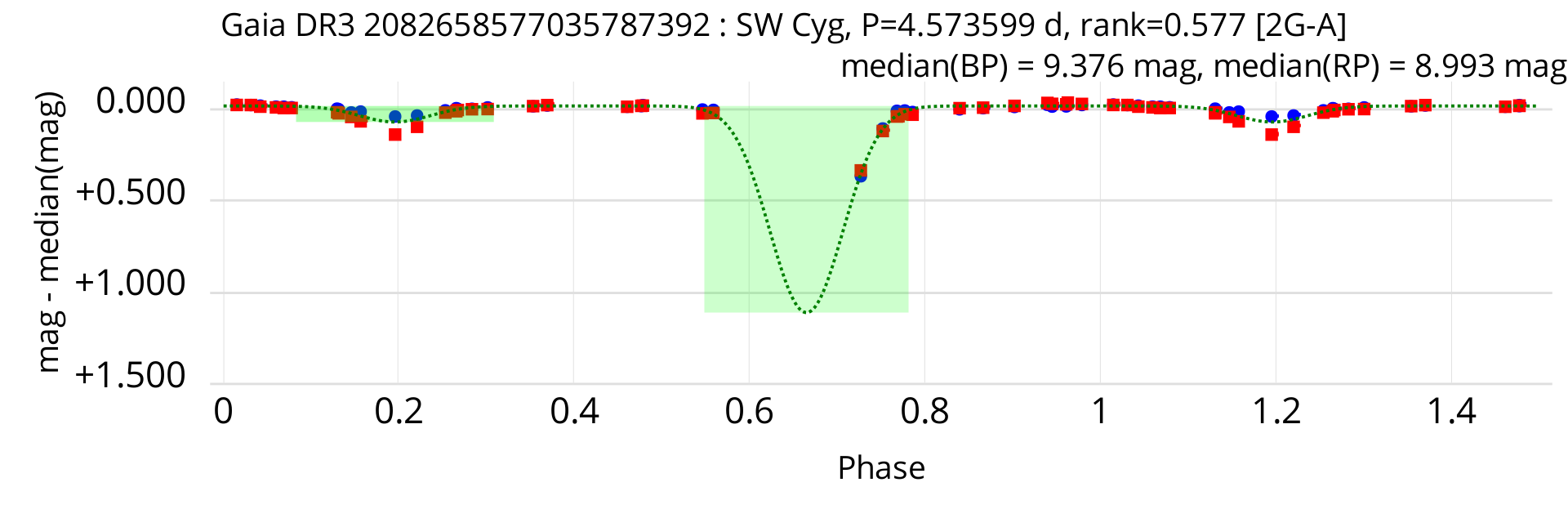}
  \vskip -0.5mm
  \includegraphics[trim={40 145 0 0},clip,width=\linewidth]{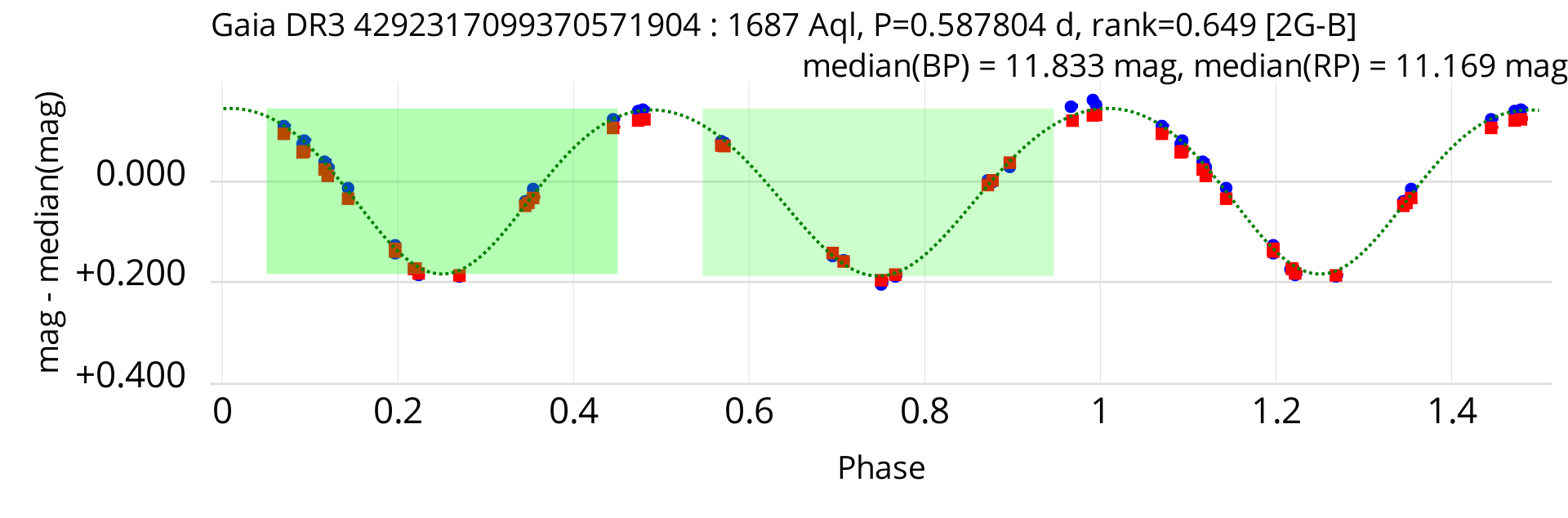}
  \vskip -0.5mm
  \includegraphics[trim={40 145 0 0},clip,width=\linewidth]{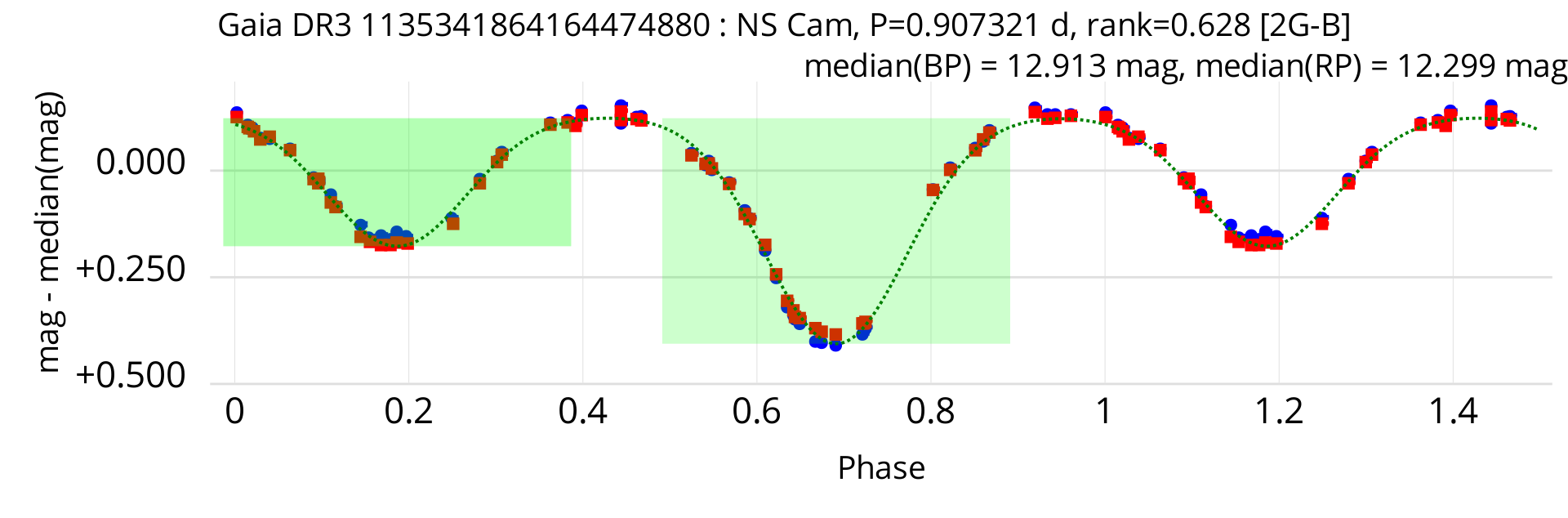}
  \vskip -0.5mm
  \includegraphics[trim={40 145 0 0},clip,width=\linewidth]{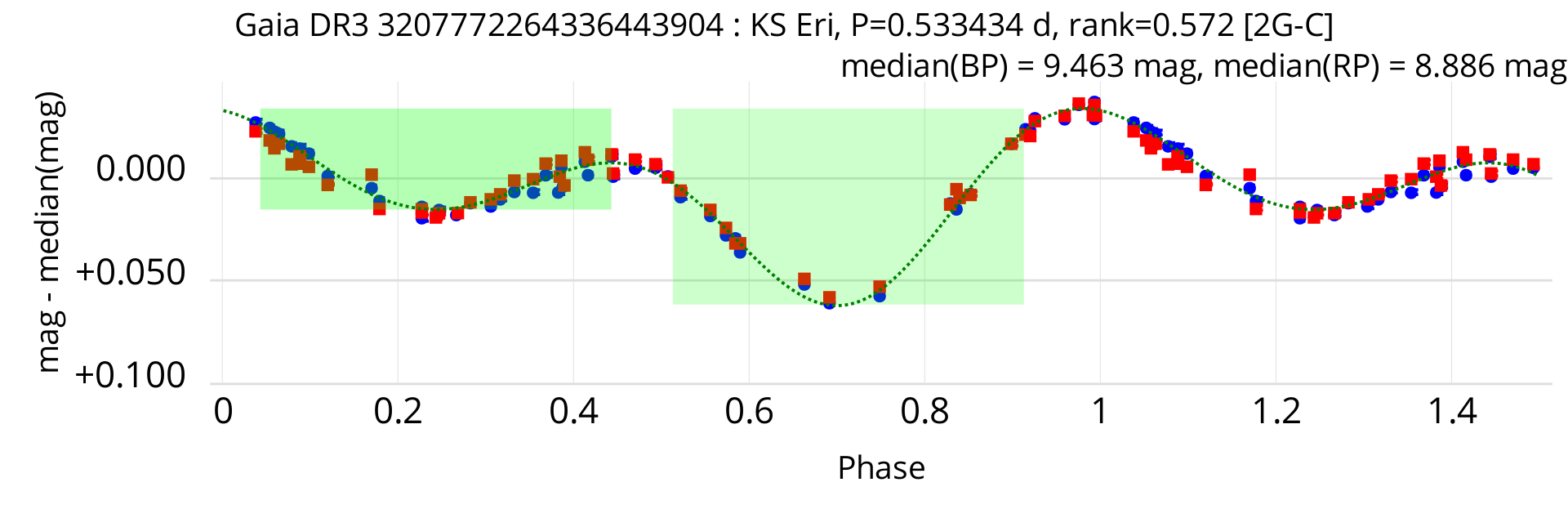}
  \vskip -0.5mm
  \includegraphics[trim={40 45 0 0},clip,width=\linewidth]{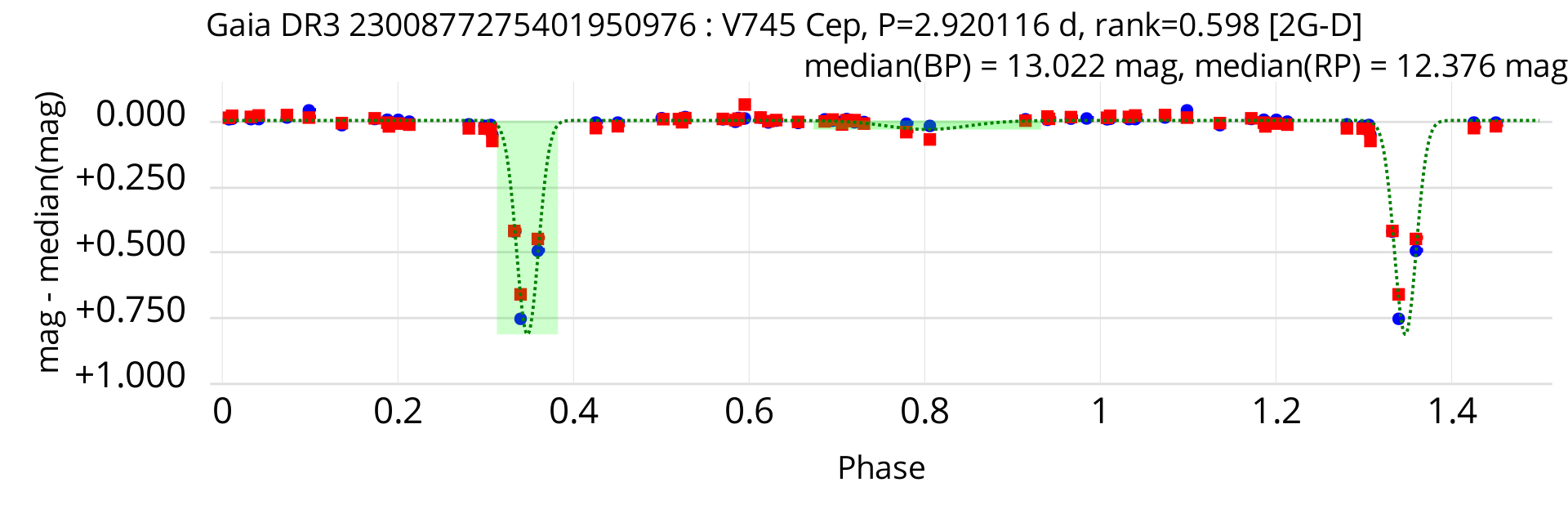}
  \caption{\gbp and \grp folded light curves of the eclipsing binaries whose \gmag are shown Fig.~\ref{Fig:lcs_examples_2G} in the main body of the article.
           The \gbp and \grp magnitudes are shifted by a value equal to the median magnitudes of their respective light curves, given in the top of each panel.
           The dotted line represents the two-Gaussian models determined from the \gmag light curves, with the green areas indicating the derived eclipse durations.
            }
\label{Fig:lcs_examples_2G_BPRP}
\end{figure}
%----
% 2082658577035787392, 4292317099370571904, 1135341864164474880, 3207772264336443904

%----
\begin{figure}
  \centering
  \includegraphics[trim={40 145 0 0},clip,width=\linewidth]{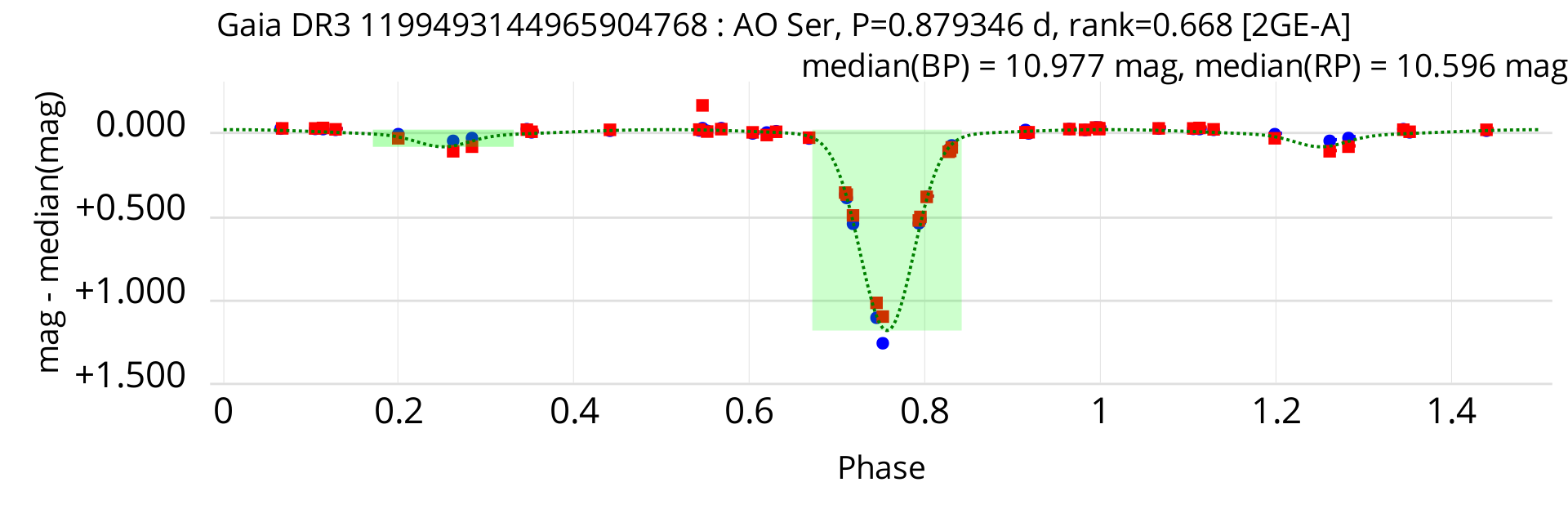}
  \vskip -0.5mm
  \includegraphics[trim={40 145 0 0},clip,width=\linewidth]{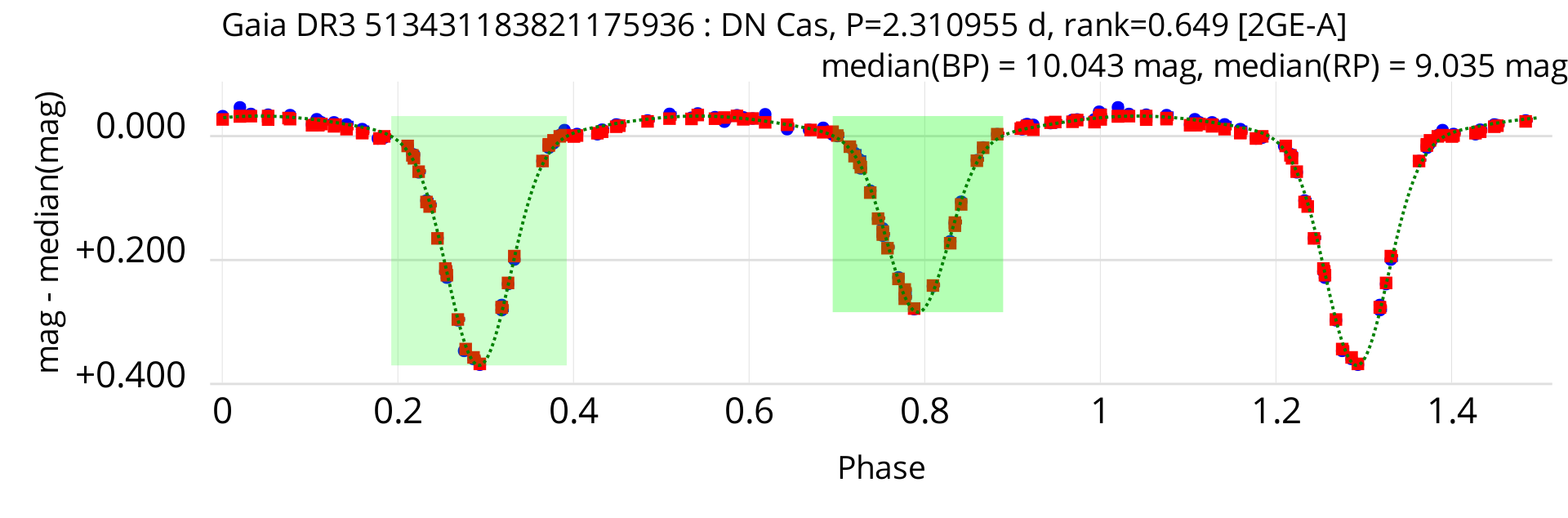}
  \vskip -0.5mm
  \includegraphics[trim={40 145 0 0},clip,width=\linewidth]{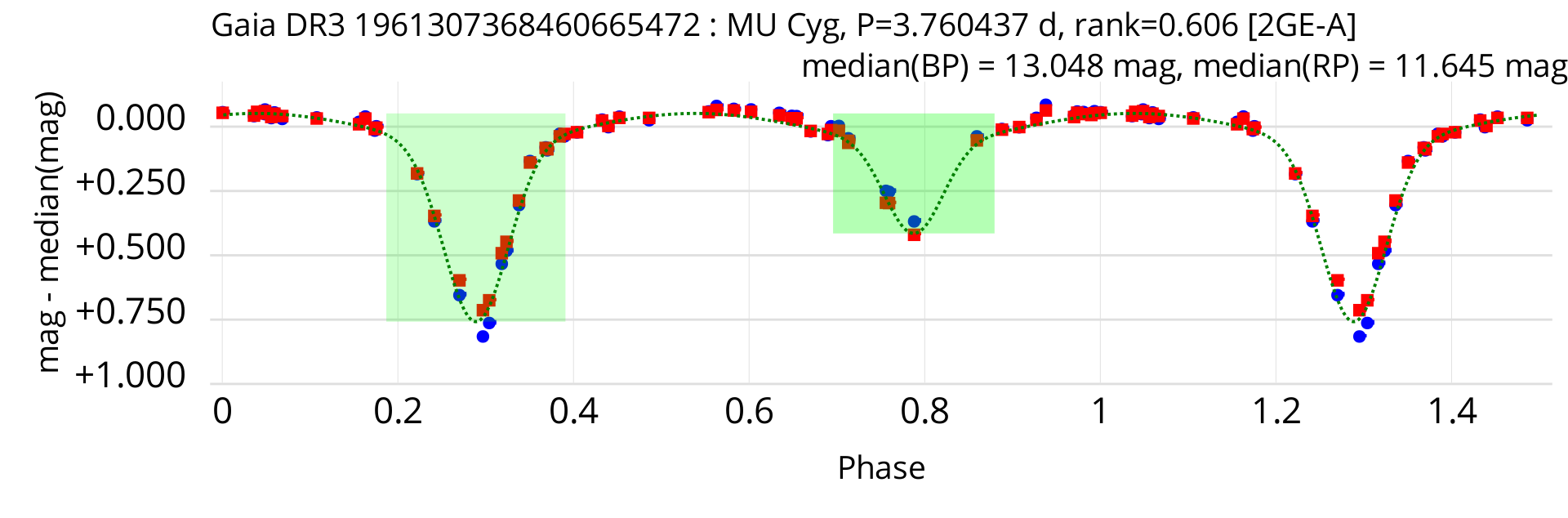}
  \vskip -0.5mm
  \includegraphics[trim={40 45 0 0},clip,width=\linewidth]{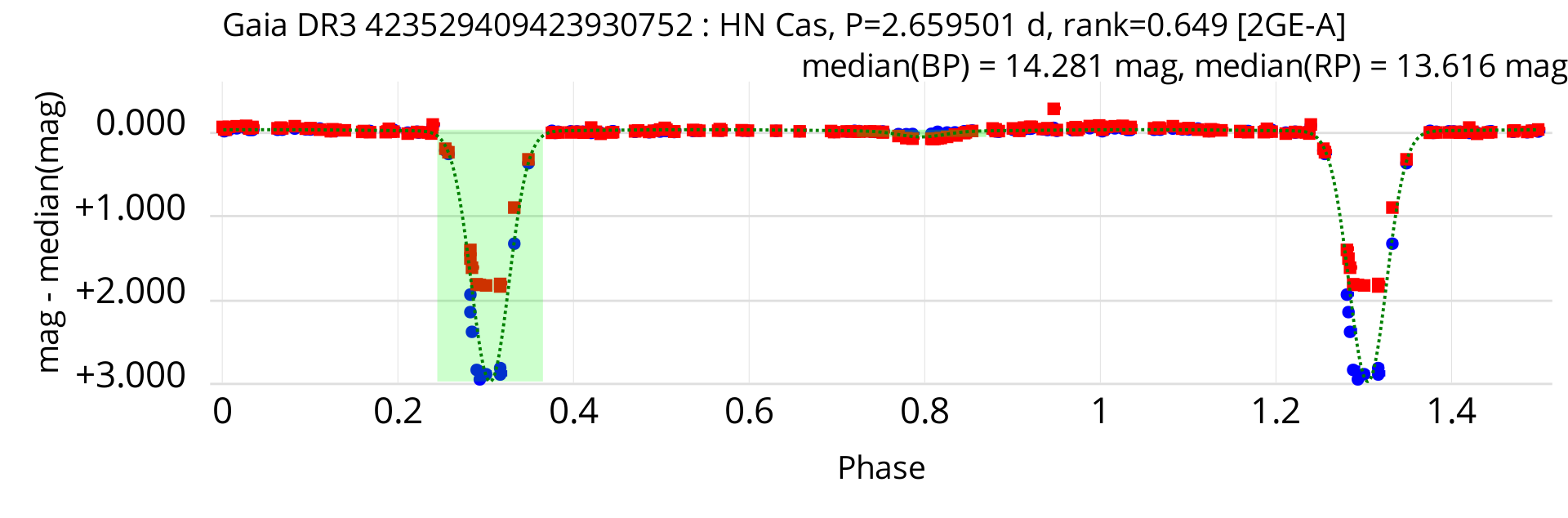}
  \caption{Same as Fig.~\ref{Fig:lcs_examples_2G_BPRP}, but for the eclipsing binaries whose \gmag light curves are shown in Fig.~\ref{Fig:lcs_examples_2GE_A} in the main body of the article.
           }
\label{Fig:lcs_examples_2GE_A_BPRP}
\end{figure}
%----

%----
\begin{figure}
  \centering
  \includegraphics[trim={40 145 0 0},clip,width=\linewidth]{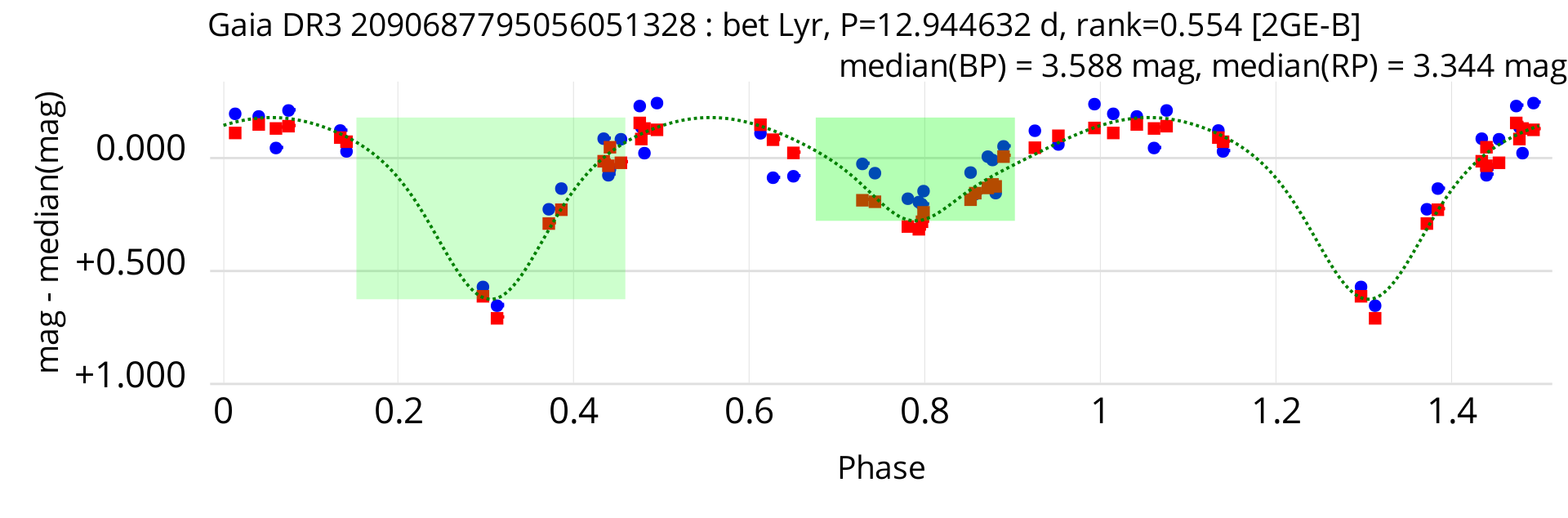}
  \vskip -0.5mm
  \includegraphics[trim={40 45 0 0},clip,width=\linewidth]{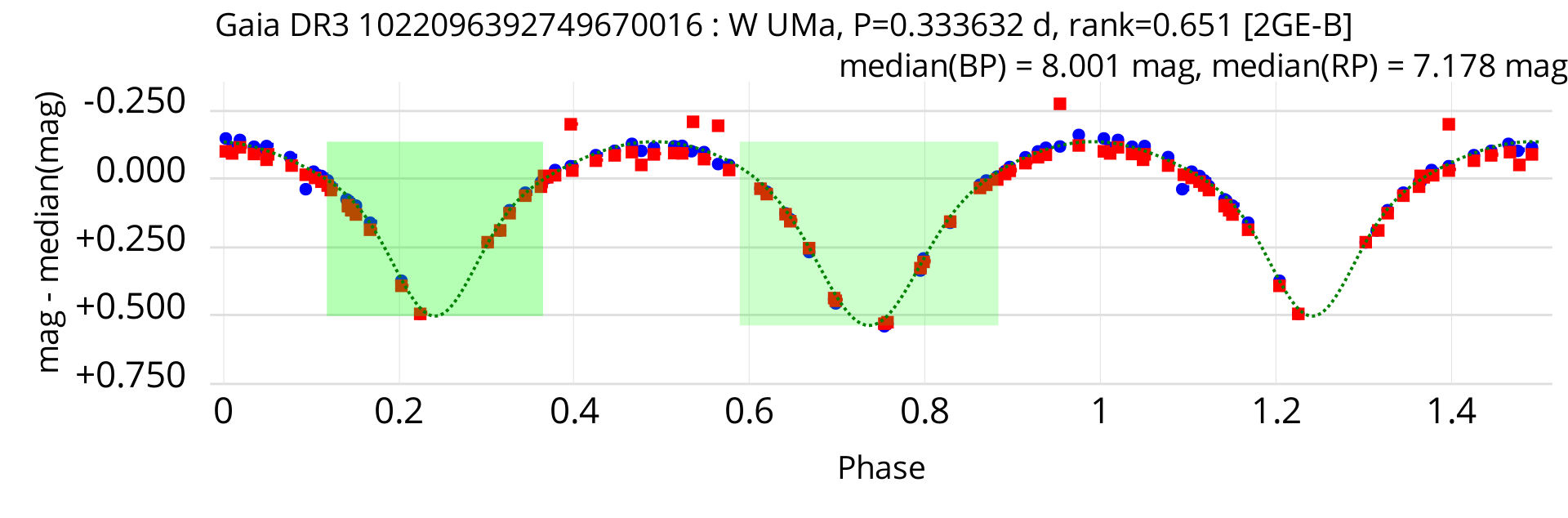}
  \caption{Same as Fig.~\ref{Fig:lcs_examples_2G_BPRP}, but for the eclipsing binaries whose \gmag light curves are shown in Fig.~\ref{Fig:lcs_examples_2GE_B} in the main body of the article.
           }
\label{Fig:lcs_examples_2GE_B_BPRP}
\end{figure}
%----

%----
\begin{figure}
  \centering
  \includegraphics[trim={40 145 0 0},clip,width=\linewidth]{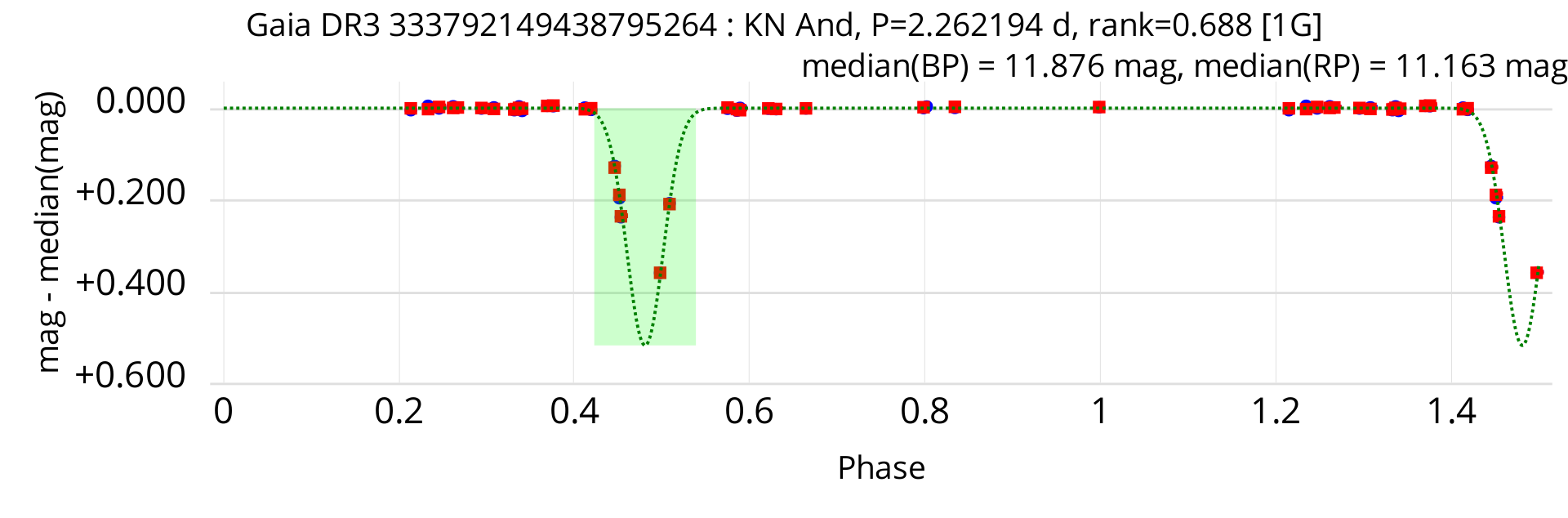}
  \vskip -0.5mm
  \includegraphics[trim={40 145 0 0},clip,width=\linewidth]{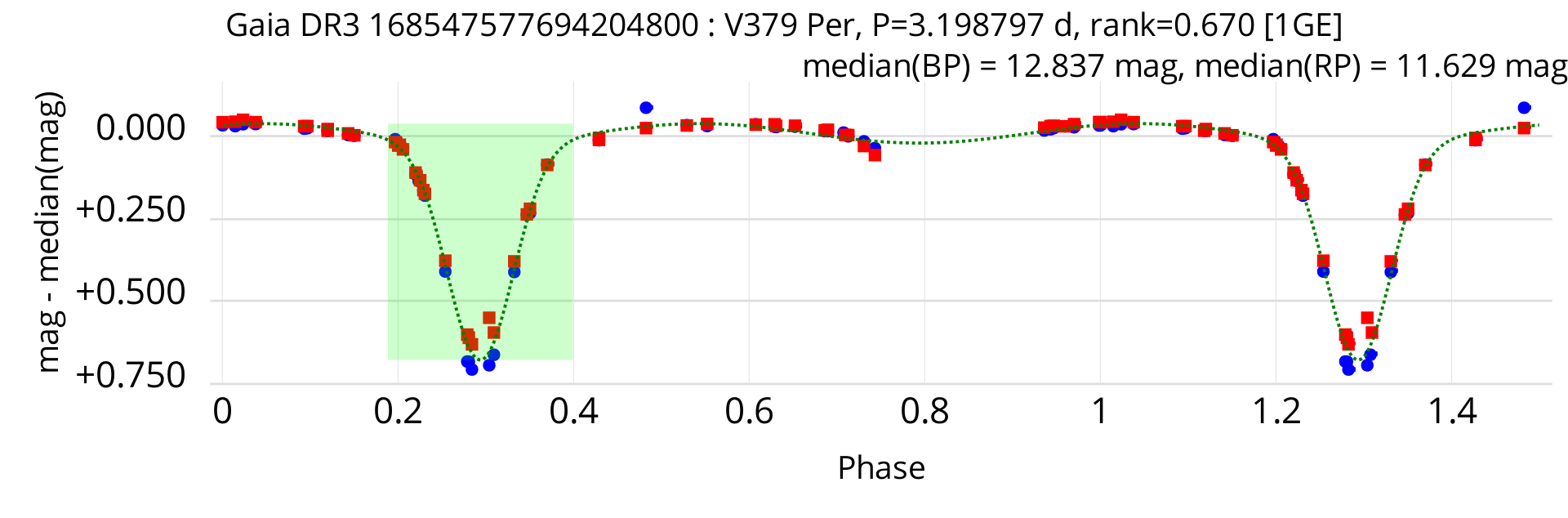}
  \vskip -0.5mm
  \includegraphics[trim={40 145 0 0},clip,width=\linewidth]{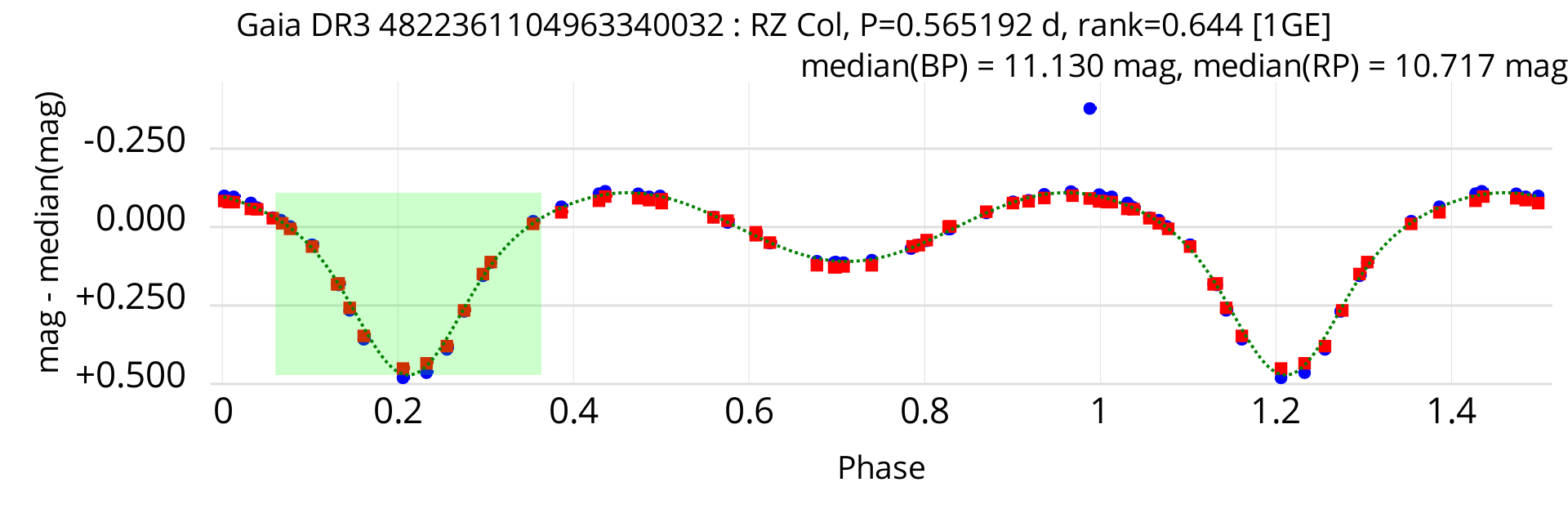}
  \vskip -0.5mm
  \includegraphics[trim={40 45 0 0},clip,width=\linewidth]{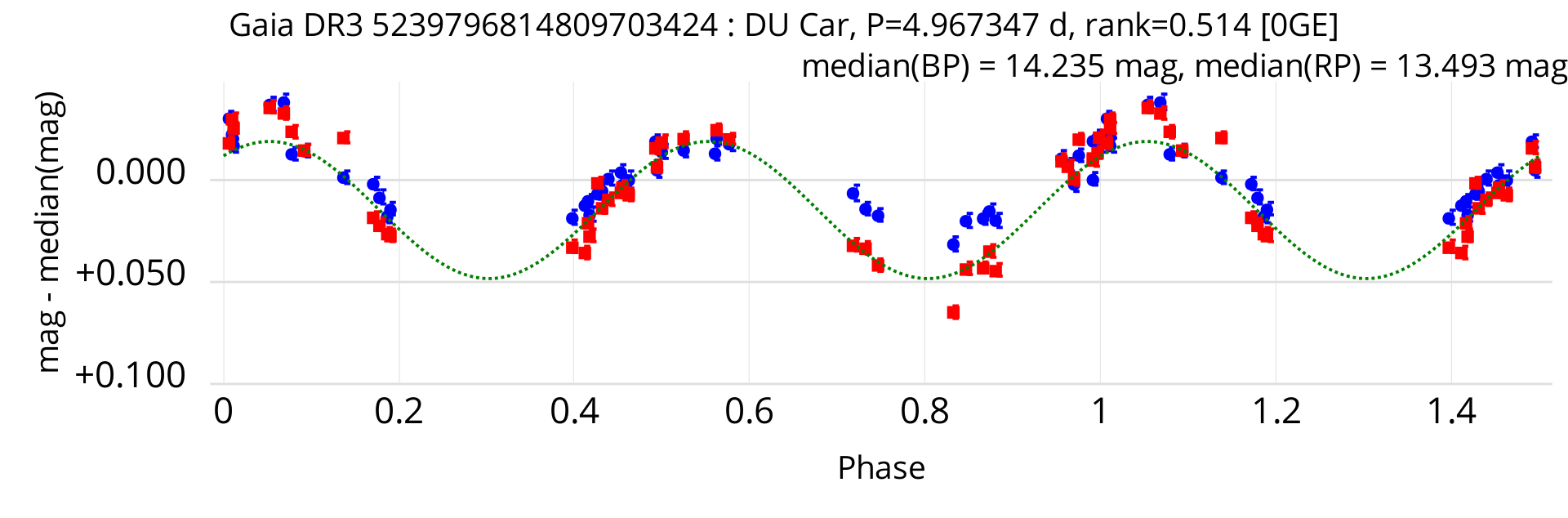}
  \caption{Same as Fig.~\ref{Fig:lcs_examples_2G_BPRP}, but for the eclipsing binaries whose \gmag light curves are shown in Fig.~\ref{Fig:lcs_examples_1G_1GE_0G} in the main body of the article. 
           }
\label{Fig:lcs_examples_1G_1GE_0G_BPRP}
\end{figure}
%----

%----
\begin{figure}
  \centering
  \includegraphics[trim={40 145 0 0},clip,width=\linewidth]{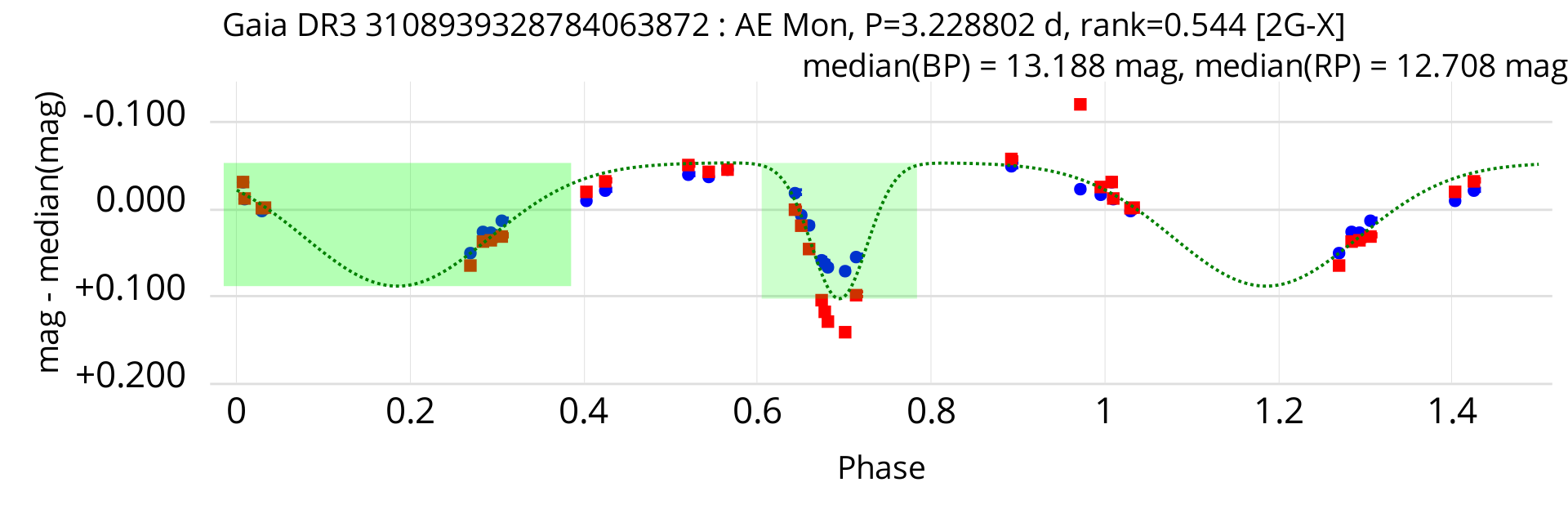}
  \vskip -0.5mm
  \includegraphics[trim={40 145 0 0},clip,width=\linewidth]{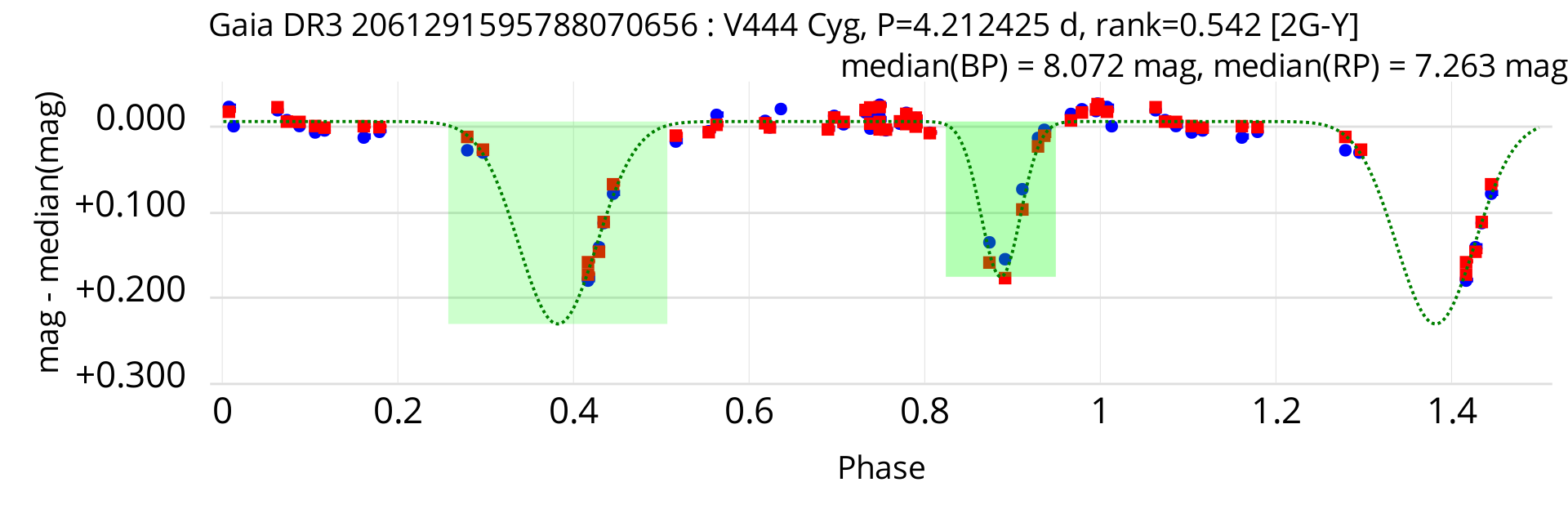}
  \vskip -0.5mm
  \includegraphics[trim={40 45 0 0},clip,width=\linewidth]{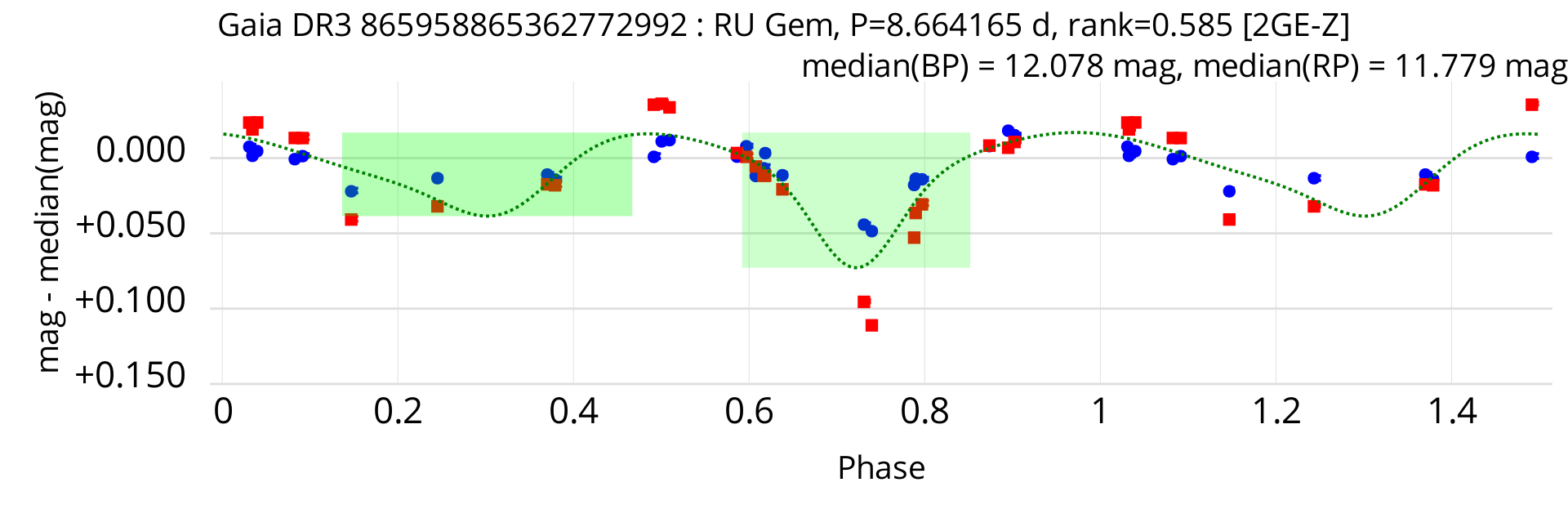}
  \caption{Same as Fig.~\ref{Fig:lcs_examples_2G_BPRP}, but for the eclipsing binaries whose \gmag light curves are shown in Fig.~\ref{Fig:lcs_examples_2G_special} in the main body of the article. 
           }
\label{Fig:lcs_examples_2G_special_BPRP}
\end{figure}
%----

%-----------
\begin{figure}
  \centering
  \includegraphics[trim={0 77 0 42},clip,width=1.0\linewidth]{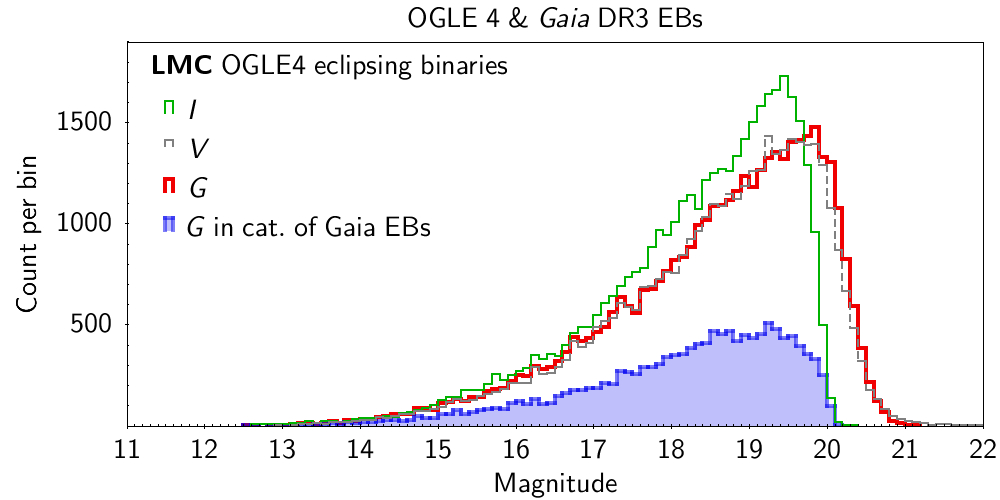}
  \vskip -0.25mm
  \includegraphics[trim={0 77 0 42},clip,width=1.0\linewidth]{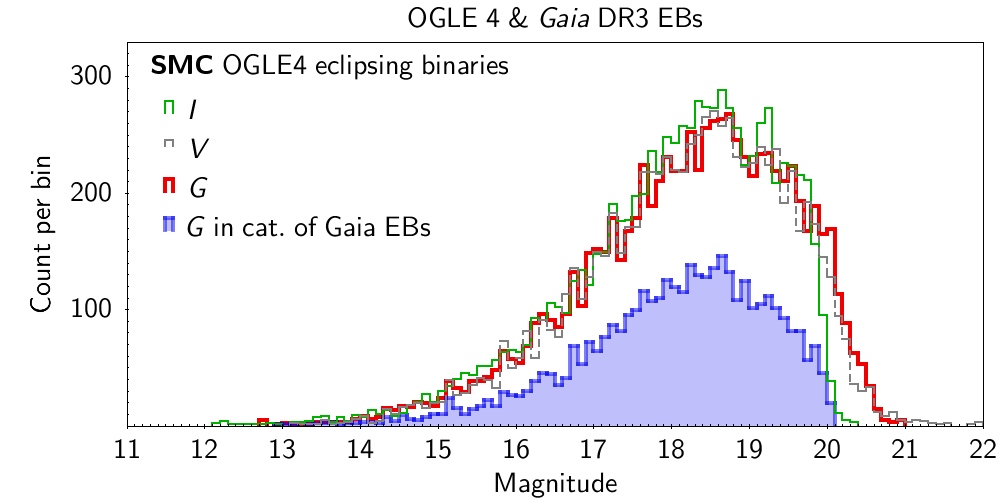}
  \vskip -0.25mm
  \includegraphics[trim={0 0 0 42},clip,width=1.0\linewidth]{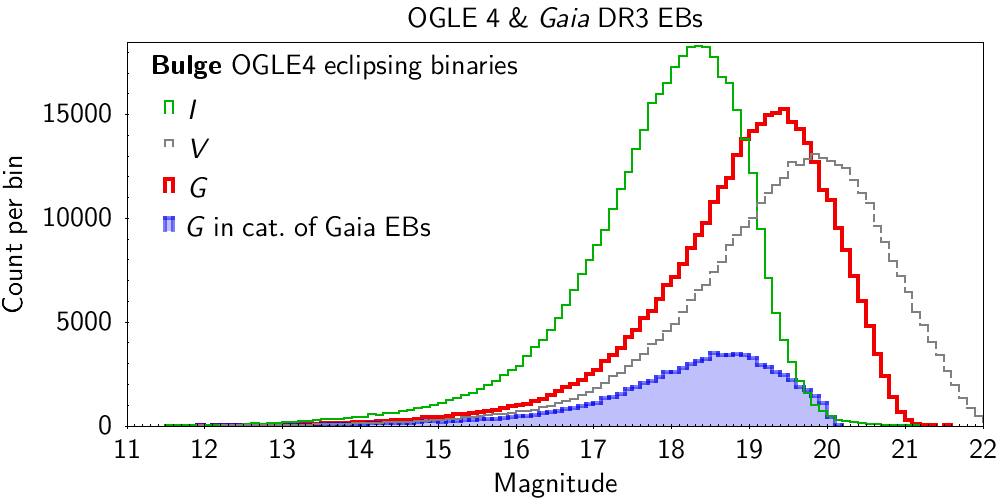}
  \caption{Magnitude distributions of OGLE4 eclipsing binaries that have a \Gaia \gmag magnitude.
           The distributions are shown for 
           The OGLE4 \OgleI and \OgleV magnitudes are shown in thin green and dashed grey, and their \Gaia \gmag magnitude in thick red.
           The \gmag distribution of the subset present in the \Gaia catalogue of eclipsing binaries is shown by the filled blue histogram.
          }
    \label{Fig:histo_mags_Ogle_DR3}
\end{figure}
%-----------

%-----------
\begin{figure}
  \centering
  \includegraphics[trim={0 0 0 42},clip,width=\linewidth]{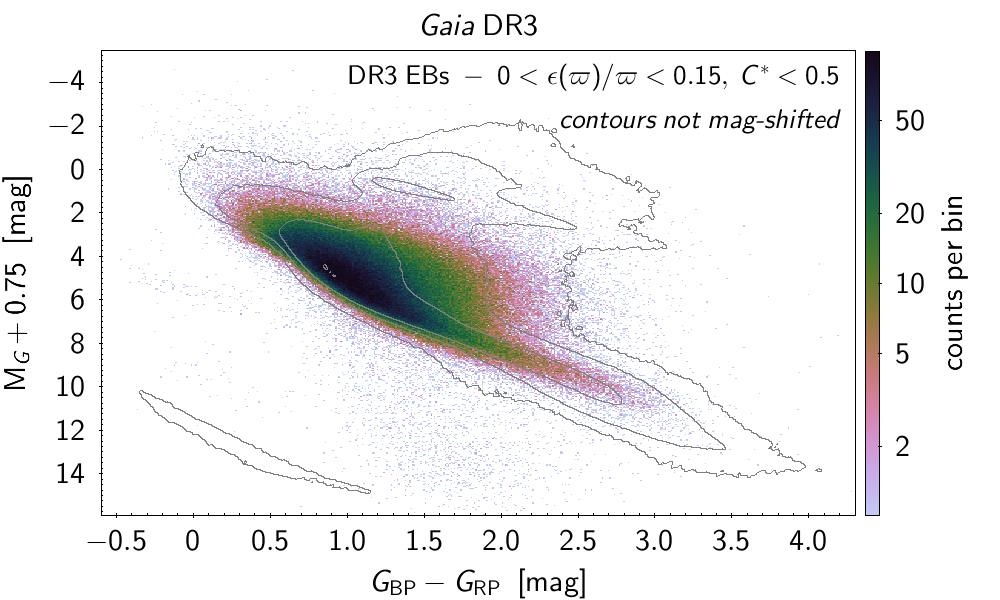}
  \caption{Same as second panel of Fig.~\ref{Fig:obsHR_15perc} in the main body of the text, but with the absolute magnitude M$_\gmag$ shifted by 0.75~mag.
           The contour lines representing the ten million random sample shown in the top panel of Fig.~\ref{Fig:obsHR_15perc} have not been shifted by 0.75~mag.
          }
\label{Fig:obsHR_15perc_Gplus0p75mag}
\end{figure}
%-----------

%-----------
\begin{figure}
  \centering
  \includegraphics[trim={40 45 0 70},clip,width=\linewidth]{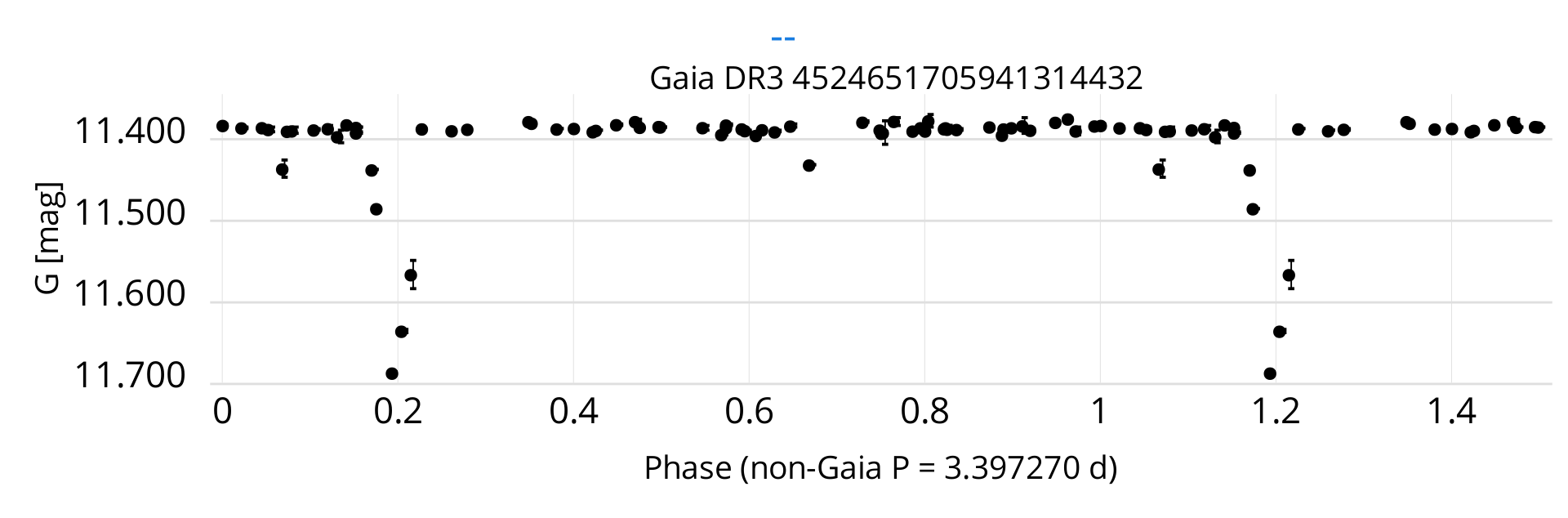}
  \vskip -0.5mm
  \includegraphics[trim={40 45 0 70},clip,width=\linewidth]{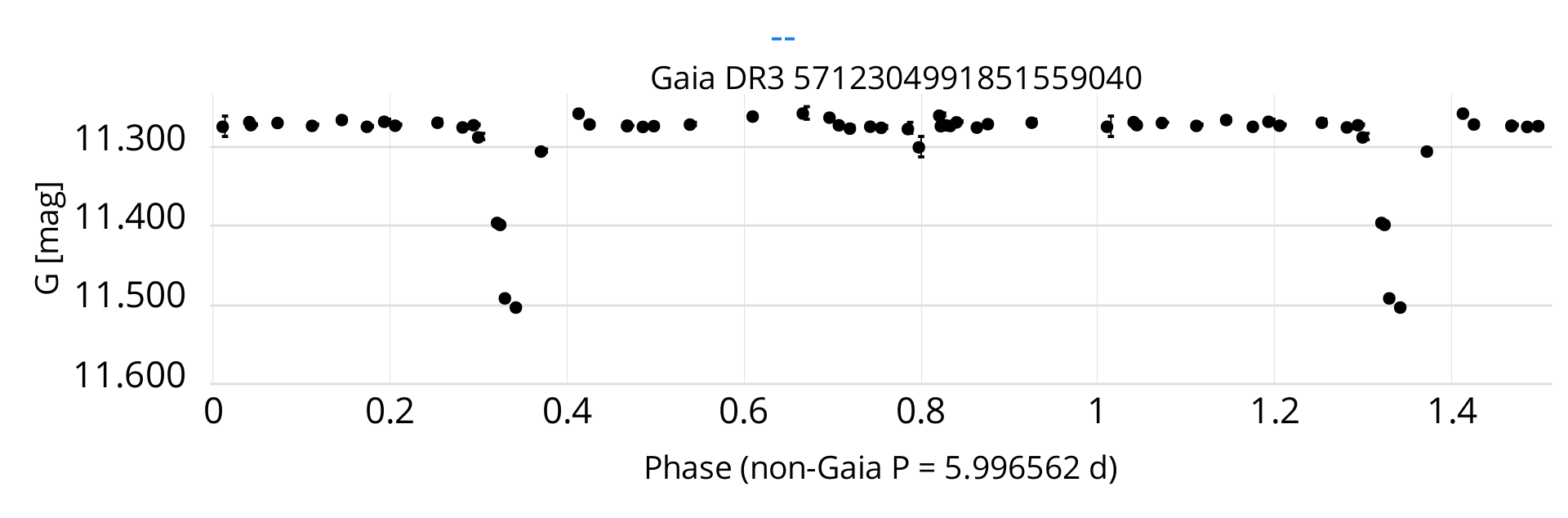}
    \caption{Light curves of \GaiaSrcIdInCaption{4524651705941314432} and \GaiaSrcIdInCaption{5712304991851559040} shown in Fig.~\ref{Fig:lcs_tooLargeEcc} in the main body of the text, folded with the periods published by the ASAS-SN survey for the respective cross-matches (\texttt{\small{ASASSN-V J075432.26-211826.4}} with P=3.3972697~d, and \texttt{\small{ASASSN-V J184156.16+192755.8}} with 5.9965616~d, respectively).
            }
\label{Fig:lcs_tooLargeEcc_ASASSN_periods}
\end{figure}
%-----------
% 4524651705941314432, 5712304991851559040

Figures \ref{Fig:lcs_examples_2G_BPRP} to \ref{Fig:lcs_examples_2G_special_BPRP} show the \gbp and \grp folded light curves of the eclipsing binaries displayed in Sect.~\ref{Sect:catalogue_usage_model}, except for V614~Ven which has its \gbp and \grp light curves already shown in Fig.~\ref{Fig:lcs_example_2GA}.

Figure~\ref{Fig:histo_mags_Ogle_DR3} shows the magnitude distributions of the OGLE4 samples of eclipsing binaries used in Sect.~\ref{Sect:quality_completeness} to estimate the completeness of the \Gaia catalogue.
The figure plots the distributions of the OGLE4 sources in the \OgleI, \OgleV and \gmag bands, separately for the OGLE4 samples towards the LMC (top panel), SMC (middle panel) and Galactic Bulge (bottom panel).
The \gmag distribution of the \Gaia -- OGLE4 crossmatches is also shown in each panel by the filled blue histograms.

Figure~\ref{Fig:obsHR_15perc_Gplus0p75mag} shows the observational diagram of the sample of eclipsing binary candidates with parallax uncertainties better than 15\%, with the absolute magnitudes M$_\gmag$ shifted by 0.75~mag to compare with the distribution of a random sample of \Gaia DR3 sources.
For this purpose, the contour lines of the random ten million sources shown in the top panel of Fig.~\ref{Fig:obsHR_15perc} (see Sect.~\ref{Sect:overview} in the main text of the article) have not been shifted by 0.75~mag.

%Figure~\ref{Fig:histo_period_15perc} compares the period distributions of various samples with good parallaxes analyzed in Sect.~\ref{Sect:overview}.
%The top panel shows the distributions for samples with parallax better than 15\% and corrected \BPRPexcess $C^*$ smaller than 0.5.
%The bottom panel shows the distributions for the same samples, but restricted to global ranking larger than 0.6 or including candidates in Sample~2G-C.

Figure~\ref{Fig:lcs_tooLargeEcc_ASASSN_periods} shows the light curves of two \Gaia candidates discussed in Sect.~\ref{Sect:overview}, folded with the ASAS-SN periods of the respective cross-matched ASAS-SN sources.

%==================================================================
\section{Acknowledgements}
\label{Appendix:acknowledgements}

This work presents results from the European Space Agency (ESA) space mission \Gaia. \Gaia data are being processed by the \Gaia Data Processing and Analysis Consortium (DPAC). Funding for the DPAC is provided by national institutions, in particular the institutions participating in the \Gaia MultiLateral Agreement (MLA). The \Gaia mission website is \url{https://www.cosmos.esa.int/gaia}. The \Gaia archive website is \url{https://archives.esac.esa.int/gaia}.

The \Gaia mission and data processing have financially been supported by, in alphabetical order by country:
\begin{itemize}
\item the Algerian Centre de Recherche en Astronomie, Astrophysique et G\'{e}ophysique of Bouzareah Observatory;
\item the Austrian Fonds zur F\"{o}rderung der wissenschaftlichen Forschung (FWF) Hertha Firnberg Programme through grants T359, P20046, and P23737;
\item the BELgian federal Science Policy Office (BELSPO) through various PROgramme de D\'{e}veloppement d'Exp\'{e}riences scientifiques (PRODEX) grants and the Polish Academy of Sciences - Fonds Wetenschappelijk Onderzoek through grant VS.091.16N, and the Fonds de la Recherche Scientifique (FNRS), and the Research Council of Katholieke Universiteit (KU) Leuven through grant C16/18/005 (Pushing AsteRoseismology to the next level with TESS, GaiA, and the Sloan DIgital Sky SurvEy -- PARADISE);  
\item the Brazil-France exchange programmes Funda\c{c}\~{a}o de Amparo \`{a} Pesquisa do Estado de S\~{a}o Paulo (FAPESP) and Coordena\c{c}\~{a}o de Aperfeicoamento de Pessoal de N\'{\i}vel Superior (CAPES) - Comit\'{e} Fran\c{c}ais d'Evaluation de la Coop\'{e}ration Universitaire et Scientifique avec le Br\'{e}sil (COFECUB);
\item the Chilean Agencia Nacional de Investigaci\'{o}n y Desarrollo (ANID) through Fondo Nacional de Desarrollo Cient\'{\i}fico y Tecnol\'{o}gico (FONDECYT) Regular Project 1210992 (L.~Chemin);
\item the National Natural Science Foundation of China (NSFC) through grants 11573054, 11703065, and 12173069, the China Scholarship Council through grant 201806040200, and the Natural Science Foundation of Shanghai through grant 21ZR1474100;  
\item the Tenure Track Pilot Programme of the Croatian Science Foundation and the \'{E}cole Polytechnique F\'{e}d\'{e}rale de Lausanne and the project TTP-2018-07-1171 `Mining the Variable Sky', with the funds of the Croatian-Swiss Research Programme;
\item the Czech-Republic Ministry of Education, Youth, and Sports through grant LG 15010 and INTER-EXCELLENCE grant LTAUSA18093, and the Czech Space Office through ESA PECS contract 98058;
\item the Danish Ministry of Science;
\item the Estonian Ministry of Education and Research through grant IUT40-1;
\item the European Commission?s Sixth Framework Programme through the European Leadership in Space Astrometry (\href{https://www.cosmos.esa.int/web/gaia/elsa-rtn-programme}{ELSA}) Marie Curie Research Training Network (MRTN-CT-2006-033481), through Marie Curie project PIOF-GA-2009-255267 (Space AsteroSeismology \& RR Lyrae stars, SAS-RRL), and through a Marie Curie Transfer-of-Knowledge (ToK) fellowship (MTKD-CT-2004-014188); the European Commission's Seventh Framework Programme through grant FP7-606740 (FP7-SPACE-2013-1) for the \Gaia European Network for Improved data User Services (\href{https://gaia.ub.edu/twiki/do/view/GENIUS/}{GENIUS}) and through grant 264895 for the \Gaia Research for European Astronomy Training (\href{https://www.cosmos.esa.int/web/gaia/great-programme}{GREAT-ITN}) network;
\item the European Cooperation in Science and Technology (COST) through COST Action CA18104 `Revealing the Milky Way with \Gaia (MW-Gaia)';
\item the European Research Council (ERC) through grants 320360, 647208, and 834148 and through the European Union?s Horizon 2020 research and innovation and excellent science programmes through Marie Sk{\l}odowska-Curie grant 745617 (Our Galaxy at full HD -- Gal-HD) and 895174 (The build-up and fate of self-gravitating systems in the Universe) as well as grants 687378 (Small Bodies: Near and Far), 682115 (Using the Magellanic Clouds to Understand the Interaction of Galaxies), 695099 (A sub-percent distance scale from binaries and Cepheids -- CepBin), 716155 (Structured ACCREtion Disks -- SACCRED), 951549 (Sub-percent calibration of the extragalactic distance scale in the era of big surveys -- UniverScale), and 101004214 (Innovative Scientific Data Exploration and Exploitation Applications for Space Sciences -- EXPLORE);
\item the European Science Foundation (ESF), in the framework of the \Gaia Research for European Astronomy Training Research Network Programme (\href{https://www.cosmos.esa.int/web/gaia/great-programme}{GREAT-ESF});
\item the European Space Agency (ESA) in the framework of the \Gaia project, through the Plan for European Cooperating States (PECS) programme through contracts C98090 and 4000106398/12/NL/KML for Hungary, through contract 4000115263/15/NL/IB for Germany, and through PROgramme de D\'{e}veloppement d'Exp\'{e}riences scientifiques (PRODEX) grant 4000127986 for Slovenia;  
\item the Academy of Finland through grants 299543, 307157, 325805, 328654, 336546, and 345115 and the Magnus Ehrnrooth Foundation;
\item the French Centre National d?\'{E}tudes Spatiales (CNES), the Agence Nationale de la Recherche (ANR) through grant ANR-10-IDEX-0001-02 for the `Investissements d'avenir' programme, through grant ANR-15-CE31-0007 for project `Modelling the Milky Way in the \Gaia era? (MOD4Gaia), through grant ANR-14-CE33-0014-01 for project `The Milky Way disc formation in the \Gaia era? (ARCHEOGAL), through grant ANR-15-CE31-0012-01 for project `Unlocking the potential of Cepheids as primary distance calibrators? (UnlockCepheids), through grant ANR-19-CE31-0017 for project `Secular evolution of galxies' (SEGAL), and through grant ANR-18-CE31-0006 for project `Galactic Dark Matter' (GaDaMa), the Centre National de la Recherche Scientifique (CNRS) and its SNO \Gaia of the Institut des Sciences de l?Univers (INSU), its Programmes Nationaux: Cosmologie et Galaxies (PNCG), Gravitation R\'{e}f\'{e}rences Astronomie M\'{e}trologie (PNGRAM), Plan\'{e}tologie (PNP), Physique et Chimie du Milieu Interstellaire (PCMI), and Physique Stellaire (PNPS), the `Action F\'{e}d\'{e}ratrice \Gaia' of the Observatoire de Paris, the R\'{e}gion de Franche-Comt\'{e}, the Institut National Polytechnique (INP) and the Institut National de Physique nucl\'{e}aire et de Physique des Particules (IN2P3) co-funded by CNES;
\item the German Aerospace Agency (Deutsches Zentrum f\"{u}r Luft- und Raumfahrt e.V., DLR) through grants 50QG0501, 50QG0601, 50QG0602, 50QG0701, 50QG0901, 50QG1001, 50QG1101, 50\-QG1401, 50QG1402, 50QG1403, 50QG1404, 50QG1904, 50QG2101, 50QG2102, and 50QG2202, and the Centre for Information Services and High Performance Computing (ZIH) at the Technische Universit\"{a}t Dresden for generous allocations of computer time;
\item the Hungarian Academy of Sciences through the Lend\"{u}let Programme grants LP2014-17 and LP2018-7 and the Hungarian National Research, Development, and Innovation Office (NKFIH) through grant KKP-137523 (`SeismoLab');
\item the Science Foundation Ireland (SFI) through a Royal Society - SFI University Research Fellowship (M.~Fraser);
\item the Israel Ministry of Science and Technology through grant 3-18143 and the Tel Aviv University Center for Artificial Intelligence and Data Science (TAD) through a grant;
\item the Agenzia Spaziale Italiana (ASI) through contracts I/037/08/0, I/058/10/0, 2014-025-R.0, 2014-025-R.1.2015, and 2018-24-HH.0 to the Italian Istituto Nazionale di Astrofisica (INAF), contract 2014-049-R.0/1/2 to INAF for the Space Science Data Centre (SSDC, formerly known as the ASI Science Data Center, ASDC), contracts I/008/10/0, 2013/030/I.0, 2013-030-I.0.1-2015, and 2016-17-I.0 to the Aerospace Logistics Technology Engineering Company (ALTEC S.p.A.), INAF, and the Italian Ministry of Education, University, and Research (Ministero dell'Istruzione, dell'Universit\`{a} e della Ricerca) through the Premiale project `MIning The Cosmos Big Data and Innovative Italian Technology for Frontier Astrophysics and Cosmology' (MITiC);
\item the Netherlands Organisation for Scientific Research (NWO) through grant NWO-M-614.061.414, through a VICI grant (A.~Helmi), and through a Spinoza prize (A.~Helmi), and the Netherlands Research School for Astronomy (NOVA);
\item the Polish National Science Centre through HARMONIA grant 2018/30/M/ST9/00311 and DAINA grant 2017/27/L/ST9/03221 and the Ministry of Science and Higher Education (MNiSW) through grant DIR/WK/2018/12;
\item the Portuguese Funda\c{c}\~{a}o para a Ci\^{e}ncia e a Tecnologia (FCT) through national funds, grants SFRH/\-BD/128840/2017 and PTDC/FIS-AST/30389/2017, and work contract DL 57/2016/CP1364/CT0006, the Fundo Europeu de Desenvolvimento Regional (FEDER) through grant POCI-01-0145-FEDER-030389 and its Programa Operacional Competitividade e Internacionaliza\c{c}\~{a}o (COMPETE2020) through grants UIDB/04434/2020 and UIDP/04434/2020, and the Strategic Programme UIDB/\-00099/2020 for the Centro de Astrof\'{\i}sica e Gravita\c{c}\~{a}o (CENTRA);  
\item the Slovenian Research Agency through grant P1-0188;
\item the Spanish Ministry of Economy (MINECO/FEDER, UE), the Spanish Ministry of Science and Innovation (MICIN), the Spanish Ministry of Education, Culture, and Sports, and the Spanish Government through grants BES-2016-078499, BES-2017-083126, BES-C-2017-0085, ESP2016-80079-C2-1-R, ESP2016-80079-C2-2-R, FPU16/03827, PDC2021-121059-C22, RTI2018-095076-B-C22, and TIN2015-65316-P (`Computaci\'{o}n de Altas Prestaciones VII'), the Juan de la Cierva Incorporaci\'{o}n Programme (FJCI-2015-2671 and IJC2019-04862-I for F.~Anders), the Severo Ochoa Centre of Excellence Programme (SEV2015-0493), and MICIN/AEI/10.13039/501100011033 (and the European Union through European Regional Development Fund `A way of making Europe') through grant RTI2018-095076-B-C21, the Institute of Cosmos Sciences University of Barcelona (ICCUB, Unidad de Excelencia `Mar\'{\i}a de Maeztu?) through grant CEX2019-000918-M, the University of Barcelona's official doctoral programme for the development of an R+D+i project through an Ajuts de Personal Investigador en Formaci\'{o} (APIF) grant, the Spanish Virtual Observatory through project AyA2017-84089, the Galician Regional Government, Xunta de Galicia, through grants ED431B-2021/36, ED481A-2019/155, and ED481A-2021/296, the Centro de Investigaci\'{o}n en Tecnolog\'{\i}as de la Informaci\'{o}n y las Comunicaciones (CITIC), funded by the Xunta de Galicia and the European Union (European Regional Development Fund -- Galicia 2014-2020 Programme), through grant ED431G-2019/01, the Red Espa\~{n}ola de Supercomputaci\'{o}n (RES) computer resources at MareNostrum, the Barcelona Supercomputing Centre - Centro Nacional de Supercomputaci\'{o}n (BSC-CNS) through activities AECT-2017-2-0002, AECT-2017-3-0006, AECT-2018-1-0017, AECT-2018-2-0013, AECT-2018-3-0011, AECT-2019-1-0010, AECT-2019-2-0014, AECT-2019-3-0003, AECT-2020-1-0004, and DATA-2020-1-0010, the Departament d'Innovaci\'{o}, Universitats i Empresa de la Generalitat de Catalunya through grant 2014-SGR-1051 for project `Models de Programaci\'{o} i Entorns d'Execuci\'{o} Parallels' (MPEXPAR), and Ramon y Cajal Fellowship RYC2018-025968-I funded by MICIN/AEI/10.13039/501100011033 and the European Science Foundation (`Investing in your future');
\item the Swedish National Space Agency (SNSA/Rymdstyrelsen);
\item the Swiss State Secretariat for Education, Research, and Innovation through the Swiss Activit\'{e}s Nationales Compl\'{e}mentaires and the Swiss National Science Foundation through an Eccellenza Professorial Fellowship (award PCEFP2\_194638 for R.~Anderson);
\item the United Kingdom Particle Physics and Astronomy Research Council (PPARC), the United Kingdom Science and Technology Facilities Council (STFC), and the United Kingdom Space Agency (UKSA) through the following grants to the University of Bristol, the University of Cambridge, the University of Edinburgh, the University of Leicester, the Mullard Space Sciences Laboratory of University College London, and the United Kingdom Rutherford Appleton Laboratory (RAL): PP/D006511/1, PP/D006546/1, PP/D006570/1, ST/I000852/1, ST/J005045/1, ST/K00056X/1, ST/\-K000209/1, ST/K000756/1, ST/L006561/1, ST/N000595/1, ST/N000641/1, ST/N000978/1, ST/\-N001117/1, ST/S000089/1, ST/S000976/1, ST/S000984/1, ST/S001123/1, ST/S001948/1, ST/\-S001980/1, ST/S002103/1, ST/V000969/1, ST/W002469/1, ST/W002493/1, ST/W002671/1, ST/W002809/1, and EP/V520342/1.
\end{itemize}

The GBOT programme  uses observations collected at (i) the European Organisation for Astronomical Research in the Southern Hemisphere (ESO) with the VLT Survey Telescope (VST), under ESO programmes
092.B-0165,
093.B-0236,
094.B-0181,
095.B-0046,
096.B-0162,
097.B-0304,
098.B-0030,
099.B-0034,
0100.B-0131,
0101.B-0156,
0102.B-0174, and
0103.B-0165;
%
% From Martin Altmann, 13 March 2019:
%  092.B-0165   01.10.13 - 31.03.14
%  093.B-0236   01.04.14 - 30.09.14
%  094.B-0181   01.10.14 - 31.03.15
%  095.B-0046   01.04.15 - 30.09.15
%  096.B-0162   01.10.15 - 31.03.16
%  097.B-0304   01.04.16 - 30.09.16
%  098.B-0030   01.10.16 - 31.03.17
%  099.B-0034   01.04.17 - 30.09.17
% 0100.B-0131   01.10.17 - 31.03.18
% 0101.B-0156   01.04.18 - 30.09.18
% 0102.B-0174   01.10.18 - 31.03.19
% 0103.B-0165   01.04.19 - 30.09.19
%
and (ii) the Liverpool Telescope, which is operated on the island of La Palma by Liverpool John Moores University in the Spanish Observatorio del Roque de los Muchachos of the Instituto de Astrof\'{\i}sica de Canarias with financial support from the United Kingdom Science and Technology Facilities Council, and (iii) telescopes of the Las Cumbres Observatory Global Telescope Network.

%==================================================================
\end{appendix}

%==================================================================
\end{document}